\pdfoutput=1 
\documentclass[rmp,aps,amssymb,twocolumn,nofootinbib,floatfix]{revtex4}
\usepackage{amsmath,amssymb,bm}
\usepackage{graphics}
\usepackage{graphicx}
\usepackage[colorlinks,linkcolor=blue,citecolor=blue]{hyperref}

\begin{document}

\title{Community detection in graphs}
\author{Santo Fortunato}
\email{fortunato@isi.it}
\affiliation{Complex Networks and Systems Lagrange Laboratory, ISI Foundation,
  Viale S. Severo 65, 10133, Torino, I-ITALY.}
 
\begin{abstract}  

The modern science of networks has brought significant advances to our understanding of complex systems.
One of the most relevant features of graphs representing real systems is community structure, or clustering, i. e.
the organization of vertices in clusters, with many edges joining vertices of the same cluster and comparatively few edges
joining vertices of different clusters. Such clusters, or communities, can be considered as fairly independent compartments 
of a graph, playing a similar role like, e. g., the tissues or the organs in the human body. 
Detecting communities is of great importance in sociology, biology and computer science,
disciplines where systems are often represented as graphs. This problem is very hard and not yet satisfactorily solved, despite the 
huge effort of a large interdisciplinary community of scientists working on it over the past few years. We will attempt a 
thorough exposition of the topic, from the definition of the main elements of the problem, to the presentation of most methods 
developed, with a special focus on techniques designed by statistical physicists, from the discussion of 
crucial issues like the significance of clustering and how methods should be tested and compared against each other, 
to the description of applications to real networks. 

\end{abstract}


\maketitle
\tableofcontents


\section{Introduction}
\label{sec0}

The origin of graph theory dates back to Euler's solution of the puzzle of 
K\"onigsberg's bridges in 1736~\cite{euler36}. Since then a lot has been learned about graphs and their mathematical properties~\cite{bollobas98}. 
In the 20th century they have also become extremely useful as representation of a wide variety of systems
in different areas. Biological, social, technological, and information networks can be studied as
graphs, and graph analysis has become crucial to understand the features of these systems.
For instance, social network analysis started in the 1930's and has become one of the most important
topics in sociology~\cite{wasserman94,scott00}. 
In recent times, the computer revolution has provided scholars with a huge amount of data
and computational resources to process and analyze these data. The size of real networks 
one can potentially handle has also grown considerably, reaching millions or even
billions of vertices. The need to deal with such a large number of units has produced
a deep change in the way graphs are approached~\cite{albert02,mendes03,newman03,pastor04,boccaletti06,barrat08}. 

\begin{figure}
\begin{center}
\includegraphics[width=\columnwidth]{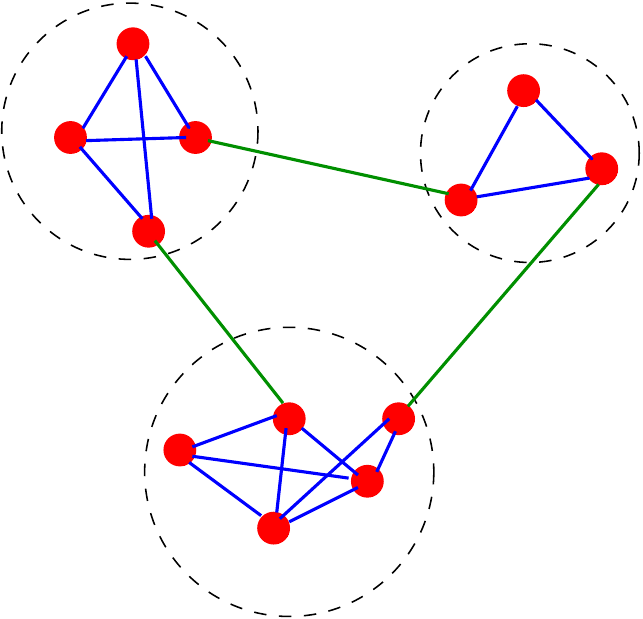}
\caption {\label{Figure1} A simple graph with three communities, enclosed by the
dashed circles. Reprinted figure with permission from Ref.~\cite{fortunato09}. \copyright 2009 by Springer.}
\end{center}
\end{figure}

Graphs representing real systems are not regular like, e. g., lattices. They are objects where 
order coexists with disorder. The paradigm of disordered graph is 
the random graph, introduced by P. Erd\"os and A. R\'enyi~\cite{erdos59}. In it, the probability 
of having an edge between a pair of vertices is equal for all possible pairs (see Appendix).
In a random graph, the distribution of edges among the vertices is highly homogeneous.
For instance, the  
distribution of the number of neighbours of a vertex, or {\it degree}, is
binomial, so most vertices have equal or similar degree.
Real networks are not random graphs, as they
display big inhomogeneities, 
revealing a high level of order and organization. The degree distribution
is broad, with a tail that often follows a power law: therefore, many vertices
with low degree coexist with some vertices with large degree. Furthermore, the 
distribution of edges is not only globally, but also locally inhomogeneous, with 
high concentrations of edges within special groups of vertices, and low concentrations between these groups.
This feature of real networks is called {\it community structure}~\cite{girvan02}, or {\it clustering}, and is the topic of this review
(for earlier reviews see Refs.~\cite{newman04,danon07,schaeffer07,fortunato09,porter09}). 
Communities, also called {\it clusters} or {\it modules}, are groups of vertices 
which probably share common properties and/or play similar roles within the graph. 
In Fig.~\ref{Figure1} a schematic example of a graph with communities is shown.

Society offers a wide variety of possible group organizations: families, working and friendship circles, villages, towns, nations.
The diffusion of Internet has also led to the creation of virtual groups, that live on the Web, like online communities.
Indeed, social communities have been 
studied for a long time~\cite{coleman64,freeman04,kottak04,moody03}. 
Communities also occur in many networked systems from biology, computer science, engineering, 
economics, politics, etc. In protein-protein interaction networks, communities are likely to group proteins 
having the same specific function within the cell~\cite{rives03,spirin03,chen06},
in the graph of the World Wide Web they
may correspond to groups of pages dealing with the same or related topics~\cite{flake02,dourisboure07}, in metabolic networks they may be related to
functional modules such as cycles and pathways~\cite{guimera05,palla05}, in food webs they may
identify compartments~\cite{pimm79,krause03}, and so on.

Communities can have concrete applications. Clustering Web clients who have similar interests and are geografically near 
to each other may improve the performance of services provided on the World Wide Web, in that each cluster of clients
could be served by a dedicated mirror server~\cite{krishnamurthy00}. Identifying clusters of customers with similar
interests in the network of purchase relationships between customers and products of online retailers (like, e. g., {\tt www.amazon.com})  
enables to set up efficient recommendation systems~\cite{reddy02}, that better guide customers through the list of items of the retailer and 
enhance the business opportunities. Clusters of large graphs can be used to create data structures  
in order to efficiently store the graph data and to handle navigational queries, like path searches~\cite{agrawal94,wu04b}.  
{\it Ad hoc networks}~\cite{perkins01}, i. e. self-configuring networks 
formed by communication nodes acting in the same region
and rapidly changing (because the devices move, for instance), 
usually have no centrally maintained routing tables that specify how nodes have to communicate to other nodes. Grouping the 
nodes into clusters enables one to generate compact routing tables while the choice of the communication 
paths is still efficient~\cite{steenstrup01}.

Community detection is important for other reasons, too. Identifying modules and their boundaries
allows for a classification of vertices, according to their structural position in the modules. So, 
vertices with a central position in their clusters, i.~e. sharing a large number of edges with the other
group partners, may have an important function of control and stability within the group; 
vertices lying at the boundaries between modules play an important role of mediation
and lead the relationships and exchanges between different communities 
(alike to Csermely's ``creative elements''~\cite{csermely08}). Such classification
seems to be meaningful in social~\cite{granovetter73,burt76,freeman77} and metabolic networks~\cite{guimera05}. 
Finally, one can study the graph where vertices are the communities and edges are set between
clusters if there are connections between some of their vertices 
in the original graph and/or if the modules overlap. In this way one attains a coarse-grained 
description of the original graph, which unveils the relationships between modules~\footnote{Coarse-graining a graph generally means
mapping it onto a smaller graph having 
similar properties, which is easier to handle. For this purpose, the vertices of the original graph 
are not necessarily grouped in communities. Gfeller and De Los Rios have proposed coarse-graining schemes 
that keep the properties of dynamic processes acting on the graph, like random walks~\cite{gfeller07} and synchronization~\cite{gfeller08}.}.  
Recent studies
indicate that networks of communities have a different degree distribution with respect to the
full graphs~\cite{palla05}; however, the origin of their structures can be explained
by the same mechanism~\cite{pollner06}. 

Another important aspect related to community structure is the hierarchical organization displayed by most networked systems in the real world.
Real networks are usually composed by communities including smaller communities, 
which in turn include smaller communities, etc. 
The human body offers a paradigmatic example of hierarchical organization: it is composed by organs, organs are composed
by tissues, tissues by cells, etc. Another example is represented by business firms, who are characterized by a pyramidal organization,
going from the workers to the president, with intermediate levels corresponding to work groups, departments and management.
Herbert A. Simon has emphasized the crucial role played by hierarchy in the structure and evolution of complex systems~\cite{simon62}. 
The generation and evolution of a system organized in interrelated stable 
subsystems are much quicker than if the system were unstructured, because 
it is much easier to assemble the smallest subparts first and use them as building blocks to get larger structures, until the whole system is assembled.
In this way it is also far more difficult that errors (mutations) occur along the process. 

The aim of community detection in graphs is to identify the modules and, possibly, their hierarchical organization, by only
using the information encoded in the graph topology. The problem has a long tradition and it has
appeared in various forms in several disciplines. 
The first analysis of community structure was carried out by Weiss and Jacobson~\cite{weiss55},
who searched for work groups within a government agency. The authors studied the matrix of 
working relationships between members of the agency, which   
were identified by means of private interviews. Work groups were separated by removing the members
working with people of different groups, which act as connectors between them. This idea of cutting the
bridges between groups is at the basis of 
several modern algorithms of community detection (Section~\ref{sec5}).
Research on communities actually started even earlier than the paper by Weiss and Jacobson.
Already in 1927, Stuart Rice looked for clusters of people in small political bodies, 
based on the similarity of their voting patterns~\cite{rice27}. Two decades later, George Homans showed that social groups could be revealed
by suitably rearranging the rows and the columns of matrices describing social ties, until they take 
an approximate block-diagonal form~\cite{homans50}. This procedure is now standard. Meanwhile, 
traditional techniques to find communities in social networks are hierarchical clustering and partitional clustering 
(Sections~\ref{sec4_2} and~\ref{sec4_3}),
where vertices are joined into groups according to their mutual similarity.

Identifying graph communities is a popular topic in computer science, too.
In parallel computing, for instance, it is crucial 
to know what is the best way to allocate tasks to processors so as to minimize the
communications between them and enable a rapid performance of the calculation. 
This can be accomplished by splitting the computer cluster into
groups with roughly the same number of processors, such that the number of physical connections
between processors of different groups is minimal. 
The mathematical formalization of this problem is called
{\it graph partitioning} (Section~\ref{sec4_1}). The first algorithms for graph partitioning 
were proposed in the early 1970's. 

In a seminal paper appeared in 2002, 
Girvan and Newman proposed a new algorithm, aiming at the identification
of edges lying between communities and their successive removal, a procedure that after some iterations
leads to the isolation of the communities~\cite{girvan02}. The intercommunity edges are detected 
according to the values of a centrality measure, the edge betweenness, that expresses the importance of the role of 
the edges in processes where signals are transmitted across the graph following paths of minimal length.
The paper triggered a big activity in the field, and many new methods have been proposed in the last years.
In particular, physicists entered the game, bringing in their tools and techniques: spin models, optimization,
percolation, random walks, synchronization, etc., became ingredients of new original algorithms. The field has also taken
advantage of concepts and methods from computer science, nonlinear dynamics, sociology, discrete mathematics.
\begin{figure*}
\begin{center}
\includegraphics[width=\textwidth]{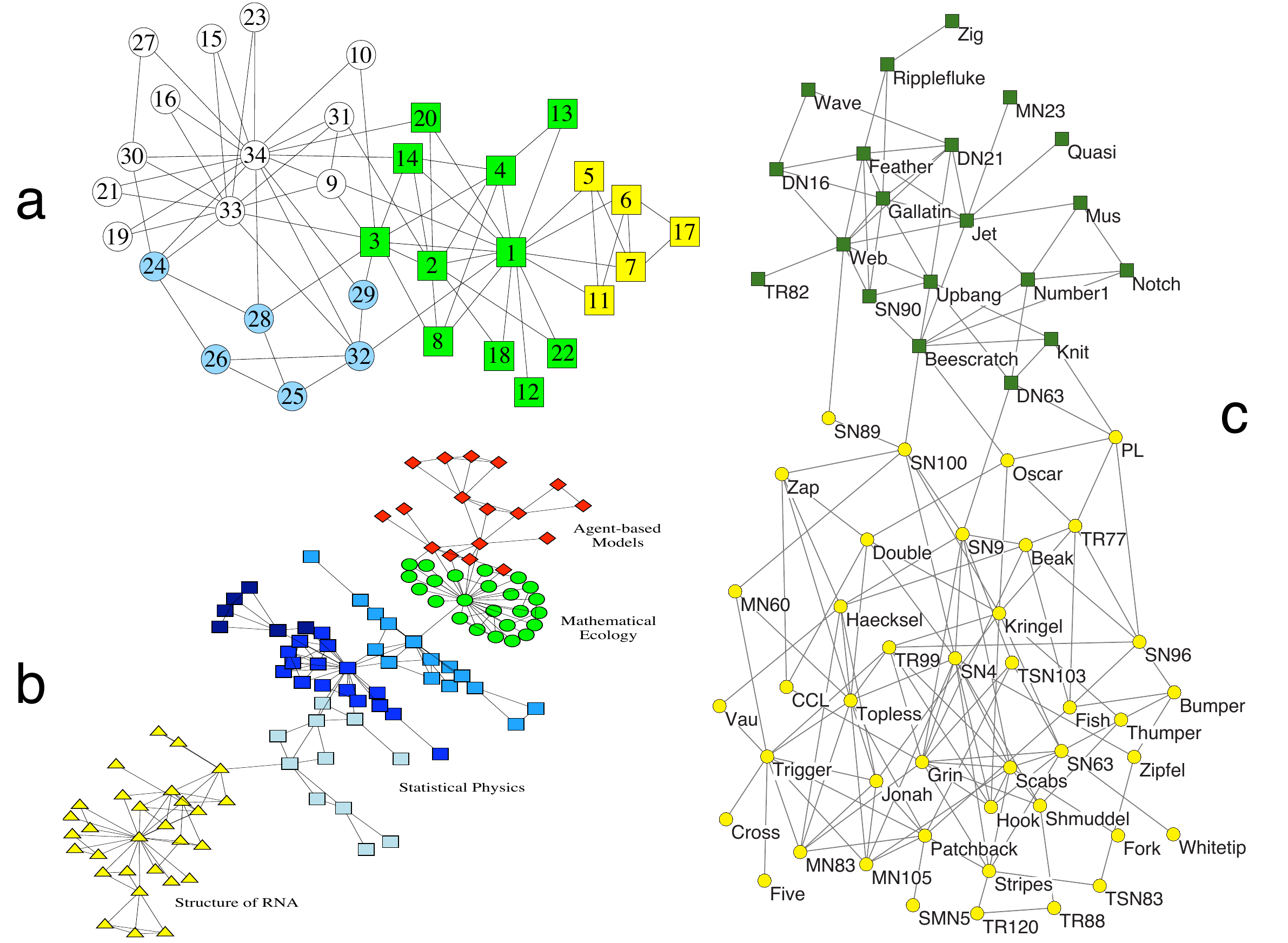}
\caption {\label{fig3} Community structure in social networks. a) Zachary's karate club, a standard benchmark in community detection. The colors correspond to the
best partition found by optimizing the modularity of Newman and Girvan (Section~\ref{sub_sec6_0}). Reprinted figure with permission from 
Ref.~\cite{donetti04}. \copyright 2004 by IOP Publishing and SISSA.
b) Collaboration network between scientists working
at the Santa Fe Institute. The colors indicate high level communities obtained by the algorithm of Girvan and Newman (Section~\ref{subsec5_1})
and correspond quite closely to research divisions of the institute. Further subdivisions
correspond to smaller research groups, revolving around project leaders. Reprinted figure with permission from 
Ref.~\cite{girvan02}. \copyright 2002 by the National Academy of Science of the USA. c) Lusseau's network
of bottlenose dolphins. The colors label the communities identified through the optimization of a modified version of
the modularity of Newman and Girvan, proposed by Arenas et al.~\cite{arenas08b} 
(Section~\ref{sub_sec6_201}). The partition matches the biological classification of the dolphins proposed by Lusseau.
Reprinted figure with permission from 
Ref.~\cite{arenas08b}. \copyright 2008 by IOP Publishing.}
\end{center}
\end{figure*}

In this manuscript we try to cover in some detail the work done in this area. We shall pay a special attention to the contributions 
made by physicists, but we shall also give proper credit to important results obtained by scholars of other disciplines.
Section~\ref{sec2} introduces communities in real networks, and is supposed to make the reader acquainted with the problem
and its relevance. In Section~\ref{sec3} we define the basic elements of community detection, i. e. the concepts of community and partition.
Traditional clustering methods in computer and social sciences, i. e. graph partitioning, hierarchical, partitional and spectral clustering
are reviewed in Section~\ref{sec4}. Modern methods, divided into categories based on the type of approach,
are presented in Sections~\ref{sec5} to~\ref{sub_sec6_2}. Algorithms to find overlapping communities, multiresolution and hierarchical techniques, are 
separately described in Sections~\ref{sec45} and \ref{sub_sec6_20}, respectively, whereas Section~\ref{sec7_1_2} is devoted
to the detection of communities evolving in time.
We stress that our categorization of the algorithms is not sharp, because many algorithms may enter more categories: we tried to classify them based on what
we believe is their main feature/purpose, even if other aspects may be present. 
Sections~\ref{sec6_4} and \ref{sec6} are devoted to the 
issues of defining when community structure is significant, and deciding about the quality of algorithms' performances. In Sections~\ref{sec7_1}
and \ref{sec7_2} we describe general properties of clusters found in real networks, and 
specific applications of clustering algorithms. Section~\ref{sec8} contains the summary of the review, along with
a discussion about future research directions in this area.
The review makes use of several concepts of graph theory, that are defined and explained in the Appendix. Readers not acquainted with 
these concepts are urged to read the Appendix first.

\section{Communities in real-world networks}
\label{sec2}

\begin{figure*}
\begin{center}
\includegraphics[width=\textwidth]{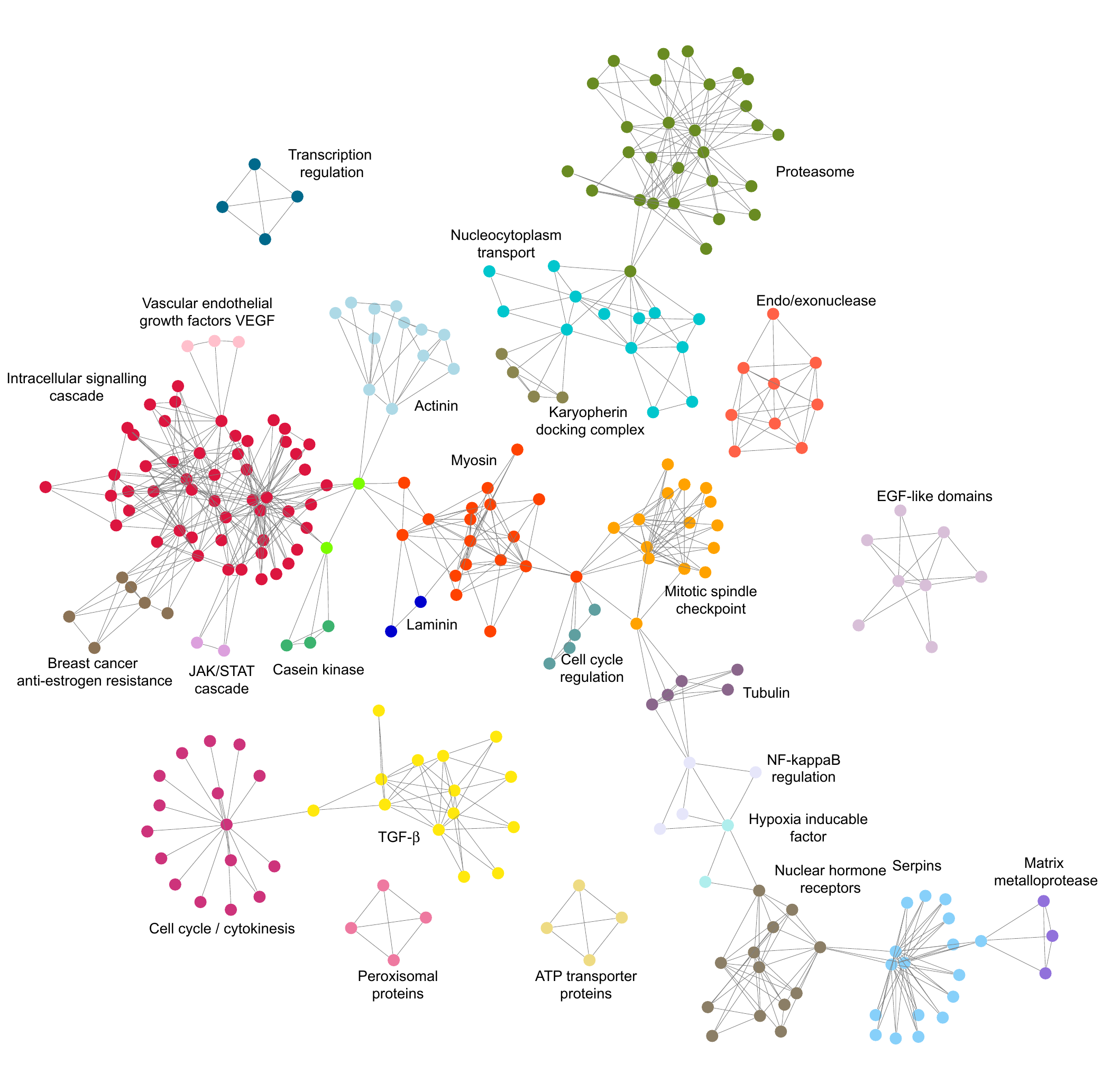}
\caption {\label{fig4} Community structure in protein-protein interaction networks. The graph pictures the interactions
between proteins in cancerous cells of a rat. Communities, labeled by colors, were detected with the Clique Percolation Method
by Palla et al. (Section~\ref{sec45_1}). Reprinted figure with permission from 
Ref.~\cite{jonsson06}. \copyright 2006 by PubMed Central.}
\end{center}
\end{figure*}
In this section we shall present some striking examples of real networks with community structure. In this way we 
shall see what communities look like and why they are important.

Social networks are paradigmatic examples of graphs with communities. The word community itself
refers to a social context. People naturally tend to form groups, within their work environment, family, friends.

In Fig.~\ref{fig3} we show some examples of social networks. The first example (Fig.~\ref{fig3}a) is Zachary's network of karate
club members~\cite{zachary77}, a well-known graph regularly used as a benchmark to test 
community detection algorithms (Section~\ref{sec6_1}). It consists of $34$
vertices, the members of a karate club in the United States, who were observed during a period of three years. Edges connect individuals
who were observed to interact outside the activities of the club. At some point, a 
conflict between the club president and the instructor led to the fission of the club in two separate 
groups, supporting the instructor and the president, respectively (indicated by squares and circles). 
The question is whether from the original network structure
it is possible to infer the composition of the two groups. Indeed, by looking at Fig.~\ref{fig3}a one can distinguish 
two aggregations, one around vertices $33$ and $34$ ($34$ is the president), the other around vertex $1$  
(the instructor). One can also identify several vertices lying between the two main structures, like $3$, $9$, $10$;
such vertices are often misclassified by community detection methods.

Fig.~\ref{fig3}b displays the largest connected component of a network of collaborations of scientists working at the Santa Fe Institute (SFI).
There are $118$ vertices, representing resident scientists at SFI and their collaborators. Edges are placed between scientists
that have published at least one paper together. The visualization layout allows to distinguish 
disciplinary groups.
In this network one observes many cliques, as authors of the same paper are all linked to each other. There are but 
a few connections between most groups. 

In Fig.~\ref{fig3}c we show the network of bottlenose dolphins living in Doubtful Sound
(New Zealand) analyzed by Lusseau~\cite{lusseau03}. There are $62$ dolphins and edges were set between animals that were seen together more 
often than expected by chance. The dolphins separated in two groups after a dolphin left the place for some time
(squares and circles in the figure). 
Such groups are quite cohesive, with several internal cliques, and easily identifiable: only six edges join vertices of different groups.
Due to this natural classification Lusseau's dolphins' network, like Zachary's karate club, is often used to test algorithms 
for community detection (Section~\ref{sec6_1}).

Protein-protein interaction (PPI) networks are subject of intense investigations in biology and bioinformatics, as the interactions between proteins 
are fundamental for each process in the cell~\cite{zhang09}. Fig.~\ref{fig4} illustrates a PPI network of the rat proteome~\cite{jonsson06}.
Each interaction is derived by homology from experimentally observed interactions in other organisms. 
In our example, the proteins interact very frequently with each other, as they belong to 
metastatic cells, which have a high motility and invasiveness with respect to normal cells.
Communities correspond to functional groups, i.~e. to proteins having the same or similar functions, which are expected to be involved
in the same processes. The modules are labeled by the overall function or the dominating protein class. 
Most communities are associated to cancer and metastasis, which indirectly shows how important  
detecting modules in PPI networks is.
\begin{figure}[t]
\begin{center}
\includegraphics[width=\columnwidth]{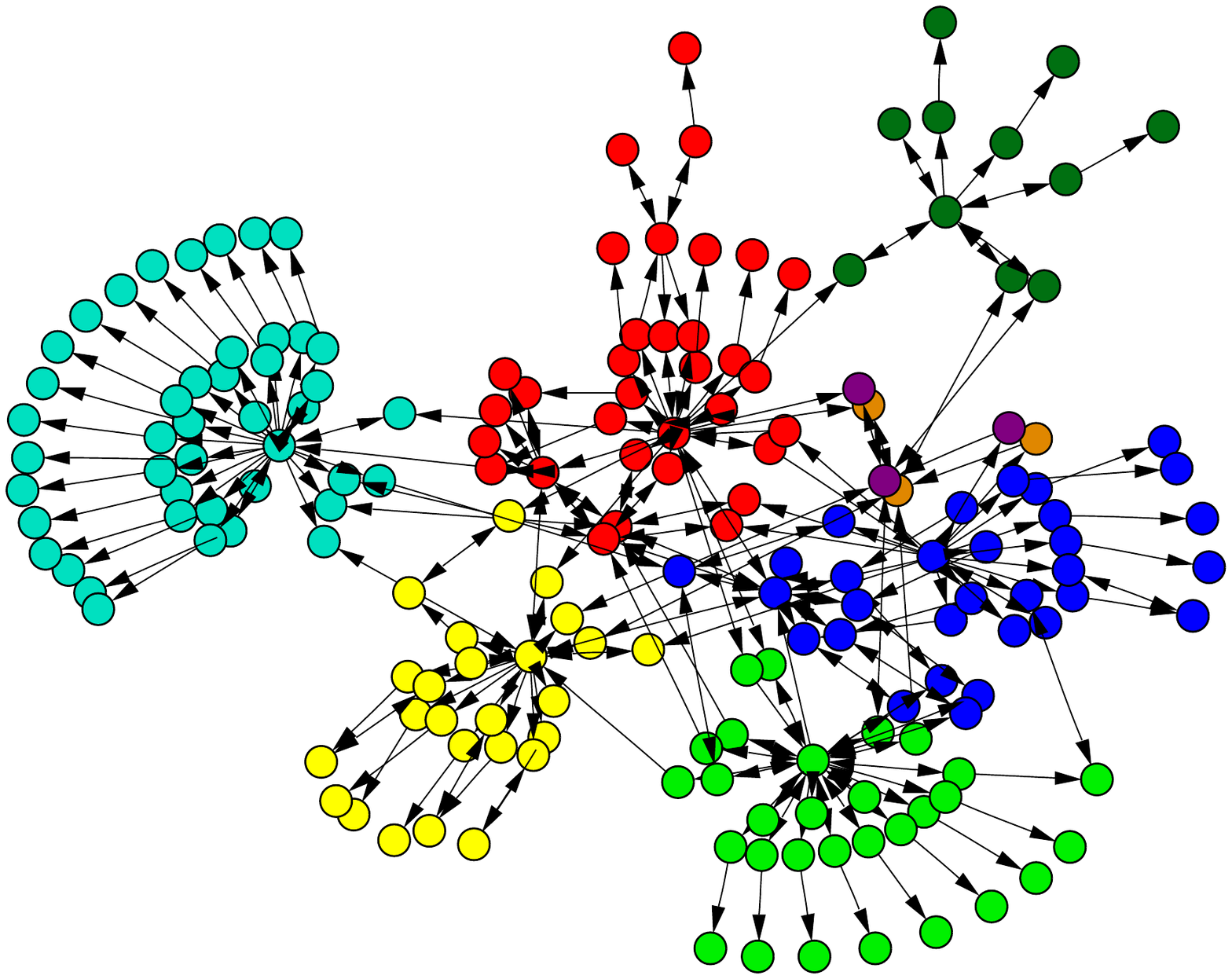}
\caption {\label{fig5} Community structure in technological networks. Sample of the web graph consisting of the pages of a web site
and their mutual hyperlinks, which are directed. Communities, indicated by the colors, were detected with the 
algorithm of Girvan and Newman (Section~\ref{subsec5_1}),
by neglecting the directedness of the edges. Reprinted figure with permission from 
Ref.~\cite{newman04b}. \copyright 2004 by the American Physical Society.}
\end{center}
\end{figure}

Relationships/interactions between elements of a system need not be
reciprocal. In many cases they have a precise direction, that needs to be taken into account 
to understand the system as a whole. As an example we can cite predator-prey relationships in food webs.
In Fig.~\ref{fig5} we see another example, taken from technology. The system is the World Wide Web, which can be seen
as a graph by representing web pages as vertices and the hyperlinks that make users move from one page to another as edges~\cite{albert99}.
Hyperlinks are directed: if one can move from page A to page B by clicking on a hyperlink of A, 
one usually does not find on B a hyperlink taking back to A. In fact, very few hyperlinks (less than $10\%$) are reciprocal.
Communities of the web graph are groups of pages having topical similarities. Detecting communities in the web graph
may help to identify the artificial clusters created by link farms in order to enhance the PageRank~\cite{brin98} value of web sites
and grant them a higher Google ranking. In this way one could discourage this unfair practice. 
One usually assumes that the existence
of a hyperlink between two pages implies that they are content-related, and that this similarity is independent
of the hyperlink direction. Therefore it is customary to neglect the directedness of the hyperlinks and to consider the graph as undirected,
for the purpose of community detection. On the other hand, taking properly into account the directedness of the edges can considerably 
improve the quality of the partition(s), as one can handle a lot of precious information about the system. Moreover, in some instances neglecting
edge directedness may lead to strange results~\cite{leicht08,rosvall08}. Developing methods of community detection for directed graphs is a hard task.
For instance, a directed graph is characterized by asymmetrical matrices (adjacency matrix, Laplacian, etc.), so spectral
analysis is much more complex.
Only a few techniques can be easily extended from the undirected to the directed
case. Otherwise, the problem must be formulated from scratch. 

\begin{figure}
\begin{center}
\includegraphics[width=\columnwidth]{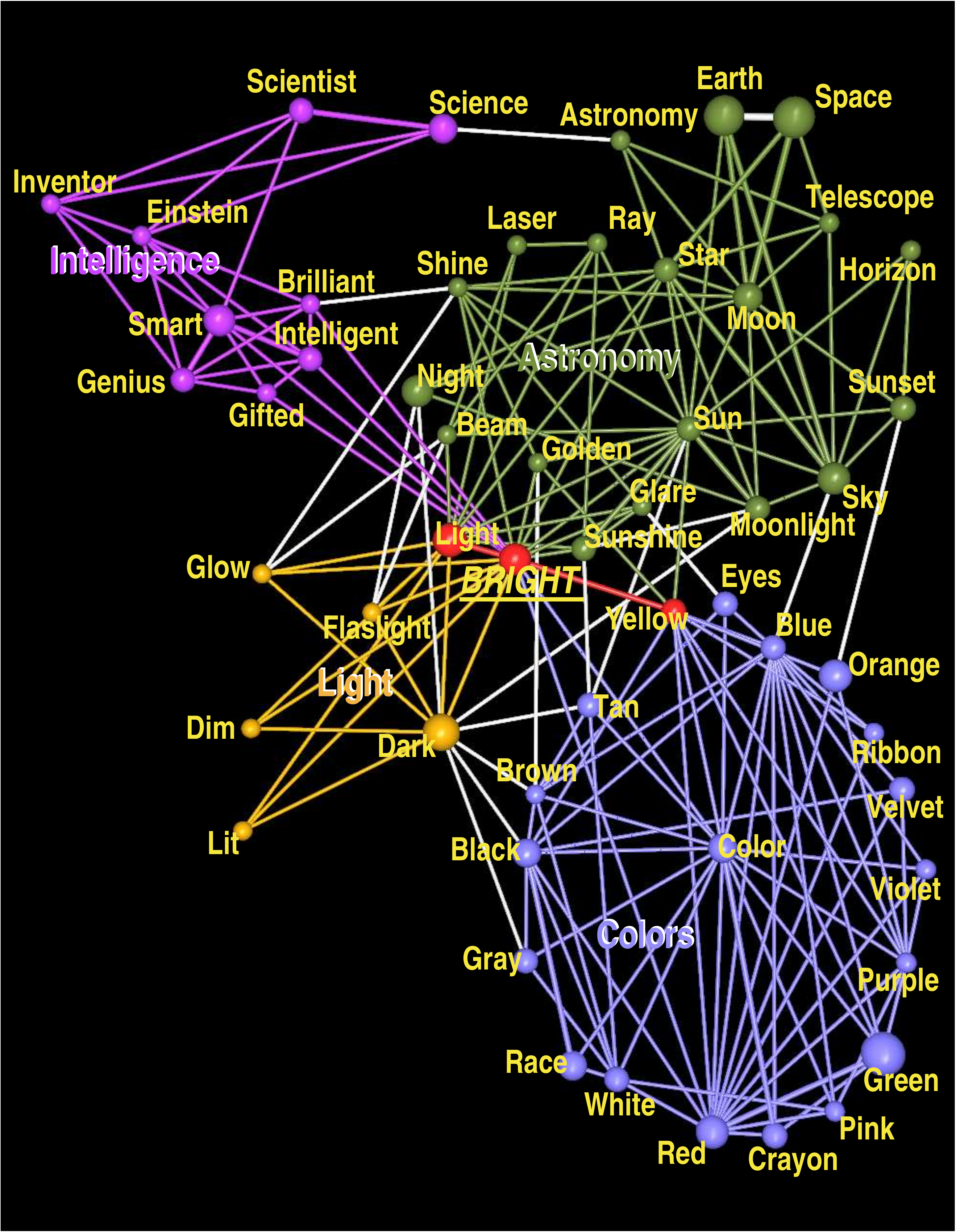}
\caption {\label{fig6} Overlapping communities in a network of word association. The groups, labeled by the colors, were detected with the
Clique Percolation Method by Palla et al. (Section~\ref{sec45_1}).
Reprinted figure with permission from 
Ref.~\cite{palla05}. \copyright 2005 by the Nature Publishing Group.}
\end{center}
\end{figure}

Edge directedness is not the only complication to deal with when facing the problem of graph clustering. In many real networks
vertices may belong to more than one group. In this case one speaks of {\it overlapping communities} and uses the term {\it cover},
rather than partition, whose standard definition forbids multiple memberships of vertices.
Classical examples are social networks, where an individual usually belongs to different 
circles at the same time, from that of work colleagues to family, sport associations, etc.. Traditional algorithms of community detection
assign each vertex to a single module. In so doing, they neglect potentially relevant information. Vertices belonging to more 
communities are likely to play an important role of intermediation between different compartments of the graph. In Fig.~\ref{fig6} we show
a network of word association derived starting from the word ``bright''. The network builds on the University of South Florida Free
Association Norms~\cite{nelson98}. An edge between words $A$ and $B$ indicates that some people associate $B$ to the word $A$.
The graph clearly displays four communities, corresponding to the categories {\it Intelligence}, {\it Astronomy}, {\it Light}
and {\it Colors}. The word ``bright'' is related to all of them by construction. Other words belong to more categories, e.g. 
``dark'' (Colors and Light). Accounting for overlapping communities introduces a further variable, the 
membership of vertices in different communities, which enormously increases the number of possible covers with respect to standard partitions.
Therefore, searching for overlapping communities is usually much more computationally demanding than detecting standard partitions. 

\begin{figure}
\begin{center}
\includegraphics[width=\columnwidth]{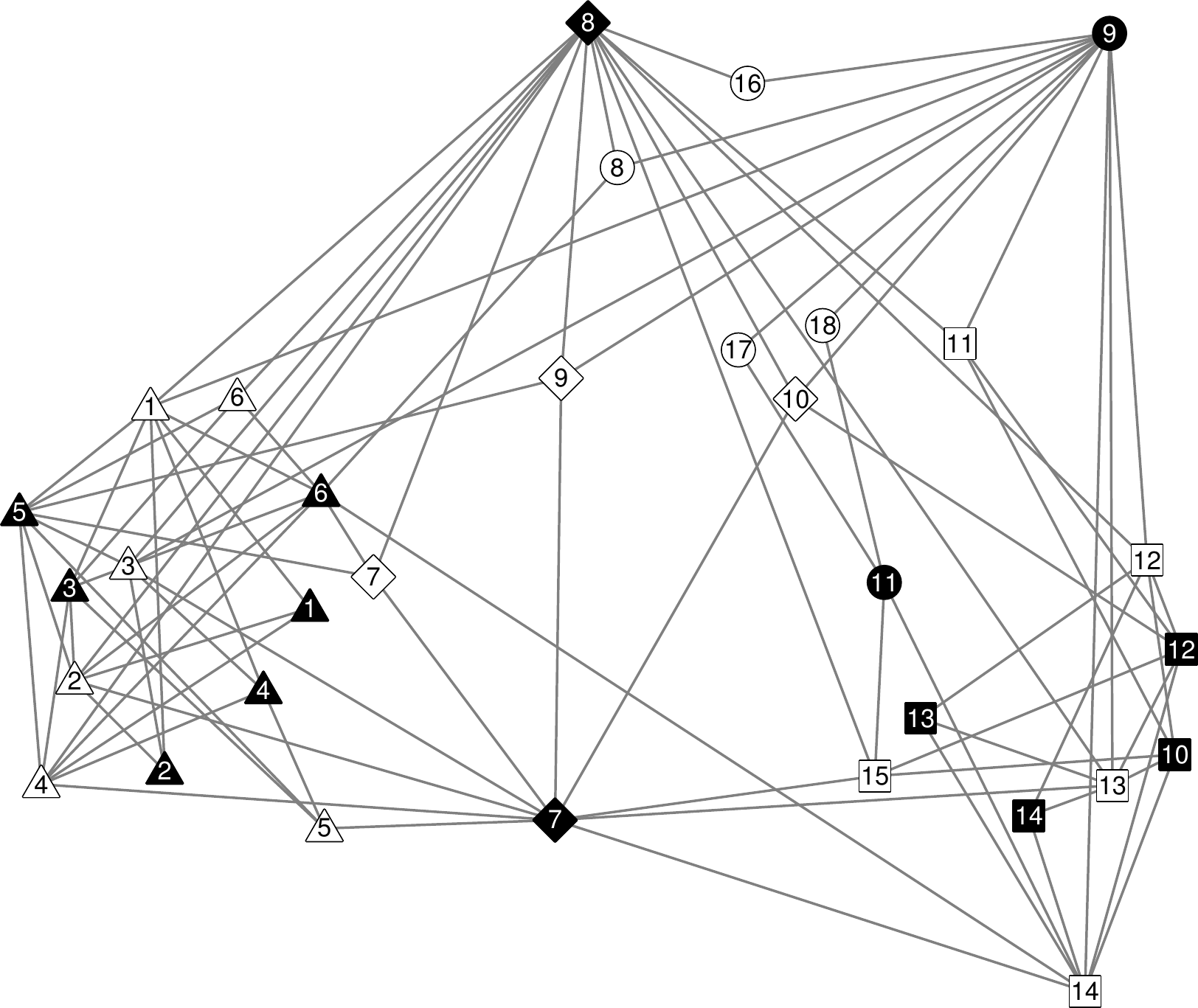}
\caption {\label{fig7} Community structure in multipartite networks. This bipartite graph refers to the Southern Women
Event Participation data set. Women are represented as open symbols with black labels, events as filled symbols
with white labels. The illustrated vertex partition has been obtained by maximizing a modified version of 
the modularity by Newman and Girvan, tailored on bipartite graphs~\cite{barber07} (Section~\ref{sub_sec6_01}).
Reprinted figure with permission from 
Ref.~\cite{barber07}. \copyright 2007 by the American Physical Society.}
\end{center}
\end{figure}

So far we have discussed examples of unipartite graphs. However, it is not uncommon to find real networks
with different classes of vertices, and edges joining only vertices of different classes. An example
is a network of scientists and papers, where edges join scientists and the papers they have authored. Here there is no 
edge between any pair of scientists or papers, so the graph is bipartite.
For a multipartite network
the concept of community does not change much with respect to the case of unipartite graphs, as it remains related to a 
large density of edges between members of the same group, with the only difference that the elements of each group
belong to different vertex classes. Multipartite graphs are usually reduced to unipartite projections of each vertex class. 
For instance, from the bipartite network of scientists and papers
one can extract a network of scientists only, who are related by coauthorship. In this way one can adopt standard techniques
of network analysis, in particular standard clustering methods, but a lot of information gets lost.
Detecting communities in multipartite networks can have interesting applications in, e.g., marketing. 
Large shopping networks, in which customers are linked to the products they have bought, 
allow to classify customers based on the 
types of product they purchase more often: this could be used both to organize targeted advertising, 
as well as to give recommendations about future purchases~\cite{adomavicius05}.  
The problem of community detection in multipartite networks is not trivial, and usually requires {\it ad hoc} methodologies.
Fig.~\ref{fig7} illustrates the famous bipartite network of Southern Women studied by Davis et al.~\cite{davis41}.
There are $32$ vertices, representing $18$ women from the area of Natchez, Mississippi, and 
$14$ social events. Edges represent the participation of the women in the events. From the figure one can see that the
network has a clear community structure.

In some of the previous examples, edges have (or can have) weights. For instance, the edges of the collaboration 
network of Fig.~\ref{fig3}b could be weighted by the number of papers coauthored by pairs of scientists.
Similarly, the edges of the word association network of Fig.~\ref{fig6} are weighted by the number of times pairs of words
have been associated by people. Weights are precious additional information on a graph, and should be considered in the analysis.
In many cases methods working on unweighted graphs can be simply extended to the weighted case. 

\section{Elements of Community Detection}
\label{sec3}

The problem of graph clustering, intuitive at first sight, is actually not well defined.
The main elements of the problem themselves, i.~e. the concepts of community and partition, 
are not rigorously defined, and require some degree of arbitrariness
and/or common sense. Indeed, some ambiguities are hidden and there are
often many equally legitimate ways of resolving them. 
Therefore, it is not surprising that there are plenty of recipes in the literature and that people
do not even try to ground the problem on shared definitions. 

It is important to stress that the identification of structural clusters
is possible only if graphs are {\it sparse}, i.~e. if
the number of edges $m$ is of the order of the number of nodes $n$ of the
graph. If $m\gg n$, the distribution
of edges among the nodes is too homogeneous for communities 
to make sense\footnote{This is not necessarily true if graphs are weighted with a heterogeneous 
distribution of weights. In such cases communities may still be identified as subgraphs with a high internal density 
of weight.}. In this case 
the problem turns into something rather different, close to data clustering~\cite{gan07}, which requires concepts and 
methods of a different nature. The main difference is that, while communities in graphs are related, explicitly or implicitly, to 
the concept of edge density (inside versus outside the community), in data clustering communities are sets of points which are ``close''
to each other, with respect to a measure of distance or similarity, defined for each pair of points. Some classical techniques for
data clustering, like {\it hierarchical}, {\it partitional} and {\it spectral clustering} 
will be discussed later in the review (Sections~\ref{sec4_2}, \ref{sec4_3} and~\ref{sec4_4}),
as they are sometimes adopted for graph clustering too. Other standard procedures for data clustering include {\it neural network clustering} techniques like, e. g.,
{\it self-organizing maps} and {\it multi-dimensional scaling} techniques like, e. g., {\it singular value decomposition} and 
{\it principal component analysis}~\cite{gan07}.

In this section we shall attempt an ordered exposition of the fundamental concepts of community detection. 
After a brief discussion of the issue of computational complexity for the algorithms, we shall review
the notions of community and partition. 

\subsection{Computational complexity}
\label{sec3_0}

The massive amount of data on real networks currently available makes the issue of the efficiency 
of clustering algorithms essential. 
The {\it computational complexity} of an algorithm is the estimate of the amount of resources required by the 
algorithm to perform a task. This involves both the number of computation steps needed and the 
number of memory units that need to be simultaneously allocated to run the computation.
Such demands are usually expressed by their scalability with the size of the system at study.
In the case of a graph, the size is typically indicated by the number of vertices $n$ and/or the number of edges $m$.
The computational complexity of an algorithm cannot always be calculated. In fact, sometimes this is a very hard task, or even impossible.
In these cases, it is however important to have at least an estimate of the {\it worst-case} complexity of the algorithm, which is the 
amount of computational resources needed to run the algorithm in the most unfavorable case for a given system size.

The notation $O(n^\alpha m^\beta)$ indicates that 
the computer time grows as a power of both the number of vertices and edges, with exponents $\alpha$ and $\beta$,
respectively. Ideally, one would like to have the lowest possible values for the exponents, which would correspond to 
the lowest possible computational demands. Samples of the Web graph, with millions of vertices and 
billions of edges, cannot be tackled by algorithms whose running time grows faster than $O(n)$ or $O(m)$.

Algorithms with polynomial complexity form the class {\bf  P}.
For some important decision and optimization problems, there are no known polynomial algorithms.
Finding solutions of such problems in the worst-case scenario may demand an exhaustive search, which takes  
a time growing faster than any polynomial function of the system size, e.g. exponentially. Problems
whose solutions can be verified in a polynomial time span the class {\bf NP}
of {\it non-deterministic polynomial time} problems, which includes {\bf P}.
A problem is {\bf NP}-hard if a solution for it can be translated into a solution for any {\bf NP}-problem.
However, a {\bf NP}-hard problem needs not be in the class {\bf NP}. If it does belong to {\bf NP} it is called
{\bf NP}-complete. The class of  
{\bf NP}-complete problems has drawn a special attention in computer science, as it includes many
famous problems like the Travelling Salesman, Boolean Satisfiability (SAT), Linear Programming, etc.~\cite{garey90,papadimitriou94}.
The fact that {\bf NP} problems have a solution which is verifiable in polynomial time does not mean that
{\bf NP} problems have polynomial complexity, i. e., that they are in {\bf P}. 
In fact, the question of whether {\bf NP}={\bf P} is the most important open problem 
in theoretical computer science. {\bf NP}-hard problems need not be in {\bf NP}
(in which case they would be {\bf NP}-complete), but they are at least as hard as {\bf NP}-complete problems, so they
are unlikely to have polynomial complexity, although a proof of that is still missing.

Many clustering algorithms or problems related to clustering are {\bf NP}-hard. In this case, it is pointless to 
use exact algorithms, which could be applied only to very small systems. 
Moreover, even if an algorithm has a polynomial complexity, it may still be too slow to tackle large systems of interest.
In all such cases it is common to use {\it approximation algorithms}, i.~e. methods that do not deliver an exact solution to the problem at hand,
but only an approximate solution, with the advantage of a lower complexity. Approximation algorithms are often non-deterministic, as they deliver 
different solutions for the same problem, for different initial conditions and/or parameters of the algorithm. The goal of such algorithms is 
to deliver a solution which
differs by a constant factor from the optimal solution. In any case, one should give provable bounds on the
goodness of the approximate solution delivered by the algorithm with respect to the optimal solution.  
In many cases it is not possible to approximate the solution within any constant, as the goodness of the approximation 
strongly depends on the specific problem at study. Approximation algorithms are commonly used for optimization problems, in which one wants to find
the maximum or minimum value of a given cost function over a large set of possible system configurations.

\subsection{Communities}
\label{sec3_1}

\subsubsection{Basics}
\label{sec3_1_1}

The first problem in graph clustering is to look for a quantitative definition of community. No definition is universally accepted.
As a matter of fact, the definition often depends on the specific system at hand and/or application one has in mind.
From intuition and the examples of Section~\ref{sec2} we get the notion that there must be more
edges ``inside'' the community than edges linking vertices of the community with the rest of the graph.
This is the reference guideline at the basis of most community definitions. But many alternative recipes are compatible with it.
Moreover, in most cases, communities are algorithmically defined, i.~e. they are just the final product of the algorithm, without a precise
{\it a priori} definition.

Let us start with a subgraph ${\cal C}$ of a graph ${\cal G}$, with $|{\cal C}|=n_c$ and $|{\cal G}|=n$ vertices,
respectively. We define the {\it internal} and {\it external} 
degree of vertex $v\in {\cal C}$, $k_{v}^{int}$ and $k_{v}^{ext}$, 
as the number of edges connecting $v$ to other vertices of ${\cal C}$ or to the rest of the graph, respectively.
If $k_{v}^{ext}=0$, the vertex has neighbors only within ${\cal C}$, which is likely to be a good cluster for $v$;
if $k_{v}^{int}=0$, instead, the vertex is disjoint from ${\cal C}$ and it should better be assigned to a different cluster.
The {\it internal degree} $k_{int}^{\cal C}$ of ${\cal C}$ is the sum of the internal degrees of its vertices. Likewise, 
the {\it external degree} $k_{ext}^{\cal C}$ of ${\cal C}$ is the sum of the external degrees of its vertices. 
The {\it total degree} $k^{\cal C}$ is the sum of the degrees of the vertices of ${\cal C}$. By definition,
$k^{\cal C}=k_{int}^{\cal C}+k_{ext}^{\cal C}$.

We define the {\it intra-cluster density} $\delta_{int}({\cal C})$ of the subgraph 
${\cal C}$ as the ratio between the number of internal edges of ${\cal C}$ and the number of all possible internal edges, i.~e.
\begin{equation} 
\delta_{int}({\cal C})=\frac{\# \mbox{ internal edges of }{\cal C}}{n_c(n_c-1)/2}.
\label{eq1}
\end{equation}
Similarly, the {\it inter-cluster density} $\delta_{ext}({\cal C})$ is the ratio between the number of edges running from
the vertices of ${\cal C}$ to the rest of the graph and the maximum number of inter-cluster edges possible, i.~e.
\begin{equation} 
\delta_{ext}({\cal C})=\frac{\# \mbox{ inter-cluster edges of }{\cal C}}{n_c(n-n_c)}.
\label{eq2}
\end{equation}
For ${\cal C}$ to be a community, we expect $\delta_{int}({\cal C})$ to be appreciably larger than the average link density $\delta({\cal G})$ of 
${\cal G}$, which is given by the ratio between the number of edges of ${\cal G}$ and the maximum number of possible edges
$n(n-1)/2$. On the other hand, $\delta_{ext}({\cal C})$ has to be much smaller than $\delta({\cal G})$. Searching
for the best tradeoff between a large $\delta_{int}({\cal C})$ and a small $\delta_{ext}({\cal C})$ is implicitly 
or explicitly the goal of most clustering algorithms. A simple way to do that is, e. g., maximizing the
sum of the differences 
$\delta_{int}({\cal C})-\delta_{ext}({\cal C})$ over all clusters of the partition\footnote{In Ref.~\cite{mancoridis98} one actually computes the inter-cluster density by summing the 
densities for each pair of clusters. Therefore the function to minimize is not exactly $\sum_{{\cal C}}[\delta_{int}({\cal C})-\delta_{ext}({\cal C})]$,
but essentially equivalent.}~\cite{mancoridis98}.

A required property of a community is 
{\it connectedness}. We expect that for ${\cal C}$ to be a community there must be a path between each pair
of its vertices, running only through vertices of ${\cal C}$. This feature simplifies the task of community
detection on disconnected graphs, as in this case one just analyzes each connected component separately, unless
special constraints are imposed on the resulting clusters. 

With these basic requirements in mind, we can now introduce the main definitions of community.
Social network analysts have devised many definitions of 
subgroups with various degrees of internal cohesion among vertices~\cite{wasserman94,scott00,moody03}. 
Many other definitions have been introduced by computer scientists and physicists.
We distinguish three classes of definitions: local, global and based on vertex similarity.
Other definitions will be given in the context of the algorithms for which they were introduced.

\subsubsection{Local definitions}
\label{sec3_1_2}

Communities are parts of the graph with a few ties with the rest of the system. To some extent, they can be 
considered as separate entities with their own autonomy. So, it makes sense to evaluate them independently of the graph as a whole.
Local definitions focus on the subgraph under study, including possibly its immediate neighborhood, but neglecting the rest of the graph.
We start with a listing of the main definitions adopted in social network analysis, for which 
we shall closely follow the exposition of Ref.~\cite{wasserman94}. There, 
four types of criteria were identified: {\it complete mutuality}, {\it reachability}, 
{\it vertex degree} and the {\it comparison of internal versus external cohesion}. The corresponding 
communities are mostly {\it maximal
subgraphs}, which cannot be enlarged with the addition of new vertices and edges without
losing the property which defines them. 

Social communities can be defined in a very strict sense as subgroups whose members are all ``friends'' to each other~\cite{luce49} (complete mutuality).
In graph terms, this corresponds to a {\it clique}, i.~e. a subset whose vertices are all adjacent to each other.
In social network analysis, a clique is a maximal subgraph, whereas in graph theory
it is common to call cliques also non-maximal subgraphs. Triangles are 
the simplest cliques, and are frequent in real networks. But larger cliques are less frequent. Moreover,
the condition is really too strict: a subgraph with all possible internal edges except one would be an extremely
cohesive subgroup, but it would not be considered a community under this recipe. Another problem is that 
all vertices of a clique are absolutely symmetric, with no differentiation between them. In many practical examples,
instead, we expect that within a community there is a whole hierarchy of roles for the vertices, with core vertices
coexisting with peripheral ones. We remark that vertices may belong to more cliques simultaneously, a property which is at the basis
of the Clique Percolation Method of Palla et al.~\cite{palla05} (see Section~\ref{sec45_1}). From a practical point of view, finding cliques in
a graph is an {\bf NP}-complete problem~\cite{bomze99}. The 
Bron-Kerbosch method~\cite{bron73} runs in a time growing exponentially with the size of the graph.

It is however possible to relax the notion of clique, defining subgroups which are still clique-like objects.
A possibility is to use properties related to reachability, i.~e. to the existence (and length) of paths between vertices.
An {\it $n$-clique} is a maximal subgraph such that the  
distance of each pair of its vertices is not larger than $n$~\cite{luce50,alba73}. For $n=1$ one recovers 
the definition of clique, as all vertices are adjacent, so each geodesic path between any pair of vertices has length $1$.
This definition, more flexible than that of clique, still has some limitations, deriving from the fact that the geodesic paths need not
run on the vertices of the subgraph at study, but may run on vertices outside the subgraph. In this way, there may be two disturbing 
consequences. First, the diameter of the subgraph may exceed $n$, even if in principle each vertex of the subgraph is less than
$n$ steps away from any of the others. Second, the subgraph may be disconnected, which is not consistent with the notion of cohesion one tries to enforce.
To avoid these problems, Mokken~\cite{mokken79} has suggested two possible alternatives, the {\it $n$-clan} and the {\it $n$-club}. 
An $n$-clan is an $n$-clique whose diameter is not larger than $n$, i.~e. a subgraph such that the distance between any 
two of its vertices, computed over shortest paths within the subgraph, does not exceed $n$. An $n$-club, instead, is a maximal subgraph
of diameter $n$. The two definitions are quite close: the difference is that an $n$-clan is maximal under the 
constraint of being an $n$-clique, whereas an $n$-club is maximal under the constraint imposed by the length of the diameter.

Another criterion for subgraph cohesion relies on the adjacency of its vertices. The idea is that a vertex
must be adjacent to some minumum number of other vertices in the subgraph. 
In the literature on social network analysis there are two complementary ways of expressing this.
A {\it $k$-plex} is a maximal subgraph in which each vertex is adjacent to all other vertices of the subgraph except at most $k$ of them~\cite{seidman78}.
Similarly, a {\it $k$-core} is a maximal subgraph in which each vertex is adjacent to at least $k$ other vertices of the subgraph~\cite{seidman83}. 
So, the two definitions impose conditions on the minimal number of absent or present edges. The corresponding clusters are more cohesive
than $n$-cliques, just because of the existence of many internal edges. In any graph there is a whole hierarchy
of cores of different order, which can be identified by means of a recent efficient algorithm~\cite{batagelj03}.
A $k$-core is essentially the same as a {\it $p$-quasi complete} subgraph, which is a subgraph such that the degree of each vertex
is larger than $p(k-1)$, where $p$ is a real number in $[0,1]$ and $k$ the order of the subgraph~\cite{matsuda99}. Determining
whether a graph includes a $1/2$-quasi complete subgraph of order at least $k$ is {\bf NP}-complete.

As cohesive as a subgraph can be, it would hardly be a community if there is a strong cohesion as well between the subgraph and the 
rest of the graph. Therefore, it is important to compare the internal and external cohesion of a subgraph. In fact, this is what is usually
done in the most recent definitions of community. The first recipe, however, is not recent and stems from social network analysis.
An {\it $LS$-set}~\cite{luccio69}, or {\it strong community}~\cite{radicchi04}, is a subgraph such that the internal degree of 
each vertex is greater than its external degree. This condition is quite strict and can be relaxed into the so-called 
{\it weak} definition of community~\cite{radicchi04}, for which it suffices that the internal degree of the subgraph exceeds its external degree.
An $LS$-set is also a weak community, while the converse is not generally true. 
Hu et al.~\cite{hu08b} have introduced alternative definitions of strong and weak communities: a community is strong
if the internal degree of any vertex of the community exceeds the number of edges that the vertex shares with any other community;
a community is weak if its total internal degree exceeds the number of edges shared by the community with the other communities.
These definitions are in the same spirit of the {\it planted partition model} (Section~\ref{sec6}). An $LS$-set is a strong community also in the sense
of Hu et al.. Likewise, a weak community according to Radicchi et al. is also a weak community for Hu et al.. In both cases the converse is not true, however.
Another definition focuses on the robustness of clusters
to edge removal and uses the concept of {\it edge connectivity}.
The edge connectivity of a pair of vertices in a graph is the minimal number of edges that need to be removed in order to disconnect the two vertices,
i.~e. such that there is no path between them. A {\it lambda set} is a subgraph such that any pair of vertices of the subgraph has a larger
edge connectivity than any pair formed by one vertex of the subgraph and one outside the subgraph~\cite{borgatti90}. 
However, vertices of a lambda-set need not be adjacent and may be quite distant from each other.

Communities can also be identified by a {\it fitness measure}, expressing to which extent 
a subraph satisfies a given property related to its cohesion. The larger the fitness, the more definite is the community.
This is the same principle behind {\it quality functions}, which give an estimate of the goodness of a graph partition (see Section~\ref{sec3_2_2}).
The simplest fitness measure for a cluster is its intra-cluster density $\delta_{int}({\cal C})$. One could assume that a subgraph ${\cal C}$ with $k$ vertices 
is a cluster if $\delta_{int}({\cal C})$ is larger than a threshold, say $\xi$. Finding such subgraphs is an 
{\bf NP}-complete problem, as it coincides with the {\bf NP}-complete Clique Problem when the threshold $\xi=1$~\cite{garey90}.
It is better to fix the size of the subgraph because, without this conditions, any clique would be one of the best possible communities,
including trivial two-cliques (simple edges). Variants of this problem focus on
the number of internal edges of the subgraph~\cite{feige01,asahiro02,holzapfel03}. Another measure of interest is the 
{\it relative density} $\rho({\cal C})$ of a subgraph ${\cal C}$, defined as the ratio between the internal and the total degree of 
${\cal C}$. Finding subgraphs of a given size with $\rho({\cal C})$ larger than a threshold is {\bf NP}-complete~\cite{sima06}.
Fitness measures can also be associated to the connectivity of the subgraph at study to the other vertices of the graph.  
A good community is expected to have a small cut size (see Section~\ref{sec1_1}), i.~e. a small number of edges joining it to the rest of the graph.
This sets a bridge between community detection and graph partitioning, which we shall discuss in Section~\ref{sec4_1}.

\subsubsection{Global definitions}
\label{sec3_1_3}

Communities can also be defined with respect to the graph as a whole. This is reasonable in those cases in which 
clusters are essential parts of the graph, which cannot be taken apart without seriously affecting the functioning of the system.
The literature offers many global criteria to identify communities. In most cases they are indirect 
definitions, in which some global property of the graph is used in an algorithm that delivers communities at the end.
However, there is a class of proper definitions, based on the idea that a graph has community structure if it is different from a random graph.
A random graph \`a la Erd\"os-R\'enyi (Section~\ref{sec1_3}), for instance, is not expected to have 
community structure, as any two vertices have the same
probability to be adjacent, so there should be no preferential linking involving special groups of vertices.
Therefore, one can define a {\it null model}, i.~e. a graph which matches the original 
in some of its structural features, but which is otherwise a random graph. The null model is used as a term of comparison,
to verify whether the graph at study displays community structure or not.
The most popular null model
is that proposed by Newman and Girvan and consists of a randomized version of the original graph, where edges are rewired
at random, under the constraint that the expected degree of each vertex matches the degree of the vertex in the original graph~\cite{newman04b}.
This null model is the basic concept 
behind the definition of {\it modularity},
a function which evaluates the goodness of partitions of a graph into clusters.
Modularity will be discussed at length in this review, because 
it has the unique privilege of being at the same time a global criterion to define a community, a quality function and 
the key ingredient of the most popular method of graph clustering.
In the standard formulation of modularity, a subgraph is a community 
if the number of edges inside the subgraph exceeds the expected number of internal edges that the same subgraph would have in the null model.
This expected number is an average over all possible realizations of the null model. Several modifications of modularity
have been proposed (Section~\ref{sub_sec6_01}).
A general class of null models, including modularity as a special case, has been designed by 
Reichardt and Bornholdt~\cite{reichardt06} (Section~\ref{sub_sec6_01}).
 
\subsubsection{Definitions based on vertex similarity}
\label{sec3_1_4}

It is natural to assume that communities are groups of vertices similar to each other. 
One can compute the similarity between each pair of vertices with respect to some reference property, local or global,
no matter whether they are connected by an edge or not. Each vertex ends up in the cluster whose vertices are most similar to it.
Similarity measures are 
at the basis of traditional methods, like hierarchical, partitional and spectral clustering, 
to be discussed in Sections~\ref{sec4_2}, \ref{sec4_3} and \ref{sec4_4}.
Here we discuss some popular measures used in the literature.

If it is possible to embed the graph vertices in an $n$-dimensional Euclidean space, by assigning a position to them, one could use the 
{\it distance} between a pair of vertices as a measure of their similarity (it is actually a measure of dissimilarity because 
similar vertices are expected to be close to each other). Given the two data points $A=(a_1,a_2,...,a_n)$ and $B=(b_1,b_2,...,b_n)$, 
one could use any norm $L_m$, like the {\it Euclidean distance} ($L_2$-norm),
\begin{equation}
d_{AB}^E=\sum_{k=1}^n\sqrt{(a_k-b_k)^2},
\label{eq3}
\end{equation}
the {\it Manhattan distance} ($L_1$-norm)
\begin{equation}
d_{AB}^M=\sum_{k=1}^n|a_k-b_k|,
\label{eq4}
\end{equation}
and the $L_{\infty}$-norm
\begin{equation}
d_{AB}^\infty=\max_{k\in [1,n]}|a_k-b_k|.
\label{eq5}
\end{equation}
Another popular spatial measure is the {\it cosine similarity}, defined as
\begin{equation}
\rho_{AB}=\mbox{arccos}\frac{{\bf a}\cdot{\bf b}}{\sqrt{\sum_{k=1}^na_k^2}\sqrt{\sum_{k=1}^nb_k^2}},
\label{eq6}
\end{equation}
where ${\bf a}\cdot{\bf b}$ is the dot product of the vectors {\bf a} and {\bf b}. The variable 
$\rho_{AB}$ is defined in the range $[0,\pi)$.

If the graph cannot be embedded in space, the similarity must be necessarily inferred from the 
adjacency relationships between vertices.
A possibility is to define a distance~\cite{burt76,wasserman94} between vertices like
\begin{equation}
d_{ij}=\sqrt{\sum_{k\neq i,j}(A_{ik}-A_{jk})^2},
\label{eq7}
\end{equation}
where {\bf A} is the adjacency matrix.
This is a dissimilarity measure, based on the concept of structural equivalence~\cite{lorrain71}. Two vertices are 
structurally equivalent if they have the same neighbors, even if they are not
adjacent themselves. If $i$ and $j$ are structurally equivalent,
$d_{ij}=0$. Vertices with large degree and different neighbours are considered very ``far'' from each other.
Alternatively, one could measure the {\it overlap} between the neighborhoods $\Gamma(i)$ and $\Gamma(j)$ 
of vertices $i$ and $j$, given by the
ratio between the intersection and the union of the neighborhoods, i.~e.
\begin{equation}
\omega_{ij}=\frac{|\Gamma(i)\cap\Gamma(j)|}{|\Gamma(i)\cup\Gamma(j)|}.
\label{eq8}
\end{equation}
Another measure related to structural equivalence is the Pearson 
correlation between columns or rows of the adjacency matrix,
\begin{equation}
C_{ij}=\frac{\sum_k (A_{ik}-\mu_i)(A_{jk}-\mu_j)}{n\sigma_i\sigma_j},
\label{eq9}
\end{equation}
where the averages $\mu_i=(\sum_jA_{ij})/n$ and the variances $\sigma_i=\sqrt{\sum_j (A_{ij}-\mu_i)^2/n}$.

An alternative measure is the number of edge- (or vertex-) independent paths between two vertices. Independent
paths do not share any edge (vertex), and their number is related to the maximum flow that can be
conveyed between the two vertices under the constraint that each edge can carry only one unit of flow
(max-flow/min-cut theorem~\cite{elias56}). The maximum flow can be calculated in a time $O(m)$, for a graph with $m$ edges,
using techniques like the augmenting path algorithm~\cite{ahuja93}. 
Similarly, one could consider all paths running between two vertices.
In this case, there is the problem that the total number of paths is infinite, but this can be avoided if 
one performs a weighted sum of the number of paths. For instance, paths of length $l$ can be weighted by 
the factor $\alpha^l$, with $\alpha<1$. Another possibility, suggested by 
Estrada and Hatano~\cite{estrada08,estrada09}, is to weigh paths of length $l$ with the inverse factorial $1/l!$.
In both cases, the contribution of long paths is strongly suppressed and the sum converges.

Another important class of measures of vertex similarity is based on properties of random walks on graphs.
One of this properties is the {\it commute-time} between a pair of vertices, which is the average number of steps needed for a random walker,
starting at either vertex, to reach the other vertex for the first time and to come back to the starting vertex.   
Saerens and coworkers~\cite{saerens04,fouss07,yen09,yen07} have extensively studied and used the 
commute-time (and variants thereof) as (dis)similarity measure: the larger the time, the farther (less similar) the vertices. 
The commute-time is closely related~\cite{chandra89} to the {\it resistance distance} introduced by Klein and Randic~\cite{klein93},
expressing the effective electrical resistance between two vertices if the graph is turned into a resistor network. 
White and Smyth~\cite{white03} and Zhou~\cite{zhou03} used instead the average first passage time, i. e. the average number of steps
needed to reach for the first time the target vertex from the source. 
Harel and Koren~\cite{harel01} proposed to build measures out of quantities like the probability to visit a target vertex 
in no more than a given number of steps after it leaves a source vertex\footnote{In the 
clustering method by Latapy and Pons~\cite{latapy05} (discussed in Section~\ref{sec44_2}) and in a recent analysis
by Nadler et al.~\cite{nadler06}, one defined a dissimilarity measure called ``diffusion distance'', which
is derived from the probability that the walker visits the 
target after a fixed number of steps.} 
and the probability that a random walker starting at a source
visits the target exactly once before hitting the source again. Another quantity used to define 
similarity measures is the {\it escape probability},
defined as the probability that the walker reaches the target vertex before coming back to the source vertex~\cite{palmer03,tong08}.
The escape probability is related to the effective conductance between the two vertices in the equivalent resistor network.
Other authors have exploited properties of modified random walks. For instance, the algorithm by Gori and Pucci~\cite{gori07} and that
by Tong et al.~\cite{tong08} used similarity measures derived from Google's PageRank process~\cite{brin98}.

\subsection{Partitions}
\label{sec3_2}

\subsubsection{Basics}
\label{sec3_2_1}

A {\it partition} is a division of a graph in clusters, such that each vertex belongs to one cluster. 
As we have seen in Section~\ref{sec2}, in real systems vertices may be shared among different communities.
A division of a graph into overlapping (or {\it fuzzy}) communities is called {\it cover}.

The number of possible partitions in $k$ clusters of a graph with $n$ vertices is the {\it Stirling number of the second kind}
$S(n,k)$~\cite{andrews76}. The total number of possible partitions is the {\it n-th Bell number}
$B_n=\sum_{k=0}^{n}S(n,k)$~\cite{andrews76}. 
In the limit of large $n$, $B_n$ has the asymptotic form~\cite{lovasz93}
\begin{equation}
B_n \sim \frac{1}{\sqrt{n}}[\lambda(n)]^{n+1/2}e^{\lambda(n)-n-1},
\label{eqpart1}
\end{equation}
where $\lambda(n)=e^{W(n)}=n/W(n)$, $W(n)$ being the {\it Lambert $W$ function}~\cite{polya98}.
Therefore, $B_n$ grows faster than exponentially with the graph size $n$, which means that an enumeration and/or 
evaluation of all partitions of a graph is impossible, unless the graph consists of very few vertices. 

\begin{figure*}
\begin{center}
\includegraphics[width=\textwidth]{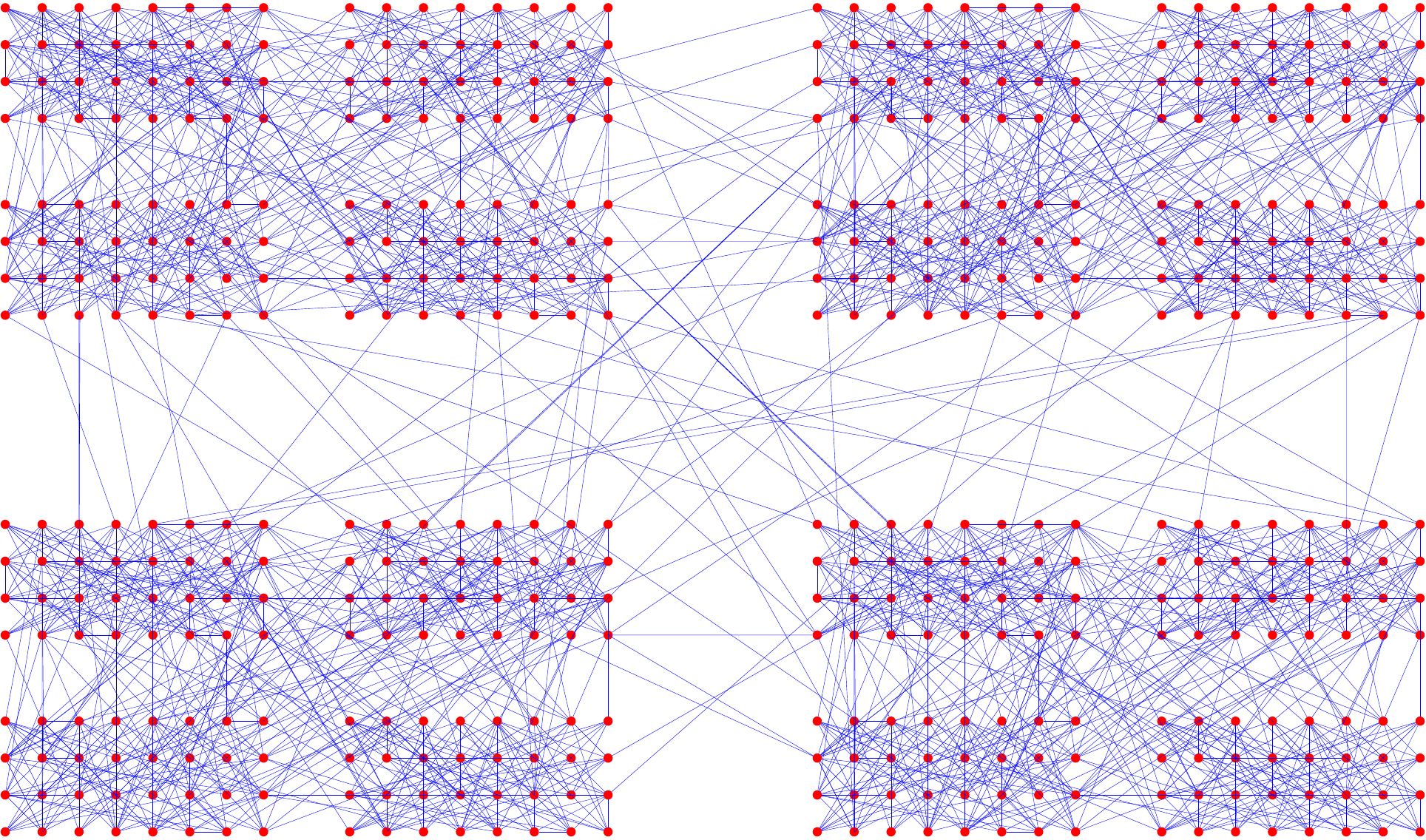}
\caption {\label{fig8} Schematic example of a hierarchical graph. Sixteen modules with $32$ vertices each
clearly form four larger clusters. All vertices have degree $64$. 
Reprinted figure with permission from 
Ref.~\cite{lancichinetti09}. \copyright 2009 by IOP Publishing.}
\end{center}
\end{figure*}

\begin{figure}
\begin{center}
\includegraphics[width=\columnwidth]{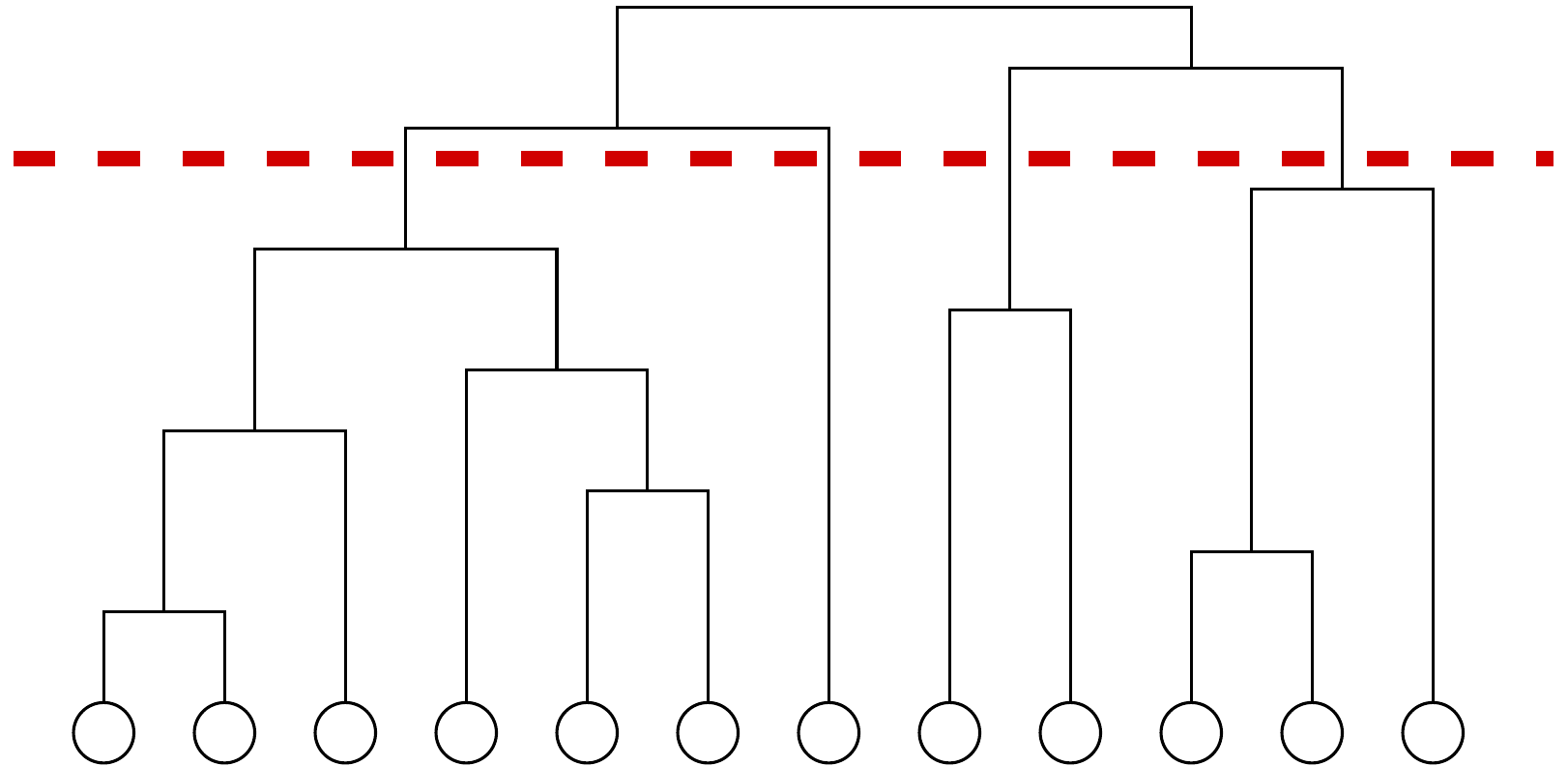}
\caption {\label{fig9} A dendrogram, or hierarchical tree. Horizontal cuts correspond to partitions of the graph in communities.
Reprinted figure with permission 
from Ref.~\cite{newman04b}. \copyright 2004 by the American Physical Society.}
\end{center}
\end{figure}

Partitions can be {\it hierarchically ordered}, when the graph has different levels of organization/structure at different scales.
In this case, clusters display in turn community structure, with smaller communities inside, which may again contain smaller communities, and so on (Fig.~\ref{fig8}).
As an example, in a social network of children living in the same town, one could group the children according to the schools they attend, but within
each school one can make a subdivision into classes. Hierarchical organization is a common feature of many real networks,
where it is revealed by a peculiar scaling of the clustering coefficient for vertices having the same degree $k$, 
when plotted as a function of $k$~\cite{ravasz02,ravasz03}.
A natural way to represent the hierarchical structure of a graph is to draw a {\it dendrogram}, like the one
illustrated in Fig.~\ref{fig9}. Here, partitions of a graph with twelve vertices are shown. At the bottom,
each vertex is its own module (the ``leaves'' of the tree). By moving upwards, groups of vertices are successively aggregated. Mergers of 
communities are represented by horizontal lines. The uppermost level represents the whole graph as a single community. 
Cutting the diagram horizontally at some height, as shown in the figure (dashed line), 
displays one partition of the graph. The diagram
is hierarchical by construction: each community belonging to a level is fully included in a community at a higher level.
Dendrograms are regularly used in sociology and biology. The technique of hierarchical clustering, described in Section~\ref{sec4_2},
lends itself naturally to this kind of representation.

\subsubsection{Quality functions: modularity}
\label{sec3_2_2}

Reliable algorithms are supposed to identify {\it good} partitions.
But what is a good clustering? In order to distinguish between ``good'' and ``bad'' clusterings,
it would be useful to require that partitions satisfy a set of basic properties, intuitive and easy to agree upon. 
In the wider context of data clustering, this issue has been studied by Jon Kleinberg~\cite{kleinberg02}, 
who has proved an important {\it impossibility theorem}. Given a set $S$ of points, a {\it distance function} $d$ is defined, 
which is positive definite and symmetric (the triangular inequality is not explicitly required). One wishes to find a clustering $f$
based on the distances between the points. Kleinberg showed that no clustering satisfies at the same time the three
following properties:
\begin{enumerate}
\item{{\it Scale-invariance}: given a constant $\alpha$, multiplying {\it any} distance function $d$ by $\alpha$ yields the same clustering.}
\item{{\it Richness}: any possible partition of the given point set can be recovered if one chooses a suitable distance function $d$.}
\item{{\it Consistency}: given a partition, any modification of the distance function that does not decrease the distance 
between points of different clusters and that does not increase the distance between points of the same cluster, yields the same clustering.}
\end{enumerate}
The theorem cannot be extended to graph clustering because the distance function cannot be in general defined for a graph which is not complete.
For weighted complete graphs, like correlation matrices~\cite{tumminello08}, it is often possible to define a distance function. On a generic
graph, except for the first property, which does not make sense without a distance function\footnote{The traditional shortest-path distance between
vertices is not suitable here, as it is integer by definition.}, the other two are quite well defined. The property
of richness implies that, given a partition, one can set edges between the vertices in such a way 
that the partition is a natural outcome of the resulting graph (e.g., it could be achieved by setting edges only between vertices of the same cluster).
Consistency here implies that deleting inter-cluster edges and adding intra-cluster edges yields the same partition.

Many algorithms are able to identify a subset of meaningful partitions, ideally one or just a few, whereas some others, like techniques based on
hierarchical clustering (Section~\ref{sec4_2}), deliver a large number of partitions. That does not mean that the partitions found
are equally good. Therefore it is helpful (sometimes even necessary) to have a quantitative {\it criterion} 
to assess the goodness of a graph partition.
A {\it quality function} is a function that assigns a number to each partition of a graph. 
In this way one can rank partitions
based on their score given by the quality function. Partitions with high scores are ``good'', so the one with the largest
score is by definition the best. Nevertheless, one should keep in mind that the question of
when a partition is better than another one is ill-posed, and the answer depends on the specific concept of community and/or quality function adopted.

A quality function $Q$ is {\it additive} if there is an elementary function $q$ such that, for any partition
$\cal P$ of a graph
\begin{equation}
Q({\cal P})=\sum_{{\cal C}\in {\cal P}}q({\cal C}),
\label{eq10}
\end{equation}
where ${\cal C}$ is a generic cluster of partition ${\cal P}$.
Eq.~\ref{eq10} states that the quality of a partition is given by the sum of the qualities of the individual clusters. 
The function $q({\cal C})$ could be any of the cluster fitness functions discussed in Section~\ref{sec3_1_2}, for instance. 
Most quality functions used in the literature are additive, although it is not a necessary requirement. 

An example of quality function is the
{\it performance} $P$, which counts the number of correctly ``interpreted'' pairs of vertices, i.~e. two vertices 
belonging to the same community and connected by an edge, or two vertices belonging to different communities and not connected by an edge. 
The definition of performance, for a partition ${\cal P}$, is
\begin{equation}
P({\cal P})=\frac{|\{(i,j) \in E, C_i=C_j\}|+|\{(i,j) \notin E, C_i\neq C_j\}|}{n(n-1)/2}.
\label{eq10_perf}
\end{equation}
By definition, $0\leq P({\cal P})\leq 1$. Another example is {\it coverage}, i.~e. the ratio of the number of
intra-community edges by the total number of edges: by definition, an ideal cluster structure, where the clusters are disconnected
from each other, yields a coverage of $1$, as all edges of the graph fall within clusters.

The most popular 
quality function is the modularity of Newman and Girvan~\cite{newman04b}. 
It is based on the idea
that a random graph is not expected to have a cluster structure, so the possible existence of clusters is revealed by 
the comparison between the actual density of edges in a subgraph and the density one would expect to have in the subgraph if the
vertices of the graph were attached regardless of community structure. This expected edge density depends on the chosen {\it null model},
i.~e. a copy of the original graph keeping some of its structural properties but without community structure. 
Modularity can then be written as follows
\begin{equation}
Q=\frac{1}{2m}\sum_{ij}\left(A_{ij}-P_{ij}\right)\delta(C_i,C_j),
\label{eq:mod_0}
\end{equation}
where the sum runs over all pairs of vertices, $A$ is the adjacency matrix,
$m$ the total number of edges of the graph, and $P_{ij}$ represents the expected number of edges between vertices $i$ and $j$ in the null model.
The $\delta$-function yields one if vertices $i$ and $j$ are in the same
community ($C_i=C_j$), zero otherwise. The choice of the null model graph is in principle arbitrary, and 
several possibilities exist. For instance, one could simply demand that the graph keeps the same number of 
edges as the original graph, and that edges are placed with the same probability between any pair of vertices. In this case (Bernoulli random graph),
the null model term in Eq.~\ref{eq:mod_0} would be a constant (i.~e. $P_{ij}=p=2m/[n(n-1)]$, $\forall i,j$). 
However this null model is not a good descriptor of real networks,
as it has a Poissonian degree distribution which is very different from the skewed distributions found in real networks. Due to the 
important implications that broad degree distributions have for the structure and function of real 
networks~\cite{albert02,boccaletti06,newman03,dorogovtsev01,pastor04,barrat08}, it is preferable to go for a null model with the same 
degree distribution
of the original graph. The standard null model of modularity
imposes that the expected degree sequence (after averaging over all possible configurations of the model) 
matches the actual degree sequence of the graph. This is a stricter constraint than merely
requiring the match of the degree distributions, and is essentially equivalent~\footnote{The difference is that the 
configuration model maintains the same degree sequence of the original graph for each realization, whereas in the null model 
of modularity the degree sequence of a realization is in general different, and only the average/expected degree sequence
coincides with that of the graph at hand. The two models are equivalent in the limit
of infinite graph size.} to the {\it configuration model}, which has been 
subject of intense investigations in the recent literature on networks~\cite{luczak92,molloy95}. 
In this null model, a vertex could be attached to any other vertex of the graph and the probability that vertices $i$ and $j$, with degrees
$k_i$ and $k_j$, are connected, can be calculated without problems. In fact, in order to form an edge between 
$i$ and $j$ one needs to join two {\it stubs} (i.~e. half-edges), incident with $i$ and $j$. The probability $p_i$ to 
pick at random a stub incident with $i$ is $k_i/2m$, as there are $k_i$ stubs incident with $i$
out of a total of $2m$. 
The probability of a connection between $i$ and $j$ is then given by the product $p_ip_j$, 
since edges are placed independently of each other. The result is 
$k_ik_j/4m^2$, which yields an expected number $P_{ij}=2mp_ip_j=k_ik_j/2m$ of edges between $i$ and $j$.
So, the final expression of modularity reads 
\begin{equation}
Q=\frac{1}{2m}\sum_{ij}\left(A_{ij}-\frac{k_ik_j}{2m}\right)\delta(C_i,C_j).
\label{eq:mod}
\end{equation}
Since the only contributions to the sum come from vertex pairs belonging to the same
cluster, we can group these contributions together and rewrite the sum over the vertex pairs as a sum over the clusters
\begin{equation}
Q=\sum_{c=1}^{n_c}\Big[\frac{l_c}{m}-\left(\frac{d_c}{2m}\right)^2\Big].
\label{eq:mod1}
\end{equation}
Here, $n_c$ is the number of clusters, $l_c$ the total number of edges joining vertices of module $c$ and 
$d_c$ the sum of the degrees of the vertices of $c$. In Eq.~\ref{eq:mod1}, the first term of each summand is the 
fraction of edges of the graph inside the module, whereas the second term represents the expected fraction of 
edges that would be there if the graph were a random graph with the same expected degree for each vertex. 

A nice feature of modularity is that it can be equivalently expressed both in terms of the intra-cluster edges, as in Eq.~\ref{eq:mod1}, and in terms of the
inter-cluster edges~\cite{djidjev07}. In fact, the maximum of modularity can be expressed as
\begin{eqnarray}
\nonumber
Q_{max}&=&\mbox{max}_{\cal P}\left\{\sum_{c=1}^{n_c}\Big[\frac{l_c}{m}-\left(\frac{d_c}{2m}\right)^2\Big]\right\}\\
&=&\nonumber
\frac{1}{m}\mbox{max}_{\cal P}\left\{\sum_{c=1}^{n_c}\Big[l_c-\mbox{Ex}(l_c)\Big]\right\}\\
&=&-\frac{1}{m}\mbox{min}_{\cal P}\left\{-\sum_{c=1}^{n_c}\Big[l_c-\mbox{Ex}(l_c)\Big]\right\},
\label{eqr11}
\end{eqnarray}
where $\mbox{max}_{\cal P}$ and $\mbox{min}_{\cal P}$ indicates the maximum and the minimum 
over all possible graph partitions ${\cal P}$ and $\mbox{Ex}(l_c)=d_c^2/4m$ indicates the expected number of links
in cluster $c$ in the null model of modularity. 
By adding and subtracting the total number of edges $m$ of the graph one finally gets
\begin{eqnarray}
\nonumber
Q_{max}&=&-\frac{1}{m}\mbox{min}_{\cal P}\Big[\Big(m-\sum_{c=1}^{n_c}l_c\Big)-\Big(m-\sum_{c=1}^{n_c}\mbox{Ex}(l_c)\Big)\Big]\\
&=&
-\frac{1}{m}\mbox{min}_{\cal P}(|\mbox{Cut}_{\cal P}|-\mbox{ExCut}_{\cal P}).
\label{eqr12}
\end{eqnarray}
In the last expression $|\mbox{Cut}_{\cal P}|=m-\sum_{c=1}^{n_c}l_c$ is the number of inter-cluster edges of partition ${\cal P}$, and 
$\mbox{ExCut}_{\cal P}=m-\sum_{c=1}^{n_c}\mbox{Ex}(l_c)$ is the expected number of inter-cluster edges of the partition in modularity's null model. 

According to Eq.~\ref{eq:mod1}, a subgraph is a module if the corresponding contribution to modularity in the sum
is positive. The more the number of internal edges of the cluster exceeds the expected number, the better defined the community.
So, large positive values of the modularity indicate good partitions\footnote{This is not necessarily true, 
as we will see in Section~\ref{sub_sec6_1}.}. 
The maximum modularity of a graph generally  
grows if the size of the graph and/or the number of (well-separated) clusters increase~\cite{good09}. Therefore, 
modularity should not be used to compare the quality of the community structure of graphs which are very different in size.
The modularity of the whole graph, taken as a single
community, is zero, as the two terms of the only summand in this case are equal and opposite. 
Modularity is always smaller than one,
and can be negative as well. For instance, the partition in which each vertex is a community is always negative: in this case
the sum runs over $n$ terms, which are all negative as the first term of each summand is zero.
This is a nice feature of the measure, implying that, if there are no partitions with positive modularity,
the graph has no community structure. On the contrary, the existence of partitions with large negative 
modularity values may hint to the existence of 
subgroups with very few internal edges and many edges lying between them ({\it multipartite structure})~\cite{newman06b}.
Holmstr\"om et al.~\cite{holmstrom09} have shown that the distribution of modularity values across the partitions of various graphs,
real and artificial (including random graphs with no apparent community structure), 
has some stable features, and that the most likely modularity values 
correspond to partitions in clusters of approximately equal size. 

Modularity has been employed as quality function in many algorithms, like some of the divisive algorithms 
of Section~\ref{sec5}. In addition, modularity optimization is itself a popular method for community detection 
(see Section~\ref{sub_sec6_0}). Modularity also allows to assess the stability of partitions~\cite{massen06} (Section~\ref{sec6_4}),
it can be used to design layouts for graph visualization~\cite{noack09} 
and to perform a sort of renormalization of a graph, by transforming a graph into a smaller one with the same community structure~\cite{arenas07c}.

\section{Traditional methods}
\label{sec4}

\subsection{Graph partitioning}
\label{sec4_1}

\begin{figure}
\begin{center}
\includegraphics[width=7cm]{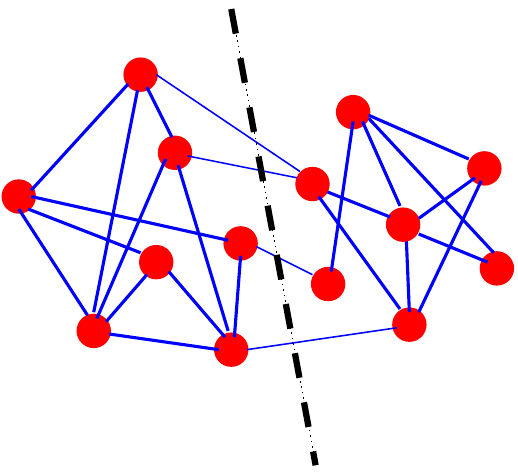}
\caption {\label{Figure5} Graph partitioning. The dashed line shows the solution of the minimum bisection
problem for the graph illustrated, i. e. the 
partition in two groups of equal size with minimal number of edges running between the groups.
Reprinted figure with permission from Ref.~\cite{fortunato09}. \copyright 2009 by Springer.}
\end{center}
\end{figure}

The problem of graph partitioning consists in dividing the vertices in $g$ groups of predefined size, 
such that the number of edges lying between the groups is minimal.
The number of edges running between clusters is called {\it cut size}.
Fig.~\ref{Figure5} presents the solution
of the problem for a graph with fourteen vertices, for $g=2$ and clusters of equal size. 

Specifying the number of clusters of the partition is necessary.
If one simply imposed
a partition with the minimal cut size, and left the number of clusters free, the solution would be trivial, corresponding to 
all vertices ending up in the same cluster, as this would yield a vanishing cut size. Specifying the size is also necessary, as
otherwise the most likely solution of the problem would consist in separating the lowest degree vertex from the rest of the graph,
which is quite uninteresting. This problem can be actually avoided by choosing a different measure to optimize for the partitioning,
which accounts for the size of the clusters. Some of these measures will be briefly introduced at the end of this section.

Graph partitioning is a fundamental issue in parallel computing, circuit partitioning and layout, and in the 
design of many serial algorithms, including techniques to solve partial differential equations and sparse
linear systems of equations.
Most variants of the graph partitioning problem are NP-hard. There are however several algorithms that 
can do a good job, even if their solutions are not necessarily optimal~\cite{pothen97a}. Many algorithms 
perform a bisection of the graph. Partitions into more than two
clusters are usually attained by iterative bisectioning. Moreover, in most cases one imposes the constraint
that the clusters have equal size. This problem is called {\it minimum bisection} and is NP-hard.

The {\it Kernighan-Lin algorithm}~\cite{kernighan70} is one of the earliest 
methods proposed and is still frequently used, often in combination
with other techniques. The authors were motivated by the problem of partitioning electronic
circuits onto boards: the nodes contained in different boards need to be linked 
to each other with the least number of connections. The procedure is an
optimization of a benefit function $Q$, which represents the difference between the number of edges inside 
the modules and the number of edges lying between them. The starting point is an initial 
partition of the graph in two clusters of the predefined size: such initial partition can be random 
or suggested by some information on the graph structure. Then, subsets consisting of 
equal numbers of vertices are swapped between the two groups, so that $Q$ has the maximal increase. 
The subsets can consist of single vertices.
To reduce the risk to be trapped in local maxima of $Q$, the procedure includes some swaps that decrease the 
function $Q$. After a series of swaps with positive and negative gains, the partition with the largest value of 
$Q$ is selected and used as starting point of a new series of iterations. The Kernighan-Lin algorithm
is quite fast, scaling as $O(n^2\log n)$ ($n$ being as usual the number of vertices), 
if only a constant number of swaps are performed at each iteration. The most expensive part is the identification of the subsets to swap,
which requires the computation of the gains/losses for any pair of candidate subsets. On sparse graphs, a slightly different heuristic
allows to lower the complexity to $O(n^2)$.
The partitions found by the procedure are strongly
dependent on the initial configuration and other algorithms can do better. It is preferable to start with a good guess
about the sought partition, otherwise the results are quite poor. Therefore the method 
is typically used to improve on the partitions found through other techniques, by using them as starting
configurations for the algorithm. The Kernighan-Lin algorithm has been extended to extract partitions in any number of
parts~\cite{suaris88}, however the run-time and storage costs increase rapidly with the number of clusters.

Another popular technique is the {\it spectral bisection method}~\cite{barnes82}, which is based on the 
properties of the spectrum of the Laplacian matrix. Spectral clustering will be discussed more thoroughly
in Section~\ref{sec4_4}, here we focus on its application to graph partitioning.

Every partition of a graph with $n$ vertices in two groups can be represented by  
an index vector ${\bf s}$, whose component ${\bf s}_i$ is $+1$ if vertex $i$ is in one group and $-1$
if it is in the other group. The cut size $R$ of the partition of the graph in the two
groups can be written as
\begin{equation}
R=\frac{1}{4}{\bf s}^T{\bf Ls},
\label{eqr}
\end{equation}
where ${\bf L}$ is the Laplacian matrix and ${\bf s}^T$ the transpose of vector ${\bf s}$.
Vector ${\bf s}$ can be written as ${\bf s}=\sum_{i}a_i{\bf v}_i$, 
where ${\bf v}_i$, $i=1,...,n$ are the eigenvectors of the Laplacian. If ${\bf s}$ is properly normalized, then
\begin{equation}
R=\sum_i a_i^2\lambda_i,
\label{eqr1}
\end{equation}
where $\lambda_i$ is the Laplacian eigenvalue corresponding to eigenvector ${\bf v}_i$.
It is worth remarking that the sum contains at most $n-1$ terms, as the Laplacian has at least one zero eigenvalue.
Minimizing $R$ equals to the minimization of the sum on the right-hand side of Eq.~\ref{eqr1}.
This task is still very hard. However, if the second lowest eigenvector
$\lambda_2$ is close enough to zero, a good approximation of the minimum can be attained by choosing ${\bf s}$
parallel to the corresponding eigenvector ${\bf v}_2$, which is called {\it Fiedler vector}~\cite{fiedler73}: 
this would reduce the sum to $\lambda_2$, which is 
a small number. But the index vector cannot be perfectly parallel to ${\bf v}_2$ by construction, because all
its components are equal in modulus, whereas the components of ${\bf v}_2$ are not. The best choice is to
match the signs of the components. So, one can set ${\bf s}_i=+1$ ($-1$) if ${\bf v}_2^i>0$ ($<0$).
It may happen that the sizes of the two corresponding groups do not match the predefined sizes one wishes to have.
In this case, if one aims at a split in $n_1$ and $n_2=n-n_1$ vertices, the best strategy is to order the components 
of the Fiedler vector from the lowest to the largest values and to put in one group the vertices corresponding 
to the first $n_1$ components from the top or the bottom, and 
the remaining vertices in the second group. 
This procedure yields two partitions: the better solution is naturally the one that gives the smaller cut size.

The spectral bisection method is quite fast. The first eigenvectors of the Laplacian can be computed by using the Lanczos
method~\cite{lanczos50}. The time required to 
compute the first $k$ eigenvectors of a matrix with the Lanczos method depends on the size of the eigengap
$|\lambda_{k+1}-\lambda_k|$~\cite{golub96}. If the eigenvalues
$\lambda_{k+1}$ and $\lambda_k$ are well separated, the running time of the algorithm is much shorter than the time required
to calculate the complete set of eigenvectors, which scales as $O(n^3)$. The method gives in general good partitions, that can be further
improved by applying the Kernighan-Lin algorithm. 

The well known max-flow min-cut theorem by Ford and Fulkerson~\cite{ford56} states that the minimum cut between any two vertices 
$s$ and $t$ of a graph, i.~e. any minimal subset of edges whose deletion would topologically separate $s$ from $t$, carries 
the maximum flow that can be transported from $s$ to $t$ across the graph. In this context edges play the role of water pipes, with
a given carrying capacity (e.g. their weights),
and vertices the role of pipe junctions.
This theorem has been used to determine minimal cuts
from maximal flows in clustering algorithms. There are several efficient routines to compute maximum flows in graphs, like the
algorithm of Goldberg and Tarjan~\cite{goldberg88}. Flake et al.~\cite{flake00, flake02} have recently used maximum flows 
to identify communities in the graph of the World Wide Web. The web graph is directed but for the purposes of the calculation
Flake et at. treated the edges as undirected.
Web communities are defined to be ``strong'' (LS-sets), i.~e. the internal degree of each vertex
must not be smaller than its external degree~\cite{radicchi04}. An artificial sink $t$ is added to the graph and one calculates the
maximum flows from a source vertex $s$ to the sink $t$: the corresponding minimum cut identifies the community of vertex $s$, provided
$s$ shares a sufficiently large number of edges with the other vertices of its community, otherwise one could get trivial separations
and meaningless clusters.

Other popular methods for graph partitioning include level-structure partitioning, the geometric algorithm, multilevel algorithms, etc.
A good description of these algorithms can be found in Ref.~\cite{pothen97a}.

Graphs can be also partitioned by minimizing measures that are affine to the cut size, like {\it conductance}~\cite{bollobas98}. 
The conductance $\Phi({\cal C})$ of the subgraph ${\cal C}$ of a graph ${\cal G}$ is defined as
\begin{equation} 
\Phi({\cal C})=\frac{c({\cal C},{\cal G}\setminus{\cal C})}{min(k_{\cal C},k_{{\cal G}\setminus{\cal C}})},
\label{eq3000}
\end{equation}
where $c({\cal C},{\cal G}\setminus{\cal C})$ is the cut size of ${\cal C}$, and $k_{\cal C}$, $k_{{\cal G}\setminus{\cal C}}$
are the total degrees of ${\cal C}$ and of the rest of the graph ${\cal G}\setminus{\cal C}$, respectively. 
Cuts are defined only between non-empty sets, otherwise the measure would not be defined (as the denominator in Eq.~\ref{eq3000} would vanish).
The minimum of the conductance is obtained in correspondence of low values of the cut size and of large values for the
denominator in Eq.~\ref{eq3000}, which peaks when the total degrees of the cluster and its complement are equal. In practical applications,
especially on large graphs, close values of the total degrees correspond to clusters of approximately equal size.
The problem of finding a cut with minimal conductance is {\bf NP}-hard~\cite{sima06}. 
Similar measures are the {\it ratio cut}~\cite{wei89} and the {\it normalized cut}~\cite{shi97,shi00}. The ratio cut of a cluster ${\cal C}$ is defined as
\begin{equation} 
\Phi_C({\cal C})=\frac{c({\cal C},{\cal G}\setminus{\cal C})}{n_{\cal C}n_{{\cal G}\setminus{\cal C}}},
\label{eq3001}
\end{equation}
where $n_{\cal C}$ and $n_{{\cal G}\setminus{\cal C}}$ are the number of vertices of the two subgraphs. The normalized cut of a cluster ${\cal C}$
is
\begin{equation} 
\Phi_N({\cal C})=\frac{c({\cal C},{\cal G}\setminus{\cal C})}{k_{\cal C}},
\label{eq3009}
\end{equation}
where $k_{\cal C}$ is again the total degree of ${\cal C}$.
As for the conductance,
minimizing the ratio cut and the normalized cut favors partitions into clusters of approximately equal size, measured
in terms of the number of vertices or edges, respectively. On the other hand, 
graph partitioning requires preliminary assumptions on the cluster sizes, whereas
the minimization of conductance, ratio cut and normalized cut
does not. The ratio cut was introduced for circuit partitioning~\cite{wei89} and its optimization is an {\bf NP}-hard problem~\cite{matula90}.
The normalized cut is frequently used in image segmentation~\cite{blake87} and its optimization
is {\bf NP}-complete~\cite{shi00}. 
The cut ratio and the normalized cut can be quite well minimized via
spectral clustering~\cite{hagen92, chan93} (Section~\ref{sec4_4}).

Algorithms for graph partitioning are not good for community detection, because it is necessary to provide as 
input the number of groups and in some cases even their sizes, about which in principle one knows nothing.
Instead, one would like an algorithm capable to produce this information
in its output. Besides, from the methodological point of view, using iterative bisectioning 
to split the graph in more pieces is not a reliable procedure. For instance, a split into three clusters 
is necessarily obtained by breaking either cluster of the original bipartition in two parts, whereas in many cases 
a minimum cut partition is obtained if the third cluster is a merger of parts of both initial clusters.

\subsection{Hierarchical clustering}
\label{sec4_2}

In general, very little is known about the community structure of a graph. It is uncommon
to know the number of clusters in which the graph is split, or other indications about the membership of the vertices.
In such cases clustering procedures like graph partitioning methods can hardly be of help, 
and one is forced to make some reasonable assumptions about the number and size of the clusters,
which are often unjustified. On the other hand, the graph may have a hierarchical structure, i.~e. may 
display several levels of grouping of the vertices, with small clusters included within large clusters, which are in turn
included in larger clusters, and so on. Social networks, for instance, often have a hierarchical structure (Section \ref{sec3_2_1}). 
In such cases, one may use {\it hierarchical clustering algorithms}~\cite{hastie01}, i.~e. 
clustering techniques that reveal the multilevel structure of the graph. Hierarchical clustering is very common in social network analysis,
biology, engineering, marketing, etc. 

The starting point of any hierarchical clustering method is the definition of a similarity measure between
vertices. After a measure is chosen, one computes the similarity for each pair of vertices, no matter 
if they are connected or not. At the end of this process, one is left with a 
new $n\times n$ matrix $X$, the similarity matrix. In Section ~\ref{sec3_1_4} we have listed several possible definitions of similarity.
Hierarchical clustering techniques aim at identifying groups of vertices with high similarity, and can be classified in two categories:
\begin{enumerate}
\item{{\it Agglomerative algorithms}, in which clusters are iteratively merged if their similarity is sufficiently high;}
\item{{\it Divisive algorithms}, in which clusters are iteratively split by removing edges connecting vertices with low similarity.}
\end{enumerate}
The two classes refer to opposite processes: agglomerative algorithms are bottom-up, as one starts from the vertices as separate clusters
(singletons) and ends up with the graph as a unique cluster; divisive algorithms are 
top-down as they follow the opposite direction. Divisive techniques have been rarely used in the past 
(meanwhile they have become more popular, see Section~\ref{sec5}), so we shall concentrate here on 
agglomerative algorithms. 

Since clusters are merged based on their mutual similarity, it is essential to define a measure that estimates how similar clusters are, out of the 
matrix $X$. This involves some arbitrariness and several prescriptions exist. 
In {\it single linkage clustering}, the similarity between two groups is the minimum element $x_{ij}$, with $i$ in one group 
and $j$ in the other. On the contrary, the maximum element $x_{ij}$ for vertices of different groups
is used in the procedure of {\it complete linkage clustering}. In {\it average linkage clustering} one has to  
compute the average of the $x_{ij}$.

The procedure can be better illustrated by means of dendrograms (Section \ref{sec3_2_1}), like the one in Fig.~\ref{fig9}.
Sometimes, stopping conditions are imposed to select a partition or a group of partitions that satisfy a special criterion, like 
a given number of clusters or
the optimization of a quality function (e.g. modularity).

Hierarchical clustering has the advantage that it does not require a preliminary knowledge on the number 
and size of the clusters. However, it does not provide a way to discriminate between the many 
partitions obtained by the procedure, and to choose that or those that better represent the community
structure of the graph. The results of the method depend on the specific similarity measure adopted.
The procedure also yields a hierarchical structure by construction, which is rather 
artificial in most cases, since the graph at hand may not have a hierarchical structure at all. 
Moreover, vertices of a community may not be correctly classified, and in many 
cases some vertices are missed even if they have a central role in their clusters~\cite{newman04}. Another problem
is that vertices with just one neighbor are often classified as separated clusters, which in most cases does not make sense. 
Finally, a major weakness of agglomerative hierarchical clustering is that it does not scale well. If points are embedded in space,
so that one can use the distance as dissimilarity measure, the computational complexity
is $O(n^2)$ for single linkage, $O(n^2\log n)$ for the complete and average linkage schemes. For graph clustering, where a distance 
is not trivially defined, the complexity can
become much heavier if the calculation of the chosen similarity measure is costly.

\subsection{Partitional clustering}
\label{sec4_3}

{\it Partitional clustering} indicates another popular class of methods to find clusters in a set of data points.
Here, the number of clusters
is preassigned, say $k$. The points are embedded in a metric space, so that each vertex is a point and a distance measure is defined between
pairs of points in the space. 
The distance is a measure of dissimilarity between vertices.
The goal is to separate the points in $k$ clusters such to maximize/minimize a given cost function based on distances between 
points and/or from points to {\it centroids}, i.~e. suitably defined positions in space. Some of the
most used functions are listed below:
\begin{itemize}
\item{{\it Minimum k-clustering}. The cost function here is the {\it diameter} of a cluster, which is the largest distance between two
points of a cluster. The points are classified such that the largest of the $k$ cluster diameters is the smallest possible. The idea is to keep the 
clusters very ``compact''.}
\item{{\it k-clustering sum}. Same as minimum $k$-clustering, but the diameter is replaced by the average distance between all pairs of points of a cluster.}
\item{{\it k-center}. For each cluster $i$ one defines a reference point $x_i$, the centroid, and computes the maximum $d_i$ of the distances
of each cluster point from the centroid.
The clusters and centroids are self-consistently chosen such to minimize the largest value of $d_i$.}
\item{{\it k-median}. Same as k-center, but the maximum distance from the centroid is replaced by the average distance.}
\end{itemize}
The most popular partitional technique in the literature is {\it $k$-means clustering}~\cite{macqueen67}. 
Here the cost function is the total intra-cluster distance, or squared error function
\begin{equation} 
\sum_{i=1}^k\sum_{{\bf x_j}\in S_i}||{\bf x_j}-{\bf c_i}||^2,
\label{eq3002}
\end{equation}
where $S_i$ indicates the subset of points of the $i$-th cluster and ${\bf c_i}$ its centroid. The $k$-means problem can be simply solved
with the Lloyd's algorithm~\cite{lloyd82}. One starts from an initial distribution of centroids such that they are 
as far as possible from each other. In the first iteration, each vertex is assigned to the nearest centroid.
Next, the centers of mass of the $k$ clusters are estimated and become a new set of 
centroids, which allows for a new classification of the vertices, and so on. 
After a small number of iterations, the positions of the centroids are stable,
and the clusters do not change any more.
The solution found is not optimal,
and it strongly depends on the initial choice of the centroids. Nevertheless, Lloyd's heuristic has remained
popular due to its quick convergence, which makes it suitable for the analysis of large data sets. 
The result can be improved by performing more runs starting from different initial conditions, and picking the 
solution which yields the minimum value of the total intra-cluster distance. Extensions of $k$-means clustering 
to graphs have been proposed by some authors~\cite{schenker03,hlaoui04,rattigan07}. 

Another popular technique, similar in spirit to $k$-means clustering, is {\it fuzzy k-means clustering}~\cite{dunn74, bezdek81}.
This method accounts for the fact that a point may belong to two or more clusters 
at the same time and is widely used in pattern recognition. The associated cost function is
\begin{equation} 
J_m=\sum_{i=1}^n\sum_{j=1}^ku_{ij}^m||{\bf x_i}-{\bf c_j}||^2,
\label{eq3003}
\end{equation}
where $u_{ij}$ is the {\it membership matrix}, which measures the degree of membership of point $i$ (with position ${\bf x_i}$) in cluster $j$, 
$m$ is a real number greater than $1$ and ${\bf c_j}$ is the center of cluster $j$
\begin{equation} 
{\bf c_j}=\frac{\sum_{i=1}^nu_{ij}^m{\bf x_i}}{\sum_{i=1}^nu_{ij}^m}.
\label{eq3004}
\end{equation}
The matrix $u_{ij}$ is normalized so that the sum of the memberships of every point in all clusters yields $1$. The membership
$u_{ij}$ is related to the distance of point $i$ from the center of cluster $j$, as it is
reasonable to assume that the
larger this distance, the lower $u_{ij}$. This can be expressed by the following relation
\begin{equation} 
u_{ij}=\frac{1}{\sum_{l=1}^k\Big(\frac{||{\bf x_i}-{\bf c_j}||}{||{\bf x_i}-{\bf c_l}||}\Big)^{\frac{2}{m-1}}}.
\label{eq3005}
\end{equation}
The cost function $J_m$ can be minimized by iterating 
Eqs.~\ref{eq3004} and ~\ref{eq3005}. One starts from some initial guess for 
$u_{ij}$ and uses Eq.~\ref{eq3004} to compute the centers, which are then
plugged back into Eqs.~\ref{eq3005}, and so on. The process stops when the corresponding elements of the membership matrix
in consecutive iterations differ from each other by less than a predefined tolerance. It can be shown that this procedure indeed 
delivers a local minimum of the cost function $J_m$ of Eq.~\ref{eq3003}. This procedure has the same problems of Lloyd's algorithm 
for $k$-means clustering, i.~e. the minimum is a local minimum, and depends on the initial choice of the matrix $u_{ij}$.

The limitation of partitional clustering is the same
as that of the graph partitioning algorithms: the number of clusters
must be specified at the beginning, the method is not able to derive it.
In addition, the embedding in a metric space can be natural for some graphs,
but rather artificial for others.

\subsection{Spectral clustering}
\label{sec4_4}

Let us suppose to have a set of $n$ objects $x_1, x_2, ..., x_n$ with a pairwise 
similarity function $S$ defined between them, which is symmetric and non-negative 
(i.~e., $S(x_i,x_j)=S(x_j,x_i)\geq 0$, $\forall i,j=1, ..n$).
{\it Spectral clustering} includes all methods and techniques that partition the set into clusters 
by using the eigenvectors of matrices, like $S$ itself or other matrices derived from it. 
In particular, the objects could be points in some metric space, or the vertices of a graph.
Spectral clustering consists of a transformation of the initial set of objects into a set of points in space,
whose coordinates are elements of eigenvectors: the set of points is then clustered via standard techniques, like 
$k$-means clustering (Section~\ref{sec4_3}).
One may wonder why it is necessary to 
cluster the points obtained through the eigenvectors, when one can directly cluster the initial set of objects, based on
the similarity matrix. 
The reason is that the change of representation induced by the eigenvectors makes the cluster properties of the initial data set
much more evident. In this way, spectral clustering is able to separate data points that could not
be resolved by applying directly $k$-means clustering, for instance, as the latter tends to deliver convex sets of points.

The first contribution on spectral clustering was a paper by Donath and Hoffmann~\cite{donath73},
who used the eigenvectors of the adjacency matrix for graph partitions.
In the same year, Fiedler~\cite{fiedler73} realized that from the
eigenvector of the second smallest eigenvalue of the Laplacian matrix it was possible to obtain a bipartition
of the graph with very low cut size, as we have explained in Section~\ref{sec4_1}. For a historical survey see 
Ref.~\cite{spielman96}. In this Section we shall follow the nice tutorial by von Luxburg~\cite{luxburg06},  
with a focus on spectral graph clustering.
The concepts and methods discussed below apply to both unweighted and weighted graphs.

The Laplacian is by far the most used matrix in spectral clustering.
In Section~\ref{sec1_2} we see that the unnormalized 
Laplacian of a graph with $k$ connected components has $k$ zero eigenvalues.
In this case the Laplacian can be written in 
block-diagonal form, i.~e. the vertices can be ordered in such a way that the Laplacian displays
$k$ square blocks along the diagonal, with (some) entries different from zero, whereas all other elements vanish.
Each block is the Laplacian of the corresponding subgraph, so it has the trivial eigenvector with components
$(1,1,1,...,1,1)$. Therefore, there are $k$ degenerate eigenvectors with equal non-vanishing components in correspondence
of the vertices of a block, whereas all other components are zero. In this way, from the components of 
the eigenvectors one can identify the connected components of the graph. For instance, let us consider the 
$n\times k$ matrix, whose columns are the $k$ eigenvectors above mentioned. The $i$-th row of this matrix 
is a vector with $k$ components representing vertex $i$ of the graph. Vectors representing vertices in the same connected component 
of the graph coincide, and their tip lies on one of the axes of a $k$-dimensional system of coordinates 
(i. e. they are all vectors of the form $(0, 0, ... 0, 1, 0, ... , 0, 0)$). So, by drawing the vertex vectors 
one would see $k$ distinct points, each on a different axis, corresponding to the graph components.

If the graph is connected, but consists of $k$ subgraphs which are weakly linked to each other, the spectrum of the unnormalized Laplacian
will have one zero eigenvalue, all others being positive. Now the Laplacian cannot be
put in block-diagonal form: even if one enumerates the vertices 
in the order of their cluster memberships 
(by listing first the vertices of one cluster, then the vertices of another cluster, etc.)
there will always be some non-vanishing entries outside of the blocks.  
However, the lowest $k-1$ non-vanishing eigenvalues are still close to zero,
and the vertex vectors of the first $k$ eigenvectors should still enable one to clearly distinguish the
clusters in a $k$-dimensional space. Vertex vectors corresponding to the same cluster are now not coincident, in general,
but still rather close to each other. So, instead of $k$ points, one would observe $k$ groups of points, 
with the points of each group localized close to each other and far from the other groups. 
Techniques like $k$-means clustering
(Section~\ref{sec4_3}) can then easily recover the clusters. 

The scenario we have described is expected from perturbation theory~\cite{stewart90,bhatia97}. 
In principle all symmetric matrices that can be put in block-diagonal form have a set of eigenvectors
(as many as the blocks), such that
the elements of each eigenvector are different from zero on the vertices of a block and zero otherwise, just like the Laplacian.
The adjacency matrix itself has the same property, for example.
This is a necessary condition for the eigenvectors to be successfully used for graph clustering, but it is not sufficient.
In the case of the Laplacian, for a graph with $k$ connected components, we know that
the eigenvectors corresponding to the $k$ lowest eigenvalues come each from
one of the components. In the case of the adjacency matrix ${\bf A}$ (or of its weighted counterpart ${\bf W}$), instead, it may
happen that large eigenvalues refer to the same component. So, if one takes the eigenvectors corresponding to the $k$ 
largest eigenvalues\footnote{Large eigenvalues of the adjacency matrix are the counterpart of the low eigenvalues of the Laplacian,
since ${\bf L}={\bf D}-{\bf A}$, where ${\bf D}$ is the diagonal matrix whose elements are the vertex degrees.}, 
some components will be overrepresented, while others will be absent. 
Therefore, using the eigenvectors of ${\bf A}$ (or ${\bf W}$) in spectral graph clustering is in general not reliable. Moreover, 
the elements of the eigenvectors corresponding to the components should be sufficiently far from zero. To understand why,
suppose that we take a (symmetric, block-diagonal) matrix, and that
one or more elements of one of the eigenvectors corresponding to the 
connected components are very close to zero. If one perturbs the graph by adding edges between 
different components, all entries of the perturbed eigenvectors will become non-zero and some may have comparable values as the 
lowest elements of the eigenvectors on the blocks. Therefore distinguishing vertices of different components may become a problem,
even when the perturbation is fairly small, and misclassifications are likely. On the other hand,
the non-vanishing elements of the (normalized) eigenvectors of the unnormalized Laplacian, for instance, are all equal to
$1/\sqrt{n_i}$, where $n_i$ is the number of vertices in the $i$-th component. In this way, there is a gap between the lowest
element (here they are all equal for the same eigenvector) and zero. This holds as well for the normalized Laplacian 
${\bf L_{rw}}$ (Section~\ref{sec1_2}). For the other normalized Laplacian ${\bf L_{sym}}$ (Section~\ref{sec1_2}), the non-zero elements
of the eigenvectors corresponding to the connected components are proportional to the square root of the degree of the corresponding
vertex. So, if degrees are very different from each other, and especially if there are vertices with very low degree, some eigenvector
elements may be quite small. As we shall see below, in the context of the technique by Ng et al.~\cite{ng01},
a suitable normalization procedure is introduced to alleviate this problem.

Now that we have explained why the Laplacian matrix is particularly suitable for spectral clustering, we proceed with the 
description of three popular methods: {\it unnormalized spectral clustering} and two {\it normalized spectral
clustering} techniques, proposed by Shi and Malik~\cite{shi97,shi00} and by Ng et al.~\cite{ng01}, respectively. 

Unnormalized spectral clustering uses the unnormalized Laplacian ${\bf L}$. The inputs are the adjacency matrix ${\bf A}$ 
(${\bf W}$ for weighted graphs) and the number $k$ of clusters to be recovered. The first step consists in computing the 
eigenvectors corresponding to the lowest $k$ eigenvalues of ${\bf L}$. Then, one builds the $n\times k$ matrix ${\bf V}$, whose 
columns are the $k$ eigenvectors. The $n$ rows of ${\bf V}$ are used to represent 
the graph vertices in a $k$-dimensional Euclidean space, through a Cartesian system of coordinates. The points 
are then grouped in $k$ clusters by using $k$-means clustering or similar techniques (Section~\ref{sec4_3}).
Normalized spectral clustering works in the same way. In the version by Shi and Malik~\cite{shi97,shi00}, 
one uses the eigenvectors of the normalized Laplacian ${\bf L_{rw}}$ (Section~\ref{sec1_2}). In the algorithm by
Ng et al.~\cite{ng01} one adopts the normalized Laplacian ${\bf L_{sym}}$ (Section~\ref{sec1_2}). Here, however, the 
matrix ${\bf V}$ is normalized by dividing the elements of each row by their sum, obtaining a new matrix ${\bf U}$, 
whose rows are then used to represent the vertices in space, as in the other methods. By doing so, it is much more unlikely that
eigenvector components for a well-separated cluster are close to zero, a scenario
which would make the classification of the corresponding vertices problematic,
as we have said above. However, if the graph has some vertices with low degree, they may still be misclassified.

Spectral clustering is closely related to graph partitioning.
Relaxed versions of the minimization of ratio cut and normalized cut (see Section~\ref{sec4_1}) can be turned
into spectral clustering problems, by following similar procedures 
as in spectral graph partitioning. The measure to minimize can be expressed in matrix form, obtaining 
similar expressions as for the cut size (see Eq.~\ref{eqr}), with index vectors defining the partition of the graph in groups
through the values of their entries. For instance, for the minimum cut bipartition of Section~\ref{sec4_1}, there is
only one index vector ${\bf s}$, whose components equal $\pm 1$, where the signs indicate the two clusters.  
The relaxation consists in performing the minimization over all possible vectors ${\bf s}$, allowing for real-valued components
as well. This version of the problem is exactly equivalent to spectral clustering. The relaxed minimization of
ratio cut for a partition in $k$ clusters yields the $n$ $k$-dimensional 
vertex vectors of unnormalized spectral clustering~\cite{luxburg06}; for 
normalized cut one obtains the $n$ $k$-dimensional vertex vectors of normalized spectral clustering, with the normalized Laplacian 
${\bf L_{rw}}$~\cite{shi97}. The problem is then to turn the resulting vectors into a partition of the graph, which can be done 
by using techniques like $k$-means clustering, as we have seen above. However, it is still unclear what is the relation between the
original minimum cut problem over actual graph partitions and the relaxed version of it, in particular how 
close one can come to the real solution via spectral clustering. 
 
Random walks on graphs are also related to spectral clustering. In fact, by minimizing 
the number of edges between clusters (properly normalized for measures like, e.~g., ratio cut
and normalized cut) one forces random walkers to spend more time within clusters and 
to move more rarely from one cluster to another. In particular, unnormalized spectral clustering with the Laplacian
${\bf L_{rw}}$ has a natural link with random walks, because  
${\bf L_{rw}}={\bf I}-{\bf D}^{-1}{\bf A}$ (Section~\ref{sec1_2}), where ${\bf D}^{-1}{\bf A}$ is the transfer matrix $\bf T$.
This has interesting consequences. For instance, Meil\u{a} and Shi have proven that 
the normalized cut for a bipartition equals the total probability that a random walker
moves from one of the clusters to the other in either sense~\cite{meila01b}. In this way, minimizing 
the normalized cut means looking for a partition minimizing the probability of transitions between clusters.

Spectral clustering requires the computation of the first $k$ eigenvectors of a Laplacian matrix. If the graph is large,
an exact computation of the eigenvectors is impossible, as it would require a time $O(n^3)$. Fortunately there are 
approximate techniques, like the power method or Krylov subspace techniques like the Lanczos method~\cite{golub96}, whose speed
depends on the size of the eigengap $|\lambda_{k+1}-\lambda_k|$, where $\lambda_{k}$ and $\lambda_{k+1}$ are the $k$-th and 
$(k+1)$-th smallest eigenvalue of the matrix. The larger the eigengap, the faster the convergence. In fact, the existence of 
large gaps between pairs of consecutive eigenvalues could suggest the number of clusters of the graph, an information which is
not delivered by spectral clustering and which has to be given as input. We know that, for a disconnected graph with $k$ components,
the first $k$ eigenvalues of the Laplacian matrix (normalized or not) are zero, whether the $(k+1)$-th is non-zero. If the clusters are 
weakly connected to each other, one expects that the first $k$ eigenvalues remain close to zero, and that the $(k+1)$-th is 
clearly different from zero. By reversing this argument, the number of clusters of a graph could be derived by
checking whether there is an integer $k$ such that the first $k$ eigenvalues are small and the $(k+1)$-th is relatively large. However, 
when the clusters are very mixed with each other, it may be hard to identify significant gaps between the eigenvalues.

The last issue we want to point out concerns the choice of the Laplacian matrix to use in the applications. 
If the graph vertices have the same or similar degrees, there is no substantial difference between the unnormalized 
and the normalized Laplacians. If there are big inhomogeneities among the vertex degrees, instead, the choice of the Laplacian
considerably affects the results. In general, normalized Laplacians are more promising because the corresponding 
spectral clustering techniques implicitly impose  
a double optimization on the set of partitions, such that the intracluster edge density is high and, at the same time,
the intercluster edge density is low. On the contrary, the unnormalized Laplacian is related to the intercluster edge density only.
Moreover, unnormalized spectral clustering does not always converge, and sometimes yields trivial partitions in which one or more clusters 
consist of a single vertex. 
Of the normalized Laplacians, ${\bf L_{rw}}$ is more reliable than ${\bf L_{sym}}$ because the eigenvectors of 
${\bf L_{rw}}$ corresponding to the lowest eigenvalues
are cluster indicator vectors, i. e., they have equal non-vanishing entries in correspondence 
of the vertices of each cluster, and zero elsewhere, if the clusters are disconnected. The eigenvectors of ${\bf L_{sym}}$, instead,
are obtained by (left-) multiplying those of ${\bf L_{rw}}$ by the matrix ${\bf D}^{1/2}$: in this way, eigenvector components
corresponding to vertices of the same cluster are no longer equal, in general, a complication that may induce artefacts in the spectral clustering
procedure.

\section{Divisive algorithms}
\label{sec5}

A simple way to identify communities in a graph is to detect the edges that connect vertices 
of different communities and remove them, so that the clusters get disconnected from each other. This is the
philosophy of divisive algorithms. The crucial point is to find a property of intercommunity edges that could 
allow for their identification. 
Divisive methods do not introduce substantial conceptual advances with respect to traditional techniques, as they
just perform hierarchical clustering on the graph at study (Section ~\ref{sec4_2}). The main difference with divisive 
hierarchical clustering is that here one removes inter-cluster edges instead of edges between pairs of vertices with low similarity and
there is no guarantee {\it a priori} that inter-cluster edges connect vertices with low similarity. In some cases vertices (with all their
adjacent edges) or whole subgraphs
may be removed, instead of single edges.
Being hierarchical clustering techniques, it is customary to
represent the resulting partitions by means of dendrograms. 

\subsection{The algorithm of Girvan and Newman}
\label{subsec5_1}

The most popular algorithm is that proposed by Girvan 
and Newman~\cite{girvan02,newman04b}. The method is historically important, because it marked the beginning
of a new era in the field of community detection and opened this topic to physicists. 
Here edges are selected according to the values of 
measures of {\it edge centrality}, estimating the importance of edges
according to some property or process running on the graph. 
The steps of the algorithm are:
\begin{enumerate}
\item{Computation of the centrality for all edges;}
\item{Removal of edge with largest centrality: in case of ties with other edges, one of them is picked at random;}
\item{Recalculation of centralities on the running graph;}
\item{Iteration of the cycle from step 2.}
\end{enumerate}
Girvan and Newman 
focused on the concept of {\it betweenness}, which is a variable expressing the frequency of the participation 
of edges to a process. They considered three alternative definitions: 
geodesic edge betweenness, random-walk edge betweenness and current-flow edge betweenness. In the following we shall refer to them as
edge betweenness, random-walk betweenness and current-flow betweenness, respectively.

\begin{figure}
\begin{center}
\includegraphics[width=\columnwidth]{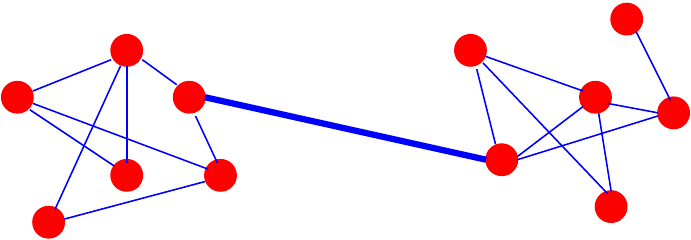}
\caption {\label{Figure6} Edge betweenness is highest for edges connecting communities. In the figure, the 
edge in the middle has a much higher betweenness than all other edges, because all shortest paths connecting vertices of the
two communities run through it. Reprinted figure with permission from Ref.~\cite{fortunato09}. \copyright 2009 by Springer.}
\end{center}
\end{figure}

Edge betweenness is the number of shortest paths between all vertex pairs that run along the edge.
It is an extension to edges of the popular concept of site betweenness, introduced by Freeman in 1977~\cite{freeman77}
and expresses the importance of edges in processes like information spreading, where information usually flows 
through shortest paths. Historically edge betweenness was introduced before site betweenness in a never published
technical report by Anthonisse~\cite{anthonisse71}. 
It is intuitive that intercommunity edges have a large value of the edge betweenness, because
many shortest paths connecting vertices of different communities will pass through them (Fig.~\ref{Figure6}). 
As in the calculation of site betweenness, if there are two or more geodesic paths with the same endpoints that run through an edge,
the contribution of each of them to the betweenness of the edge must be divided by the multiplicity of the paths, as one assumes that the signal/information 
propagates equally along each geodesic path. 
The betweenness of all edges of the graph can be calculated in a time that scales as $O(mn)$, or $O(n^2)$ on a sparse graph, with 
techniques based on breadth-first-search~\cite{newman04b,brandes01,zhou06}. 

In the context of information spreading, one could imagine that signals flow across random rather than geodesic paths.
In this case the betweenness of an edge is given by the frequency of the passages across the edge of a random walker running on the graph
(random-walk betweenness). A random walker moving from a vertex follows each adjacent edge with equal probability.
A pair of vertices is chosen at random, $s$ and $t$. The walker starts at $s$ and keeps moving until
it hits $t$, where it stops. One computes the probability that each edge was crossed by the walker, and averages over all
possible choices for the vertices $s$ and $t$. It is meaningful to compute the {\it net} crossing probability, which is 
proportional to the number of times the walk crossed the edge in one direction. In this way one neglects back and forth passages that 
are accidents of the random walk and tell nothing about the centrality of the edge.
Calculation of random-walk betweenness requires
the inversion of an $n\times n$ matrix (once), followed by obtaining and
averaging the flows for all pairs of nodes. The first task requires a time $O(n^3)$, the second $O(mn^2)$, for a
total complexity $O[(m+n)n^2]$, or $O(n^3)$ for a sparse matrix. The complete calculation requires a time $O(n^3)$ on a sparse graph.

Current-flow betweenness is defined by considering the graph a resistor network, with edges having 
unit resistance. If a voltage difference is applied 
between any two vertices, each edge carries some amount of current, that can be calculated by solving 
Kirchoff's equations. The procedure is repeated for all possible vertex pairs: the current-flow
betweenness of an edge is the average value of the current 
carried by the edge. It is possible to show that this measure is equivalent to 
random-walk betweenness, as the voltage differences and the random walks net flows across the edges satisfy the same equations~\cite{newman05}.
Therefore, the calculation of current-flow betweenness has the same complexity $O[(m+n)n^2]$, or $O(n^3)$ for a sparse graph.

Calculating edge betweenness is much faster than current-flow or random walk betweenness 
[$O(n^2)$ versus $O(n^3)$ on sparse graphs].
In addition, in practical applications the Girvan-Newman algorithm with edge
betweenness gives better results than adopting the other centrality measures~\cite{newman04b}.
Numerical studies show that the
recalculation step 3 of Girvan-Newman algorithm is essential to detect meaningful communities.
This introduces an additional factor $m$ in the running time of the algorithm: consequently, the edge betweenness
version scales as $O(m^2n)$, or
$O(n^3)$ on a sparse graph. On graphs with strong community structure, that quickly break into communities, the recalculation 
step needs to be performed only within the connected component including the last removed edge (or the two components bridged by it if
the removal of the edge splits a subgraph), as the edge betweenness of all other edges remains the same. This 
can help saving some computer time, although 
it is impossible to give estimates of the gain since it depends on the specific graph at hand. 
Nevertheless, the algorithm is quite slow, and applicable to sparse
graphs with up to $n\sim 10000$ vertices, with current computational resources.  In the original version of 
Girvan-Newman's algorithm~\cite{girvan02}, the authors had to deal with the whole hierarchy of partitions, as they 
had no procedure to say which partition is the best. In a successive refinement~\cite{newman04b}, they 
selected the partition with the largest value of modularity (see Section~\ref{sec3_2_2}), a criterion that has been
frequently used ever since. 
The method can be simply extended to the case of weighted graphs, by suitably generalizing 
the edge betweenness. The betweenness of a weighted edge equals the betweenness of the edge in the corresponding unweighted graph,
divided by the weight of the edge~\cite{newman04d}.
There have been countless applications of the Girvan-Newman method: the algorithm  
is now integrated in well known libraries of network analysis programs.

Tyler et al. proposed a modification of the Girvan-Newman
algorithm, to improve the speed of the calculation~\cite{tyler03, wilkinson04}. The gain in speed
was required by the analysis of graphs of gene co-occurrences, which are too large to be analyzed by the algorithm
of Girvan and Newman. Algorithms computing site/edge betweenness
start from any vertex, taken as center, and compute the contribution to betweenness from all paths
originating at that vertex; the procedure is then repeated for all vertices~\cite{newman04b,brandes01,zhou06}. 
Tyler et al. proposed to calculate the contribution to edge betweenness only from a limited number of centers, chosen at random,
deriving a sort of Monte Carlo estimate. Numerical tests indicate that, for each connected subgraph, 
it suffices to pick a number of centers growing as the logarithm
of the number of vertices of the component. For a given choice of the centers, the algorithm proceeds just like that of Girvan and
Newman. The stopping criterion is different, though, as it does not require the calculation of modularity on the resulting partitions, but
relies on a particular definition of community. According to such definition, a connected subgraph with $n_0$ vertices 
is a community if the edge betweenness of any of its edges does not exceed $n_0-1$. Indeed, if the subgraph consists of two 
parts connected by a single edge, the betweenness value of that edge would be greater than or equal to $n_0-1$, 
with the equality holding only if one of the two parts consists of a single vertex. Therefore, the condition on the betweenness of the edges
would exclude such situations, although other types of cluster structures might still be compatible with it. In this way, in the method of 
Tyler et al., edges are removed until all connected components of the partition are ``communities'' in the sense explained above.
The Monte Carlo sampling of the edge betweenness necessarily induces statistical errors.
As a consequence, the partitions are in general different for different
choices of the set of center vertices. However, the authors showed that, by repeating the
calculation many times, the method gives
good results on a network of gene co-occurrences~\cite{wilkinson04}, with a substantial gain of computer time. 
The technique has been also applied to a network of people corresponding via email~\cite{tyler03}.
In practical examples, only 
vertices lying at the boundary between communities may not be clearly classified, and be assigned 
sometimes to a group, sometimes to another. This is actually a nice feature of the method, as it allows to identify 
overlaps between communities, as well as the degree of membership of overlapping vertices in the clusters they belong to.
The algorithm of Girvan and Newman, which is deterministic, is unable to accomplish this\footnote{It may happen that, at a given iteration,
two or more edges of the graph have the same value of maximal betweenness. In this case one can pick any of them at random, which
may lead in general to (slightly) different partitions at the end of the computation.}.
Another fast version of the Girvan-Newman algorithm has been proposed by Rattigan et al.~\cite{rattigan07}. Here, 
a quick approximation of the edge betweenness values is carried out 
by using a {\it network structure index}, which consists of a set of vertex annotations combined with a distance measure~\cite{rattigan06}.
Basically one divides the graph into regions and computes the distances of every vertex from each region. In this way Rattigan et al.
showed that it is possible to lower the complexity of the algorithm to $O(m)$, by keeping a fair accuracy in the estimate of the 
edge betweenness values. This version of the Girvan-Newman algorithm 
gives good results on the benchmark graphs proposed by Brandes et al.~\cite{brandes03} 
(see also Section~\ref{sec6_1}), as well as 
on a collaboration network of actors and on a citation network.

Chen and Yuan
have pointed out that counting all possible shortest paths in the calculation of the 
edge betweenness may lead to unbalanced partitions, with communities of very different size,
and proposed to count only {\it non-redundant} paths, i.~e. paths whose endpoints are all different from each other: the 
resulting betweenness yields better results than standard edge betweenness 
for mixed clusters on the benchmark graphs of Girvan and Newman~\cite{chen06}. Holme et al. have used a modified version of the algorithm
in which vertices, rather than edges, are removed~\cite{holme03}. A centrality measure for the vertices, 
proportional to their site betweenness, and inversely proportional to their indegree, is chosen to
identify boundary vertices, which are then iteratively removed with all their edges. This modification,
applied to study the hierarchical organization of biochemical networks, is motivated by the need to 
account for reaction kinetic information, that simple site betweenness does not include. The indegree of a vertex 
is solely used because it indicates the
number of substrates to a metabolic reaction involving that vertex; 
for the purpose of clustering the graph is considered undirected, as usual. 

The algorithm of Girvan and Newman is unable to find 
overlapping communities, as each vertex is assigned to a single cluster. Pinney and Westhead have proposed a modification of the algorithm 
in which vertices can be split between communities~\cite{pinney06}. To do that, they also compute the betweenness of all vertices of the graph. Unfortunately 
the values of edge and site betweenness cannot be simply compared, due to their different normalization, 
but the authors remarked that the two endvertices of an inter-cluster edge should have
similar betweenness values, as the shortest paths crossing one of them are likely to reach the other one as well through the edge. So they take the edge 
with largest betweenness and remove it only if the ratio of the betweenness values of its endvertices is between $\alpha$ and $1/\alpha$, with $\alpha=0.8$.
Otherwise, the vertex with highest betweenness (with all its adjacent edges) 
is temporarily removed. When a subgraph is split by vertex or edge removal, all deleted vertices  
belonging to that subgraph are ``copied'' in each subcomponent, along with all their edges. Gregory~\cite{gregory07} has proposed 
a similar approach, named CONGA (Cluster Overlap Newman-Girvan Algorithm), in which 
vertices are split among clusters if their site betweenness exceeds the maximum value of the betweenness of the edges. A vertex is split by 
assigning some of its edges to one of its duplicates, and the rest to the other. There are several possibilities to do that, Gregory
proposed to go for the split that yields the maximum of a new centrality measure, called {\it split betweenness}, which is the number of shortest
paths that would run between two parts of a vertex if the latter were split. 
The method has a worst-case complexity $O(m^3)$, or $O(n^3)$ on a sparse graph,
like the algorithm of Girvan and Newman. The code can be found at {\tt http://www.cs.bris.ac.uk/$\sim$steve/networks/index.}\\{\tt html}.

\subsection{Other methods}
\label{subsec5_2}

\begin{figure}
\begin{center}
\includegraphics[width=\columnwidth]{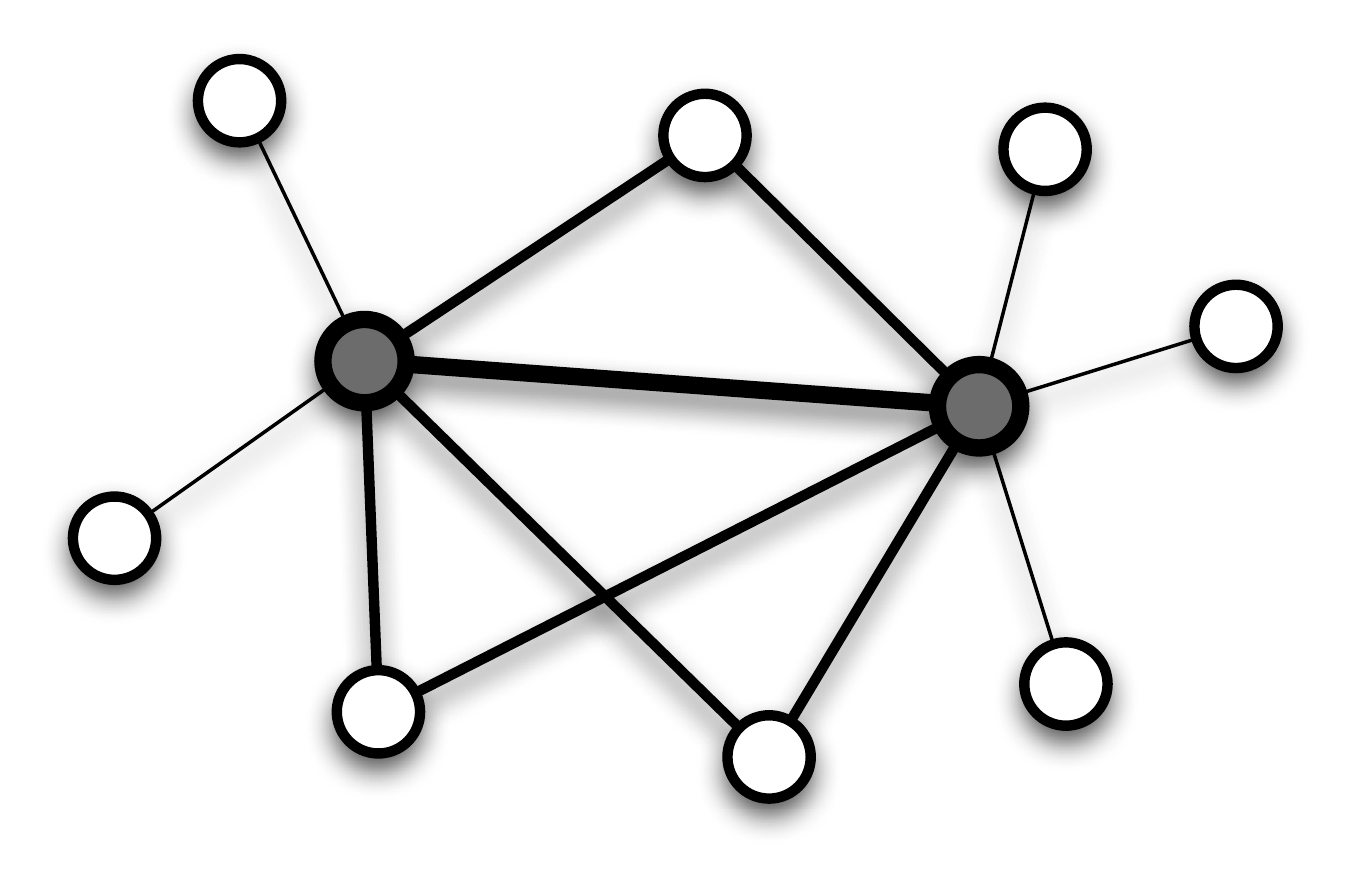}
\caption {\label{edgeclus} Schematic illustration of the edge clustering coefficient introduced by Radicchi et al.~\cite{radicchi04}.
The two grey vertices have five and six other neighbors, respectively. Of the five possible triangles based on the edge connecting
the grey vertices, three are actually there, yielding an edge clustering coefficient $C^3=3/5$. Courtesy by F. Radicchi.}
\end{center}
\end{figure}

Another promising track to detect inter-cluster edges is related to the presence of cycles, i.~e. closed non-intersecting paths,
in the graph. Communities are characterized by a high density of edges, so it is reasonable to expect that such edges form 
cycles. On the contrary, edges lying between communities will hardly be 
part of cycles. Based on this intuitive idea, Radicchi et al. proposed a new measure, the edge clustering coefficient, such that 
low values of the measure are likely to correspond to intercommunity edges~\cite{radicchi04}. The edge clustering coefficient generalizes to edges the
notion of clustering coefficient introduced by Watts and Strogatz for vertices~\cite{watts98} (Fig.~\ref{edgeclus}). 
The clustering coefficient of a vertex
is the number of triangles including the vertex divided by the number of possible triangles that can be formed (Section~\ref{sec1_1}).
The edge clustering coefficient is defined as
\begin{equation} 
\tilde{C}_{i,j}^{(g)}=\frac{z_{i,j}^{(g)}+1}{s_{i,j}^{(g)}},
\label{eq3010}
\end{equation}
where $i$ and $j$ are the extremes of the edge, $z_{i,j}^{(g)}$ the number of cycles of length $g$ built upon edge $ij$ and 
$s_{i,j}^{(g)}$ the possible number of cycles of length $g$ that one could build based on the existing edges of $i$, $j$ and their neighbors.
The number of actual cycles in the numerator is augmented by $1$ to enable a ranking among edges without cycles, which would all
yield a coefficient $\tilde{C}_{i,j}^{(g)}$ equal to zero, independently of the degrees of the extremes $i$ and $j$ and their neighbors. 
Usually, cycles of length $g=3$ (triangles) or $4$ 
are considered. The measure is (anti)correlated with edge betweenness: edges with low edge clustering coefficient usually have high betweenness and vice versa,
although the correlation is not perfect.
The method works as the algorithm by Girvan and Newman.
At each iteration, the edge with smallest clustering coefficient
is removed, the measure is recalculated again, and so on. If the removal of an edge leads to a split of a subgraph in two parts,
the split is accepted only if both clusters are LS-sets (``strong'') or 
``weak'' communities (see Section~\ref{sec3_1_2}). The verification of the community condition on the clusters 
is performed on the full adjacency matrix of the initial graph.
If the condition were satisfied only for one of the two clusters, the initial subgraph 
may be a random graph, as it can be easily seen that by cutting a random graph \'a la Erd\"os and R\'enyi in two parts, the larger of them is
a strong (or weak) community with very high probability, whereas the smaller part is not. Enforcing the community condition on both clusters, 
it is more likely that the subgraph to be split indeed has a cluster structure. Therefore, the algorithm stops when all clusters
produced by the edge removals are communities in the strong or weak sense, and further splits would violate this condition.
The authors suggested to use the same stopping criterion for the algorithm of Girvan and Newman, 
to get structurally well-defined clusters.
Since the edge clustering coefficient is a local measure, involving
at most an extended neighborhood of the edge, 
it can be calculated very quickly. The running time of the algorithm to completion
is $O(m^4/n^2)$, or $O(n^2)$ on a sparse graph, if $g$ is small, so it is much shorter than the running time of the Girvan-Newman method.
The recalculation step becomes slow if $g$ is not so small, as in this case the number of edges whose 
coefficient needs to be recalculated may reach a sizeable fraction of the edges of the graph; likewise, counting the number of cycles
based on one edge becomes lengthier. 
If $g\sim 2d$, where $d$ is the diameter of the graph (which is usually a small number for real networks),
the cycles span the whole graph and the measure becomes global and no more local. 
The computational complexity in this case exceeds that of the algorithm of Girvan and Newman, but it can come close to it for practical 
purposes even at lower values of $g$. 
So, by tuning $g$ one can smoothly interpolate between a local
and a global centrality measure. The software of the algorithm can be found in
{\tt http://filrad.homelinux.org/Data/}.
In a successive paper~\cite{castellano04} the authors
extended the method to the case of weighted networks, by modifying the edge clustering coefficient of Eq.~\ref{eq3010},
in that the number of cycles $z_{i,j}^{(g)}$ is multiplied by the weight of the edge $ij$. 
The definitions of strong and weak communities can be trivially extended to weighted graphs by replacing the internal/external degrees of the vertices/clusters
with the corresponding strengths. More recently, the method has been extended to bipartite networks~\cite{zhang_peng07}, where only cycles 
of even length are possible ($g=4$, $6$, $8$, etc.). 
The algorithm by Radicchi et al. may give poor results when the graph has few cycles, as it happens in some social and many non-social networks. 
In this case, in fact, the edge clustering coefficient is small and fairly similar for most edges, and the algorithm 
may fail to identify the bridges between communities.

An alternative measure of centrality for edges is 
information centrality. It is based on the concept of efficiency~\cite{latora01}, which estimates how easily information
travels on a graph according to the length of shortest paths between
vertices. The efficiency of a network is defined as the average of the inverse distances between all pairs of vertices. If the vertices are
``close'' to each other, the efficiency is high. 
The information centrality of an edge is the relative variation of the efficiency of the graph
if the edge is removed. 
In the algorithm by Fortunato et al.~\cite{fortunato04}, edges are removed
according to decreasing values of information centrality. The method is analogous to that of Girvan and Newman. 
Computing the information centrality of an edge requires the calculation of the distances between all pairs of vertices, which can be done
with breadth-first-search in a time $O(mn)$. So, in order to compute the information centrality of all edges one requires a time $O(m^2n)$. At this point one
removes the edge with the largest value of information centrality and recalculates the information centrality of all remaining edges with respect to
the running graph. Since the procedure is iterated until there are no more edges in the network, the final complexity is $O(m^3n)$, or $O(n^4)$ on a sparse graph.
The partition with the largest value of modularity is chosen as most representative of the community structure of the graph.
The method is much slower than the algorithm of Girvan and Newman. Partitions obtained with both techniques are rather consistent, mainly because 
information centrality has a strong correlation with edge betweenness.
The algorithm by Fortunato et al. gives better results when 
communities are mixed, i.~e. with a high degree of interconnectedness, but it tends to isolate leaf vertices and small loosely bound subgraphs.

A measure of vertex centrality based on loops, similar to the clustering coefficient by Watts and Strogatz~\cite{watts98}, has
been introduced by Vragovi\^{c} and Louis~\cite{vragovic06}. 
The idea is that neighbors of a vertex well inside a community are ``close'' to each other, even in the absence of the vertex, 
due to the high density of intra-cluster edges. Suppose that $j$ and $k$ are neighbors of  
a vertex $i$: $d_{jk/i}$ is the length of a shortest path between $j$ and $k$, if $i$ is removed from the graph.
Naturally, the existence of alternative paths to $j-i-k$ implies the existence of loops in the graph.
Vragovi\^{c} and Louis defined the {\it loop coefficient} of $i$ as the average of $1/d_{jk/i}$ over all pairs of neighbors of $i$, somewhat reminding of 
the concept of information centrality used in the method by Fortunato et al.~\cite{fortunato04}. High values of the loop coefficient
are likely to identify core vertices of communities, whereas low values correspond to vertices lying at the boundary between communities.
Clusters are built around the vertices with highest values of the loop coefficient. The method has time complexity 
$O(nm)$; its results are not so accurate, as compared to popular clustering techniques.

\section{Modularity-based methods}
\label{sec6_0}

Newman-Girvan modularity $Q$ (Section~\ref{sec3_2_2}), originally introduced to define a stopping criterion for the 
algorithm of Girvan and Newman, has rapidly become an essential element of many clustering methods. Modularity is by far 
the most used and best known quality function. It represented one of the first attempts to achieve a first principle
understanding of the clustering problem, and it embeds in its compact form all essential ingredients and questions, from the 
definition of community, to the choice of a null model, to the expression of the ``strength'' of communities and partitions.  
In this section we shall focus on all clustering techniques that require modularity, directly and/or indirectly. We will examine 
fast techniques that can be used on large graphs, but which do not find 
good optima for the measure~\cite{newman04c,clauset04,danon06,pujol06,du07,wakita07,blondel08,schuetz08,schuetz08b,noack09b,xiang09,mei09}; 
more accurate methods, which are computationally demanding~\cite{guimera04,massen05,medus05}; 
algorithms giving a good tradeoff between high accuracy and low complexity~\cite{duch05,white05,newman06,lehmann07,ruan07}.
We shall also
point out other properties of modularity, discuss some extensions/modifications of it, as well as highlight its limits.

\subsection{Modularity optimization}
\label{sub_sec6_0}

By assumption, high values of modularity indicate good partitions\footnote{This is not true in general, as we shall discuss in
Section~\ref{sub_sec6_1}.}. So,
the partition corresponding to its maximum value 
on a given graph should be the best, or at least a very good one. 
This is the main motivation for modularity maximization,
by far the most popular class of methods to detect communities in graphs.
An exhaustive optimization of $Q$ is 
impossible, due to the huge number of ways in which it is possible to partition a graph, even when the latter is small. 
Besides, the true maximum is out of reach, as it has been recently proved that modularity optimization is 
an NP-complete problem~\cite{brandes08}, so it is probably impossible to find the solution in a time 
growing polynomially with the size of the graph.
However, there are currently several algorithms able to find fairly good approximations of 
the modularity maximum in a reasonable time.

\subsubsection{Greedy techniques}
\label{sub_sec6_0_1}

The first algorithm devised to maximize modularity 
was a greedy method of Newman~\cite{newman04c}.
It is an agglomerative hierarchical clustering method, where groups of vertices are successively joined to form larger communities such that modularity 
increases after the merging. One starts from $n$ clusters, each containing a single vertex. Edges are not initially 
present, they are added one by one during the procedure. However, the modularity of partitions explored during the procedure is always 
calculated from the full topology of the graph, as we want to find the modularity maximum on the space of partitions of the full graph. 
Adding a first edge to the set of disconnected 
vertices reduces the number of groups from $n$ to
$n-1$, so it delivers a new partition of the graph. The edge is chosen such that this partition gives the maximum 
increase (minimum decrease) of modularity with respect to the previous configuration. 
All other edges are added based on the same principle. If the insertion of an edge does not change the
partition, i.~e. the edge is internal to one of the clusters previously formed, modularity stays the same.
The number of partitions
found during the procedure is $n$, each with a different number of clusters, from $n$ to $1$. 
The largest value of modularity in this subset of partitions is the approximation of the modularity 
maximum given by the algorithm. At each iteration step, one needs to compute the variation $\Delta Q$ of modularity
given by the merger of any two communities of the running partition, so that one can choose the best merger. 
However, merging communities between which there are no edges
can never lead to an increase of $Q$, so one has to check only the pairs of communities which are connected by edges, of which there cannot
be more than $m$. Since the calculation of each $\Delta Q$ can be done in constant time, this part of the calculation requires a time $O(m)$.
After deciding which communities are to be merged, one needs to update the matrix $e_{ij}$ expressing the fraction of edges
between clusters $i$ and $j$ of the running partition (necessary to compute $Q$), 
which can be done in a worst-case time $O(n)$. Since the algorithm requires $n-1$ iterations (community mergers) to run to completion,
its complexity is $O((m+n)n)$, or $O(n^2)$ on a sparse graph, so it enables one to perform a clustering analysis on much larger networks than 
the algorithm of Girvan and Newman (up to an order of $100000$ vertices with current computers).  
In a later paper~\cite{clauset04}, Clauset et al. pointed out that the update of the matrix $e_{ij}$ in
Newman's algorithm involves a large number of useless operations, due to the sparsity of the adjacency matrix.
This operation can be performed more efficiently by using data structures for sparse matrices, like
{\it max-heaps}, which rearrange the data in the form of binary trees.
Clauset et al. maintained the matrix of modularity variations
$\Delta Q_{ij}$, which is also sparse, a max-heap containing the largest elements of each row of the matrix $\Delta Q_{ij}$ as well as the labels
of the corresponding communities,
and a simple array whose elements are the sums of the elements of each row of the old matrix $e_{ij}$. 
The optimization of modularity can be carried out using these three data structures, whose update is much quicker
than in Newman's technique.
The complexity of the algorithm is $O(md\log n)$, where $d$ is the depth of the dendrogram describing 
the successive partitions found during the execution of the algorithm, which 
grows as $\log n$ for graphs with a strong hierarchical structure. For those graphs, the running time of the
method is then $O(n\log^2n)$, which allows to analyse the community structure
of very large graphs, up to $10^6$ vertices. The greedy optimization of Clauset et al. is currently one of the few algorithms that can be used
to estimate the modularity maximum on such large graphs. The code can be freely downloaded 
from {\tt http://cs.unm.edu/$\sim$aaron/research/fastmodulari}\\{\tt ty.htm}.

This greedy optimization of modularity tends to form quickly large communities at the expenses of small ones, 
which often yields poor values of the modularity maxima. Danon et al. suggested to normalize the modularity variation
$\Delta Q$ produced by the merger of two communities by the fraction of edges incident to one of the two communities, 
in order to favor small clusters~\cite{danon06}.
This trick leads to better modularity optima as compared to the original recipe of Newman, especially when communities are very
different in size. 
Wakita and Tsurumi~\cite{wakita07} have noticed that, due to the bias towards large communities, the fast algorithm by Clauset et al.
is inefficient, because it yields very unbalanced dendrograms, for which the  
relation $d\sim \log n$ does not hold, and as a consequence the method often runs at its worst-case complexity.
To improve the situation they proposed a modification
in which, at each step, one seeks the community merger delivering the largest value of the product of
the modularity variation $\Delta Q$ times a factor ({\it consolidation ratio}), that peaks for communities of equal size. In this way 
there is a tradeoff between the gain in modularity and the balance of the communities to merge,
with a big gain in the speed of the procedure, that enables 
the analysis of systems with up to $10^7$ vertices. Interestingly, this modification often leads to better modularity maxima than those
found with the version of Clauset et al., at least on large social networks. The code can be found at 
{\tt http://www.is.titech.ac.jp/$\sim$wakita/en/software}\\{\tt/community-analysis-software/}.
Another trick to avoid the formation of large communities was proposed 
by Schuetz and Caflisch and consists in allowing for the merger of more community pairs, instead of one, at each iteration~\cite{schuetz08,schuetz08b}. 
This generates several ``centers'' around which communities are formed, which grow simultaneously so that
a condensation into a few large clusters is unlikely. This modified version of the greedy algorithm is combined with a simple
refinement procedure in which single vertices are moved to the neighboring community that yields the maximum increase of modularity. The method 
has the same complexity
of the fast optimization by Clauset et al., but comes closer to the modularity maximum. The software is available at 
{\tt http://www.biochem-caflisch.uzh.ch/public/5/net}\\{\tt work-clusterization-algorithm.html}. The accuracy of the 
greedy optimization can be significantly improved if the hierarchical agglomeration is started from some 
reasonable intermediate configuration, rather than from the individual vertices~\cite{pujol06,du07}. Xiang et al.
suggested to start from a configuration obtained by merging the original isolated vertices into larger subgraphs, 
according to the values of a measure of topological similarity between subgraphs~\cite{xiang09}. A similar approach has been 
described by Ye et al.~\cite{ye08}:
here the initial partition is such that no single vertex can be moved from its cluster to another without decreasing $Q$.  
Higher-quality modularities
can be also achieved by applying refinement strategies based on local search at various steps of the greedy agglomeration~\cite{noack09b}. 
Such refinement procedures are similar to the technique proposed by Newman to improve the results of 
his spectral optimization of modularity (\cite{newman06} and Section~\ref{sub_sec6_0_4}).
Another good strategy consists in alternating greedy optimization with stochastic perturbations of the 
partitions~\cite{mei09}.

\begin{figure*}
\begin{center}
\includegraphics[width=\textwidth]{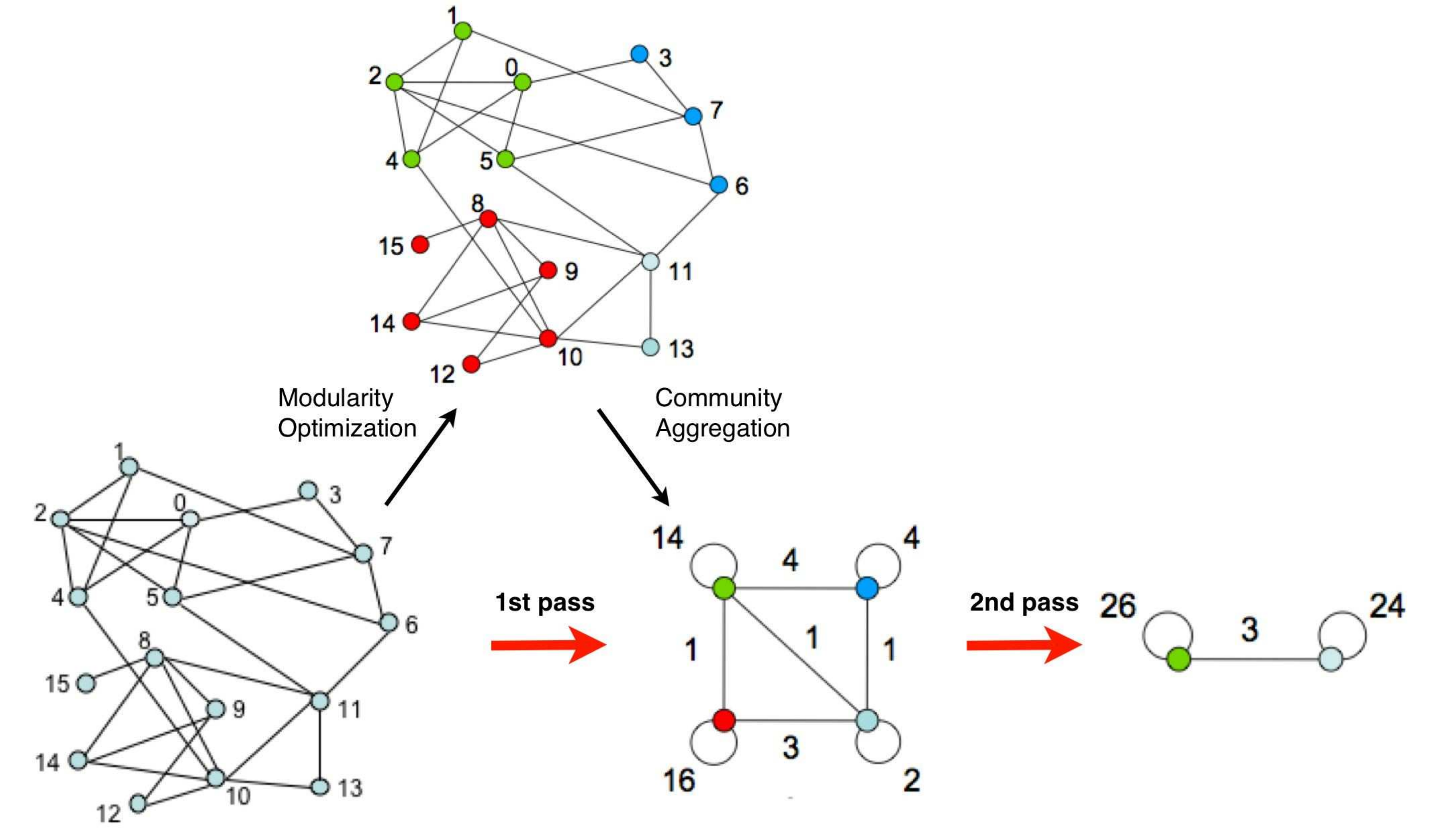}
\caption {\label{blondelmethod} Hierarchical optimization of modularity by Blondel et al.~\cite{blondel08}. The diagram shows two
iterations of the method, starting from the graph on the left. Each iteration
consists of a step, in which every vertex is assigned to the (local) cluster that produces the
largest modularity increase, followed by a successive transformation of the clusters into vertices of a smaller (weighted) graph, representing the
next higher hierarchical level. Reprinted figure with permission from Ref.~\cite{blondel08}.
\copyright 2008 by IOP Publishing and SISSA.}
\end{center}
\end{figure*}

A different greedy approach has been introduced by Blondel et al.~\cite{blondel08}, for the general case of weighted graphs.
Initially, all vertices of the graph are put in different communities.
The first step consists of a sequential sweep over all vertices. Given a vertex $i$, one computes the gain in
weighted modularity (Eq.~\ref{eq:mod2}) coming from putting $i$ in the community of its neighbor $j$ 
and picks the community of the neighbor that yields the largest increase of $Q$,
as long as it is positive. At the end of the sweep, one obtains the first level partition. In the second step communities are replaced
by supervertices, and two supervertices are connected if there is at least an edge between vertices of the corresponding communities. In this case,
the weight of the edge between the supervertices is the sum of the weights of the edges between the represented communities
at the lower level. The two steps of the algorithm are then repeated, yielding new hierarchical 
levels and supergraphs (Fig.~\ref{blondelmethod}). 
We remark that 
modularity is always computed from the initial graph topology: operating on supergraphs enables one to consider
the variations of modularity for partitions of the original graph 
after merging and/or splitting of groups of vertices. Therefore, at some iteration, modularity
cannot increase anymore, and the algorithm stops.
The technique is more limited by storage demands than by computational time. The latter grows like $O(m)$, so the algorithm is extremely fast and 
graphs with up to $10^9$ edges can be analyzed in a reasonable time on current computational resources. 
The software can be found at {\tt http://findcommunities.googlepages.com/}.
The modularity maxima found by the method are better than those found 
with the greedy techniques by 
Clauset et al.~\cite{clauset04} and Wakita and Tsurumi~\cite{wakita07}. 
However, closing communities 
within the immediate neighborhood of vertices may be inaccurate and yield spurious partitions in practical cases. So, it is not clear whether
some of the intermediate partitions could correspond to meaningful hierarchical levels of the graph.
Moreover, the results of the algorithm depend on the order of the sequential sweep over the vertices. 

We conclude by stressing that, despite the improvements and refinements of the last years, the accuracy of greedy optimization 
is not that good, as compared with other techniques.

\subsubsection{Simulated annealing}
\label{sub_sec6_0_2}

Simulated annealing~\cite{kirkpatrick83} is a
probabilistic procedure for global optimization used in different fields and problems.
It consists in performing an exploration of the space of possible states, looking for the
global optimum of a function $F$, say its maximum. Transitions from one state to another occur with probability
$1$ if $F$ increases after the change, otherwise
with a probability $\exp(\beta\Delta F)$, where $\Delta F$ is the decrease of the function and 
$\beta$ is an index of stochastic noise, a sort of inverse temperature, which increases after
each iteration. 
The noise reduces the risk that the system gets trapped in local optima. At some stage, 
the system converges to a stable state, which can be an arbitrarily good approximation of the
maximum of $F$, depending on how many states were explored and how slowly $\beta$ is varied.
Simulated annealing was first employed for modularity optimization by Guimer\`a et al.~\cite{guimera04}.
Its standard implementation~\cite{guimera05} combines two types of ``moves'': local moves, where a single 
vertex is shifted from one cluster to another, taken at random; global moves, consisting 
of mergers and splits of communities. Splits can be carried out in several distinct ways. 
The best performance is achieved if one optimizes the modularity of a bipartition of the cluster, taken as
an isolated graph. This is done again with simulated annealing, where one considers only 
individual vertex movements, and the temperature is decreased until it reaches the running value for the global optimization. 
Global moves reduce the risk of getting trapped in local minima and they have proven
to lead to much better optima than using simply local moves~\cite{massen05,medus05}.
In practical applications, one typically combines $n^2$ local moves with 
$n$ global ones in one iteration. 
The method can potentially come very close to the 
true modularity maximum, but it is slow. The actual complexity cannot be estimated, as it heavily depends on the
parameters chosen for the optimization (initial temperature, cooling factor), not only on the graph size. 
Simulated annealing can be used for small graphs, with up to about $10^4$ vertices. 

\subsubsection{Extremal optimization}
\label{sub_sec6_0_3} 

Extremal optimization (EO) is a heuristic search procedure proposed
by Boettcher and Percus~\cite{boettcher01}, in order to achieve an accuracy comparable with simulated
annealing, but with a substantial gain in computer time. It is based on the optimization of 
local variables, expressing the contribution of each unit of the system to the global function at study.
This technique was used for modularity optimization by Duch and Arenas~\cite{duch05}.
Modularity can be indeed written as a sum over the vertices: the local modularity of a vertex
is the value of the corresponding term in this sum. A fitness measure for each vertex is obtained
by dividing the local modularity of the vertex by its degree, as in this case the measure does not depend
on the degree of the vertex and is suitably normalized. One starts from a random partition of the graph 
in two groups with the same number of vertices. At each iteration, the vertex with the lowest fitness is shifted to the other cluster.
The move changes the partition, so the local fitnesses of many vertices need to be recalculated. The process continues 
until the global modularity $Q$ cannot be improved any more by the procedure. 
This technique reminds one of the Kernighan-Lin~\cite{kernighan70} algorithm for graph partitioning (Section ~\ref{sec4_1}), 
but here the sizes of the communities are determined
by the process itself, whereas in graph partitioning they are fixed from the beginning.
After the bipartition, each cluster is 
considered as a graph on its own and the procedure is repeated, as long as $Q$ increases for the 
partitions found. The procedure, as described, proceeds deterministically from the given initial partition, as one 
shifts systematically the vertex with lowest fitness, and is likely to get trapped in local optima.
Better results can be obtained if one introduces a probabilistic selection, 
in which vertices are ranked based on their fitness values and one picks the vertex of rank $q$ with the probability
$P(q)\sim q^{-\tau}$ ($\tau$-EO, \cite{boettcher01}). 
The algorithm finds very good estimates of the modularity maximum, and performs very well
on the benchmark of Girvan and Newman~\cite{girvan02} (Section~\ref{sec6_1}) . 
Ranking the fitness values has a cost $O(n\log n)$, which can be reduced
to $O(n)$ if heap data structures are used. Choosing the vertex to be shifted can be done with a binary search, which amounts to 
an additional factor $O(\log n)$. Finally, the number of steps needed to verify whether the running modularity maximum can be improved or not
is also $O(n)$. The total complexity of the method is then
$O(n^2\log n)$. We conclude that EO represents a good tradeoff between accuracy and speed, although 
the use of recursive bisectioning may lead to poor results on large networks with many communities.  

\subsubsection{Spectral optimization}
\label{sub_sec6_0_4} 

\begin{figure}
\begin{center}
\includegraphics[width=\columnwidth]{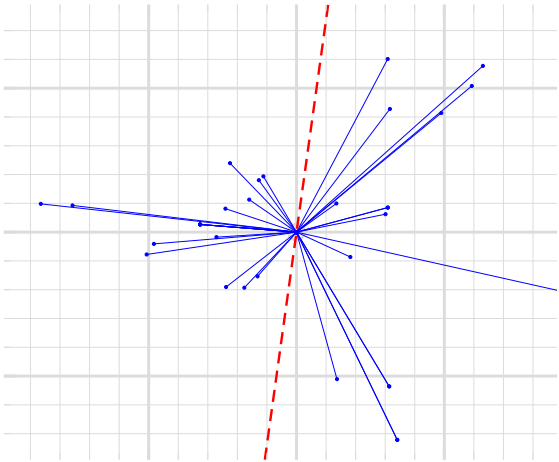}
\caption{\label{Qspectra_plane} Spectral optimization of modularity by Newman~\cite{newman06,newman06b}. By using the first two eigenvectors of 
the modularity matrix, vertices can be represented as points on a plane. By cutting the plane with a line passing through the origin (like the dashed line
in the figure) one obtains bipartitions of the graph with possibly high modularity values. Reprinted figure with permission from 
Ref.~\cite{newman06b}. \copyright 2006 by the American Physical Society.}
\end{center}
\end{figure}

Modularity can be
optimized using the eigenvalues and eigenvectors of a special matrix, the modularity matrix ${\bf B}$, whose elements are
\begin{equation}
B_{ij}=A_{ij}-\frac{k_ik_j}{2m}.
\label{eqr4}
\end{equation}
Here the notation is the same used in Eq.~\ref{eq:mod}. Let ${\bf s}$ be the vector representing any partition
of the graph in two clusters ${\cal A}$ and ${\cal B}$: $s_i=+1$ if vertex $i$ belongs to ${\cal A}$, $s_i=-1$
if $i$ belongs to ${\cal B}$. Modularity can be written as
\begin{eqnarray}
\nonumber
Q&=&\frac{1}{2m}\sum_{ij}\left(A_{ij}-\frac{k_ik_j}{2m}\right)\delta(C_i,C_j)\\\nonumber
&=&\frac{1}{4m}\sum_{ij}\left(A_{ij}-\frac{k_ik_j}{2m}\right)(s_is_j+1)\\
&=&
\frac{1}{4m}\sum_{ij}B_{ij}s_is_j=\frac{1}{4m}{\bf s}^T{\bf B s}.
\label{eqr5}
\end{eqnarray}
The last expression indicates standard matrix products. The vector ${\bf s}$ can be decomposed 
on the basis of eigenvectors ${\bf u}_i$ ($i=1,...,n$) of the modularity matrix 
${\bf B}$: ${\bf s}=\sum_ia_i{\bf u}_i$, with $a_i={\bf u}_i^T\cdot {\bf s}$. By plugging this expression of ${\bf s}$ into
Eq.~\ref{eqr5} one finally gets
\begin{equation}
Q=\frac{1}{4m}\sum_{i}a_i{\bf u}_i^T{\bf B}\sum_j a_j{\bf u}_j=\frac{1}{4m}\sum_{i=1}^{n}({\bf u}_i^T\cdot{\bf s})^2\beta_i,
\label{eqr6}
\end{equation}
where $\beta_i$ is the eigenvalue of ${\bf B}$ corresponding to the eigenvector ${\bf u}_i$. Eq.~\ref{eqr6} is analogous
to Eq.~\ref{eqr1} for the cut size of the graph partitioning problem. This suggests that one can optimize modularity on 
bipartitions via spectral bisection (Section~\ref{sec4_1}), by replacing the Laplacian matrix with the modularity matrix~\cite{newman06,newman06b}.
Like the Laplacian matrix, ${\bf B}$ has always the trivial eigenvector $(1,1,...,1)$ with eigenvalue zero, because the sum of the elements
of each row/column of the matrix vanishes.
From Eq.~\ref{eqr6} we see that, if ${\bf B}$ has no positive eigenvalues, the maximum coincides with the trivial partition consisting of the 
graph as a single cluster (for which $Q=0$), i.~e. it has no community structure.
Otherwise, one has to look for the eigenvector of $B$ with largest (positive) eigenvalue, ${\bf u}_1$,
and group the vertices according to the signs of the components of ${\bf u}_1$, just like in Section~\ref{sec4_1}. Here, however, one does
not need to specify the sizes of the two groups: the vertices with positive components are all in one group, the others in the other group.
If, for example, the component of ${\bf u}_1$ corresponding to vertex $i$ is positive, but we set $s_i=-1$, the modularity is lower than by setting $s_i=+1$.  
The values of the components of ${\bf u}_1$ are also informative, as they indicate the level of the participation of the vertices to their communities.
In particular, components whose values are close to zero lie at the border between the two clusters and can be well considered as belonging to both of them.
The result obtained from the spectral bipartition can be further 
improved by shifting single vertices from one community to the other, such to have the highest increase (or lowest decrease) of 
the modularity of the resulting graph partition. This refinement technique, similar to the Kernighan-Lin algorithm (Section~\ref{sec4_1}),
can be also applied to improve the results of other optimization techniques (e.g. greedy algorithms, extremal optimization, etc.). 
The procedure is repeated for each of the clusters separately, and the number of communities
increases as long as modularity does. At variance with graph partitioning, 
where one needs to fix the number of clusters and their size beforehand, here there is a clear-cut
stopping criterion, represented by the fact that cluster subdivisions are admitted only if they lead to a modularity increase. 
We stress that modularity needs to be always computed from the full adjacency matrix 
of the original graph\footnote{Richardson et al.~\cite{richardson09} have actually shown that if one instead seeks the optimization of 
modularity for each cluster, taken as an independent graph, the combination of spectral bisectioning and the post-processing technique
may yield better results for the final modularity optima.}.
The drawback of the method is similar as for spectral bisection, i.~e. the algorithm gives the best results 
for bisections, whereas it is less accurate when the number of communities is larger than two. Recently, Sun et al.~\cite{sun09} have 
added a step after each bipartition of a cluster, in that single vertices can be moved from one cluster to another and even form the seeds 
of new clusters. We remark that the procedure is different from the Kernighan-Lin-like refining steps, as here the number of clusters can change.
This variant, which does not increase the complexity of the original spectral optimization, leads to better modularity maxima.
Moreover, one does not need to stick to bisectioning, if
other eigenvectors with positive eigenvalues of the modularity matrix are used. 
Given the first $p$ eigenvectors, one can construct $n$ $p$-dimensional vectors, each corresponding to a vertex, 
just like in spectral partitioning (Section~\ref{sec4_4}). The components of the 
vector of vertex $i$ are proportional to the $p$ entries of the eigenvectors in position $i$. Then one can define {\it community vectors},
by summing the vectors of vertices in the same community. It is possible to show that, if the vectors of two communities 
form an angle larger that $\pi/2$, keeping the communities separate yields larger modularity than if they are merged (Fig.~\ref{Qspectra_plane}). 
In this way, in
a $p$-dimensional space the modularity maximum corresponds to a partition in at most $p+1$ clusters.
Community vectors were used by 
Wang et al. to obtain high-modularity partitions into a number of communities smaller than a given maximum~\cite{wang08}.  
In particular, if one takes the eigenvectors corresponding
to the two largest eigenvalues, one can obtain a split of the graph in three clusters: in a recent work, Richardson et al. presented
a fast technique to obtain graph tripartitions with large modularity along these lines~\cite{richardson09}.  
The eigenvectors with the most negative eigenvalues can also be used to extract useful information, like the presence of
a possible multipartite structure of the graph, as they give the most relevant 
contribution to the modularity minimum. 

The spectral optimization of modularity is quite fast. The leading eigenvector of the modularity matrix can be computed with the power method, by 
repeatedly multiplying ${\bf B}$ by an arbitrary vector (not orthogonal to ${\bf u}_1$). The number of required iterations to reach convergence 
is $O(n)$. Each multiplication seems to require a time $O(n^2)$, as ${\bf B}$ is a complete matrix, but the peculiar form of 
${\bf B}$ allows for a much quicker calculation, taking time $O(m+n)$. So, a graph bipartition requires a time $O[n(m+n)]$, or $O(n^2)$ on a sparse
graph. To find the modularity optimum one needs a number of subsequent bipartitions that equals the depth $d$ of the resulting hierarchical tree.
In the worst-case scenario, $d=O(n)$, but in practical cases the procedure usually stops much before reaching the leaves of the dendrogram, so
one could go for the average value $\langle d\rangle \sim \log n$, for a total complexity of $O(n^2\log n)$. The algorithm
is faster than extremal optimization and it is also slightly more accurate, especially for large graphs. The modularity
matrix and the corresponding spectral optimization can be trivially extended to weighted graphs.

A different spectral approach had been previously proposed by White and Smyth~\cite{white05}. 
Let ${\bf W}$ indicate the weighted adjancency matrix of a graph ${\cal G}$.
A partition of ${\cal G}$ in $k$ clusters can be
described through an $n\times k$ assignment matrix ${\bf X}$, where $x_{ic}=1$ if vertex $i$ belongs to cluster $c$, otherwise
$x_{ic}=0$. It can be easily shown that, up to a multiplicative constant, 
modularity can be rewritten in terms of the matrix ${\bf X}$ as 
\begin{equation}
Q\propto \mbox{tr}[{\bf X}^T({\bf \cal W}-{\bf \cal D}){\bf X}]= -\mbox{tr}[{\bf X}^T{\bf {L}_Q}{\bf X}],
\label{eqr7}
\end{equation}
where ${\bf {\cal W}}$ is a diagonal matrix with identical elements, equal to the sum of all edge weights, 
and the entries of ${\bf \cal D}$ are ${\bf \cal D}_{ij}=k_ik_j$, where $k_i$ is the 
degree of vertex $i$. The matrix ${\bf L}_Q={\bf \cal D}-{\bf \cal W}$
is called the $Q$-Laplacian.
Finding the assignment matrix ${\bf X}$ 
that maximizes $Q$ is an $NP$-complete problem, but one can get a good approximation by relaxing the constraint that the elements of ${\bf X}$ 
have to be discrete. By doing so $Q$ becomes a sort of continuous functional of ${\bf X}$ and one can determine the extremes of $Q$ by setting its first 
derivative (with respect to ${\bf X}$) to zero. This leads to the eigenvalue problem
\begin{equation}
{\bf L_Q X}={\bf X \Lambda}.
\label{eqr8}
\end{equation}
Here ${\bf \Lambda}$ is a diagonal matrix. Eq.~\ref{eqr8} turns modularity maximization into a 
spectral graph partitioning problem (Section~\ref{sec4_4}), using the $Q$-Laplacian matrix.  
A nice feature of the $Q$-Laplacian is that, for graphs which are not too small, it can be 
approximated (up to constant factors) 
by the transition matrix ${\bf \tilde{\cal W}}$, obtained by normalizing ${\bf W}$ such that the sum of the elements of
each row equals one. Eq.~\ref{eqr8} is at the basis of the algorithms developed by
White and Smyth, which search for partitions with at most $K$ clusters, where $K$ is a predefined input
parameter that may be suggested by preliminary information on the graph cluster structure. 
The first $K-1$ eigenvectors 
of the transition matrix ${\bf \tilde{\cal W}}$ (excluding the trivial eigenvector with all equal components)  
can be computed with a variant 
of the Lanczos method~\cite{demmel00}.  Since the eigenvector components
are not integer, the eigenvectors do not correspond directly to a partition of the graph in clusters. 
However, as usual in spectral graph partitioning,
the components of the eigenvectors can be used as coordinates of the graph vertices in an Euclidean space
and $k$-means clustering is applied to obtain the desired partition. White and Smyth proposed two methods to derive
the clustering after embedding the graph in space. Both methods have a worst-case complexity $O(K^2n+Km)$, which is essentially linear
in the number of vertices of the graph if the latter is sparse and $K\ll n$. However, on large and sparse graphs, $K$ could scale with the 
number of vertices $n$, so the procedure might become quite slow.
In order to speed up calculations without losing
much accuracy in the final estimates of the maximum modularity, Ruan and Zhang~\cite{ruan07} 
have proposed an algorithm, called {\it Kcut}, that applies 
recursively the method by White and Smyth, in a slightly modified form: after a first application to the full graph, in the next iteration the method
is applied to all clusters of the first partition, treated as independent networks, and so on. The procedure goes on
as long as the modularity of the resulting partitions keeps growing. The advantage of {\it Kcut} is that one can play with low values
for the (maximal) number of clusters $\ell$ at each iteration; if 
partitions are balanced, after $a$ levels of recursions, the number of clusters of the partition is approximately $K=\ell^a$. Therefore the   
complexity of {\it Kcut} is $O[(n+m)\log K]$ for a final partition in (at most) $K$ clusters, which is much lower than the complexity of the algorithm
by White and Smyth. Ruan and Zhang tested {\it Kcut}
on artificial graphs generated with the planted $\ell$-partition model (Section~\ref{sec6}), and on real networks including
Zachary's karate club~\cite{zachary77}, the American college football network~\cite{girvan02} and two 
collaboration networks of Jazz musicians~\cite{gleiser03} and physicists~\cite{newman01}: the accuracy of 
{\it Kcut} is comparable to that of the algorithm by White and Smyth, though generally lower.

\subsubsection{Other optimization strategies}
\label{sub_sec6_001}

Agarwal and Kempe have suggested to maximize modularity within the framework of mathematical programming~\cite{agarwal08}.
In fact, modularity optimization can be formulated both as a linear and as a quadratic program. In the first case, 
the variables are defined on the links: $x_{ij}=0$ if $i$ and $j$ are in the same cluster, otherwise $x_{ij}=1$. The modularity
of a partition, up to a multiplicative constant, can then be written as
\begin{equation}
Q\propto \sum_{ij}B_{ij}(1-x_{ij}),
\label{eqr9}
\end{equation}
where ${\bf B}$ is the modularity matrix defined by Newman (see Section ~\ref{sub_sec6_0_4}).
Eq.~\ref{eqr9} is linear in the variables $\{ x\}$, which must obey the constraint $x_{ij}\leq x_{ik}+x_{kj}$, because, 
if $i$ and $j$ are in the same cluster, and so are $i$ and $k$, then $j$ and $k$ must be in that cluster too. Maximizing 
the expression in Eq.~\ref{eqr9} under the above constraint is $NP$-hard, if the variables have to be integer as required. 
However, if one relaxes this condition by using real-valued $\{ x\}$, the problem can be solved in polynomial time~\cite{karloff91}. On the other hand,
the solution does not correspond to an actual partition, as the $x$ variables are fractional. To get clusters out of the $\{ x\}$ one needs a rounding step.
The values of the $x$ variables are used as sort of distances in a metric space (the triangular inequality is satisfied by construction): clusters of 
vertices ``close'' enough to each other (i.~e. whose mutual $x$ variables are close to zero) are formed and removed until each vertex is assigned to a cluster.
The resulting partition is further refined with the same post-processing technique used by Newman for the spectral optimization of modularity, 
i.~e. by a sequence of steps similar to those of the algorithm by Kernighan and Lin
(see Section ~\ref{sub_sec6_0_4}). Quadratic programming can be used to get bisections of graphs with high modularity, that can be iterated 
to get a whole hierarchy of partitions as in Newman's spectral optimization. One starts from one of the identities in Eq.~\ref{eqr5}
\begin{equation}
Q=\frac{1}{4m}\sum_{ij}B_{ij}(1+s_is_j),
\label{eqr10}
\end{equation}
where $s_i=\pm 1$, depending on whether the vertex belongs to the first or the second cluster. 
Since the optimization of the expression in Eq.~\ref{eqr10} is $NP$-complete,
one must relax again the constraint on the variables $s$ being integer. A possibility is to transform each $s$
into an $n$-dimensional vector $\bf s$ and each product in the scalar product between vectors. The vectors are normalized so that
their tips lie on the unit-sphere of the $n$-dimensional space. This vector 
problem is polynomially solvable, but one needs a method to associate a bipartition to the set of $n$ vectors of the solution.
Any $(n-1)$-dimensional hyperplane centered at the origin cuts the space in two halves, separating the vectors in two subsets.
One can then choose multiple random hyperplanes and pick the one which delivers the partition with highest modularity.
As in the linear program, a post-processing technique \'a la Newman (see Section ~\ref{sub_sec6_0_4}) 
is used to improve the results of the procedure.
The two methods proposed by Agarwal and Kempe
are strongly limited by their high computational complexity, due mostly to the large storage demands, 
making graphs with more than $10^4$ vertices intractable. On the other hand, 
the idea of applying mathematical programming to graph clustering is promising. The code of the algorithms can be downloaded from
{\tt http://www-scf.usc.edu/$\sim$gaurava/}. 
In a recent work~\cite{xu07}, Xu et al. have 
optimized modularity using mixed-integer mathematical programming, with both integer and continuous variables, obtaining
very good approximations of the modularity optimum, at the price of high computational costs. Chen et al. have used integer
linear programming to transform the initial graph into an optimal target graph consisting of disjoint cliques, which effectively 
yields a partition~\cite{chen08}.
Berry et al. have formulated the problem
of graph clustering as a {\it facility location problem}~\cite{hillier04}, consisting in the 
minimization of a cost function based on a local variation of modularity~\cite{berry07}.

Lehmann and Hansen~\cite{lehmann07} optimized modularity via 
mean field annealing~\cite{peterson87}, a deterministic alternative to simulated annealing~\cite{kirkpatrick83}. The method uses Gibbs probabilities to compute the 
conditional mean value for the variable of a vertex, which indicates its community membership. By making a mean field approximation on the variables 
of the other vertices in the Gibbs probabilities one derives a self-consistent set of non-linear equations, that can be solved by iteration 
in a time $O[(m+n)n]$. The method yields better modularity maxima than the spectral optimization by Newman (Section~\ref{sub_sec6_0_4}), at least on artificial 
graphs with built-in community structure, similar to the benchmark graphs by Girvan and Newman (Section~\ref{sec6_1}). 

Genetic algorithms~\cite{holland92} have also been used to optimize modularity. In a standard genetic algorithm one has a set of candidate solutions
to a problem, which are numerically encoded as chromosomes, and an objective function to be optimized on the space of solutions. The objective 
function plays the role of biological fitness for the chromosomes. One usually starts 
from a random set of candidate solutions, which are progressively changed through manipulations inspired by 
biological processes regarding real chromosomes, like point mutation (random variations of some parts of the chromosome) and 
crossing over (generating new chromosomes by merging parts of existing chromosomes). Then, the fitness of the new
pool of candidates is computed and the chromosomes with the highest fitness have the greatest chances to survive in the next generation. 
After several iterations only solutions with large fitness survive.
In a work by Tasgin et al.~\cite{tasgin07}, partitions are the chromosomes and modularity is the fitness function. 
With a suitable choice of the algorithm parameters, like the number of chromosomes and the rates of mutation and crossing over,
Tasgin et al. could obtain results of comparative quality as greedy modularity optimization on Zachary's karate club~\cite{zachary77}, 
the college football network~\cite{girvan02} and the
benchmark by Girvan and Newman (Section~\ref{sec6_1}). Genetic algorithms were also adopted by Liu et al.~\cite{liu07}. Here  
the maximum modularity partition is obtained via successive bipartitions of the graph, where each bipartition is determined by   
applying a genetic algorithm to each subgraph (starting from the original graph itself), 
which is considered isolated from the rest of the graph. A bipartition is accepted only if
it increases the total modularity of the graph. 

In Section~\ref{sec3_2_2} we have seen that the modularity maximum is obtained for the partition that minimizes the difference
between the cut size and the expected cut size of the partition (Eq.~\ref{eqr12}). In the complete weighted graph ${\cal G}_w$ such that
the weight $w_{ij}$ of an edge is $1-k_ik_j/2m$, if $i$ and $j$ are connected in ${\cal G}$, and $-k_ik_j/2m$ if they are not, the difference
$|\mbox{Cut}_{\cal P}|-\mbox{ExCut}_{\cal P}$ is simply the cut size of partition ${\cal P}$. 
So, maximizing modularity for $\cal G$ is equivalent to the problem of
finding the partition with minimal cut size of the weighted graph ${\cal G}_w$, i.~e. to a graph partitioning problem. 
The problem can then be efficiently solved by using
existing software for graph partitioning~\cite{djidjev07}.

\subsection{Modifications of modularity}
\label{sub_sec6_01}

In the most recent literature on graph clustering several modifications and extensions of modularity can be found. 
They are usually motivated by specific classes of clustering problems and/or graphs that one may want to analyze. 

Modularity can be easily extended to graphs with weighted edges~\cite{newman04d}. One needs to replace the
degrees $k_i$ and $k_j$ in Eq.~\ref{eq:mod} with the strengths $s_i$ and $s_j$ of vertices $i$ and $j$. We remind that 
the strength of a vertex is the
sum of the weights of edges adjacent to the vertex (Section~\ref{sec1_1}). For a proper normalization, the number of 
edges $m$ in Eq.~\ref{eq:mod} has to be replaced by the sum $W$ of the weights of all edges. The product $s_is_j/2W$ is 
now the expected weight of the edge $ij$ in the null model of modularity, which has to be compared with the actual weight 
$W_{ij}$ of that edge in the original graph. This can be understood if we consider the case in which all weights are multiples
of a unit weight, so they can be rewritten as integers. The weight of the connection between two nodes can then be replaced by
as many edges between the nodes as expressed by the number of weight units. For the resulting multigraph we can use the same procedure 
as in the case of unweighted graphs, which leads to the formally identical expression 
\begin{equation}
Q_w=\frac{1}{2W}\sum_{ij}\left(W_{ij}-\frac{s_is_j}{2W}\right)\delta(C_i,C_j),
\label{eq:mod2}
\end{equation}
which can be also written as a sum over the modules
\begin{equation}
Q=\sum_{c=1}^{n_c}\Big[\frac{W_c}{W}-\left(\frac{S_c}{2W}\right)^2\Big],
\label{eq:mod20}
\end{equation}
where $W_c$ is the sum of the weights of the internal edges of module $c$ and $S_c$ is the sum of the strengths of the vertices of $c$.
If edge weights are not mutually commensurable, one can always represent them as integers with good approximation, provided
a sufficiently small weight unit is adopted, so the expressions for weighted modularity of Eqs.~\ref{eq:mod2}, \ref{eq:mod20} are generally valid.
In principle, weights can be assigned to the edges of an undirected graph, by using any measure of similarity/correlation 
between the vertices (like, e. g.,
the measures introduced in Section~\ref{sec3_1_4}). In this way, one could derive the corresponding weighted modularity and use it 
to detect communities, with a potentially better exploitation of the structural 
information of the graph as compared to standard modularity~\cite{feng07, ghosh08}. 

\begin{figure}
\begin{center}
\includegraphics[width=9cm]{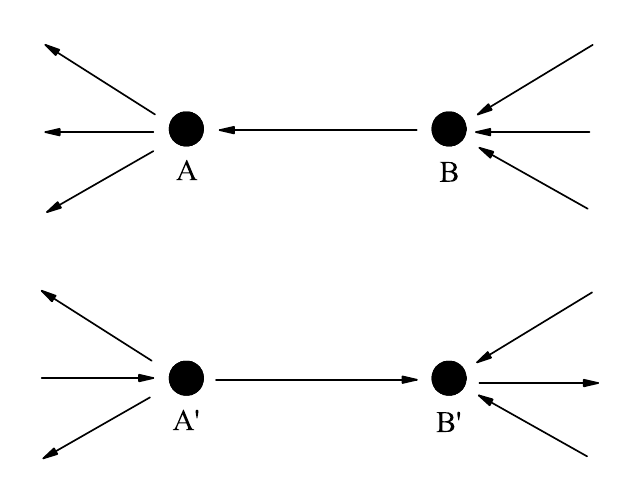}
\caption {\label{directedQ} Problem of the directed modularity introduced by Arenas et al.~\cite{arenas07c}. The two situations illustrated
are equivalent for modularity, as vertices $A$ and $A^\prime$, as well as $B$ and $B^\prime$, have identical
indegrees and outdegrees. In this way, the optimization
of directed modularity is not able to distinguish a situation in which there is directed flow (top) or not (bottom). Reprinted figure with permission from 
Ref.~\cite{kim09}.}
\end{center}
\end{figure}

Modularity has also a straightforward extension to the case of directed graphs~\cite{arenas07c,leicht08}. If an edge is directed,
the probability that it will be oriented in either of the two possible directions depends on the in- and out-degrees of the 
end-vertices. For instance, taken two vertices $A$ and $B$, where $A$ ($B$) has a high (low) 
indegree and low (high) outdegree, in the null model of modularity an edge will be much more likely to point from $B$ to $A$ than from $A$ to $B$.
Therefore, the expression of modularity for directed graphs reads
\begin{equation}
Q_d=\frac{1}{m}\sum_{ij}\left(A_{ij}-\frac{k_i^{out}k_j^{in}}{m}\right)\delta(C_i,C_j),
\label{eq:mod3}
\end{equation}
where the factor $2$ in the denominator of the second summand has been dropped because the sum of the 
indegrees (outdegrees) equals $m$, whereas the sum of the degrees of the vertices of an undirected graph equals $2m$; the factor
$2$ in the denominator of the prefactor has been dropped because the number of non-vanishing elements of the adjacency matrix is $m$,
not $2m$ as in the symmetric case of an undirected graph. If a graph is both directed and weighted, formulas 
\ref{eq:mod2} and \ref{eq:mod3} can be combined as
\begin{equation}
Q_{gen}=\frac{1}{W}\sum_{ij}\left(W_{ij}-\frac{s_i^{out}s_j^{in}}{W}\right)\delta(C_i,C_j),
\label{eq:mod4}
\end{equation}
which is the most general (available) expression of modularity~\cite{arenas07c}. Kim et al.~\cite{kim09} have remarked that the 
directed modularity of Eq.~\ref{eq:mod3} may not properly account for the directedness of the edges (Fig.~\ref{directedQ}), and
proposed a different definition based on diffusion on directed graphs, inspired by Google's PageRank algorithm~\cite{brin98}. 
Rosvall and Bergstrom raised similar objections~\cite{rosvall08}.

If vertices may belong to more clusters, it is not obvious how to find a proper generalization of modularity.
In fact, there is no unique recipe.
Shen et al.~\cite{shen09}, for instance, suggested the simple definition
\begin{equation}
Q=\frac{1}{2m}\sum_{ij}\frac{1}{O_iO_j}\left(A_{ij}-\frac{k_ik_j}{2m}\right)\delta(C_i,C_j).
\label{eq:mod04}
\end{equation}
Here $O_i$ is the number of communities including vertex $i$. The contribution of each edge to modularity is then
the smaller, the larger the number of communities including its endvertices. 
Nicosia et al.~\cite{nicosia09} have made some more general considerations on the problem of extending modularity to 
the case of overlapping communities. They considered the case of directed unweighted networks,
starting from the following general expression
\begin{equation}
Q_{ov}=\frac{1}{m}\sum_{c=1}^{n_c}\sum_{i,j}\Big[r_{ijc}A_{ij}-s_{ijc}\frac{k_i^{out}k_j^{in}}{m}\Big],
\label{eq:mod8}
\end{equation}
where $k_i^{in}$ and $k_j^{out}$ are the indegree and outdegree of vertices $i$ and $j$, the index $c$ labels the communities and 
$r_{ijc}$, $s_{ijc}$ express the contributions to the sum corresponding to the edge 
$ij$ in the network and in the null model, due to the multiple memberships of $i$ and $j$. If there is no overlap between the communities,
$r_{ijc}=s_{ijc}=\delta_{c_ic_jc}$, where $c_i$ and $c_j$ correspond to the communities of $i$ and $j$. In this case, the edge 
$ij$ contributes to the sum only if $c_i=c_j$, as in the original definition of modularity. For overlapping communities, the coefficients
$r_{ijc}$, $s_{ijc}$ must depend on the membership coefficients $\alpha_{i,c}$, $\alpha_{j,c}$ of vertices $i$ and $j$. One can assume that
$r_{ijc}={\cal F}(\alpha_{i,c},\alpha_{j,c})$, where ${\cal F}$ is some function. The term $s_{ijc}$ is related to the null model of modularity,
and it must be handled with care. In modularity's original null model edges are formed by joining two random stubs, so one needs to define the membership of 
a random stub in the various communities.
If we assume that there is no correlation {\it a priori} between the membership coefficients of any two vertices,
we can assign to a stub originating from a vertex $i$ in community $c$ the average membership corresponding to all edges
which can be formed with $i$. On a directed graph we have to distinguish between outgoing and incoming stubs, so one has
\begin{equation}
\beta_{i\rightarrow,c}^{out}=\frac{\sum_{j}{\cal F}(\alpha_{i,c},\alpha_{j,c})}{n},
\label{eq:mod9}
\end{equation}
\begin{equation}
\beta_{i\leftarrow,c}^{in}=\frac{\sum_{j}{\cal F}(\alpha_{j,c},\alpha_{i,c})}{n},
\label{eq:mod10}
\end{equation}
and one can write the following general expression for modularity
\begin{equation}
Q_{ov}=\frac{1}{m}\sum_{c=1}^{n_c}\sum_{i,j}\Big[r_{ijc}A_{ij}-\frac{\beta_{i\rightarrow,c}^{out}k_i^{out}\beta_{j\leftarrow,c}^{in}k_j^{in}}{m}\Big].
\label{eq:mod11}
\end{equation}
The question now concerns the choice of the function ${\cal F}(\alpha_{i,c},\alpha_{j,c})$. If the formula of Eq.~\ref{eq:mod11} 
is to be an extension of modularity to the case of overlapping communities, it has to satisfy some general properties of classical modularity.
For instance, the modularity value of a cover consisting of the whole network as a single cluster should be zero. It turns out that a large class of functions
yield an expression for modularity that fulfills this requirement. Otherwise, the choice of ${\cal F}$ is rather arbitrary and good choices can 
be only tested {\it a posteriori}, based on the results of the optimization. Membership coefficients are also present in an extension of modularity
to overlapping communities
proposed by Shen et al.~\cite{shen09b}. Here the membership coefficient of vertex $v$ in community $c$ 
is a sum over the edges of $v$ belonging to $c$, where each edge has a weight
proportional to the number of maximal cliques of $c$ containing the edge.

Gaertler et al. have introduced quality measures based on modularity's principle of the comparison between
a variable relative to the original graph and the corresponding variable of a null model~\cite{gaertler07}. They 
remark that modularity is just the difference between the coverage of a partition and the expected coverage
of the partition in the null model. We remind that the coverage of a partition is the ratio between the number 
of edges within clusters and the total number of edges (Section~\ref{sec3_2_2}). Based on this observation, Gaertler et al.
suggest that the comparison between the two terms can be done with other binary operations as well. For instance, one could consider the 
ratio
\begin{equation}
S_{cov}^{\div}=\frac{\sum_{c=1}^{n_c}l_c/m}{\sum_{c=1}^{n_c}(d_c/2m)^2},
\label{eqmodrat}
\end{equation}
where the notation is the same as in Eq.~\ref{eq:mod1}. This can be done as well for any variable other than coverage. By using performance, for instance,
(Section~\ref{sec3_2_2}) one obtains two new quality functions $S_{perf}^{-}$ and $S_{perf}^{\div}$, corresponding to taking the difference 
or the ratio between performance and its null model expectation value, respectively. Gaertler et al. compared the results obtained with  
the four functions $S_{cov}^{-}=Q$, $S_{cov}^{\div}$, $S_{perf}^{-}$ and $S_{perf}^{\div}$, on a class of benchmark graphs with built-in 
cluster structure (Section~\ref{sec6_1}) and social networks. They found that the ``absolute'' variants $S_{cov}^{-}$ and $S_{perf}^{-}$
are better than the ``relative'' variants $S_{cov}^{\div}$ and $S_{perf}^{\div}$ on the artificial benchmarks, whereas $S_{perf}^{\div}$ is better
on social networks\footnote{The comparison was done by computing the values
of significance indices like coverage and performance on the final partitions.}. 
Furthermore $S_{perf}^{-}$ is better than the standard modularity $S_{cov}^{-}$.

Modifications of modularity's null model have been introduced by Massen and Doye~\cite{massen05}
and Muff et al.~\cite{muff05}. Massen and Doye's null model
is still a graph with the same expected degree sequence as the original,
and with edges rewired at random among the vertices, but one imposes the additional constraint that
there can be neither multiple edges between a pair of vertices nor edges joining a vertex with itself (loops or self-edges).
This null model is more realistic, as multiple edges and loops are usually absent in real graphs. The maximization of the corresponding
modified modularity yields partitions with smaller average cluster size than standard modularity. The latter tends to disfavor 
small communities, because the actual densities of edges inside small communities hardly exceed the null model densities, 
which are appreciably enhanced by
the contributions from multiple connections and loops. 
Muff et al. proposed a local version of modularity, in which the expected number of edges within 
a module is not calculated with respect to the full graph, but considering just a portion of it, namely the subgraph 
including the module and its neighbouring modules. Their motivation is the fact that modularity's null 
model implicitly assumes that each vertex could be attached
to any other, whereas in real cases a cluster is usually connected to few other clusters.
On a directed graph, their {\it localized modularity} $LQ$ reads  
\begin{equation}
LQ=\sum_{c=1}^{n_c}\left[\frac{l_c}{L_{c_n}}- \frac{d_c^{in}d_c^{out}}{L_{c_n}^2}\right].
\label{eqr16}
\end{equation}
In Eq.~\ref{eqr16} $l_c$ is the number of edges inside cluster $c$, $d_c^{in}$ ($d_c^{out}$) the total internal (external) degree of 
cluster $c$ and $L_{c_n}$ the total number of edges in the subgraph comprising cluster $c$ and its neighbor clusters. The localized modularity is 
not bounded by $1$, but can take any value. Its maximization delivers more accurate partitions than standard modularity optimization
on a model network describing the social interactions between children in a school (school network) and on the metabolic and protein-protein
interaction networks of {\it E. coli}. 

Reichardt and Bornholdt have shown that it is possible to reformulate the problem of community detection as the problem 
of finding the ground state of a spin glass model~\cite{reichardt06}. Each vertex $i$ is labeled by a Potts 
spin variable $\sigma_i$, which indicates the 
cluster including the vertex. The basic principle of the model is that edges should connect vertices of the same class (i.~e. same spin state), 
whereas vertices of different classes (i.~e. different spin states) should be disconnected (ideally). So,
one has to energetically favor edges between vertices in the same class,
as well as non-edges between vertices in different classes, and penalize edges between
vertices of different classes, along with non-edges between vertices in the same class.
The resulting Hamiltonian of the spin model is
\begin{equation}
{\cal H}(\{\sigma\})=-\sum_{i<j}J_{ij}\delta(\sigma_i,\sigma_j)=-\sum_{i<j}J(A_{ij}-\gamma p_{ij})\delta(\sigma_i,\sigma_j),
\label{eqr13}
\end{equation}
where $J$ is a constant expressing the coupling strength, 
$A_{ij}$ are the elements of the adjacency matrix of the graph, $\gamma>0$ a parameter expressing the relative contribution to the energy from  
existing and missing edges, and $p_{ij}$ is the expected number of links connecting $i$ and $j$ for a null model graph with the same total number of 
edges $m$ of the graph considered. The system is a spin glass~\cite{mezard87}, as the couplings $J_{ij}$ 
between spins are both ferromagnetic (on the edges of the graph,
provided $\gamma p_{ij}<1$) and antiferromagnetic (between disconnected vertices, as $A_{ij}=0$ and $J_{ij}=-J\gamma p_{ij}<0$). The multiplicative costant $J$
is irrelevant for practical purposes, so in the following we set $J=1$.
The range of the spin-spin interaction is 
infinite, as there is a non-zero coupling between any pair of spins. Eq.~\ref{eqr13} bears a strong resemblance with the expression of modularity of 
Eq.~\ref{eq:mod}.
In fact, if $\gamma=1$ and $p_{ij}=k_ik_j/2m$ we recover exactly modularity, up to a factor $-1/m$. In this case, finding the spin configuration for which
the Hamiltonian is minimal is equivalent to maximizing modularity.
Eq.~\ref{eqr13} is much more general than modularity, though, as both the null model
and the parameter $\gamma$ can be arbitrarily chosen. In particular, the value of $\gamma$ determines the importance
of the null model term $p_{ij}$ in the quality function. Eq.~\ref{eqr13} can be rewritten as
\begin{eqnarray}
\nonumber
{\cal H}(\{\sigma\})&=&-\sum_{s}\left[l_{s}-\gamma(l_{s})_{p_{ij}}\right]=-\sum_{s=1}c_{ss}\\
&=&\sum_{s<r}\left[l_{rs}-\gamma(l_{rs})_{p_{ij}}\right]=\sum_{s<r}a_{rs}.
\label{eqr14}
\end{eqnarray}
Here, the sums run over the clusters: $l_{s}$ and $l_{rs}$ indicate the number of edges within cluster $s$ and between clusters $r$ and $s$, respectively;
$(l_{s})_{p_{ij}}$ and $(l_{rs})_{p_{ij}}$ are the corresponding null model expectation values. Eq.~\ref{eqr14} defines the coefficients $c_{ss}$ of {\it cohesion}
and $a_{rs}$ of {\it adhesion}. If a subset of a cluster $s$
has a larger coefficient of adhesion with another cluster $r$ than with its complement in $s$, the energy can be reduced by merging the subset with cluster $r$.
In the particular case in which the coefficient of adhesion of a subset $\cal G^\prime$ of a cluster $s$
with its complement in the cluster exactly matches the coefficient of adhesion of $\cal G^\prime$ with another cluster $r$, the partitions in 
which $\cal G^\prime$ stays within $s$ or is merged with $r$ have the same energy. In this case one can say that clusters $r$ and $s$ are overlapping.
In general, the partition with minimum energy has the following properties: 1) 
every subset of each cluster has a coefficient of adhesion with its complement in the cluster not smaller
than with any other cluster; 2) every cluster has non-negative coefficient of cohesion; 3) the coefficient of adhesion between any two clusters is non-positive.

By tuning the parameter $\gamma$ one can vary the number of clusters in the partition with minimum energy, going from
a single cluster comprising all vertices ($\gamma=0$), to $n$ clusters with a single vertex ($\gamma\rightarrow\infty$). So, $\gamma$ is 
a resolution parameter that allows to explore the cluster structure of a graph at different scales (see Section~\ref{sub_sec6_1}). 
The authors used
single spin heatbath simulated annealing algorithms to find the ground state of the Hamiltonian of Eq.~\ref{eqr13}.

Another generalization of modularity was recently suggested by Arenas et al.~\cite{arenas08}. They remarked that the 
fundamental unit to define modularity is the edge, but that high edge densities inside clusters usually imply the existence of 
long-range topological correlations between vertices, which are revealed by the presence of {\it motifs}~\cite{milo02}, i.~e. connected undirected subgraphs,
like cycles (Section~\ref{sec1_1}). For instance, a high edge density inside
a cluster usually means that there are also several triangles in the cluster, and comparatively few between clusters, a criterion that has inspired on its own
popular graph clustering algorithms~\cite{radicchi04,palla05}.   
Modularity can then be simply generalized by comparing the density of motifs inside clusters with the expected density in modularity's null model
({\it motif modularity}).
As a particular case, the {\it triangle modularity} of a partition ${\cal C}$ reads
\begin{equation}
  Q_{\triangle}({\cal C}) =
    \frac{\displaystyle
      \sum_{ijk} A_{ij}({\cal C}) A_{jk}({\cal C}) A_{ki}({\cal C})
    }{\displaystyle
      \sum_{ijk} A_{ij} A_{jk} A_{ki}
    }
    -
    \frac{\displaystyle
      \sum_{ijk} n_{ij}({\cal C}) n_{jk}({\cal C}) n_{ki}({\cal C})
    }{\displaystyle
      \sum_{ijk} n_{ij} n_{jk} n_{ki}
    }\,
\label{eqr15}
\end{equation}
where $A_{ij}({\cal C})=A_{ij}\delta(C_i,C_j)$ ($C_i$ is the label of the cluster $i$ belongs to), $n_{ij}=k_ik_j$ ($k_i$ is the degree of vertex $i$) and 
$n_{ij}({\cal C})=n_{ij}\delta(C_i,C_j)$. If one chooses as motifs paths with even length, and removes the constraint that all vertices of the motif/path
should stay inside the same cluster, maximizing motif modularity could 
reveal the existence of multipartite structure. For example, if a graph is bipartite, one expects to see many $2$-paths starting 
from one vertex class and returning to it from the other class. Motif modularity can be trivially extended to the case of weighted graphs.

Several graphs representing real systems are built out of correlation data between elements. 
Correlation matrices are very common in the study of complex systems: 
well-known examples are the correlations of price returns, which 
are intensively studied by economists and econophysicists~\cite{mantegna00}.
Correlations may be positive as well as negative, so the corresponding weighted edges indicate both attraction and repulsion
between pairs of vertices.
Usually the correlation values are filtered or otherwise transformed such to eliminate the weakest correlations and anticorrelations
and to maintain strictly positive weights for the edges, yielding graphs that can be treated with standard techniques.  
However, ignoring negative correlations means to give up useful information on the relationships between vertices.
Finding clusters in a graph with both positive and negative weights is called {\it correlation clustering problem}~\cite{bansal04}.
According to intuition, one expects that vertices of the same cluster are linked by positive edges, whereas vertices 
of different clusters are linked by negative edges. The best cluster structure is the partition that
maximizes the sum of the strengths (in absolute value) of positive edges within clusters and negative edges between clusters, or, equivalently,
the partition that minimizes the sum of the strengths  (in absolute value) of positive edges between clusters and negative edges within clusters.
This can be formulated by means of modularity, if one accounts for the contribution of the negative edges. A natural way to proceed is 
to create two copies of the graph at study: in one copy only the weights of the positive edges are kept, 
in the other only the weights of the negative edges (in absolute value). By applying Eq.~\ref{eq:mod2}
to the same partition of both graphs, one derives the contributions $Q^+$ and $Q^-$ to the modularity of that partition for 
the original graph. G\'omez et al.
define the global modularity as a linear combination of $Q^+$ and $Q^-$, that accounts for the relative total strengths of positive
and negative edge weights~\cite{gomez09}. Kaplan and Forrest~\cite{kaplan08} have proposed a similar expression, with two important differences. First,
they have used the total strength of the graph, i.~e. the sum of the absolute values of all weights, 
to normalize $Q^+$ and $Q^-$; G\'omez et al. instead have used the positive and the negative strengths, for $Q^+$ and $Q^-$, respectively, which
seems to be the more natural choice looking at Eq.~\ref{eq:mod2}. Second, Kaplan and Forrest have given equal weight to the contributions
of $Q^+$ and $Q^-$ to their final expression of modularity, which is just the difference $Q^+-Q^-$. In another work, Traag and 
Bruggeman~\cite{traag09} have introduced negative links in the general spin glass formulation of modularity of Reichardt and Bornholdt~\cite{reichardt06}.
Here the relative importance of the contribution of positive and negative edge weights is a free parameter, the tuning of which allows to 
detect communities of various sizes and densities of positive/negative edges.

Some authors have pointed out that the original expression of modularity is not ideal to detect communities in bipartite graphs, which describe 
several real systems, like food webs~\cite{williams00}, scientific~\cite{newman01} and artistic~\cite{gleiser03} collaboration networks, etc.. 
Expressions of modularity for bipartite graphs were suggested by Guimer\`a et al.~\cite{guimera07b} and Barber~\cite{barber07,barber08}. 
Guimer\`a et al. call the two classes of vertices actors and teams, and indicate with $t_i$ the degree of actor $i$ and $m_a$ the degree of team $a$.
The null model graphs are random graphs with the same expected degrees for the vertices, as usual. 
The bipartite modularity ${\cal M}_B({\cal P})$ for a partition ${\cal P}$
(of the actors) has the following expression
\begin{equation}
{\cal M}_B({\cal P})=\sum_{c=1}^{n_c}\left[\frac{\sum_{i\neq j\in c}c_{ij}}{\sum_{a}m_a(m_a-1)}-\frac{\sum_{i\neq j\in c}t_it_j}{\left(\sum_{a}m_a\right)^2}\right].
\label{eqr17}
\end{equation}
Here, $c_{ij}$ is the number of teams in which actors $i$ and $j$ are together and the sum ${\sum_{a}m_a(m_a-1)}$ gives the number of ordered pairs of actors 
in the same team. The second ratio of each summand is the null model term, indicating the expected (normalized) number of teams for pairs of actors in cluster $c$.
The bipartite modularity can also be applied to (unipartite) directed graphs: 
each vertex can be duplicated and assigned to both classes, based on its twofold role of 
source and target for the edges.

Another interesting alternative was introduced by Barber~\cite{barber07,barber08} and is a simple extension of Eq.~\ref{eq:mod}.
Let us suppose that the two vertex classes (red and blue) are made out of $p$ and $q$ vertices, respectively. The degree of a red vertex
$i$ is indicated with $k_i$, that of a blue vertex $j$ with $d_j$. The adjacency matrix ${\bf A}$ of the graph
is in block off-diagonal form, as there are edges only between red and blue vertices. Because of that, Barber assumes that the null model matrix 
${\bf P}$, whose element $P_{ij}$ indicates as usual the expected number of edges between vertices $i$ and $j$ in the null model, also has the block
off-diagonal form
\begin{equation}
{\bf P}=
\begin{bmatrix} {\bf O}_{p\times p}&{\bf\tilde{P}}_{p\times q} \\ {\bf\tilde{P}^T}_{q\times p}&{\bf O}_{q\times q}  
\end{bmatrix},
\label{eqr18}
\end{equation}
where the ${\bf O}$ are square matrices with all zero elements and $\tilde{P}_{ij}=k_id_j/m$, as in the null model 
of standard modularity (though other choices are possible).
The modularity maximum can be computed through the modularity matrix ${\bf B}={\bf A}-{\bf P}$, as we have seen in Section~\ref{sub_sec6_0_4}.
However, spectral optimization of modularity gives excellent results for bipartitions, while its performance worsens
when the number of clusters is unknown, as it is usually the case. Barber has proposed a different optimization technique,
called Bipartite Recursively Induced Modules (BRIM), based on the bipartite nature of the graph. The algorithm is based on the
special expression of modularity for the bipartite case, for which 
once the partition of the red or the blue vertices is known, it is easy to get the partition of the other 
vertex class that yields the maximum modularity. Therefore, one starts from an arbitrary 
partition in $c$ clusters of, say, the blue vertices, and recovers the partition of the 
red vertices, which is in turn used as input to get a better partition of the blue vertices, and so on until modularity converges. BRIM does not 
predict the number of clusters $c$ of the graph, but one can obtain good estimates for it by exploring different values with a simple bisection approach.
Typically, for a given $c$ the algorithm needs a few steps to converge, each step having a complexity $O(m)$. An expression
of the number of convergence steps in terms of $n$ and/or $m$ still needs to be derived.

\subsection{Limits of modularity}
\label{sub_sec6_1}

In this Section we shall discuss some features of modularity, which are crucial to identify the 
domain of its applicability and
ultimately to assess the issue of the reliability of the measure for the problem of graph clustering.

An important question concerns the value of the maximum modularity $Q_{max}$ for a graph. We know that it must be 
non-negative, as there is always at least a partition with zero modularity, consisting in a single cluster with all vertices (Section~\ref{sec3_2_2}).
However,
a large value for the modularity maximum does not necessarily mean that a graph has community structure.
Random graphs are supposed to have no community structure, as the linking probability between vertices is either constant or
a function of the vertex degrees, so there is no bias {\it a priori} towards 
special groups of vertices. Still, random graphs
may have partitions with large modularity values~\cite{guimera04, reichardt06}. This is due to 
fluctuations in the distribution of edges in the graph, which in many graph realizations is not homogeneous even if 
the linking probability is constant, like in Erd\"os-R\'enyi graphs. The fluctuations determine concentrations of links in 
some subsets of the graph, which then appear like communities.
According to the definition of modularity, a graph has community structure with respect to a random graph with equal
size and expected degree sequence. 
Therefore, the modularity maximum
of a graph reveals a significant community structure only if it is appreciably larger than the modularity maximum
of random graphs of the same size and expected degree sequence. The significance of the modularity maximum $Q_{max}$ for a graph can be estimated by calculating
the maximum modularity for many realizations of the null model, obtained from the original graph by randomly rewiring 
its edges. One then computes the average $\langle Q\rangle_{NM}$ and the standard deviation $\sigma_Q^{NM}$ of the results.
The statistical significance of $Q_{max}$ is indicated by the distance of $Q_{max}$ from the null model average $\langle Q\rangle_{NM}$
in units of the standard deviation $\sigma_Q^{NM}$, i.~e. by the $z$-score
\begin{equation}
z=\frac{Q_{max}-\langle Q\rangle_{NM}}{\sigma_Q^{NM}}.
\label{eq:zscore}
\end{equation}
If $z\gg 1$, $Q_{max}$ indicates strong community structure. Cutoff values of $2-3$ for the $z$-scores are customary. This approach has problems, though.
It can generate both false positives and false negatives: a few graphs that most people would consider without a significant community structure have a large
$z$-score; on the other hand, some graphs that are agreed to display cluster structure have very low values for the $z$-score. Besides, 
the distribution of the maximum modularity values of the null model, though peaked, 
is not Gaussian. Therefore, one cannot attribute to the values of the $z$-score the significance  
corresponding to a Gaussian distribution, and one would need instead to compute the statistical significance for the right distribution.

Reichardt and Bornholdt have studied the issue of the modularity values for random graphs in some depth~\cite{reichardt06b,reichardt07}, using their general 
spin glass formulation of the clustering problem (Section~\ref{sub_sec6_01}). 
They considered the general case of a random graph with arbitrary degree distribution $P(k)$ and without degree-degree correlations. 
They set $\gamma=1$, so that the 
energy of the ground state coincides with modularity (up to a constant factor). For modularity's null model graphs,
the modularity maximum corresponds to an equipartition of the graph, i.~e. the magnetization of the ground state of the 
spin glass is zero, a result confirmed by numerical simulations~\cite{reichardt06b,reichardt07}. This is because
the distribution of the couplings has zero mean, and the mean is only coupled to
magnetization~\cite{fu86}. For a partition of any graph with $n$ vertices and $m$ edges 
in $q$ clusters with equal numbers of vertices, 
there is a simple linear relation between the cut size ${\cal C}_q$ of the partition
and its modularity $Q_q$: ${\cal C}_q=m[(q-1)/q-Q_q]$. We remind that the cut size ${\cal C}_q$ is the total number of 
inter-cluster edges of the partition (Section~\ref{sec4_1}). In this way, the partition with maximum modularity is also the one
with minimum cut size, and community detection becomes equivalent to graph partitioning.
Reichardt and Bornholdt
derived analytically the ground state energy for Ising spins ($q=2$), which corresponds to the following
expression of the expected maximum modularity $Q^{max}_2$ for a bipartition~\cite{reichardt07}
\begin{equation}
Q^{max}_2=U_0J\frac{\langle k^{1/2}\rangle}{\langle k\rangle}.
\label{eqr19}
\end{equation}
Here $\langle k^\alpha\rangle=\int P(k)k^\alpha dk$ and $U_0$ is the ground state energy of the Sherrington-Kirkpatrick model~\cite{sherrington75}.
The most interesting feature of Eq.~\ref{eqr19} is the simple scaling with ${\langle k^{1/2}\rangle}/{\langle k\rangle}$. 
Numerical calculations show that this scaling holds for both Erd\"os-R\'enyi and scale-free graphs (Section~\ref{sec1_3}). Interestingly, the result
is valid for partitions in $q$ clusters, where $q$ is left free, not only for $q=2$. The number of 
clusters of the partition with maximum modularity decreases if the average degree $\langle k\rangle$ increases, and
tends to $5$ for large values of $\langle k\rangle$, regardless
of the degree distribution and the size of the graph. From Eq.~\ref{eqr19} we also see that the expected maximum modularity for a random graph
increases when $\langle k\rangle$ decreases, i. e. if the graph gets sparser.
So it is particularly hard to detect communities in sparse graphs
by using modularity optimization. As we shall see in Section~\ref{sec6_4}, the sparsity of a graph is generally a serious obstacle for graph clustering methods,
no matter if one uses modularity or not.

A more fundamental issue, raised by Fortunato and Barth\'elemy~\cite{fortunato07}, concerns the capability of modularity to detect ``good'' partitions.
If a graph has a clear cluster structure, one expects that the maximum modularity of the graph reveals it.
The null model of modularity assumes that any vertex $i$ ``sees'' any other vertex $j$, and the expected number of edges between them is
$p_{ij}=k_ik_j/2m$. Similarly, the expected number of edges between two clusters ${\cal A}$ and ${\cal B}$ with total degrees $K_{\cal A}$ and $K_{\cal B}$, 
respectively, is $P_{\cal AB}=K_{\cal A}K_{\cal B}/2m$. The variation
of modularity determined by the merger of ${\cal A}$ and ${\cal B}$ with respect to the partition in which they are separate clusters
is $\Delta Q_{\cal AB}=l_{\cal AB}/m-K_{\cal A}K_{\cal B}/2m^2$, with $l_{\cal AB}$ number of edges connecting ${\cal A}$ to ${\cal B}$.
If $l_{\cal AB}=1$, i.~e. there is a single edge joining ${\cal A}$ to ${\cal B}$, we expect that  
the two subgraphs will often be kept separated. Instead,
if $K_{\cal A}K_{\cal B}/2m<1$, $\Delta Q_{\cal AB}>0$. Let us suppose for simplicity that $K_{\cal A}\sim K_{\cal B}=K$, i.~e.
that the two subgraphs are of about the same size, measured in terms of edges. 
We conclude that, if $K<\sim\sqrt{2m}$ and the two subgraphs ${\cal A}$ and ${\cal B}$ are connected, 
modularity is greater if they are in the same cluster~\cite{fortunato07}. 
The reason is intuitive: if there are more edges than expected between ${\cal A}$ and ${\cal B}$, there is
a strong topological correlation between the subgraphs. If the subgraphs are sufficiently small (in degree), the expected number of 
edges for the null model can be smaller than one, so even the weakest possible connection (a single edge) suffices to keep the subgraphs together.
Interestingly, this result holds independently of the structure of the subgraphs. In particular it remains true if the subgraphs are cliques,
which are the subgraphs with the largest possible density of internal edges, and represent the strongest possible communities.
\begin{figure}
\begin{center}
\includegraphics[width=\columnwidth]{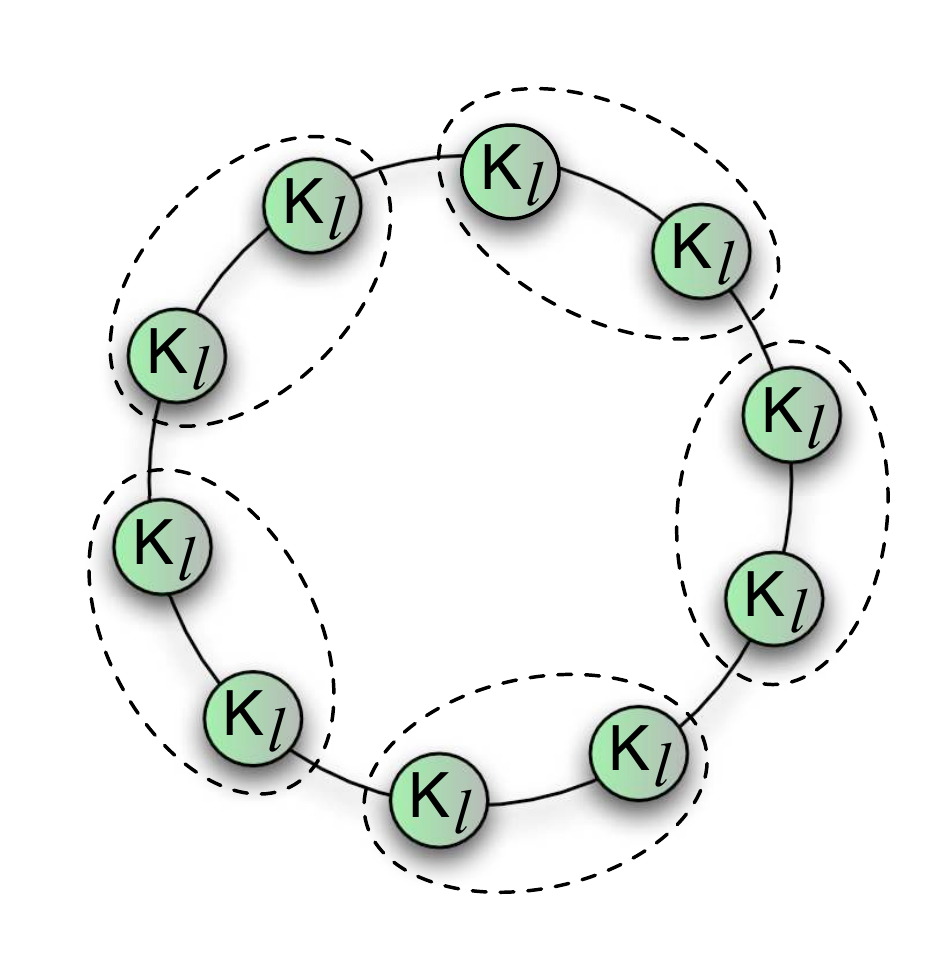}
\caption {\label{Figure7} Resolution limit of modularity optimization. The natural community
structure of the graph, represented by the individual cliques (circles), is not recognized by
optimizing modularity, if the cliques are smaller than a scale depending on the size of the graph. In this case,
the maximum modularity corresponds to a partition whose clusters include two or more cliques (like the groups 
indicated by the dashed contours). 
Reprinted figure with permission from Ref.~\cite{fortunato07}. \copyright 2007 from the National Academy of Science of the USA.}
\end{center}
\end{figure}
In Fig.~\ref{Figure7} a graph is made out of $n_c$ identical cliques,
with $l$ vertices each, connected by single edges.
It is intuitive to think that the clusters of the best partition are the individual cliques: instead,
if $n_c$ is larger than about $l^2$, modularity would be higher for partitions in which 
the clusters are groups of cliques (like the clique pairs indicated by the dashed lines in the figure). 

The conclusion is striking: modularity optimization has a resolution limit that may prevent it from detecting 
clusters which are comparatively small with respect to the graph as a whole, even when they are well defined communities like cliques.
So, if the partition with maximum modularity includes clusters with total degree of the order of $\sqrt{m}$ or smaller,
one cannot know {\it a priori} whether the clusters are single communities or combinations of smaller weakly interconnected communities.
This resolution problem has a large impact in practical applications. Real graphs with community structure usually contain communities
which are very diverse in size~\cite{palla05,guimera03,danon05,clauset04}, so many (small) communities may remain undetected.
Besides, modularity is extremely sensitive to even individual connections. Many real graphs, in biology and in the social
sciences, are reconstructed through experiments and surveys, so edges may occasionally be false positives: if two small subgraphs
happen to be connected by a few false edges, modularity will put them in the same cluster, inferring a relationship between 
entities that in reality may have nothing to do with each other.

The resolution limit comes from the very definition of modularity, in particular from its null model. The weak point of the null model is
the implicit assumption that each vertex can interact with every other vertex, which implies that each part of the graph knows about everything else. 
This is however questionable, and certainly wrong for large systems like, e.g., the Web graph. It is certainly more reasonable to assume
that each vertex has a limited horizon within the graph, and interacts just with a portion of it. However, nobody knows yet
how to define such local territories for the graph vertices. The null model of the localized modularity of Muff et al. (Section~\ref{sub_sec6_01}) 
is a possibility, since it
limits the horizon of a vertex to a local neighborhood, comprising the cluster of the vertex and the clusters linked to it by at least 
one edge (neighboring clusters). However,
there are many other possible choices.
In this respect, the null model of Girvan and Newman, though unrealistic, is the simplest one can think of, which partly explains its success.
Quality functions that, like modularity, are based on a null model such that the horizon of vertices is of the order of the size of the whole graph, 
are likely to be affected by a resolution limit~\cite{fortunato07c}. The problem is more general, though.
For instance, Li et al.~\cite{li08} have introduced a quality function, called {\it modularity density},
which consists in the sum over the clusters of the ratio between the difference of the internal and external degrees of the cluster and the
cluster size. The modularity density does not require a null model, and delivers better results than modularity optimization (e. g. it correctly 
recovers the natural partition of the graph in Fig.~\ref{Figure7} for any number/size of the cliques). However, it is still affected 
by a resolution limit. To avoid that, Li et al. proposed a more general definition of their measure, including a tunable parameter
that allows to explore the graph at different resolutions, in the spirit of the methods of Section~\ref{sub_sec6_20}.

A way to go around the resolution limit problem could be to perform further subdivisions of the clusters obtained 
from modularity optimization, in order to eliminate possible artificial mergers of communities. 
For instance, one could recursively optimize modularity for each single cluster, 
taking the cluster as a separate entity~\cite{fortunato07,ruan08}. However, this is not a reliable procedure, for two reasons:
1) the local modularities used to find partitions within the clusters have different null models, as they depend on the cluster sizes, so they 
are inconsistent with each other; 
2) one needs to define a criterion to decide when one has to stop partitioning a cluster, but there is no obvious prescription, 
so any choice is necessarily based on arbitrary assumptions\footnote{Ruan and Zhang~\cite{ruan08} propose a stopping criterion based on the statistical
significance of the maximum modularity values of the subgraph. The maximum modularity of a subgraph is compared with
the expected maximum modularity for a random graph with the same size and expected degree sequence of the subgraph. If the corresponding $z$-score 
is sufficiently high, the subgraph is supposed to have community structure and one accepts the partition in smaller pieces. The procedure
stops when none of the subgraphs of the running partitions has significant community structure, based on modularity.}.

Resolution limits arise as well in the more general formulation of community detection by Reichardt and Bornholt~\cite{kumpula07}.
Here the limit scale for the undetectable clusters is $\sqrt{\gamma m}$. We remind that $\gamma$ weighs the contribution of the null model term
in the quality function.
For $\gamma=1$ one recovers the resolution limit of modularity.
By tuning the parameter $\gamma$ it is possible to arbitrarily vary the resolution scale of the corresponding quality function.
This in principle solves the problem of the resolution limit, as one could adjust the resolution of the method to
the actual scale of the communities to detect. The problem is that usually one has no information about the community sizes,
so it is not possible to decide {\it a priori} the proper value(s) of $\gamma$ for a specific graph. In the most recent literature
on graph clustering quite a few {\it multiresolution methods} have been introduced, addressing this problem in several ways.
We will discuss them in some detail in Section~\ref{sub_sec6_20}. 

The resolution limit can be easily extended to the case of weighted graphs. In a recent paper~\cite{berry09}, Berry et al.
have considered the special case in which intracluster edges have weight $1$, whereas intercluster edges have weight $\epsilon$. By repeating the
same procedure as in Ref.~\cite{fortunato07}, they conclude that clusters with internal strength (i. e. sum of all weights
of internal edges) $w_s$ may remain undetected if $w_s<\sqrt{W\epsilon/2}-\epsilon$, where $W$ is the total strength of the graph.
So, the resolution limit decreases when $\epsilon$ decreases. Berry et al. use this result to show that,
by properly weighting the edges of a given unweighted graph, it becomes possible 
to detect clusters with very high resolution by still using modularity optimization.   
\begin{figure}
\begin{center}
\includegraphics[width=\columnwidth]{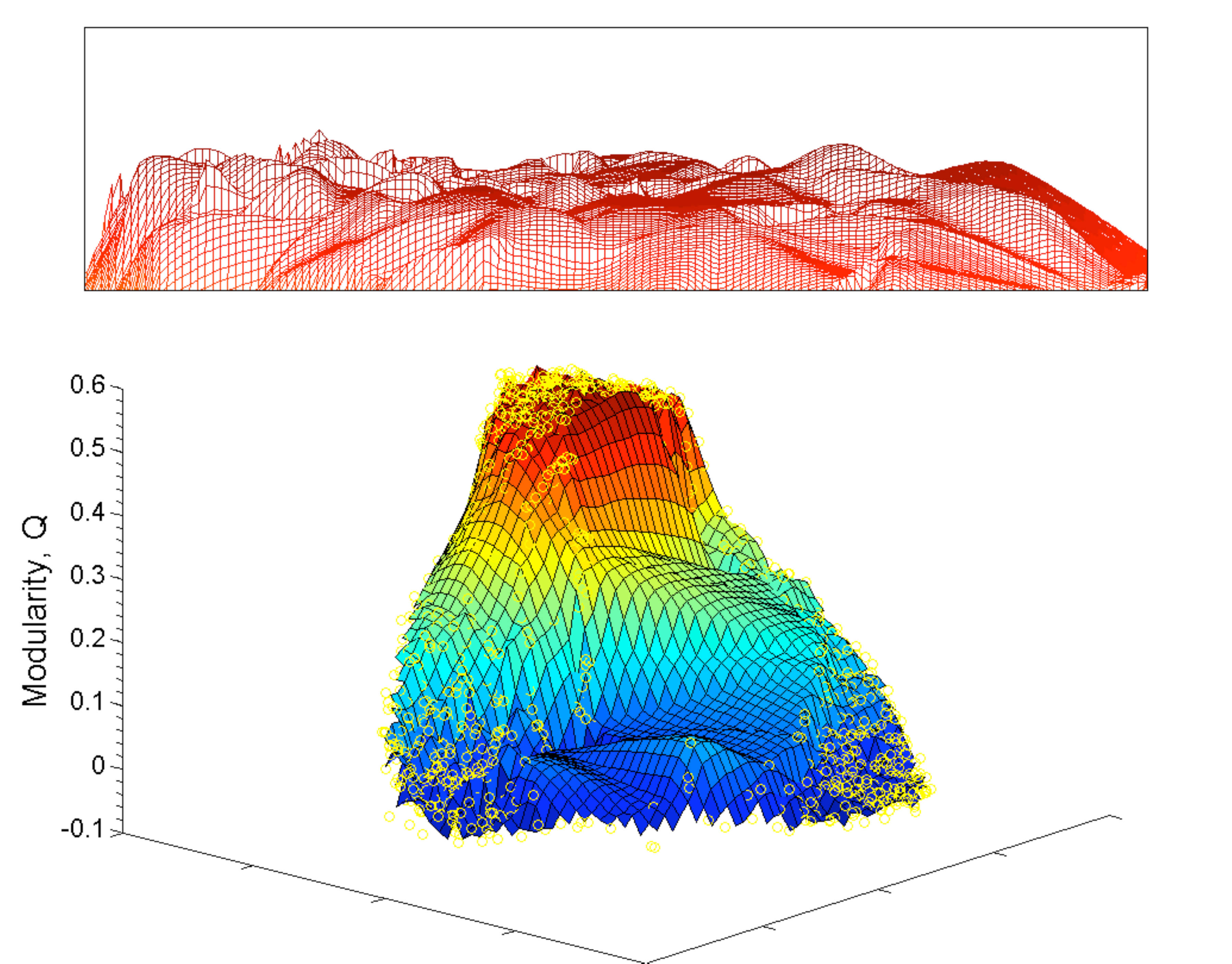}
\caption {\label{figterr} Low-dimensional visualization of the modularity landscape for the metabolic network of the 
spirochete {\it Treponema pallidum}. The big degeneracy
of suboptimal high-modularity partitions is revealed by the plateau (whose shape is detailed in the inset), 
which is large and very irregular. Modularity values in the plateau
are very close to the absolute maximum, although they may correspond to quite different partitions.
Reprinted figure with permission from Ref.~\cite{good09}.}
\end{center}
\end{figure}

Very recently, Good et al.~\cite{good09} have made a careful analysis of modularity and its performance. They discovered
that the modularity landscape is characterized by an exponential number of distinct states/partitions, whose modularity 
values are very close to the global maximum (Fig.~\ref{figterr}). This problem is particularly dramatic if a graph has a hierarchical community
structure, like most real networks. Such enormous number of solutions explains why
many heuristics are able to come very close to modularity's global maximum, but it also implies
that the global maximum is basically impossible to find.
In addition, high-modularity partitions are not necessarily similar to each other,
despite the proximity of their modularity scores. The optimal partition from a topological point of view, which 
usually does not correspond to the modularity maximum due to the resolution limit, may however have a large modularity score.
Therefore the optimal partition is basically indistinguishable from a huge number of 
high-modularity partitions, which are in general structurally dissimilar from it.  The large structural inhomogeneity of the 
high-modularity partitions implies that one cannot rely on any of them, at least in principle, in the absence of additional information 
on the particular system at hand and its structure.

\section{Spectral Algorithms}
\label{sec43}

In Sections~\ref{sec4_1} and \ref{sec4_4} we have learned that spectral properties of graph matrices are frequently used to find 
partitions. A paradigmatic example is spectral graph clustering, which makes use of the eigenvectors of
Laplacian matrices (Section~\ref{sec4_4}). We have also seen that Newman-Girvan modularity
can be optimized by using the eigenvectors of the modularity matrix (Section~\ref{sub_sec6_0_4}).
Most spectral methods have been introduced and developed in computer science and generally focus  
on data clustering, although applications to graphs are often possible as well.  
In this section we shall review recent spectral techniques proposed mostly by physicists explicitly for graph clustering.

Early works have shown that the eigenvectors of the {\it transfer matrix} ${\bf T}$ (Section~\ref{sec1_2}) can be used 
to extract useful information on community structure. The transfer matrix acts as a time propagator for the process
of random walk on a graph. Given the
eigenvector ${\bf c}^{\alpha}$ of the transposed transfer matrix ${\bf T^\dagger}$, corresponding to the eigenvalue $\lambda_{\alpha}$, $c^{\alpha}_i$ is
the outgoing current flowing from vertex $i$, corresponding to the eigenmode $\alpha$. The {\it participation ratio} (PR) 
\begin{equation}
\chi_{\alpha}=\left[\sum_{i=1}^n(c^{\alpha}_i)^4\right]^{-1}
\label{eqm19}
\end{equation}
indicates the effective number of vertices contributing to eigenvector ${\bf c}^{\alpha}$. 
If $\chi_{\alpha}$ receives contributions only from vertices of the same cluster, i.~e. eigenvector ${\bf c}^{\alpha}$ is ``localized'', the value of
$\chi_{\alpha}$ indicates the size of that cluster~\cite{eriksen03,simonsen04}. 
The significance of the cluster can be assessed by comparing $\chi_{\alpha}$ with the corresponding participation ratio for a random graph 
with the same expected degree sequence as the original graph. Eigenvectors of the adjacency matrix may be localized as well
if the graph has a clear community structure~\cite{slanina05}. A recent comprehensive analysis of spectral properties 
of modular graphs has been carried out by Mitrovi\'{c} and Tadi\'{c}~\cite{mitrovic09}.

\begin{figure}
\begin{center}
\includegraphics[width=\columnwidth]{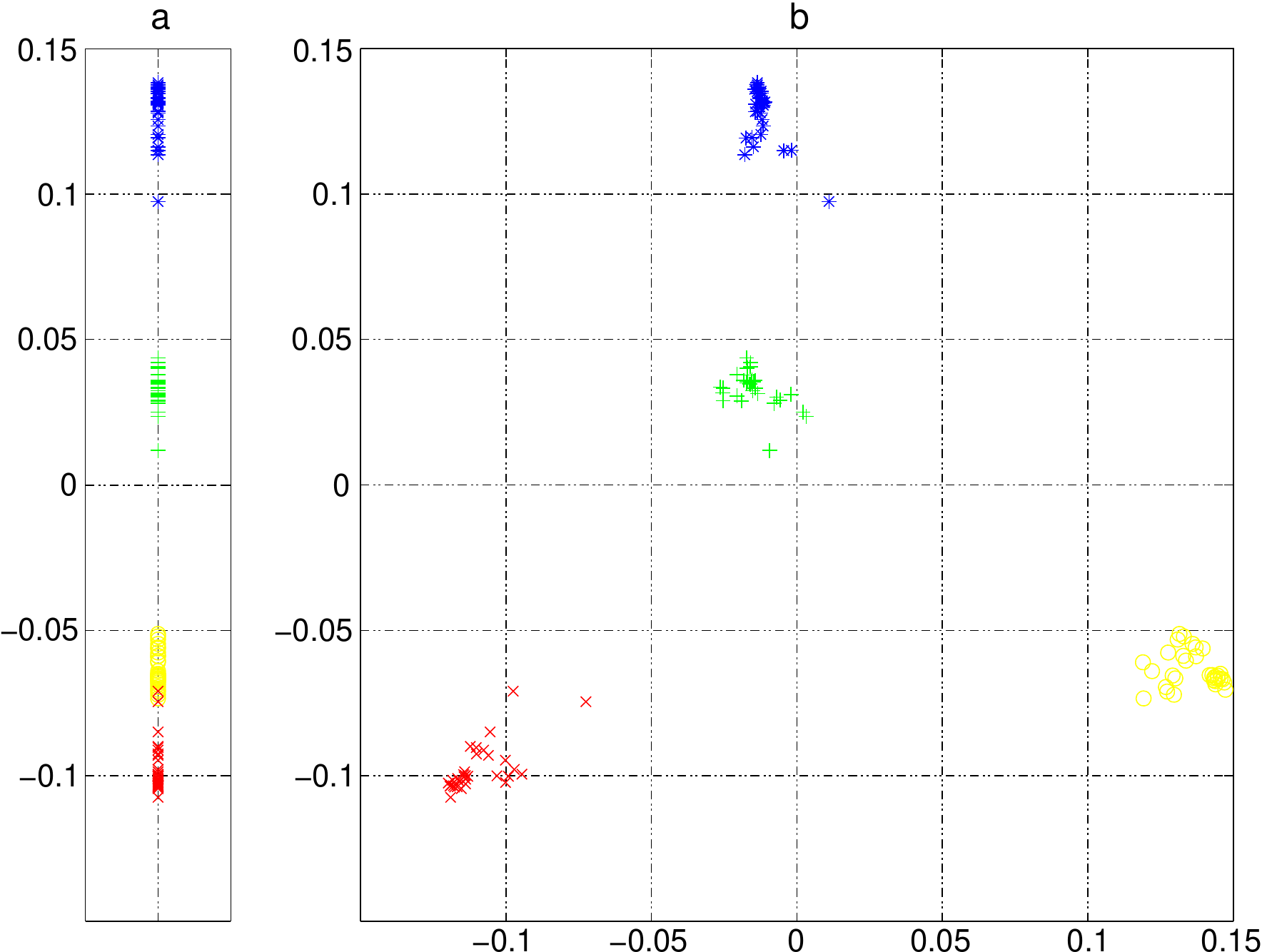}
\caption {\label{Figure8} Spectral algorithm by Donetti and Mu\~noz. Vertex $i$ is represented by the values of
the $i$th components of Laplacian eigenvectors. In this example, the graph has an ad-hoc division in four communities,
indicated by the colours. The communities are better separated in two dimensions (b) than in one (a). 
Reprinted figure with permission from Ref.~\cite{donetti04}.
\copyright 2004 by IOP Publishing and SISSA.}
\end{center}
\end{figure}

Donetti and Mu\~noz have devised an elegant method based on the eigenvectors of 
the Laplacian matrix~\cite{donetti04}. The idea is the same as in spectral graph clustering
(Section~\ref{sec4_4}): since
the values of the eigenvector components are close for vertices in the same community, one can use them 
as coordinates, such that vertices turn into points in a metric space. So, if one uses $M$ eigenvectors,
one can embed the vertices in an $M$-dimensional space. Communities appear as groups of points well separated
from each other, as illustrated in Fig.~\ref{Figure8}. The separation is the more visible, the larger the number
of dimensions/eigenvectors $M$.
The originality of the method consists in the procedure to group the points and to extract the partition. Donetti and Mu\~noz
used hierarchical clustering (see Section~\ref{sec4_2}), with the constraint that only pairs of 
clusters which have at least one interconnecting edge in the original graph are merged.
Among all partitions of the resulting dendrogram, the one with largest modularity is chosen.
For the similarity measure between vertices, Donetti and Mu\~noz used both the Euclidean distance
and the angle distance. The angle distance between two points 
is the angle between the vectors going from the origin of the
$M$-dimensional space to either point. Tests on the benchmark by Girvan and Newman (Section~\ref{sec6_1}) 
show that the best results are obtained with complete-linkage
clustering. The most computationally expensive part of the algorithm is the calculation of the Laplacian eigenvectors.
Since a few eigenvectors suffice to get good partitions, one can determine them with the Lanczos method~\cite{lanczos50}.
The number $M$ of eigenvectors that are needed to have a clean separation of the clusters is 
not known {\it a priori}, but one can compute a number $M_0>1$ of them and search for the highest modularity partition
among those delivered by the method for all $1\leq M\leq M_0$. In a related work, Simonsen has embedded graph vertices in space
by using as coordinates the components of the eigenvectors of the right stochastic matrix~\cite{simonsen05}.

Eigenvalues and eigenvectors of the Laplacian matrix have been used by Alves to compute the effective conductances for 
pairs of vertices in a graph, assuming that the latter is an electric network with edges of unit resistance~\cite{alves07}.
The conductances enable one to compute the transition probabilities for a random walker moving on the graph, 
and from the transition probabilities one builds a similarity matrix between vertex pairs. Hierarchical clustering is applied to join 
vertices in groups. The method can be trivially extended to the case of weighted graphs.
The algorithm by Alves is rather slow, as one needs to compute the whole spectrum of the Laplacian, which 
requires a time $O(n^3)$. Moreover, there is no criterion to select which partition(s) of the dendrogram is (are) the best.

\begin{figure}
\begin{center}
\includegraphics[width=\columnwidth]{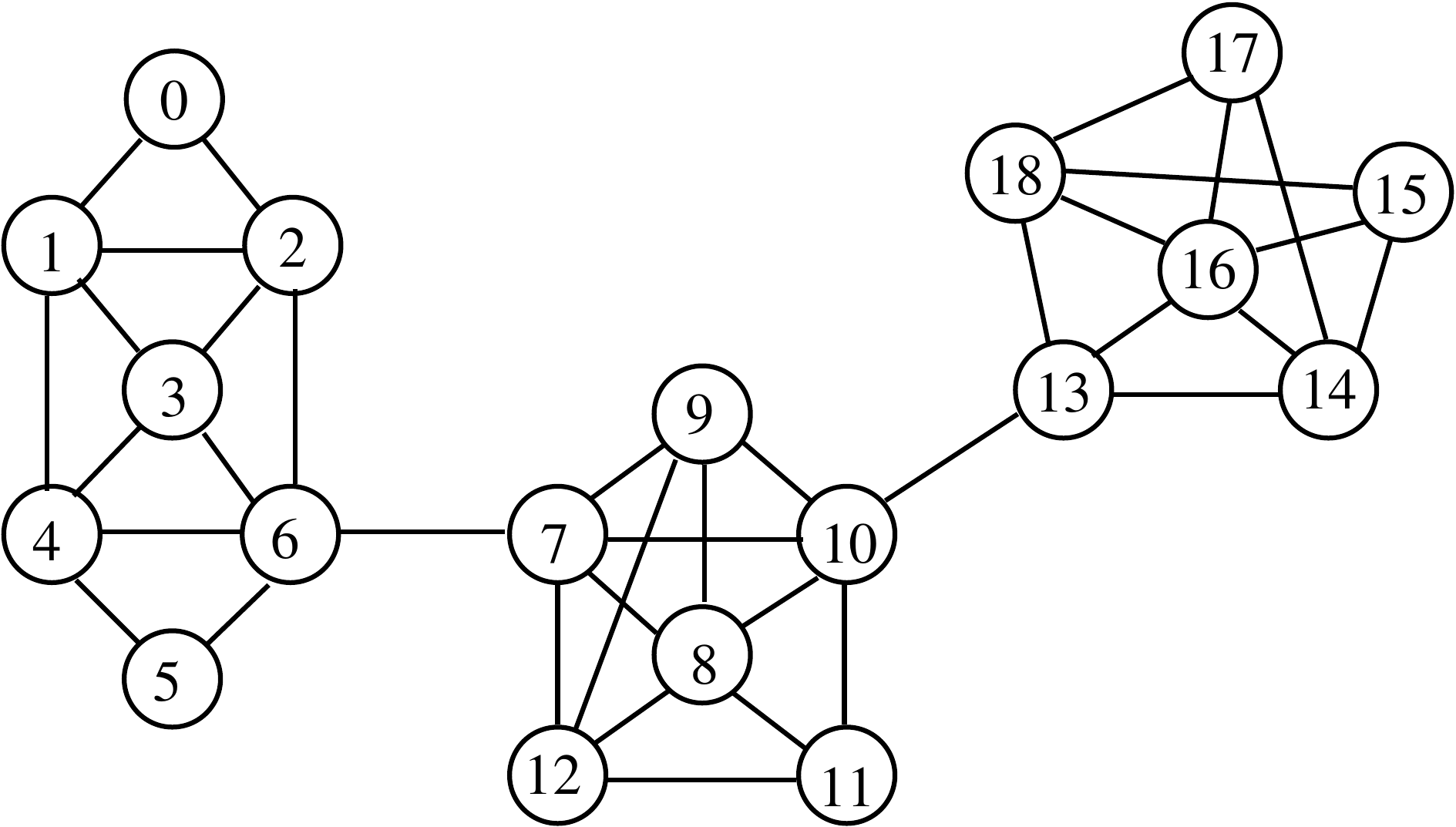}
\vskip0.4cm
\includegraphics[width=\columnwidth]{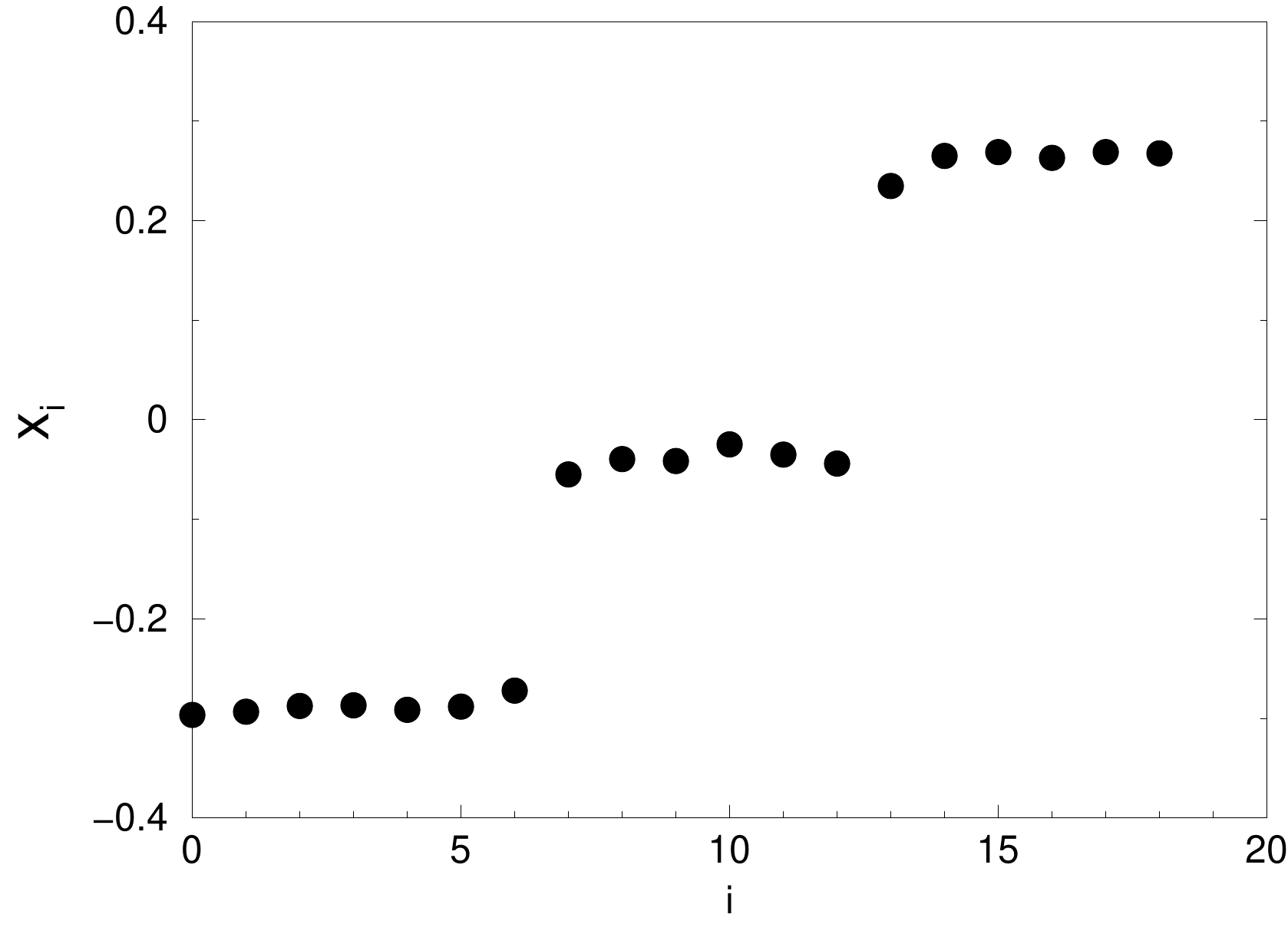}
\caption {\label{capoccifig} Basic principle of the spectral algorithm by Capocci et al.~\cite{capocci05}. The bottom diagram shows the
values of the components of the second eigenvector of the right stochastic matrix for the graph drawn on the top. The three plateaus of the 
eigenvector components correspond to the three evident communities of the graph. Reprinted figures with permission from 
Ref.~\cite{capocci05}. \copyright 2005 by Elsevier.}
\end{center}
\end{figure}

Capocci et al.~\cite{capocci05} used eigenvector components of the
right stochastic matrix $\bf R$ (Section~\ref{sec1_2}), that is derived from the adjacency matrix by dividing each row by the sum of its elements.
The right stochastic matrix has similar properties as the Laplacian. 
If the graph has $g$ connected components, the largest $g$ eigenvalues are equal to $1$, with eigenvectors characterized by having
equal-valued components for vertices belonging to the same component. In this way,
by listing the vertices according to the connected components they belong to, 
the components of any eigenvector of $\bf R$, corresponding to eigenvalue $1$, display a step-wise profile, 
with plateaus indicating vertices in the same connected component. For connected graphs with cluster structure, one can still see plateaus, if  
communities are only loosely connected to each other (Fig.~\ref{capoccifig}). 
Here the communities can be 
immediately deducted by an inspection of the components of any eigenvector with eigenvalue $1$. 
In practical cases, plateaus are not clearly visible, and one eigenvector is not enough. However, one expects that there should be 
a strong correlation between eigenvector components corresponding to vertices in the same cluster. Capocci et al.
derived a similarity matrix, where the similarity between vertices $i$ and $j$ is the
Pearson correlation coefficient between their corresponding eigenvector components, averaged
over a small set of eigenvectors.
The eigenvectors can be calculated by performing a constrained optimization
of a suitable cost function. The method can be extended to weighted and directed 
graphs. It is useful to estimate vertex similarities, however it does
not provide a well-defined partition of the graph.

Yang and Liu~\cite{yang08} adopted a recursive bisectioning procedure. Communities are subgraphs such that
the external degree of each vertex does not exceed the internal degree ({\it strong communities} 
or {\it $LS$-sets}, see Section~\ref{sec3_1_2}).
In the first step of the algorithm, the adjacency matrix of the graph is put in
approximately block-diagonal form. This is done by computing a 
new centrality measure for the vertices, called {\it clustering centrality}. This measure is similar to Bonacich's eigenvector
centrality~\cite{bonacich72,bonacich87}, which is given by the eigenvector of the adjacency matrix corresponding to the largest eigenvalue.
The clustering centrality of a vertex basically measures the probability that a random walker starting at that vertex hits a given target.
Such probability is larger if the origin and the target vertices belong to the same cluster than if they do not. If the graph has 
well-separated communities, the values of the clustering centrality would be similar for vertices in the same cluster.  
In this way, one can rearrange the original adjacency matrix by listing the vertices in non-decreasing order of their clustering centralities,
and blocks would be visible. The blocks are then identified by iterative bisection: each cluster found at some step is split in two  
as long as the resulting parts
are still communities in the strong sense, otherwise the procedure stops. The worst-case complexity of the method is $O[Kt(n\log n+m)]$, where
$K$ is the number of clusters of the final partition and $t$ the (average) number of iterations required to compute
the clustering centrality with the power method~\cite{golub96}. Since $t$ is fairly independent of the graph size, 
the method scales quite well on sparse graphs [$O(n\log n)$]. The main limit of this technique is the assumption that communities 
are defined in the strong sense, which is too restrictive. On the other hand, one could think of using alternative definitions.

\section{Dynamic Algorithms}
\label{sec44}

This Section describes methods employing processes running on the graph,
focusing on spin-spin interactions, random walks and synchronization.

\subsection{Spin models}
\label{sec44_1}

The Potts model is among the most popular models in 
statistical mechanics~\cite{wu82}. It describes a system of spins that can be in $q$ 
different states. The interaction is ferromagnetic, i.~e. 
it favours spin alignment, so at zero temperature all spins are in the same 
state. If antiferromagnetic interactions are also present, the ground state of the system may 
not be the one where all spins are aligned, but a state where different spin values coexist, in homogeneous clusters.
If Potts spin variables are assigned to the vertices of a graph with community structure, and the interactions 
are between neighbouring spins, it is likely that the structural clusters could be recovered from 
like-valued spin clusters of the system, as there are many more interactions inside communities than outside.
Based on this idea, inspired by an earlier paper by Blatt et al.~\cite{blatt96}, Reichardt and Bornholdt
proposed a method to detect communities that maps the graph onto a zero-temperature $q$-Potts model with nearest-neighbour 
interactions~\cite{reichardt04}. The Hamiltonian of the model, i.~e. its energy, reads
\begin{equation}
{\cal H}=-J\sum_{i,j}A_{ij}\delta(\sigma_i,\sigma_j)+\gamma\sum_{s=1}^q \frac{n_s(n_s-1)}{2},
\label{eqspin1}
\end{equation}
where $A_{ij}$ is the element of the adjacency matrix, $\delta$ is Kronecker's function, $n_s$ the number of spins in state $s$, $J$ and 
$\gamma$ are coupling parameters.
The energy $\cal H$ is the sum of two 
competing terms: the first is the classical ferromagnetic Potts model energy, and favors spin alignment; the second term instead peaks when
the spins are homogeneously distributed. The ratio $\gamma/J$ expresses the relative importance of the two terms: by tuning 
$\gamma/J$ one can explore different levels of modularity of the system, from the whole graph seen as a single cluster to clusters
consisting of individual vertices. If $\gamma/J$ is set to the value $\delta({\cal G})$ of the average
density of edges of the graph $\cal G$, the energy of the system is smaller if spins align within subgraphs such that their
internal edge density exceeds $\delta({\cal G})$, whereas the external edge density is smaller than $\delta({\cal G})$, i.~e.
if the subgraphs are clusters (Section~\ref{sec3_1_1}). The minimization of $\cal H$ is carried out via
simulated annealing (\cite{kirkpatrick83} and Section~\ref{sub_sec6_0_2}), starting from 
a configuration where spins are randomly assigned to the vertices and the number of states $q$ is very high.
The procedure is quite fast and the results do not depend on $q$ (provided $q$ is sufficiently high). The method also allows
to identify vertices shared between communities, from the comparison of partitions corresponding to 
global and local energy minima. The Hamiltonian $\cal H$
can be rewritten as 
\begin{equation}
{\cal H}=\sum_{i<j}\delta(\sigma_i,\sigma_j)(\gamma-A_{ij}),
\label{eqspin2}
\end{equation}
which is the energy of an infinite-range Potts spin glass, as all pairs of spins are interacting (neighboring or not) 
and there may be both positive and negative couplings. 
The method can be simply extended to the analysis of weighted graphs, by introducing spin couplings proportional to the edge weights,
which amounts to replacing the adjacency matrix ${\bf A}$ with the weight matrix ${\bf W}$ in Eq.~\ref{eqspin1}.
Ispolatov et al.~\cite{ispolatov06} have adopted a similar Hamiltonian as in Eq.~\ref{eqspin1}, with a tunable 
antiferromagnetic term interpolating
between the corresponding term of Eq.~\ref{eqspin1} and the entropy term (proportional to $n_s\log n_s$) 
of the free energy, whose minimization is equivalent to finding the states of the finite-temperature Potts model
used by Blatt et al.~\cite{blatt96}.
Eq.~\ref{eqspin2} is at the basis of the successive generalization
of modularity with arbitrary null models proposed by Reichardt and Bornholdt, that we have discussed in Section~\ref{sub_sec6_01}.

In another work~\cite{son06}, Son et al. have presented a clustering technique based on the Ferromagnetic 
Random Field Ising Model (FRFIM). Given a weighted graph with weight matrix ${\bf W}$,
the Hamiltonian of the FRFIM on the graph is
\begin{equation}
{\cal H}=-\frac{1}{2}\sum_{i,j}W_{ij}\sigma_i\sigma_j-\sum_{i}B_i\sigma_i.
\label{eqspin3}
\end{equation}
In Eq.~\ref{eqspin3} $\sigma_i=\pm 1$ and $B_i$ are 
the spin and the random magnetic field of vertex $i$, respectively. The FRFIM has been studied to understand the nature of the 
spin glass phase transition~\cite{middleton02} and the disorder-driven roughening transition of interfaces in disordered media~\cite{noh01,noh02}.
The behavior of the model depends on the choice of the magnetic fields. Son et al. set to zero the magnetic fields of all vertices but two, say $s$ and $t$,
for which the field has infinite strength and opposite signs. This amounts to fix the spins of $s$ and $t$ to opposite values, introducing frustration
in the system. The idea is that,
if $s$ and $t$ are central vertices of different communities, they impose their spin state to the other community members. So, the state of minimum
energy is a configuration in which the graph is polarized into a subgraph with all positive spins and a subgraph with all negative spins, coinciding with 
the communities, if they are well defined. Finding the minimum of $\cal H$ is equivalent to solving a maximum-flow/minimum-cut problem, which can be done 
through well known techniques of combinatorial optimization, like the augmenting path algorithm~\cite{ahuja93}. For a given choice of $s$ and $t$,
many ground states can be found. The vertices that end up in the same cluster in all ground states represent the cores of the clusters, which are called 
{\it coteries}. Possible vertices not belonging to the coteries indicate that the two clusters overlap. In the absence of information about the 
cluster structure of the graph, one needs to repeat the procedure for any pair of vertices $s$ and $t$. Picking vertices of the same
cluster, for instance, would not give meaningful partitions. Son et al. distinguish relevant clusters if they are of about the same size. The procedure can be
iteratively applied to each of the detected clusters, considered as a separate graph, until all clusters have no community structure any more.
On sparse graphs, the algorithm has complexity $O(n^{2+\theta})$, where $\theta\sim 1.2$, so it is very slow and can be currently used for graphs of 
up to few thousands vertices. If one happens to know which are the important vertices of the clusters, e.g. by computing appropriate centrality values
(like degree or site betweenness~\cite{freeman77}), the choices for $s$ and $t$ are constrained and the complexity can become as low as 
$O(n^{\theta})$, which enables the analysis of systems with millions of vertices. Tests on Barab\'asi-Albert graphs (Section~\ref{sec1_3})
show that the latter have no community structure, as expected.

\subsection{Random walk}
\label{sec44_2}

Random walks~\cite{hughes95} can also be useful to find communities. If a graph has a strong community structure,
a random walker spends a long time inside a community due to the high density of internal edges and 
consequent number of paths that could be followed. Here we describe the most popular clustering algorithms based on 
random walks. All of them can be trivially extended to the case of weighted graphs.

Zhou used random walks 
to define a distance between pairs of vertices~\cite{zhou03}: the distance
$d_{ij}$ between $i$ and $j$ is the average number of edges that a random
walker has to cross to reach $j$ starting from $i$. Close vertices are likely to belong
to the same community. Zhou defines a ``global attractor'' of a vertex $i$ to be a
closest vertex to $i$
(i.~e. any vertex lying at the smallest distance from $i$), 
whereas the ``local attractor'' of $i$ is its closest
neighbour. Two types of communities are defined, 
according to local or global attractors: a vertex $i$ has to be put in the same 
community of its attractor and of all other vertices for which $i$ is an attractor. Communities 
must be minimal subgraphs, i.~e. they cannot include smaller subgraphs which are communities according
to the chosen criterion. Applications to real networks, like Zachary's karate club~\cite{zachary77} and the college football network
compiled by Girvan and Newman~\cite{girvan02} (Section~\ref{sec6_1}), along with artificial graphs like the benchmark by 
Girvan and Newman~\cite{girvan02} (Section~\ref{sec6_1}),
show that the method can find meaningful partitions. 
The method can be refined, in that vertex $i$ is associated to its attractor $j$ only with a probability proportional to
$\exp(-\beta d_{ij})$, $\beta$ being a sort of inverse temperature. The computation of the distance matrix 
requires solving $n$ linear-algebra equations (as many as the vertices), 
which requires a time $O(n^3)$. On the other hand, an exact computation of the distance
matrix is not necessary, as
the attractors of a vertex can be identified by considering only a localized portion of the graph around the vertex; therefore the
method can be applied to large graphs as well.  
In a successive paper~\cite{zhou03b}, Zhou introduced a measure of dissimilarity between vertices based on 
the distance defined above. The measure resembles the definition of distance based on structural equivalence of 
Eq.~\ref{eq7}, where the elements of the adjacency matrix are replaced by the corresponding distances.
Graph partitions are obtained with a divisive procedure that, starting from the graph as a single community,
performs successive splits based on the criterion that vertices in the same cluster must be less dissimilar than
a running threshold, which is decreased during the process. The hierarchy of partitions derived by the method
is representative of actual community structures for several real and artificial graphs, including
Zachary's karate club~\cite{zachary77}, the college football network~\cite{girvan02} and the benchmark by 
Girvan and Newman~\cite{girvan02} (Section~\ref{sec6_1}). The time complexity of the procedure is
again $O(n^3)$. The code of the algorithm can be downloaded from 
{\tt http://www.mpikg-golm.mpg.de/theory/people/zhou}\\{\tt /networkcommunity.html}.

In another work~\cite{zhou04}, Zhou and Lipowsky adopted biased random walkers, where the bias is
due to the fact that walkers move preferentially towards vertices sharing a large number of neighbours
with the starting vertex. They defined a proximity index, which indicates how close a pair of vertices
is to all other vertices. Communities are detected with a procedure called {\it NetWalk}, which is an agglomerative
hierarchical clustering method (Section~\ref{sec4_2}), where the similarity between vertices is expressed by their proximity.
The method 
has a time complexity $O(n^3)$: however, the proximity index of a pair of vertices can be computed with good approximation by 
considering just a small portion of the graph around the two vertices, with a considerable gain in time. 
The performance of the method is comparable with that of the algorithm of Girvan and Newman (Section~\ref{subsec5_1}).

A different distance measure between vertices based on
random walks was introduced by Latapy and Pons~\cite{latapy05}. The distance is calculated from the
probabilities that the random walker moves from a vertex to another in a fixed number of steps. The number of 
steps has to be large enough to explore a significant portion of the graph, but not too long, as otherwise 
one would approach the stationary limit in which transition probabilities trivially depend on the vertex degrees. 
Vertices are then grouped into communities through an agglomerative hierarchical clustering technique based on Ward's method~\cite{ward63}. 
Modularity (Section~\ref{sec3_2_2}) is used to select the best partition of the resulting dendrogram. The algorithm runs
to completion in a time $O(n^2d)$ on a sparse graph, where $d$ is the depth of the dendrogram. Since $d$ is often 
small for real graphs [$O(\log n)$], the expected complexity in practical computations is $O(n^2\log n)$. The software of the algorithm can
be found at {\tt http://www-rp.lip6.fr/$\sim$latapy/PP/walktrap.html}.

Hu et al.~\cite{hu08} designed a graph clustering technique based on a signaling process between vertices, somewhat resembling diffusion. Initially a
vertex $s$ is assigned one unit of signal, all the others have no signal. In the first step, the source vertex $s$ sends 
one unit of signal to each of its neighbors. Next, all vertices send as many units of signals they have to each of their neighbors.
The process is continued until a given number of iterations $T$ is reached. The intensity of the signal at vertex $i$, normalized by the total
amount of signal, is the $i$-th
entry of a vector ${\bf u_s}$, representing the source vertex $s$. The procedure is then repeated by choosing each vertex
as source. In this way one can associate an $n$-dimensional vector to each vertex, which correspons to a point in an
Euclidean space. 
The vector ${\bf u_s}$ is actually the 
$s$-th column of the matrix $({\bf I}+{\bf A})^T$, where ${\bf I}$ and ${\bf A}$ are the identity and adjacency matrix, respectively.
The idea is that the vector ${\bf u_s}$ describes the influence that vertex $s$ exerts on the graph through signaling.
Vertices of the same community are expected to have similar influence on the graph and thus to correspond to vectors which are ``close'' in space.
The vectors are finally grouped via fuzzy $k$-means clustering (Section~\ref{sec4_3}). The optimal number of clusters corresponds to  
the partition with the shortest average distance between vectors in the same community and the largest average distance between
vectors of different communities. The signaling process is similar to diffusion, but with the important difference that here 
there is no flow conservation, as the amount of signal at each vertex is not distributed among its neighbors but transferred entirely to each neighbor
(as if the vertex sent multiple copies of the same signal). The complexity of the algorithm is $O[T(\langle k\rangle+1)n^2]$, where $\langle k\rangle$
is the average degree of the graph. Like in the previous algorithm by Latapy and Pons~\cite{latapy05}, finding an optimal value for 
the number of iterations $T$ is non-trivial.

Delvenne et al.~\cite{delvenne08} have shown that random walks enable one to introduce a general quality function, 
expressing the persistence of clusters in time. A cluster is persistent with respect to a random walk after $t$ time steps if the 
probability that the walker escapes the cluster before $t$ steps is low. 
Such probability is computed via the {\it clustered autocovariance matrix} ${\bf R}_t$, which, for a partition of the graph in $c$ clusters, is defined as
\begin{equation}
{\bf R}_t={\bf H}^T({\bf \Pi}{\bf M}^t-{\pi}^T{\bf \pi}){\bf H}. 
\label{eqbarahona}
\end{equation}
Here, ${\bf H}$ is the $n\times c$ membership matrix, whose element $H_{ij}$ equals one if vertex $i$ is in cluster $j$, zero otherwise;
${\bf M}$ is the transition matrix of the random walk; ${\bf \Pi}$ the diagonal matrix whose elements are the stationary probabilities of the random walk, i. e.
$\Pi_{ii}=k_i/2m$, $k_i$ being the degree of vertex $i$; ${\bf \pi}$ is the vector whose entries are the diagonal elements of ${\bf \Pi}$.
The element $(R_t)_{ij}$ expresses the probability for the walk to start in cluster $i$ and end up in cluster $j$ after $t$ steps, minus the 
stationary probability that two independent random walkers are in $i$ and $j$. In this way, the persistence of a cluster $i$ is related to the diagonal element
$(R_t)_{ii}$. Delvenne et al. defined the {\it stability of the clustering} 
\begin{equation}
r(t;{\bf H})=\min_{0\leq s\leq t}\sum_{i=1}^c(R_s)_{ii}=\min_{0\leq s\leq t} {\mbox{trace}}[R_s].
\label{eqbarahona1}
\end{equation}
The aim is then, for a given time $t$, finding the partition with the largest value for $r(t;{\bf H})$.
For $t=0$, the most stable partition is that in which all vertices are their own clusters. Interestingly, for $t=1$, 
maximizing stability is equivalent to maximizing Newman-Girvan modularity (Section~\ref{sec3_2_2}). The cut size of the partition
(Section~\ref{sec4_1}) equals $[r(0)-r(1)]$, so it is also a one-step measure.
In the limit $t\rightarrow\infty$, the most stable partition coincides with the Fiedler partition~\cite{fiedler73,fiedler75}, i. e.
the bipartition where vertices are put in the same class according to the signs of the corresponding component of the Fiedler eigenvector (Section~\ref{sec4_1}).
Therefore, the measure $r(t;{\bf H})$ is very general, and gives a unifying interpretation in the framework of the random walk
of several measures that were defined in different contexts. In particular, modularity has a natural interpretation 
in this dynamic picture~\cite{lambiotte08}.
Since the size of stable clusters increases with $t$, time can be considered as a resolution
parameter. Resolution can be fine tuned by taking time as a continuous variable 
(the extension of the formalism is straightforward); the linearization of the stability
measure at small (continuous) times delivers multiresolution versions of modularity~\cite{reichardt06,arenas08b} (Section~\ref{sub_sec6_201}).

In a method by Weinan et al.~\cite{weinan08}, the best partition of a graph in $k$ clusters is such that the Markov chain
describing a random walk on the meta-graph, whose vertices are the clusters of the original graph, gives the best approximation of the 
full random walk dynamics on the whole graph. 
The quality of the approximation is given by the distance between the 
left stochastic matrices of the two processes, which thus needs to be minimized. The minimization is performed by using a variant of the 
$k$-means algorithm (Section~\ref{sec4_3}), and the result is the best obtained out of $l$ runs starting from different initial conditions, a strategy
that considerably improves the quality of the optimum. The time complexity is
$O[tlk(n+m)]$, where $t$ is the number of steps required to reach convergence. The optimal number of clusters
could in principle be determined by analyzing how the quality of the approximation varies with $k$, but the authors do not give 
any general recipe. The method is rather accurate on the benchmark by Girvan and Newman~\cite{girvan02} (Section~\ref{sec6_1}) 
and on Zachary's karate club network.
The algorithm by Weinan et al. is asymptotically equivalent to spectral graph partitioning (Section~\ref{sec4_4}) when
the Markov chain describing the random walk presents a sizeable spectral gap between some of the largest eigenvalues
of the transfer matrix (Section~\ref{sec1_2}), approximately equal to one,  
and the others.

We conclude this section by describing the {\it Markov Cluster Algorithm (MCL)}, which was 
invented by Van Dongen~\cite{vandongen00}. 
This method simulates a peculiar process of flow diffusion in a graph. 
One starts from the {\it transfer matrix} of the graph $\bf T$ (Section~\ref{sec1_2}).
The element $T_{ij}$ of the
transfer matrix gives the probability that a random walker, sitting at vertex $j$, moves to $i$. The sum of the elements
of each column of ${\bf T}$ is one.
Each iteration of the algorithm consists of two steps.
In the first step, called {\it expansion}, the transfer matrix of the graph is raised to an integer power $p$
(usually $p=2$). The
entry $M_{ij}$ of the resulting matrix gives the probability that a random walker, starting
from vertex $j$, reaches $i$ in $p$ steps (diffusion flow).
The second step, which has no physical counterpart,
consists in raising each single entry of the matrix ${\bf M}$ to
some power $\alpha$, where $\alpha$ is now real-valued. This operation, called {\it inflation}, enhances the weights
between pairs of vertices with large values of the diffusion flow, which are likely to be in the same community.
Next, the elements of each column must be divided by their sum, such that
the sum of the elements of the column equals one and a new transfer matrix is recovered.
After some iterations, the process delivers a stable matrix, with some remarkable properties. Its elements are 
either zero or one, so it is a sort of adjacency matrix. Most importantly, the graph described
by the matrix is disconnected, and its connected components are the communities of the original graph.
The method is really simple to implement, which is the main reason of its success: as of now, the MCL is 
one of the most used clustering algorithms in bioinformatics. The code can be downloaded from 
{\tt http://www.micans.org/mcl/}. Due to the matrix multiplication of the expansion step,
the algorithm should scale as $O(n^3)$, even if the graph is sparse, as the running matrix becomes quickly 
dense after a few steps of the algorithm. However, while computing the matrix multiplication, MCL keeps only
a maximum number $k$ of non-zero elements per column, where $k$ is usually much smaller than $n$.
So, the actual worst-case running time of the algorithm is $O(nk^2)$ on a sparse graph. 
A problem of the method is the fact that the final partition is sensitive to the parameter $\alpha$
used in the inflation step. Therefore several different partitions can be obtained, and it is not clear 
which are the most meaningful or representative.

\subsection{Synchronization}
\label{sec44_3}

\begin{figure*}
\begin{center}
\includegraphics[width=\textwidth]{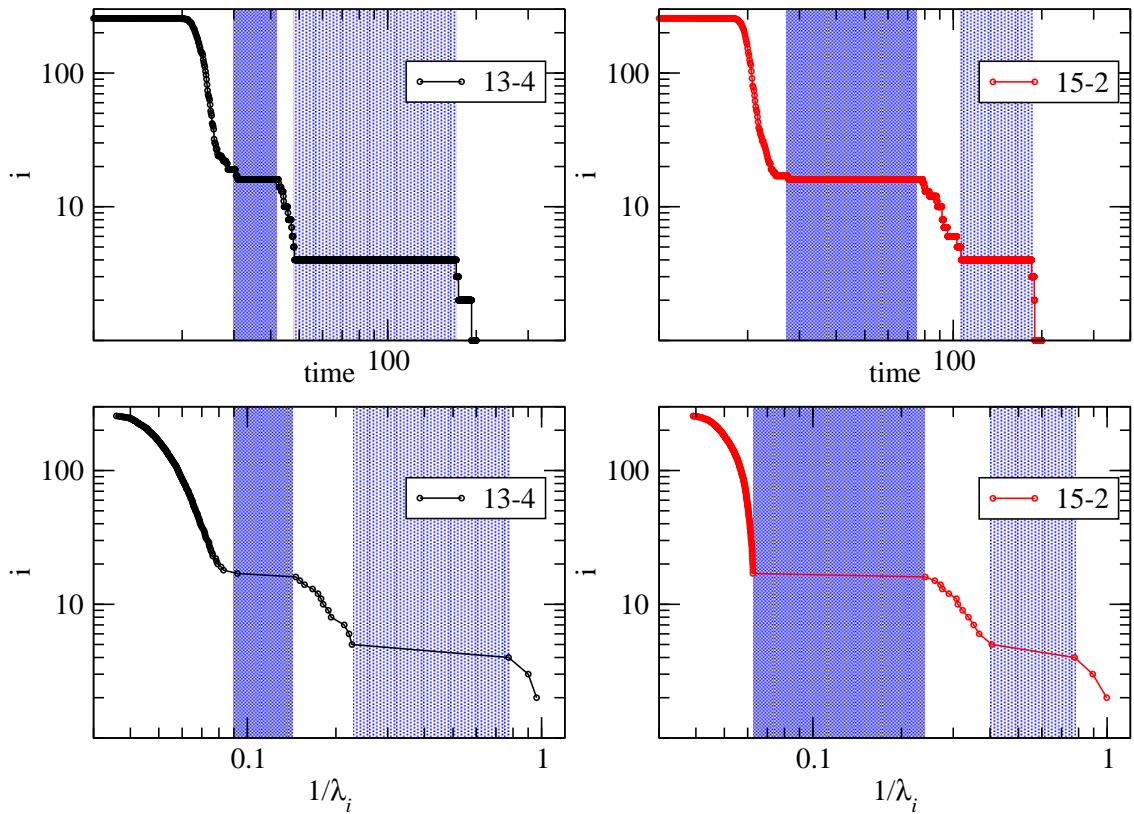}
\caption {\label{syncarenas} Synchronization of Kuramoto oscillators on graphs with 
two hierarchical levels of communities. (Top) The number of different synchronized components is plotted versus time for two 
graphs with different densities of edges within the clusters. (Bottom) The rank index of the eivenvalues of the 
Laplacian matrices of the same two graphs of the upper panels is plotted versus the inverse eigenvalues
(the ranking goes from the largest to the smallest eigenvalue). The two types of communities are revealed by the plateaus.
Reprinted figure with permission from 
Ref.~\cite{arenas06}. \copyright 2006 by the American Physical Society.}
\end{center}
\end{figure*}

Synchronization~\cite{pikovsky01} is an emergent phenomenon occurring in systems of interacting units and is ubiquitous in nature, society and technology.
In a synchronized state, the units of the system are in the same or similar state(s) at every time. Synchronization has also been applied 
to find communities in graphs. 
If oscillators are placed at the vertices, with initial random phases, and have nearest-neighbour interactions, 
oscillators in the same community synchronize first, whereas a full synchronization requires a longer time.
So, if one follows the time evolution of the process, states with synchronized clusters of vertices 
can be quite stable and long-lived, so they can be easily recognized.
This was first shown by Arenas, D\'iaz-Guilera and P\'erez-Vicente~\cite{arenas06}. They used 
Kuramoto oscillators~\cite{kuramoto84}, which are coupled two-dimensional vectors endowed with a proper 
frequency of oscillations. In the Kuramoto model, the phase $\theta_i$ of oscillator 
$i$ evolves according to the following dynamics
\begin{equation}
\frac{d\theta_i}{dt}=\omega_i+\sum_{j}K\sin(\theta_j-\theta_i),
\label{eqsync1}
\end{equation}
where $\omega_i$ is the natural frequency of $i$, $K$ the strength of the coupling between oscillators and the sum
runs over all oscillators (mean field regime). If the interaction coupling exceeds a threshold, depending on the width of the
distribution of natural frequencies, the dynamics leads to
synchronization. If the dynamics runs on a graph, each oscillator is coupled only to
its nearest neighbors. In order to reveal the effect of local synchronization, 
Arenas et al. introduced the local order parameter
\begin{equation}
\rho_{ij}(t)=\langle \cos[\theta_i(t)-\theta_j(t)]\rangle,
\label{eqsync2}
\end{equation}
measuring the average correlation between oscillators $i$ and $j$. The average is computed over different initial conditions.
By visualizing the correlation matrix ${\bf \rho}(t)$ at a given time $t$, one may distinguish groups of vertices that synchronize 
together. The groups can be identified by means of the {\it dynamic connectivity matrix} ${\bf {\cal D}_t}(T)$, which is a binary matrix
obtained from ${\bf \rho}(t)$ by thresholding its entries. The dynamic connectivity matrix embodies information about both 
the synchronization dynamics and the underlying graph topology. 
From the spectrum of ${\bf {\cal D}_t}(T)$ it is possible to derive the number of 
disconnected components at time $t$. By plotting the number of components as a function of time, plateaus may appear at some
characteristic time scales, indicating structural scales of the graph with robust communities (Fig.~\ref{syncarenas}).
Partitions corresponding to long plateaus are characterized by high values of the modularity of Newman and Girvan (Section~\ref{sec3_2_2}) on graphs
with homogeneous degree distributions, whereas such correlation is poor in the presence of hubs~\cite{arenas07b}.
Indeed, it has been proven that the stability (Eq.~\ref{eqbarahona1}) 
of the dynamics associated to the standard Laplacian matrix, which describes the convergence towards synchronization of the Kuramoto model
with equal intrinsic frequencies, coincides with modularity only for graphs whose vertices have the same degree~\cite{lambiotte08}.
The appearance of plateaus at different time scales hints to a hierarchical organization of the graph. After a sufficiently long $t$ 
all oscillators are synchronized and the whole system behaves as a single component.
Interestingly, Arenas et al. found that the structural scales revealed by
synchronization correspond to groups of eigenvalues of the Laplacian matrix of the graph, separated by gaps (Fig.~\ref{syncarenas}). 

Based on the same principle, Boccaletti et al. designed a community detection method based on synchronization~\cite{boccaletti07}.
The synchronization dynamics is a variation of Kuramoto's model, the opinion changing rate (OCR) model~\cite{pluchino05}. Here the interaction
coupling between adjacent vertices is weighted by a term proportional to a (negative) power 
of the betweenness of the edge connecting the vertices (Section~\ref{subsec5_1}), with exponent $\alpha$. 
The evolution equations of the model are solved by decreasing the value of $\alpha$ during the evolution of the dynamics, starting from a configuration
in which the system is fully synchronized ($\alpha=0$). The graph tends to get split into clusters of synchronized elements, because the
interaction strengths across inter-cluster edges get suppressed due to their
high betweenness scores. By varying $\alpha$, different partitions
are recovered, from the graph as a whole until the vertices as separate communities: the partition with the largest value of modularity is taken
as the most relevant.
The algorithm scales in a time $O(mn)$, or $O(n^2)$ on sparse graphs, and gives good results in practical examples, 
including Zachary's karate club~\cite{zachary77} and the benchmark by 
Girvan and Newman~\cite{girvan02} (Section~\ref{sec6_1}).
The method can be refined by homogeneizing the natural frequencies of the oscillators during the evolution of the system. In this way, the system
becomes more stable and partitions with higher modularity values can be recovered.

In a recent paper by Li et al.~\cite{li08b}, it was shown that
synchronized clusters in modular networks are characterized by interfacial vertices, whose oscillation frequency is intermediate between those of two
or more clusters, so that they do not belong to a specific community. Li et al. used this result to devise a technique able to detect
overlapping communities.

Synchronization-based algorithms may not be reliable when communities are very different in size; tests in this direction
are still missing.
%
%
\section{Methods based on statistical inference}
\label{sec_statinf}

{\it Statistical inference}~\cite{mackay03} aims at deducing properties of data sets, starting from a set of 
observations and model hypotheses. If the data set is a graph, the model, based on hypotheses on how vertices are connected to each other,
has to {\it fit} the actual graph topology. In this section we review those clustering techniques attempting to find 
the best fit of a model to the graph, where the model assumes that vertices have some sort of classification, based on
their connectivity patterns. We mainly focus on methods adopting {\it Bayesian inference}~\cite{winkler03}, 
in which the best fit is obtained through the maximization of
a likelihood ({\it generative models}), but we also discuss related techniques, based on {\it blockmodeling}~\cite{doreian05}, 
{\it model selection}~\cite{burnham02} and {\it information theory}~\cite{mackay03}.

\subsection{Generative models}
\label{subsec_statinf1}

Bayesian inference uses observations to estimate the probability that 
a given hypothesis is true. It consists of two ingredients: the evidence, expressed by the information $D$ one has about the system 
(e.g., through measurements); a statistical model with parameters $\{\theta\}$. Bayesian inference starts by writing the likelihood
$P(D|\{\theta\})$ that the observed evidence is produced by the model for a given set of parameters $\{\theta\}$. The aim is to determine 
the choice of $\{\theta\}$ that maximizes the
posterior distribution $P(\{\theta\}|D)$ of the parameters given the model and the evidence. By using Bayes' theorem one has
\begin{equation}
P(\{\theta\}|D)=\frac{1}{Z}P(D|\{\theta\})P(\{\theta\}),
\label{eqstinf0}
\end{equation}
where $P(\{\theta\})$ is the prior distribution of the model parameters and
\begin{equation}
Z=\int P(D|\{\theta\})P(\{\theta\})d\theta.
\label{eqstinf01}
\end{equation}
Unfortunately, computing the integral~\ref{eqstinf01} is a major challenge. Moreover, the choice of the prior distribution $P(\{\theta\})$ is 
non-obvious. Generative models differ from each other by the choice of the model and the
way they address these two issues.

Bayesian inference is frequently used in the analysis and modeling of real graphs, including
social~\cite{handcock07,koskinen07,rhodes07} and biological networks~\cite{rowicka04,berg06}. 
Graph clustering can be considered a specific example of inference problem. Here, the
evidence is represented by the graph structure
(adjacency or weight matrix) and there is an additional ingredient, represented by
the classification of the vertices in groups, which is a {\it hidden} (or {\it missing}) information that 
one wishes to infer along with the parameters of the model which is supposed to be responsible for
the classification. This idea is at the basis of several recent papers, which we discuss here.
In all these works, one essentially maximizes the likelihood $P(D|\{\theta\})$ that the model is consistent with the observed graph structure,
with different constraints. We specify the set of parameters $\{\theta\}$ as the triplet $(\{q\},\{\pi\},k)$, where $\{q\}$ indicates the
community assignment of the vertices, $\{\pi\}$ the model parameters, and $k$ the number of clusters. In the following we shall stick 
to the notation of the papers, so the variables above may be indicated by different symbols. However, to better 
show what each method specifically does we shall refer to our general notation at the end of the section.

Hastings~\cite{hastings06} chooses as a model of network with communities the {\it planted partition model} (Section~\ref{sec6}).
In it, $n$ vertices are assigned to $q$ groups: vertices of the same group are linked with a probability $p_{in}$,
while vertices of different groups are linked with a probability $p_{out}$. If $p_{in}>p_{out}$, the model graph has a
built-in community structure. The vertex classification is indicated
by the set of labels $\{q_i\}$. The probability that, given a graph,
the classification $\{q_i\}$ is the right one according to the model
is\footnote{The actual likelihood includes an additional factor expressing the 
{\it a priori} probability of the community sizes. Hastings assumes that this probability is constant.}
\begin{equation}
p(\{q_i\})\propto \{\exp[-\sum_{\langle ij\rangle}J\delta_{q_iq_j}-\sum_{i\neq j}J^\prime\delta_{q_iq_j}/2]\}^{-1},
\label{eqstinf1}
\end{equation}
where $J=\log\{[p_{in}(1-p_{out})]/[p_{out}(1-p_{in})]\}$, $J^\prime=\log[(1-p_{in})/(1-p_{out})]$ and the first sum runs over nearest neighboring vertices.
Maximizing $p(\{q_i\})$ is equivalent to minimizing the argument of the exponential, which is the Hamiltonian of a Potts model 
with short- and long-range interactions. For $p_{in}>p_{out}$, $J>0$ and $J^{\prime}<0$, so the model is a spin glass 
with ferromagnetic nearest-neighbor interactions and antiferromagnetic long-range interactions, similar to the 
model proposed by Reichardt and Bornholdt to generalize Newman-Girvan modularity~\cite{reichardt06} (Section~\ref{sub_sec6_01}).
Hastings used belief propagation~\cite{gallager63} to find the ground state of the spin model. 
On sparse graphs, the complexity of the algorithm is expected to be $O(n\log^\alpha{n})$, where $\alpha$ needs to be estimated numerically.
In principle one needs to input the 
parameters $p_{in}$ and $p_{out}$, which are usually unknown in practical applications. However, it turns out that they can be chosen rather
arbitrarily, and that bad choices can be recognized and corrected. 

Newman and Leicht~\cite{newman07} have recently proposed a similar method based on a mixture model and the expectation-maximization 
technique~\cite{dempster77}. The method bears some resemblance with an {\it a posteriori} blockmodel 
previously introduced by Snijders and Nowicki~\cite{snijders97,nowicki01}. 
They start from a directed graph with $n$ vertices, whose vertices
fall into $c$ classes. The group of vertex $i$ is indicated by $g_i$, $\pi_r$ the fraction of vertices in group $r$,
and $\theta_{ri}$ the probability that there is a directed edge from vertices of group $r$ to vertex $i$. By definition,
the sets $\{\pi_i\}$ and $\{\theta_{ri}\}$ satisfy the normalization conditions $\sum_{r=1}^c\pi_r=1$ and $\sum_{i=1}^n\theta_{ri}=1$.
Apart from normalization, the probabilities $\{\theta_{ri}\}$ are assumed to be independent of each other.
The best classification of the vertices corresponds to the maximum of the average log-likelihood $\bar{\cal L}$ that the model,
described by the values of the parameters $\{\pi_i\}$ and $\{\theta_{ri}\}$ 
fits the adjacency matrix ${\bf A}$ of the graph. The expression of the average log-likelihood $\bar{\cal L}$ requires the 
definition of the probability $q_{ir}=Pr(g_i=r|A, \pi, \theta)$, that vertex $i$ belongs to group $r$. 
By applying Bayes' theorem the probabilities $\{q_{ir}\}$
can be computed in terms of the $\{\pi_i\}$ and the $\{\theta_{ri}\}$, as
\begin{equation}
q_{ir}=\frac{\pi_r\prod_j\theta_{rj}^{A_{ij}}}{\sum_s\pi_s\prod_j\theta_{sj}^{A_{ij}}},
\label{eqstinf2}
\end{equation}
while the maximization of the average log-likelihood $\bar{\cal L}$, under the normalization constraints of the model variables
$\{\pi_i\}$ and $\{\theta_{ri}\}$, yields the relations
\begin{equation}
\pi_r=\frac{1}{n}\sum_iq_{ir},\,\,\,\,\,\,\,\theta_{rj}=\frac{\sum_iA_{ij}q_{ir}}{\sum_ik_{i}q_{ir}},
\label{eqstinf3}
\end{equation}
where $k_i$ is the outdegree of vertex $i$. Equations~\ref{eqstinf2} and~\ref{eqstinf3} are self-consistent, and can be solved by iterating them 
to convergence, starting from a suitable set of initial conditions.
Convergence is fast, so the algorithm could be applied to fairly large graphs, with up to about $10^6$ vertices.

The method, designed for directed graphs, can be easily extended to the undirected case, whereas
an extension to weighted graphs is not straightforwad. 
A nice feature of the method is that it does not require any preliminary indication
on what type of structure to look for; the resulting structure is the most likely classification
based on the connectivity patterns of the vertices. Therefore, various types of structures can be detected, not necessarily communities. 
For instance, multipartite structure could be uncovered, or mixed patterns where 
multipartite subgraphs coexist with communities, etc..
In this respect, it is more powerful than most methods of community detection, 
which are bound to focus only on proper communities, i.~e. subgraphs
with more internal than external edges. 
In addition, since partitions are defined by assigning probability values to the vertices, 
expressing the extent of their membership in a group,
it is possible that some vertices are not clearly assigned to a group, but to more groups, so the method is able to deal with 
overlapping communities. 
The main drawback of the algorithm is the fact that one needs to specify the number of 
groups $c$ at the beginning of the calculation, a number that is typically unknown for real networks. It is possible to derive this
information self-consistently by maximizing the probability that the data are reproduced
by partitions with a given number of clusters. But this procedure involves some degree 
of approximation, and the results are often not good. 

In a recent study it has been shown that the method by Newman and Leicht
enables one to rank vertices based on their degree of influence on other vertices, which allows to identify the
vertices responsible for the group structure and its stability~\cite{mungan08}. A very similar 
technique has also been applied by V\'azquez~\cite{vazquez08} to
the problem of population stratification, where animal populations and their attributes are represented as hypergraphs (Section~\ref{sec1_1}). V\'azquez
also suggested an interesting criterion to decide the 
optimal number of clusters, namely picking the number $\bar{c}$ whose solution has the greatest similarity with
solutions obtained at different values of $c$. The similarity between two partitions can be estimated in various ways, for instance
by computing the normalized mutual information (Section~\ref{sec6}). In a successive paper~\cite{vazquez09}, 
V\'azquez showed that better results are obtained if the classification likelihood is maximized
by using Variational Bayes~\cite{jordan99,beal03}.

\begin{figure}
\begin{center}
\includegraphics[width=\columnwidth]{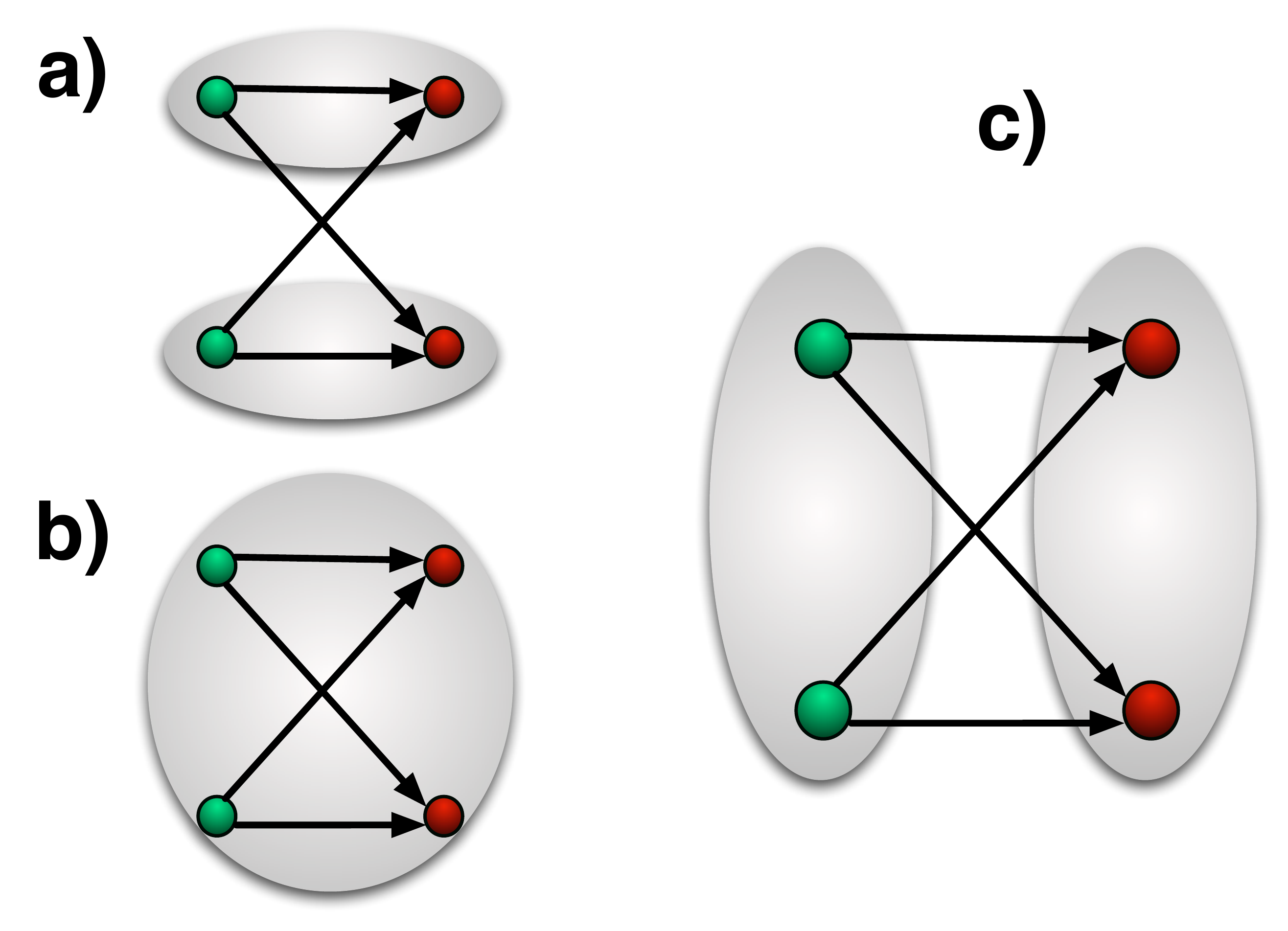}
\caption {\label{Figramasco} Problem of method by Newman and Leicht. By applying the method to 
the illustrated complete bipartite graph (colors indicate the vertex classes)
the natural group structure c) is not recovered; instead, the most likely classifications are a) and b).
Reprinted figure with permission from Ref.~\cite{ramasco08}.\copyright 2008 by the American Physical Society.}
\end{center}
\end{figure}

Ramasco and Mungan~\cite{ramasco08} remarked that the normalization condition on the probabilities $\{\theta_{ri}\}$
implies that each group $r$ must have non-zero outdegree and that therefore the method fails to detect the intuitive 
group structure of (directed) bipartite graphs (Fig.~\ref{Figramasco}). 
To avoid this problem, 
they proposed a modification, that consists in introducing three sets for the 
edge probabilities $\{\theta_{ri}\}$, relative to edges going from group $r$ to vertex $i$ (as before), from $i$ to $r$ and 
in both directions, respectively. Furthermore, they used the average entropy of the 
classification $S_q=-(\sum_{i,r}q_{ir}\ln\,q_{ir})/n$, where the $q_{ir}$ are the analogs of the probabilities 
in Eq.~\ref{eqstinf2}, to infer the optimal number of groups, that the method of Newman and Leicht is unable to 
provide. Another technique similar to that by Newman and Leicht has been designed by
Ren et al.~\cite{ren09}. The model is based on the group fractions $\{\pi_i\}$, defined as above, 
and a set of probabilities $\{\beta_{r,i}\}$, expressing the relevance of vertex $i$ for group $r$; 
the basic assumption is that the probability that two vertices of the same group are connected by an edge 
is proportional to the product of the relevances of the two vertices. In this way, there is an explicit relation
between group membership and edge density, and the method can only detect community structure. The community assignments
are recovered through an expectation-maximization procedure that closely follows that by Newman and Leicht. 

Maximum likelihood estimation has been used by \v{C}opi\v{c} et al. to define an axiomatization of the 
problem of graph clustering and its related concepts~\cite{copic05}. The starting point is again the planted partition model (Section~\ref{sec6}),
with probabilities $p_{in}$ and $p_{out}$. A novelty of the approach is the introduction of the {\it size matrix} ${\bf S}$, whose element
$S_{ij}$ indicates the maximum strength of interaction between vertices $i$ and $j$. For instance, in a graph with unweighted connections,
all elements of ${\bf S}$ equal $1$. In this case, the probability that the graph conceals a community structure coincides with the
expression (\ref{eqstinf1}) by Hastings. \v{C}opi\v{c} et al. used this probability as a quality function to define rankings
between graph partitions ({\it likelihood rankings}). The authors show that the likelihood rankings satisfy a number of general properties, which 
should be satisfied by any reasonable ranking. They also propose an algorithm to find the maximum likelihood partition, by using the auxiliary concept
of {\it pseudo-community structure}, i. e. a grouping of the graph vertices in which it is specified which pairs of vertices stay in the same 
community and which pairs instead stay in different communities. A pseudo-community may not be a community because the transitive property 
is not generally valid, as the focus is on pairwise vertex relationships: it may happen that $i$ and $j$ are classified in the same group, and that $j$ and $k$ 
are classified in the same group, but that $i$ and $k$ are not classified as belonging to the same group. We believe that the work by 
\v{C}opi\v{c} et al. is an important first step towards a more rigorous formalization of the problem of graph clustering.

Zanghi et al.~\cite{zanghi08} have designed a clustering technique that lies somewhat in between the method by Hastings and that by Newman and Leicht.
As in Ref.~\cite{hastings06}, they use the planted partition model to represent a graph with community structure; as in
Ref.~\cite{newman07}, they maximize the classification likelihood using an expectation-maximization algorithm~\cite{dempster77}.
The algorithm runs for a fixed number of clusters $q$, like that by Newman and Leicht; however, the optimal number of clusters can be determined by
running the algorithm for a range of $q$-values and
selecting the solution that maximizes the Integrated Classification Likelihood introduced by Biernacki et al.~\cite{biernacki00}. The time complexity
of the algorithm is $O(n^2)$.

Hofman and Wiggins have proposed a general Bayesian approach to the problem of graph clustering~\cite{hofman08}. 
Like Hastings~\cite{hastings06}, they model a graph 
with community structure as in the planted partition problem (Section~\ref{sec6}), in that
there are two probabilities $\theta_c$ and $\theta_d$ that there is an edge between vertices of the same or different
clusters, respectively. The unobserved community structure is indicated by the set of labels $\vec{\sigma}$ for the vertices; 
$\pi_r$ is again the fraction of vertices in group $r$. The conjugate prior distributions $p(\vec{\theta})$ and $p(\vec{\pi})$
are chosen to be Beta and Dirichlet distributions. The most probable number of clusters $K^*$ maximizes the conditional probability
$p(K|{\bf A})$ that there are $K$ clusters, given the matrix $\bf A$. 
Like Hastings, Hofman and Wiggins assume that the prior probability $p(K)$
on the number of clusters is a smooth function, therefore maximizing $p(K|{\bf A})$ amounts to 
maximizing the Bayesian evidence $p({\bf A}|K)\propto p(K|{\bf A})/p(K)$, obtained by integrating the
joint distribution $p({\bf A},\vec{\sigma}| \vec{\pi}, \vec{\theta}, K)$, which is factorizable, over the model parameters
$\vec{\theta}$ and $\vec{\pi}$. The integration can be performed exactly only for small graphs. 
Hofman and Wiggins used Variational Bayes~\cite{jordan99,beal03}, in order to 
compute controlled approximations of $p({\bf A}|K)$. 
The complexity of the algorithm was estimated numerically on synthetic graphs, yielding $O(n^\alpha)$, with $\alpha=1.44$. In fact,
the main limitation comes from high memory requirements. The method is more powerful than the one by Hastings~\cite{hastings06}, in that the 
edge probabilities $\vec{\theta}$ are inferred by the procedure itself and need not be specified (or guessed) at the beginning.
It also includes the expectation-maximization approach by Newman and Leicht~\cite{newman07} as a special case, with the big advantage
that the number of clusters need not be given as an input, but is an output of the method. The software of the algorithm can be 
found at {\tt http://www.columbia.edu/$\sim$chw2/}.

We conclude with a brief summary of the main techniques described above, coming back to our notation at the beginning of the section.
In the method by Hastings, one maximizes the likelihood 
$P(D|\{q\},\{\pi\},k)$ over the set of all possible community assignments $\{q\}$, given the number of clusters $k$ and the 
model parameters (i.~e. the linking probabilities $p_{in}$ and $p_{out}$). Newman and Leicht maximize the likelihood 
$P(D|\{q\},\{\pi\},k)$ for a given number of clusters, 
over the possible choices for the model parameters and community assignments, by deriving the optimal choices for both 
variables with a self-consistent procedure. Hofman and Wiggins maximize the likelihood 
$P_{HW}(k)=\sum_{\{q\}}\int P(D|\{q\},\{\pi\},k)P(\{q\}|\{\pi\})P(\{\pi\})d\pi$ over the possible choices for the number of clusters.

\subsection{Blockmodeling, model selection and information theory}
\label{subsec_statinf2}

Block modeling is a common approach in statistics and social network analysis to decompose a graph in classes of 
vertices with common properties. In this way, a simpler description of the graph is attained. Vertices are usually grouped
in classes of equivalence. There are two main definitions of topological equivalence for vertices: 
{\it structural equivalence}~\cite{lorrain71} (Section~\ref{sec3_1_4}),
in which vertices are equivalent if they have the same 
neighbors\footnote{More generally, if they have the same ties/edges to the same vertices, 
as in a social network there may be different types of ties/edges.}; 
{\it regular equivalence}~\cite{white83, everett94}, in which 
vertices of a class have similar connection patterns to vertices of the other classes (ex. parents/children). 
Regular equivalence does not require that ties/edges
are restricted to specific target vertices, so it is a more general concept than structural equivalence.
Indeed, vertices which are structurally equivalent
are also regularly equivalent, but the inverse is not true. The concept of structural equivalence can be generalized to probabilistic
models, in which one compares classes of graphs, not single graphs, characterized by a set of linking probabilities
between the vertices. In this case, vertices are organized in classes such that the linking probabilities of a vertex
with all other vertices of the graph are the same for vertices in the same class, 
which are called {\it stochastically equivalent}~\cite{fienberg81,holland83}. 
 
A thorough discussion of blockmodeling is beyond
the scope of this review: we point the reader to Ref.~\cite{doreian05}. Here we discuss a recent work by Reichardt and White~\cite{reichardt07b}.
Let us suppose to have a directed graph with $n$ vertices and $m$ edges. 
A classification of the graph is indicated by the set of labels $\{\sigma\}$, where $\sigma_i=1, 2, ..., q$ is the class of vertex $i$.
The corresponding blockmodel, or {\it image graph}, is expressed by a $q\times q$ adjacency matrix ${\bf B}$: 
$B_{q_1q_2}=1$ if edges between classes $q_1$ and $q_2$ are allowed, otherwise it is zero. The aim is finding the classification $\{\sigma\}$ and 
the matrix $B$ that best fits the adjacency matrix ${\bf A}$ of the graph. The goodness of the fit is expressed by 
the quality function
\begin{equation}
{\cal Q}^B(\{\sigma\})=\frac{1}{m}\sum_{i\neq j}[a_{ij}A_{ij}B_{\sigma_i\sigma_j}+b_{ij}(1-A_{ij})(1-B_{\sigma_i\sigma_j})],
\label{eqstinf4}
\end{equation}
where $a_{ij}$ ($b_{ij}$) reward the presence (absence) of edges between vertices if there
are edges (non-edges) between the corresponding classes, and $m$ is the number of edges of the graph, as usual.
Eq.~\ref{eqstinf4} can be rewritten as a sum over the classes
\begin{equation}
{\cal Q}^B(\{\sigma\})=\sum_{r,s}^{q}(e_{rs}-[e_{rs}])B_{rs},
\label{eqstinf5}
\end{equation}
by setting $e_{rs}=(1/m)\sum_{i\neq j}(a_{ij}+b_{ij})A_{ij}\delta_{\sigma_ir}\delta_{\sigma_js}$ and 
$[e_{rs}]=(1/m)\sum_{i\neq j}b_{ij}\delta_{\sigma_ir}\delta_{\sigma_js}$. If one sets $a_{ij}=1-p_{ij}$ and $b_{ij}=p_{ij}$,
$p_{ij}$ can be interpreted as the linking probability between $i$ and $j$, in some null model. Thereof, $e_{rs}$ becomes the number of edges 
running between vertices of class $r$ and $s$, and $[e_{rs}]$ the expected number of edges in the null model. Reichardt and White
set $p_{ij}=k^{out}_ik^{in}_j/m$, which defines the same null model of Newman-Girvan modularity for directed graphs (Section~\ref{sub_sec6_01}). 
In fact, if the image graph has only self-edges, i.~e. $B_{rs}=\delta_{rs}$, the quality function ${\cal Q}^B(\{\sigma\})$ exactly matches modularity.
Other choices for the image graph are possible, however. For instance, a matrix $B_{rs}=1-\delta_{rs}$ describes the classes of a $q$-partite graph (Section~\ref{sec1_1}).
From Eq.~\ref{eqstinf5} we see that, for a given classification $\{\sigma\}$, the image graph that yields the largest value
of the quality function ${\cal Q}^B(\{\sigma\})$ is that in which $B_{rs}=1$ when the term $e_{rs}-[e_{rs}]$ is non-negative,
and $B_{rs}=0$ when the term $e_{rs}-[e_{rs}]$ is non-positive. So, the best classification is the one maximizing the quality function
\begin{equation}
{\cal Q}^*(\{\sigma\})=\frac{1}{2}\sum_{r,s}^{q}||e_{rs}-[e_{rs}]||,
\label{eqstinf6}
\end{equation}
where all terms of the sum are taken in absolute value. The function ${\cal Q}^*(\{\sigma\})$ is maximized via simulated annealing.
The absolute maximum $Q_{max}$ is obtained by construction when $q$ matches the number $q^*$ of structural equivalence classes of the graph. 
However, the absolute maximum $Q_{max}$ does not have a meaning by itself, as one can achieve fairly high values of ${\cal Q}^*(\{\sigma\})$
also for null model instances of the original graph, i. e. if one randomizes the graph by keeping the same expected indegree and outdegree sequences.
In practical 
applications, the optimal number of classes is determined by comparing the ratio $Q^*(q)/Q_{max}$ [$Q^*(q)$ is the maximum of ${\cal Q}^*(\{\sigma\})$
for $q$ classes] with the expected ratio for the null model.
Since classifications for different $q$-values are not 
hierarchically ordered, overlaps between classes may be detected. The method can be trivially extended to the case of weighted graphs. 

Model selection~\cite{burnham02} aims at finding models which are at the same time 
simple and good at describing a system/process. A basic example of a model selection problem is curve fitting.
There is no clear-cut recipe to select a model, but a bunch of heuristics, like
Akaike Information Criterion (AIC)~\cite{akaike74}, 
Bayesian Information Criterion (BIC)~\cite{schwarz78}, Minimum Description Length 
(MDL)~\cite{rissanen78,grunwald05}, Minimum Message Length (MML)~\cite{wallace68}, etc..

\begin{figure*}
\begin{center}
\includegraphics[width=\textwidth]{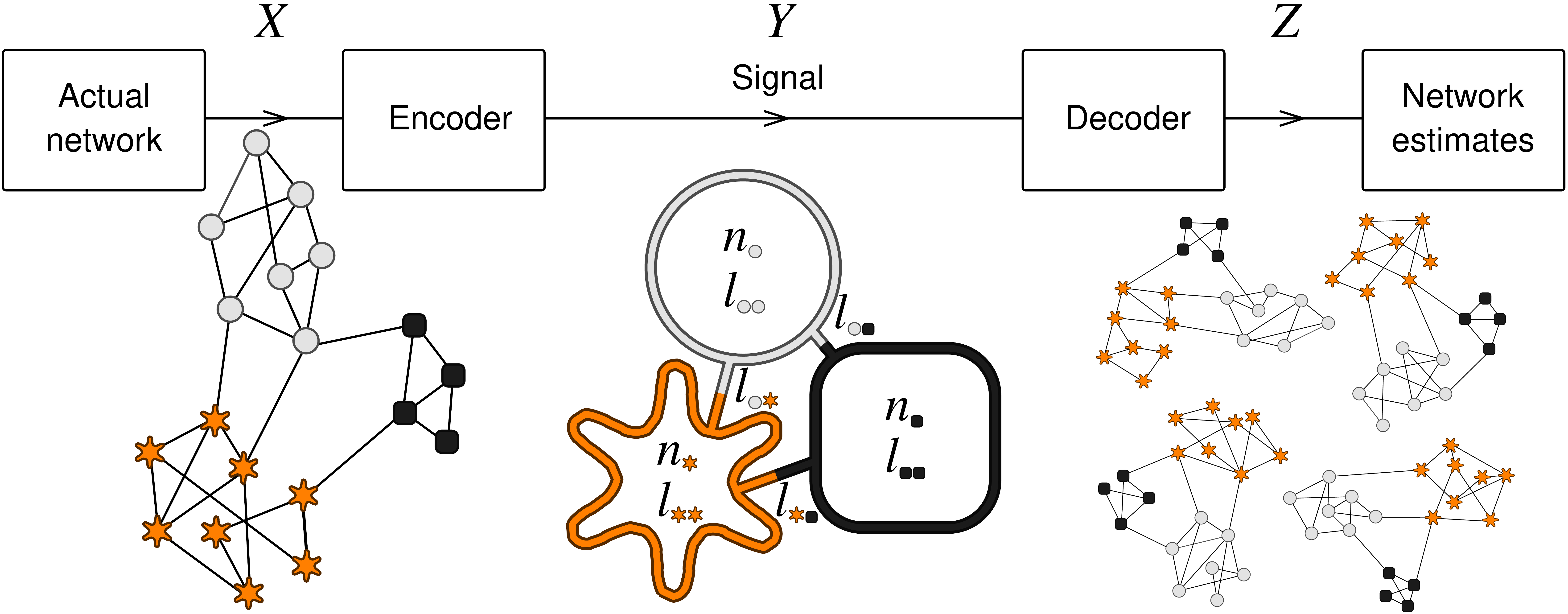}
\caption{\label{rosvall1} Basic principle of the method by Rosvall and Bergstrom~\cite{rosvall07}. An encoder sends
to a decoder a compressed information about the topology of the graph on the left. The information gives a coarse description of the graph,
which is used by the decoder to deduce the original graph structure. Reprinted figure with permission from 
Ref.~\cite{rosvall07}. \copyright 2007 by the National Academy of Science of the USA.}
\end{center}
\end{figure*}

The modular structure of a graph can be considered as a compressed description
of the graph to approximate the whole information contained in its adjacency matrix.
Based on this idea, Rosvall and Bergstrom~\cite{rosvall07} envisioned
a communication process in which a partition of a graph in communities represents a
synthesis $Y$ of the full structure that a signaler sends to a receiver, who tries to infer the original
graph topology $X$ from it (Fig.~\ref{rosvall1}). The same idea is at the basis of an earlier method by Sun et al.~\cite{sun07}, which 
was originally designed for bipartite graphs evolving in time and will be described in Section~\ref{sec7_1_2}. 
The best partition corresponds to the signal $Y$ that contains the most information about
$X$. This can be quantitatively assessed by the minimization of the conditional information $H(X|Y)$ of $X$ given $Y$,
\begin{equation}
H(X|Y)=\log\left[\prod_{i=1}^q
\binom{n_i(n_i-1)/2}{l_{ii}}
\prod_{i>j}\binom{n_in_j}{l_{ij}}\right],
\label{eqstinf7}
\end{equation}
where $q$ is the number of clusters, $n_i$ the number of vertices in cluster $i$, $l_{ij}$ the number of edges between clusters $i$ and $j$.
We remark that, if one imposes no constraints on $q$, $H(X|Y)$ is minimal in the trivial case in which $X=Y$ ($H(X|X)=0$). This solution is not
acceptable because it does not correspond to a compression of information with respect to the original data set. One has to look for the ideal 
tradeoff between a good compression and a small enough information $H(X|Y)$. The Minimum Description Length (MDL) principle~\cite{rissanen78,grunwald05} 
provides a solution to this problem, which amounts to the minimization of a function given by $H(X|Y)$ plus a function of the number $n$ of 
vertices, $m$ of edges and $q$ of clusters.
The optimization is performed by simulated annealing, so the method is rather slow and can 
be applied to graphs with up to about $10^4$ vertices. However, faster techniques may in principle be used, even if they imply a loss
in accuracy. 
The method appears superior than modularity optimization, especially when communities are of different sizes. This 
comes from tests performed on the benchmark of Girvan and Newman~\cite{girvan02} (Section~\ref{sec6_1}), both in its original version
and in asymmetric versions, proposed by the authors, where the clusters have different sizes or different average degrees. 
In addition, it can detect other types of vertex classifications than communities, as in Eq.~\ref{eqstinf7} there are
no constraints on the relative importance of the edge densities within communities with respect to the edge densities between communities.
The software of the algorithm can be found at {\tt http://www.tp.umu.se/$\sim$rosvall/code.html}.

In a recent paper~\cite{rosvall08}, Rosvall and Bergstrom pursued the same idea of describing a graph by using less information
than that encoded in the full adjacency matrix. The goal is to optimally compress the information needed to describe
the process of information diffusion across the graph. Random walk is chosen as a proxy of information diffusion. A two-level description, 
in which one gives unique names to important structures of the graph and to vertices within the same structure, but the vertex names
are recycled among different structures, leads to a more compact description than by simply coding all vertices with different names.
This is similar to the procedure usually adopted in geographic maps, where the structures are cities and one usually chooses the same names
for streets of different cities, as long as there is only one street
with a given name in the same city. 
Huffman coding~\cite{huffman52} 
is used to name vertices. For the random walk, the above-mentioned structures are communities, as 
it is intuitive that walkers will spend a lot of time within them, so they play a crucial role in the process of 
information diffusion. 
Graph clustering turns then into the following coding problem: finding the partition that yields the minimum 
description length of an infinite random walk. Such description length consists of two terms, expressing the Shannon 
entropy of the random walk within and between clusters. Every time the walker 
steps to a different cluster, one needs to use the codeword of that cluster in the description, to inform the decoder 
of the transition\footnote{Instead, for a one-level description, in which all vertices have different names, it is enough to specify 
the codeword of the vertex reached at every step to completely define the process, but this may be costly.}.
Clearly, if clusters are well separated from each other, 
transitions of the random walker between clusters will be unfrequent, so it is advantageous to use the map, with the clusters
as regions, because in the description of the random walk
the codewords of the clusters will not be repeated many times,
while there is a considerable saving in the description due to the 
limited length of the codewords used to denote the vertices. Instead, if there are no well-defined clusters and/or if the partition is not
representative of the actual community structure of the graph, transitions between the clusters of the partition will be 
very frequent and there will be little or no gain by using the two-level description of the map.
The minimization of the description length is carried out by combining greedy search with simulated annealing.
In a successive paper~\cite{rosvall09}, the authors adopted the fast greedy technique designed by Blondel et al. for modularity
optimization~\cite{blondel08}, with some refinements.
The method can be applied to weighted graphs, both undirected and directed. In the latter case, the random walk process is modified
by introducing a teleportation probability $\tau$, to guarantee ergodicity, just like in Google's PageRank algorithm~\cite{brin98}.
The partitions of directed graphs 
obtained by the method differ from those derived by optimizing the directed version of 
Newman-Girvan modularity (Section~\ref{sub_sec6_01}): this is due to the fact that modularity focuses on pairwise relationships between vertices,
so it does not capture flows.
The code of the method is available at {\tt http://www.tp.umu.se/$\sim$rosvall/code.html}.

Chakrabarti~\cite{chakrabarti04} has applied the MDL principle to put the adjacency matrix of a graph into the (approximately) block diagonal
form representing the best tradeoff between having a limited number of blocks, for a good compression of the graph topology, and 
having very homogeneous blocks, for a compact description of their structure. The total encoding cost $T$ includes the information on the
total number of vertices of the graph, on the number of blocks and the number of vertices and edges in each block, along with the adjacency matrices
of the blocks. The minimization of $T$ is carried out by starting from the partition in which the graph is 
a single cluster. At each step, one operates a bipartition of the cluster of the partition with the maximum Shannon entropy per vertex.
The split is carried out in order to remove from the original cluster those vertices carrying the highest contribution to
the entropy per vertex of the cluster. Then, starting from the resulting partition, which has one more cluster than the previous one,
$T$ is optimized among those partitions with the same number of clusters. The procedure continues until one
reaches a number of clusters $k^\star$, for which $T$ cannot be further decreased.
The method by Chakrabarti has complexity
$O[I(k^\star)^2m]$, where $I$ is the number of iterations required for the convergence of the optimization 
for a given number of clusters, which is usually small ($I\leq 20$ in the 
experiments performed by the author). Therefore the algorithm can be applied to fairly large graphs.

Information theory has also been used to detect communities in graphs. Ziv et al.~\cite{ziv05} have designed a method 
in which the information contained in the graph topology is compressed such to preserve some predefined information.
This is the basic principle of the information bottleneck method~\cite{tishby99}. To understand this criterion, we need to introduce
an important measure, the {\it mutual information} $I(X,Y)$~\cite{mackay03} of two random variables $X$ and $Y$. It is defined as
\begin{equation}
I(X,Y)=\sum_{x}\sum_{y}P(x,y)\log\frac{P(x,y)}{P(x)P(y)},
\label{eqstinf8}
\end{equation}
where $P(x)$ indicates the probability that $X=x$ (similarly for $P(y)$) and $P(x,y)$ is the joint probability 
of $X$ and $Y$, i. e. $P(x,y)=P(X=x,Y=y)$. 
The measure $I(X,Y)$ tells how much we learn about $X$ if we know $Y$, and viceversa. If $X$ is the input variable, $Z$ the variable 
specifying the partition and $Y$ the variable encoding the information we want to keep, which is called {\it relevant variable},
the goal is to minimize the mutual information between $X$ and $Z$ (to achieve the largest possible data compression), under the constraint 
that the information on $Y$ extractable from $Z$ be accurate. 
The optimal tradeoff between the values of $I(X,Z)$ and $I(Y,Z)$ (i.~e. compression versus accuracy) 
is expressed by the minimization of a functional, where the 
relative weight of the two contributions is given by a parameter playing the role of a temperature.
In the case of graph clustering, the question is what to choose as relevant information variable. Ziv et al. proposed to 
adopt the structural information encoded in the process of diffusion on the graph. They also introduce the concept of 
{\it network modularity}, which characterizes the graph as a whole, not a specific partition like the modularity by Newman and Girvan
(Section~\ref{sec3_2_2}). The network modularity is defined as the area under the {\it information curve}, which essentially 
represents the relation between
the extent of compression and accuracy for all solutions found by the method and all possible numbers of clusters.
The software of the algorithm by Ziv et al. can be found at {\tt http://www.columbia.edu/$\sim$chw2/}.

\section{Alternative methods}
\label{sub_sec6_2}

In this section we describe some algorithms that do not fit in the previous categories, although 
some overlap is possible.

Raghavan et al.~\cite{raghavan07} have designed a simple and fast method based on {\it label propagation}.
Vertices are initially given unique labels (e.g. their vertex labels). At each iteration, a sweep over all vertices, in random sequential order,
is performed: each vertex takes the 
label shared by the majority of its neighbors. If there is no unique majority, one of the majority labels is picked at random.
In this way, labels propagate across the graph: most labels will disappear, others will dominate. The process reaches convergence
when each vertex has the majority label of its neighbors. Communities are defined as groups of vertices having identical labels at convergence.
By construction, each vertex has more neighbors in its community than in any other community. This resembles the 
strong definition of community we have discussed in Section~\ref{sec3_1_2}, although the latter is stricter, in that
each vertex must have more neighbors in its community than in the rest of the graph. The algorithm does not deliver a unique solution.
Due to the many ties encountered along the process it is possible to derive different partitions starting from the same initial condition,
with different random seeds. Tests on real graphs show that all partitions found are similar to each other, though. The most precise
information that one can extract from the method is contained by aggregating the various partitions obtained, which can be done
in various ways. The authors proposed to label each vertex with the set of all labels it has in different partitions.
Aggregating partitions enables one to detect possible overlapping communities.
The main advantage of the method is the fact that it does not need any information on the number and the size of the clusters.
It does not need any parameter, either.
The time complexity of each iteration of the algorithm is $O(m)$, the number of iterations to convergence
appears independent of the graph size, or growing very slowly with it. So the technique is really fast and could be used for the analysis of large
systems. In a recent paper~\cite{tibely08}, Tib\'ely and Kert\'esz showed that the method is equivalent to finding the local energy
minima of a simple zero-temperature kinetic Potts model, 
and that the number of such energy minima is considerably larger than the number of 
vertices of the graph. Aggregating partitions as Raghavan et al. suggest leads to a fragmentation of the 
resulting partition in clusters that are the smaller,
the larger the number of aggregated partitions. This is 
potentially a serious problem of the algorithm by Raghavan et al., especially when large graphs are investigated.
In order to eliminate undesired solutions, Barber and Clark introduced some constraints in the optimization process~\cite{barber09}.
This amounts to adding some terms to the objective function ${\cal H}$ whose maximization is equivalent to
the original label propagation algorithm\footnote{${\cal H}=1/2\sum_{ij}A_{ij}\delta_{ij}$, where ${\bf A}$ is the adjacency matrix of the graph and
$\delta$ is Kronecker's function. It is just the negative of the energy of a zero-temperature Potts model,
as found by Tib\'ely and Kert\'esz~\cite{tibely08}}. Interestingly, if one imposes the constraint
that partitions have to be balanced, i. e. that clusters have similar total degrees, the objective function
becomes formally equivalent to Newman-Girvan modularity $Q$ (Section~\ref{sec3_2_2}), so the corresponding 
version of the label propagation algorithm is essentially based on a local optimization of modularity.
Leung et al. have found that the original algorithm by Raghavan et al., applied on 
online social networks, often yields partitions with one giant community together with much smaller ones~\cite{leung09}.
In order to avoid this disturbing feature, which is an artefact of the algorithm, Leung et al. proposed to modify the method by introducing
a score for the labels, which decreases as the label propagates far from the vertex to which the label was originally assigned. 
When choosing the label of a vertex, the labels of its neighbors are weighted by their scores, therefore
a single label cannot span too large portions of the graph (as its weight fades away with the distance from the origin), 
and no giant communities can be recovered.
Tests of the modified algorithm 
on the LFR benchmark~\cite{lancichinetti08} (Section~\ref{sub_sec6_201}) give good results and encourage further
investigations. 

Bagrow and Bollt designed an agglomerative technique, called {\it L-shell method} \cite{bagrow05}.
It is a procedure that finds the community of any vertex, although the authors also presented a more general procedure 
to identify the full community structure of the graph.
Communities are defined locally, based on a simple criterion involving the number of edges inside and outside a group
of vertices. One starts from a vertex-origin and keeps adding vertices lying on successive shells, where a shell is defined as
a set of vertices at a fixed geodesic distance from the origin. The first shell includes 
the nearest neighbours of the origin, the second the next-to-nearest neighbours, and so on.
At each iteration, one calculates the number of edges connecting vertices of the new 
layer to vertices inside and outside the running cluster. If the ratio of these two numbers (``emerging degree'')
exceeds some predefined threshold, the vertices of the new shell are added to the cluster, otherwise the process stops.
The idea of closing a community by expanding a shell has been previously introduced by Costa~\cite{costa04}, in which 
shells are centered on hubs. However, in this procedure the number of clusters is preassigned and no cluster can contain more than one hub.
Because of the local nature of the process, the L-shell method is very fast and can identify communities
very quickly. Unfortunately the method works well only when the source vertex is approximately equidistant from
the boundary of its community. To overcome this problem, Bagrow and Bollt 
suggested to repeat the process starting from every vertex and derive a {\it membership matrix} $M$: the element
$M_{ij}$ is one if vertex $j$ belongs to the community of vertex $i$, otherwise it is zero.
The membership matrix can be rewritten by suitably permutating rows and columns based on their mutual distances.
The distance between two rows (or columns) is defined as the number of entries whose elements differ.
If the graph has a clear community structure, the membership matrix takes a block-diagonal 
form, where the blocks identify the communities. The method enables one to detect overlaps between communities as well~\cite{porter07}.
Unfortunately, the rearrangement of the matrix 
requires a time $O(n^3)$, so it is quite slow. A variant of the algorithm by Bagrow and Bollt, in which boundary vertices 
are examined separately and both first and second nearest neighbors of the running community 
are simultaneously investigated, was suggested by Rodrigues et al.~\cite{rodrigues07}.

A recent methodology introduced by Papadopoulos et al.~\cite{papadopoulos09}, called
{\it Bridge Bounding}, is similar to the L-shell algorithm, but here the cluster around a vertex grows until one ``hits''
the boundary edges. Such edges can be recognized from the values of various measures, like betweenness~\cite{girvan02} or 
the edge clustering coefficient~\cite{radicchi04}. The problem is that there are often no clear gaps in the distributions of the 
values of such measures, so one is forced to set a threshold to automatically 
identify the boundary edges from the others, and there is no obvious way to do it. The best
results of the algorithm are obtained by using a measure consisting of a weighted sum of the edge clustering coefficient over
a wider neighborhood of the given edge. This version of the method has a time complexity $O(\langle k\rangle^2m+\langle k\rangle n)$, where
$\langle k\rangle$ is the average degree of the graph.

\begin{figure}
\begin{center}
\includegraphics[width=9cm]{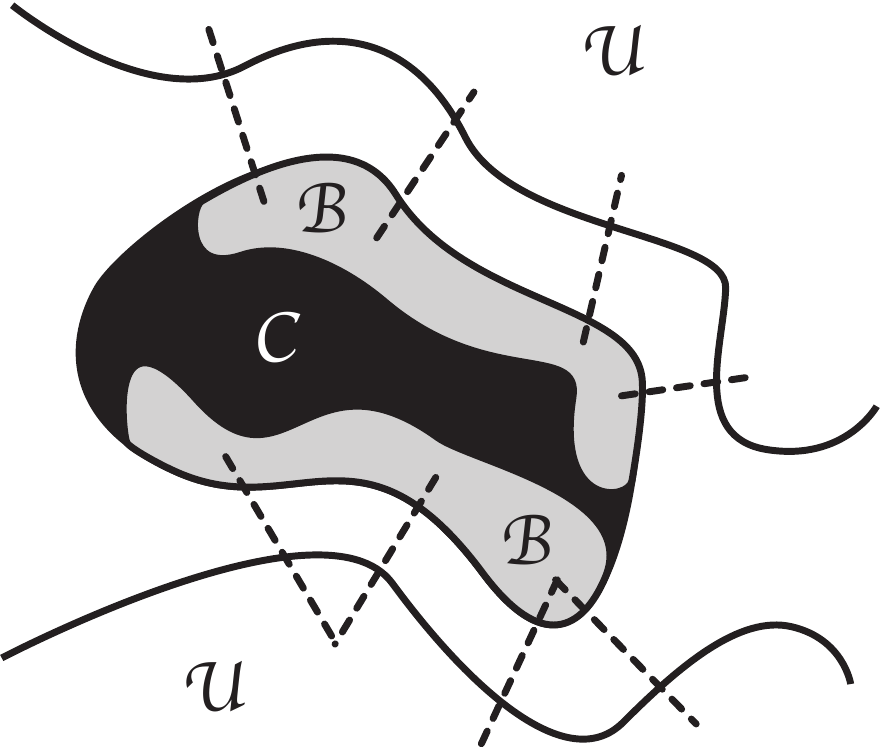}
\caption {\label{clausetloc} Schematic picture of a community ${\cal C}$
used in the definition of local modularity by Clauset~\cite{clauset05}. The black area indicates the subgraph of ${\cal C}$
including all vertices of ${\cal C}$, whose neighbors are also in ${\cal C}$. The boundary ${\cal B}$ entails 
the vertices of ${\cal C}$ with at least one neighbor outside the community.  
Reprinted figure with permission from 
Ref.~\cite{clauset05}. \copyright 2005 by the American Physical Society.}
\end{center}
\end{figure}

In another algorithm by Clauset, local communities 
are discovered through greedy maximization of a local modularity measure~\cite{clauset05}. Given a community ${\cal C}$,
the boundary ${\cal B}$ is the set of vertices of ${\cal C}$ with at least one neighbor outside ${\cal C}$ (Fig.~\ref{clausetloc}). 
The local modularity $R$ by Clauset
is the ratio of the number of edges having both endpoints in ${\cal C}$ (but at least one in ${\cal B}$), with the number of edges 
having at least one endpoint in ${\cal B}$. It is a measure of the sharpness of the community boundary. 
Its optimization consists of a local exploration of the community starting from a source vertex: at each step
the neighboring vertex yielding the largest increase (smallest decrease) of $R$ is added, until the community has reached a predefinite
size $n_c$. This greedy optimization takes a time $O(n_c^2\langle k\rangle)$, where $\langle k\rangle$ is the average degree of the graph.  
The local modularity $R$ has been used in a paper by Hui et al.~\cite{hui07}, where methods to find communities in 
networks of mobile devices are designed.

Another method, where communities are defined based on 
a local criterion, was presented by Eckmann and Moses~\cite{eckmann02}. 
The idea is to use the clustering coefficient~\cite{watts98} of a vertex as a 
quantity to distinguish tightly connected groups of vertices. Many edges mean many loops inside a community,
so the vertices of a community are likely to have a large clustering coefficient. The latter 
can be related to the average distance between pairs of neighbours of the vertex.
The possible values of the distance are $1$ (if neighbors are connected)
or $2$ (if they are not), so the average distance lies between $1$ and $2$. The more triangles there are in
the subgraph, the shorter the average distance. Since each vertex always has distance $1$ from its neighbours,
the fact that the average distance between its neighbours is different from $1$ reminds what happens when
one measures segments on a curved surface. Endowed with a metric, represented by the geodesic distance between
vertices/points, and a curvature, the graph can be embedded in a geometric space. Communities appear as
portions of the graph with a large curvature. The algorithm was applied to the graph representation of 
the World Wide Web, where vertices are web pages and edges are the hyperlinks that take users from
a page to the other. The authors found that communities correspond to web pages dealing with the same topic.

Long et al. have devised an interesting technique that is able to detect various types of vertex groups, not
necessarily communities~\cite{long07}. The method is based on graph approximation, as it tries to 
match the original graph topology onto a coarse type of graph, the {\it community prototype graph}, which 
has a clear group structure (block-diagonal for clusters, block-off-diagonal for classes of multipartite graphs, etc.).
The goal is to determine the community prototype graph that best approximates the graph at study, where the 
goodness of the approximation is expressed by the distance between the corresponding matrices. In this way the original problem 
of finding graph subsets becomes an optimization problem. Long et al. called this procedure {\it Community Learning by Graph
Approximation} (CLGA). Sometimes  
the minimization of the matrix distance can be turned into the maximization of the trace of a matrix.
Measures like cut size or ratio cut can be also formulated as the trace of matrices
(see for instance Eq.~\ref{eqr}). In fact, CLGA includes traditional graph partitioning as a special case (Section~\ref{sec4_1}).
Long et al. designed three algorithms for CLGA: two of them seek for 
divisions of the graph into overlapping or non-overlapping groups, respectively; 
in the third one an additional constraint is introduced to produce groups of comparable size. 
The complexity of these algorithms is $O(tn^2k)$, where $t$ is the number of iterations until the optimization converges and 
$k$ the number of groups. The latter has to be given as an input, which is a serious limit of CLGA.
  
A fast algorithm by Wu and Huberman
identifies communities based on the properties of resistor networks~\cite{wu04}. 
It is essentially a method for partitioning graphs in two parts, similar to spectral bisection,   
although partitions in an arbitrary number of communities can be obtained by iterative applications.
The graph is transformed into a resistor network where each edge has unit resistance.
A unit potential difference is set between two randomly chosen vertices. The idea is that, if 
there is a clear division in two communities of the graph, there will be a visible
gap between voltage values for vertices at the borders between the clusters.
The voltages are calculated by solving Kirchoff's equations: an exact solution would be too 
time consuming, but it is possible to find a reasonably good approximation in a linear time for a sparse graph with
a clear community structure, so the more time consuming part of the algorithm is the sorting of the voltage values, which
takes time $O(n\log n)$. Any possible vertex pair can be chosen to set the initial potential difference, so the 
procedure should be repeated for all possible vertex pairs. The authors showed that this is not necessary, and that
a limited number of sampling pairs is sufficient to get good results, so the algorithm scales as $O(n\log n)$ and is very fast.
An interesting feature of the method is that it can quickly find the natural community of any vertex,
without determining the complete partition of the graph. For that, one uses the vertex
as source voltage and places the sink at an arbitrary vertex.
The same feature is present 
in an older algorithm by Flake et al.~\cite{flake02}, where one uses max-flow instead of current flow (Section~\ref{sec4_1}). 
An algorithm by Orponen and Schaeffer~\cite{orponen05} is based on the same principle, but it does not need the specification of target
sources as it is based on diffusion in an unbounded medium.
The limit of such methods is the fact that one has to give as input the number of clusters, which is usually not known 
beforehand.

Ohkubo and Tanaka~\cite{ohkubo06} pointed out that, since communities are rather compact structures, they should have a small volume,
where the volume of a community is defined as the ratio of the number of vertices by the internal edge density of 
the community. Ohkubo and Tanaka assumed that the sum $V_{total}$ of the volumes of the communities of a partition 
is a reliable index of the goodness of the partition. So, the most relevant partition is the one minimizing $V_{total}$.
The optimization is carried out with simulated annealing.

Zarei and Samani~\cite{zarei09} remarked that there is a symmetry between community structure and anti-community (multipartite)
structure, when one considers a graph and its complement, whose edges are the missing edges of the original graph.
In fact, if a graph has a well identified communities, the same groups of vertices would be strong anti-communities in the complement graph, i. e.
they should have a few intra-cluster edges and many inter-cluster edges. Based on this remark, 
the communities of a graph can be identified by looking for anticommunities in the complement graph, which can sometimes be easier.
Zarei and Samani devised a spectral method using matrices of the complement graph. The results of this technique appear good as compared to
other spectral methods on artificial graphs generated with the planted $\ell$-partition model~\cite{condon01}, as well as on 
Zachary's karate club~\cite{zachary77}, Lusseau's dolphins' network~\cite{lusseau03} and a network of protein-protein interactions.
However, the authors have used very small graphs for testing. Communities make sense on sparse graphs, but
the complements of large sparse graphs would not be sparse, but very dense, and their community (multipartite) structure basically invisible.

Gudkov and Montealegre detected communities by means of 
dynamical simplex evolution~\cite{gudkov08}. Graph vertices are represented as points in an $(n-1)$-dimensional space. 
Each point initially sits on the $n$ vertices of a simplex, and then moves in space due to forces exerted by the other points.
If vertices are neighbors, the mutual force acting on their representative points is 
attractive, otherwise it is repulsive. If the graph has a clear community structure, the corresponding spatial
clusters repel each other because of the few connections between them (repulsion dominates over attraction). If communities
are more mixed with each other, clusters are not well separated and they could be mistakenly aggregated in larger structures. To avoid that,
Gudkov and Montealegre defined  
clusters as groups of points such that the distance between each pair of points does not
exceed a given threshold, which can be arbitrarily tuned, to reveal structures at different resolutions (Section~\ref{sub_sec6_201}).
The algorithm consists in solving first-order differential equations, describing the dynamics of mass points moving in a viscous medium.
The complexity of the procedure is $O(n^2)$. Differential equations are also at the basis 
of a recent method designed by Krawczyk and Ku{\l}akowski~\cite{krawczyk07,krawczyk08}. Here the equations describe a dynamic process,
in which the original graph topology evolves to a disconnected graph, whose components are the clusters of the original graph.

Despite the significant improvements in computational complexity, it is still problematic to apply 
clustering algorithms to many large networks available today.
Therefore Narasimhamurthy et al.~\cite{narasimhamurthy08} proposed a two-step procedure: first, the graph at study is 
decomposed in smaller pieces by a fast graph partitioning technique; then, a clustering method is applied to each of the smaller subgraphs
obtained [Narasimhamurthy et al. used the Clique Percolation Method (Section~\ref{sec45_1})].
The initial decomposition of the graph is carried out through the multilevel
method by Dhillon et al.~\cite{dhillon07}. It is crucial to verify that the initial partitioning does not
split the communities of the graph among the various subgraphs of the decomposition. 
This can be done by comparing, on artificial graphs,
the final clusters obtained with the two-step method with those detected 
by applying the chosen clustering technique to the entire graph.

\section{Methods to find overlapping communities}
\label{sec45}

Most of the methods discussed in the previous sections aim at detecting standard partitions, i.~e. partitions in which 
each vertex is assigned to a single community. However, in real graphs vertices are often 
shared between communities (Section~\ref{sec2}), and the issue of detecting 
overlapping communities has become quite popular in the last few years. 
We devote this section to the main techniques to detect
overlapping communities. 

\subsection{Clique percolation}
\label{sec45_1}

\begin{figure}
\begin{center}
\includegraphics[width=\columnwidth]{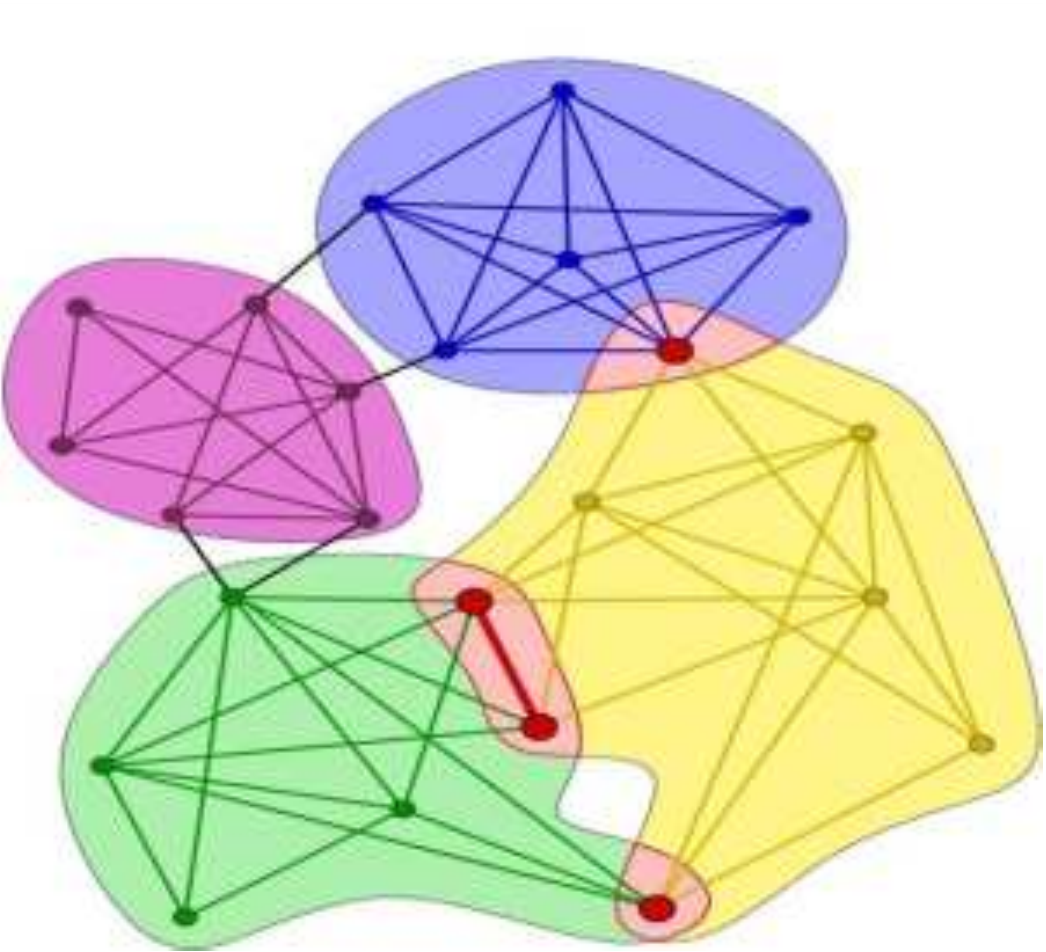}
\caption {\label{Figure10} Clique Percolation Method. The example shows communities spanned by 
adjacent 4-cliques. Overlapping vertices are shown by the bigger dots. Reprinted figure with 
permission from Ref.~\cite{palla05}. \copyright 2005 by the Nature
Publishing Group.}
\end{center}
\end{figure}

The most popular technique is the Clique Percolation Method (CPM)
by Palla et al.~\cite{palla05}. It is based on the concept that the internal edges of a community are likely
to form cliques due to their high density. On the other hand, it is unlikely that intercommunity edges
form cliques: this idea was already used in the divisive method of Radicchi et al. (Section~\ref{subsec5_2}).
Palla et al. use the term $k$-clique to indicate a complete graph with $k$ vertices\footnote{In graph theory
the k-clique by Palla et al. is simply called clique, or complete graph, with $k$ vertices (Section~\ref{sec1_1}).}.
Notice that a $k$-clique is different from the
$n$-clique (see Section~\ref{sec3_1_2}) used in social science.
If it were possible for a clique to move on a graph, in some way, it would probably get trapped
inside its original community, as it could not cross the bottleneck formed by the intercommunity edges.
Palla et al. introduced a number of concepts to implement this idea.
Two $k$-cliques are {\it adjacent} if
they share $k-1$ vertices. The union of adjacent $k$-cliques is called {\it $k$-clique chain}.
Two $k$-cliques are connected if they are part of a $k$-clique chain. Finally, a {\it $k$-clique
community} is the largest connected subgraph obtained by the union of a $k$-clique and 
of all $k$-cliques which are connected to it. Examples of $k$-clique communities are shown in Fig.~\ref{Figure10}.
One could say that a $k$-clique community is identified by making  
a $k$-clique ``roll'' over adjacent $k$-cliques, where rolling 
means rotating a $k$-clique about the $k-1$ vertices it shares with any adjacent $k$-clique.
By construction, $k$-clique communities can share vertices, so they can be overlapping.
There may be vertices belonging to non-adjacent $k$-cliques, which could be reached 
by different paths and end up in different clusters. Unfortunately, there are also vertices 
that cannot be reached by any $k$-clique, like, e. g. vertices with degree one (``leaves'').
In order to find $k$-clique communities,
one searches first for maximal cliques. Then a clique-clique overlap matrix ${\bf O}$ is built~\cite{everett98}, 
which is an $n_c\times n_c$ matrix, $n_c$ being the number of cliques; $O_{ij}$ is the number of vertices
shared by cliques $i$ and $j$. To find $k$-cliques, one needs simply to keep the entries of ${\bf O}$
which are larger than or equal to $k-1$, set the others to zero and find the connected components of the resulting matrix.
Detecting maximal cliques is known to require a running time
that grows exponentially with the size of the graph. However, the authors found that, for 
the real networks they analyzed, the procedure is quite fast, due to the fairly limited number of cliques,
and that (sparse) graphs
with up to $10^5$ vertices can be analyzed in a reasonably short time. The actual scalability of the algorithm 
depends on many factors, and cannot be expressed in closed form. 
An interesting aspect of $k$-clique communities is that they allow to make a clear distinction between random graphs and 
graphs with community structure. This is a rather delicate issue: we have seen in Section~\ref{sub_sec6_1} that Newman-Girvan modularity
can attain large values on random graphs. 
Der\'enyi et al.~\cite{derenyi05} have studied the percolation properties of $k$-cliques on random graphs, when the edge probability
$p$ varies. They found that 
the threshold $p_c(k)$ for the emergence of a giant $k$-clique community, i.~e. a community occupying a macroscopic portion of the graph, is 
$p_c(k)=[(k-1)n]^{-1/(k-1)}$, $n$ being the number of vertices of the graph, as usual. For $k=2$, for which the $k$-cliques reduce to edges,
one recovers the known expression for the emergence of a giant connected component in Erd\"os-R\'enyi graphs (Section~\ref{sec1_3}). 
This percolation transition is quite sharp: if the edge probability $p<p_c(k)$, $k$-clique communities are rather small; if $p>p_c(k)$ there is 
a giant component and many small communities. To assess the significance of the clusters found with the CPM, one 
can compare the detected cover\footnote{We remind that {\it cover} is the equivalent of partition for overlapping communities.} 
with the cover found on a null model graph, which is random but preserves
the expected degree sequence of the original graph. The modularity of Newman and Girvan is based on the same null model (Section~\ref{sec3_2_2}).
The null models of real graphs seem to display the same two scenarios found for Erd\"os-R\'enyi graphs, characterized by the 
presence of very small $k$-clique communities, with or without a giant cluster. Therefore, covers with $k$-clique communities of large or appreciable size
can hardly be due to random fluctuations. Palla and coworkers~\cite{adamcsek06} have designed a software package implementing the CPM, called
{\it CFinder}, which is freely available ({\tt www.cfinder.org}). 

The algorithm has been extended 
to the analysis of weighted, directed and bipartite graphs. For weighted graphs, in principle one can 
follow the standard procedure of thresholding the weights, and apply the method on the resulting graphs, treating them as unweighted.
Farkas et al.~\cite{farkas07} proposed instead to threshold the weight of cliques, defined as the geometric mean of the weights 
of all edges of the clique. The value of the threshold is chosen slightly above the critical value at which a giant
$k$-clique community emerges, in order to get the richest possible variety of clusters. On directed graphs, 
Palla et al. defined {\it directed $k$-cliques} as complete graphs with $k$ vertices, such that there is 
an ordering among the vertices, and each edge goes from a vertex with higher order to one with lower order. The ordering is determined 
from the {\it restricted outdegree} of the vertex, expressing the fraction of outgoing edges pointing to the other vertices of the clique
versus the total outdegree. The method has been extended to bipartite graphs by Lehmann et al.~\cite{lehmann08}. In this case one 
uses bipartite cliques, or {\it bicliques}: a subgraph $K_{a,b}$ is a biclique if each of $a$ vertices of one class are connected
with each of $b$ vertices of the other class. Two cliques $K_{a,b}$ are adjacent if they share a clique
$K_{a-1,b-1}$, and a $K_{a,b}$ clique community is the union of all $K_{a,b}$ cliques that can be reached from each other
through a path of adjacent $K_{a,b}$ cliques. Finding all $N_c$ bicliques of a graph is an ${\bf NP}$-complete problem~\cite{peeters03},
mostly because the number of bicliques tends to grow exponentially with the size of the graph.  
The algorithm designed by Lehmann et al. to find biclique communities
is similar to the original CPM, and has a total complexity of $O(N_c^2)$.
On sparse graphs, $N_c$ often grows linearly with the number of edges $m$, yielding a time complexity $O(m^2)$. 
Bicliques are also the main ingredients of {\it BiTector}, a recent algorithm to detect community structure in 
bipartite graphs~\cite{du08}.

Kumpula et al. have developed a fast implementation of the CPM, called Sequential Clique Percolation algorithm (SCP)~\cite{kumpula08}.
It consists in detecting $k$-clique communities by sequentially inserting the edges of the graph at study, one by one, 
starting from an initial empty graph. Whenever a new edge is added, one checks whether new $k$-cliques are formed, by searching for
$(k-2)$-cliques in the subset of neighboring vertices of the endpoints of the inserted edge.
The procedure requires to build a graph $\Gamma^{*}$, in which the vertices are $(k-1)$-cliques and edges are set between
vertices corresponding to $(k-1)$-cliques which are subgraphs of the same $k$-clique. At the end of the process, the connected components of
$\Gamma^{*}$ correspond to the searched $k$-clique communities. The technique has a time complexity which is linear in the number of $k$-cliques
of the graph, so it can vary a lot in practical applications. Nevertheless, it turns out to be much faster than the original implementation of the 
CPM. The big advantage of the SCP, however, consists of its implementation for weighted graphs. By inserting edges in decreasing order of weight,
one recovers in a single run the community structure of the graph for all possible weight thresholds, by storing every cover detected 
after the addition of each edge. The standard CPM, instead, needs to be applied once for each threshold. If, instead of 
edge weight thresholding, one performs $k$-clique weight thresholding, as
prescribed by Farkas et al.~\cite{farkas07}, the SCP remains much faster than the CPM, 
if one applies a simple modification to it, consisting in detecting and storing all
$k$-cliques on the full graph, sorting them based on their weights, and finding the communities by sequentially adding the
$k$-cliques in decreasing order of weight.

The CPM has the same limit as the algorithm of Radicchi et al.~\cite{radicchi04} (Section~\ref{subsec5_2}): 
it assumes that the graph has a large number of cliques, so it may fail to give meaningful covers
for graphs with just a few cliques, like technological networks and some social networks. On the other hand,
if there are many cliques, the method may deliver trivial community structure, like a cover consisting of the 
whole graph as a single cluster. A more fundamental issue is the fact that the method does not look for 
actual communities, consistent with the shared notion of dense subgraphs, but for subgraphs ``containing'' many cliques,
which may be quite different objects than communities (for instance, they could be ``chains'' of cliques with low internal edge density). 
Another big problem is that on real networks
there is a considerable fraction of vertices that are left out of the communities, like leaves.
One could think of some post-processing procedure to include them in the communities, but for that it is necessary
to introduce a new criterion, outside the framework that inspired the method. 
Furthermore it is not clear {\it a priori}
which value of $k$ one has to choose to identify meaningful structures.
Finally, the criterion to choose the threshold 
for weighted graphs and the definition of directed $k$-cliques are rather arbitrary.

\subsection{Other techniques}
\label{sec45_2}

One of the first methods to find overlapping communities was designed by Baumes et al.~\cite{baumes05}. 
A community is defined as a subgraph which locally optimizes a given function $W$, typically some measure related to the edge density
of the cluster\footnote{Community definitions based on local 
optimization are adopted in other algorithms as well, like that by Lancichinetti et al.~\cite{lancichinetti09}
(Section~\ref{sub_sec6_201}).}. Different overlapping subsets 
may all be locally optimal, so vertices can be shared between communities. Detecting the cluster structure of a graph
amounts to finding the set of all locally optimal clusters. Two efficient heuristics are proposed, called Iterative Scan (${\tt IS}$)
and Rank Removal (${\tt RaRe}$).
${\tt IS}$ performs a greedy optimization of the function $W$. One starts from a random seed vertex/edge and adds/deletes vertices one by one as long as
$W$ increases. Then another seed is randomly picked and the procedure is repeated. The algorithm stops when, by picking any seed, one recovers
a previously identified cluster. ${\tt RaRe}$ consists in removing important vertices such to disconnect the graphs in small components
representing the cores of the clusters. The importance of vertices is determined by their centrality scores 
(e.g. degree, betweenness centrality~\cite{freeman77}), PageRank~\cite{brin98}). Vertices are removed until one fragments the graph into 
components of a given size. After that, the removed vertices are added again to the graph, and are associated to those clusters for which
doing so increases the value of the function $W$. The complexity of ${\tt IS}$ and ${\tt RaRe}$ is $O(n^2)$ on sparse graphs. The best performance is achieved
by using ${\tt IS}$ to refine results obtained from ${\tt RaRe}$. In a successive paper~\cite{baumes05a}, Baumes et al. further improved
such two-step procedure, in that the removed vertices in ${\tt RaRe}$ are reinserted in decreasing order of their centrality scores, and the 
optimization of $W$ in ${\tt IS}$ is only extended to neighboring vertices of the running cluster. The new 
recipe maintains time complexity $O(n^2)$, but on sparse graphs it requires a time lower by an order of magnitude than the old one, while the
quality of the detected clustering is comparable.

A different method, combining spectral mapping, fuzzy clustering and the optimization of a quality function, has been 
presented by Zhang et al.~\cite{zhang07}. The membership of vertex $i$ in cluster $k$  
is expressed by $u_{ik}$, which is a number between $0$ and $1$. The sum of the $u_{ik}$ over all communities $k$ of a cover
is $1$, for every vertex. This normalization is suggested by the fact that the entry $u_{ik}$ can be thought of as the
probability that $i$ belongs to community $k$, so the 
sum of the $u_{ik}$ represents the probability that the vertex belongs to any community of the cover, which is necessarily $1$.
If there were no overlaps, $u_{ik}=\delta_{k_ik}$, where $k_i$ represents the unique community of
vertex $i$. The algorithm consists of three phases: 1) embedding vertices in Euclidean space; 2) grouping the corresponding vertex points in
a given number $n_c$ of clusters; 3) maximizing a modularity function over the set of covers found in step 2), corresponding to different values of $n_c$.
This scheme has been used in other techniques as well, like in the algorithm of Donetti and Mu\~noz~\cite{donetti04} (Section~\ref{sec43}). 
The first step builds upon a spectral technique introduced by White and Smyth~\cite{white05}, that we have
discussed in Section~\ref{sub_sec6_0_4}. Graph vertices are embedded in a $d$-dimensional Euclidean space by using the top $d$ 
eigenvectors of the right stochastic matrix ${\bf W}$ (Section~\ref{sec1_2}), 
derived from the adjacency matrix ${\bf A}$ by dividing each element by the sum of the elements 
of the same row. The spatial coordinates of vertex $i$ are the $i$-th components of the eigenvectors. In the second step, the vertex points 
are associated to $n_c$ clusters by using fuzzy $k$-means clustering~\cite{dunn74, bezdek81} (Section~\ref{sec4_3}).  
The number of clusters $n_c$ varies from $2$ to a maximum $K$, so one obtains $K-1$ covers. The best cover is the one that yields the largest value of 
the modularity $Q_{ov}^{zh}$, defined as
\begin{equation}
Q_{ov}^{zh}=\sum_{c=1}^{n_c}\Big[\frac{\bar{W}_c}{W}-\left(\frac{\bar{S}_c}{2W}\right)^2\Big],
\label{eq:mod5}
\end{equation}
where
\begin{equation}
\bar{W}_c=\sum_{i,j\in V_c}\frac{u_{ic}+u_{jc}}{2}w_{ij}, 
\label{eq:mod6}
\end{equation}
and
\begin{equation}
\bar{S}_c=\bar{W}_c+\sum_{i\in V_c, j\in V\setminus V_c}\frac{u_{ic}+(1-u_{jc})}{2}w_{ij}. 
\label{eq:mod7}
\end{equation}
The sets $V_c$ and $V$ include the vertices of module $c$ and of the whole network, respectively. 
Eq.~\ref{eq:mod5} is an extension of the weighted modularity in Eq.~\ref{eq:mod20}, obtained by weighing the 
contribution of the edges' weights to the sums in $W_c$ and $S_c$ 
by the (average) membership coefficients of the vertices of the edge. The determination of the 
eigenvectors is the most computationally expensive part of the method, so the time complexity is the same
as that of the algorithm by White and Smyth (see Section~\ref{sub_sec6_0_4}), i.~e. $O(K^2n+Km)$, which is essentially linear in $n$
if the graph is sparse and $K \ll n$.

Nepusz et al. proposed a different approach based on vertex similarity~\cite{nepusz08}. One starts from the membership matrix
${\bf U}$, defined as in the previous method by Zhang et al. 
From ${\bf U}$ a matrix ${\bf S}$ is built, where $s_{ij}=\sum_{k=1}^{n_c}u_{ik}u_{jk}$, expressing the 
similarity between vertices ($n_c$ is the number of clusters). 
If one assumes to have information about the actual vertex similarity, corresponding to the matrix ${\bf \tilde{S}}$,
the best cover is obtained by choosing ${\bf U}$ such that ${\bf S}$ approximates as closely as possible ${\bf \tilde{S}}$. This 
amounts to minimize the function
\begin{equation}
D_{\cal G}({\bf U})=\sum_{i=1}^n\sum_{j=1}^nw_{ij}(\tilde{s}_{ij}-s_{ij})^2,
\label{eqoverl1}
\end{equation}
where the $w_{ij}$ weigh the importance of the approximation for each entry of the similarity matrices. In the absence of 
any information on the community structure of the graph, 
one sets $w_{ij}=1$, $\forall i,j$ (equal weights) and ${\bf \tilde{S}}$ equal to the adjacency matrix ${\bf A}$,
by implicitly assuming that vertices are similar if they are neighbors, dissimilar otherwise. On weighted graphs, one can 
set the $w_{ij}$ equal to the edge weights. Minimizing $D_{\cal G}({\bf U})$ is a nonlinear constrained optimization problem, that can be solved 
with a gradient-based iterative optimization method, like simulated annealing. The optimization procedure adopted by Nepusz et al., for a 
fixed number of clusters $n_c$, has a time complexity
$O(n^2n_ch)$, where $h$ is the number of iterations leading to convergence, so the method can only be applied to fairly
small graphs. If $n_c$ is unknown, as it usually happens, the best cover 
is the one corresponding to the largest value of the modularity
\begin{equation}
Q=\frac{1}{2m}\sum_{ij}\left(A_{ij}-\frac{k_ik_j}{2m}\right)s_{ij}.
\label{eqoverl2}
\end{equation}
Eq.~\ref{eqoverl2} is very similar to
the expression of Newman-Girvan modularity (Eq.~\ref{eq:mod_0}): the difference is that the
Kronecker's $\delta$ is replaced by the vertices' similarity, to account for overlapping communities.
Once the best cover is identified, one can use the entries of the partition matrix ${\bf U}$ to evaluate
the participation of each vertex in the $n_c$ clusters of the cover. 
Nepusz et al. defined the {\it bridgeness} $b_{i}$ of a vertex $i$ as
\begin{equation}
b_i=1-\sqrt{\frac{n_c}{n_c-1}\sum_{j=1}^c\left(u_{ji}-\frac{1}{n_c}\right)^2}.
\label{eqoverl3}
\end{equation}
If $i$ belongs to a single cluster, $b_i=0$. 
If, for a vertex $i$, $u_{ik}=1/n_c$, $\forall k$, $b_i=1$ and $i$ is a perfect {\it bridge}, as it lies exactly between all clusters.
However, a vertex with low $b_i$ may be simply an outlier, not belonging to any cluster. Since real bridges are usually rather central vertices,
one can identify them by checking for large values of the {\it centrality-corrected bridgeness}, obtained by multiplying the bridgeness 
of Eq.~\ref{eqoverl3} by the centrality of the vertex (expressed by, e.g., degree, betweenness~\cite{freeman77}, etc.).
A variant of the algorithm by Nepusz et al. can be downloaded from 
{\tt http://www.cs.rhul.ac.uk/home/tamas/assets/file}\\{\tt s/fuzzyclust-static.tar.gz}.

\begin{figure}
\begin{center}
\includegraphics[width=7cm]{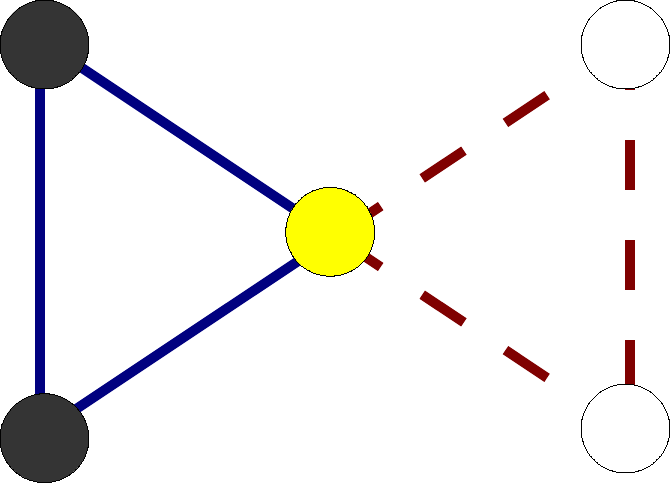}
\caption {\label{figlink} Communities as sets of edges. In the figure,
the graph has a natural division in two triangles, with the central vertex shared between them. If communities are identified by their internal edges,
detecting the triangles and their overlapping vertex becomes easier than by using methods that group vertices.
Reprinted figure with permission from Ref.~\cite{evans09}. \copyright 2009 by the American Physical Society.}
\end{center}
\end{figure}

In real networks it is often easier to discriminate between intercluster and intracluster edges than recognizing overlapping vertices.
For instance, in social networks, even though many people may belong to more groups, their social ties within each group can be easily spotted. 
Besides, it may happen that communities are joined to each other through their overlapping vertices (Fig.~\ref{figlink}), without intercluster edges.
For these reasons, it has been recently suggested that defining clusters as sets of edges, rather than vertices, may be a promising
strategy to analyze graphs with overlapping communities~\cite{evans09,ahn09}. One has to focus on the {\it line graph}~\cite{balakrishnan97}, 
i. e. the graph whose vertices are the 
edges of the original graph; vertices of the line graph are linked if the corresponding edges in the original graph are adjacent, i. e. if they
share one of their endvertices. Partitioning the line graph means grouping the edges of the starting graph\footnote{Ideally one wants to put together
only the edges lying within clusters, and exclude the others. Therefore partitioning 
does not necessarily mean assigning each vertex of the line graph to a group, as standard clustering techniques would do.}.
Evans and Lambiotte~\cite{evans09} introduced a set of quality functions, similar to Newman-Girvan modularity (Eq.~\ref{eq:mod_0}), 
expressing the stability of partitions against random walks taking place on the graph, 
following the work of Delvenne et al.~\cite{delvenne08} [Section~\ref{sec44_2}]. They considered a projection of 
the traditional random walk on the line graph, along with two other diffusion processes, 
where walkers move between adjacent edges (rather than between neighboring vertices).
Evans and Lambiotte optimized the three corresponding modularity functions to look for partitions in two real networks,
Zachary's karate club~\cite{zachary77} (Section~\ref{sec6_1}) and the network of word associations derived from the University of South Florida Free
Association Norms~\cite{nelson98} (Section~\ref{sec2}). The optimization was carried out with the hierarchical technique by Blondel et al.~\cite{blondel08}
and the multi-level algorithm by Noack and Rotta~\cite{noack09b}. While the results for the word association network are reasonable, the test 
on the karate club yields partitions in more than two clusters. However, the modularities used by Evans et Lambiotte can be modified to include
longer random walks (just like in Ref.~\cite{delvenne08}), and the length of the walk represents a resolution parameter that can be tuned to get better results.
Ahn et al.~\cite{ahn09} proposed to group edges with an agglomerative hierarchical clustering technique, called
{\it hierarchical link clustering} (Section~\ref{sec4_2}). 
They use a similarity measure for a pair of (adjacent) edges that expresses the size of the overlap between the neighborhoods of the non-coincident endvertices,
divided by the total number of (different) neighbors of such endvertices. Groups of edges are merged pairwise in descending order of similarity, until
all edges are together in the same cluster. The resulting dendrogram provides the most complete information on the community structure of the graph.
However, as usual, most of this information is redundant and is an artefact of the procedure itself. So, Ahn et al. introduced a quality
function to select the most meaningful partition(s), called {\it partition density}, which is essentially the average edge density 
within the clusters. The method is able to find meaningful clusters in biological networks, like protein-protein and metabolic networks,
as well as in a social network of mobile phone communications. It can also be extended to multipartite and weighted graphs.

The idea of grouping edges is surely interesting. However it is not {\it a priori}
better than grouping vertices. In fact, the two situations are somewhat symmetric. Edges connecting vertices of different clusters
are ``overlapping'', but they will be assigned just to one cluster (or else the clusters would be merged).

The possibility of having overlapping communities makes most standard clustering methods inadequate, and enforces 
the design of new {\it ad hoc} techniques, like the ones we have described so far. On the other hand, if it 
were possible to identify the overlapping vertices and ``separate'' them among the clusters they belong to, the overlaps would be 
removed and one could then apply any of the traditional clustering methods to the resulting graph. This idea is at the basis
of a recent method proposed by Gregory~\cite{gregory09}. It is a three-stages procedure: first, one transforms the graph into a larger graph without
overlapping vertices; second, a clustering technique is applied to the resulting graph; third, one maps the partition obtained into a cover
by replacing the vertices with those of the original graph. The transformation step, called {\it Peacock}, is performed by 
identifying the vertices with highest {\it split betweenness} (Section~\ref{subsec5_1}) and splitting them in multiple parts, connected by edges.
This is done as long as the split betweenness of the vertices is sufficiently high, which is determined by a parameter $s$. 
In this way, most vertices of the resulting graph are exactly the same one had initially, the others are multiple copies of the overlapping vertices
of the initial graph. The overlaps of the final cover are obtained by checking if copies of the same initial vertex end up in different disjoint clusters.
The complexity is dominated by the Peacock algorithm, if one computes the exact values of the split betweenness for the vertices, which
requires a time $O(n^3)$ on a sparse graph\footnote{The split betweenness needs to be recalculated 
after each vertex split, just as one does for the edge betweenness in the Girvan-Newman algorithm~\cite{girvan02}.
Therefore both computations have the same complexity.}. Gregory proposed
an approximate local computation, which scales as $O(n\log n)$: in this way 
the total complexity of the method becomes competitive, if one chooses a fast algorithm for the identification of the clusters. 
The goodness of the results depends on the specific method one uses to 
find the clusters after the graph transformation. The software of the version of the method used by Gregory in his applications
can be found at {\tt http://www.cs.bris.ac.uk/$\sim$steve/networks/peaco}\\{\tt ckpaper/}.
The idea of Gregory is interesting, as it allows to exploit traditional methods 
even in the presence of overlapping communities. The choice of the parameter $s$, which determines whether a 
vertex is overlapping or not, does not seem to affect significantly the results, as long 
as $s$ is taken sufficiently small. 

\section{Multiresolution methods and cluster hierarchy}
\label{sub_sec6_20}

The existence of a resolution limit for Newman-Girvan modularity (Section~\ref{sub_sec6_1}) implies that the straight optimization of 
quality functions yields a coarse description of the cluster structure of the graph, at a scale which has {\it a priori} nothing to do with
the actual scale of the clusters. 
In the absence of information on the cluster sizes of the graph, a method 
should be able to explore all possible scales, to make sure that it will eventually identify the right communities. Multiresolution methods
are based on this principle. However, many real graphs display {\it hierarchical} cluster structures, with clusters inside other clusters~\cite{simon62}.
In these cases, there are more levels of organization of vertices in clusters, and more relevant scales. 
In principle, clustering algorithms should be able to identify them. Multiresolution methods can do the trick, in principle, 
as they scan continuously the range of possible cluster scales.
Recently other methods have been developed, where partitions are by construction hierarchically nested in each other. In this section we discuss both classes 
of techniques.

\subsection{Multiresolution methods}
\label{sub_sec6_201}

In general, multiresolution methods have a freely tunable parameter, that allows to set the characteristic size of the clusters to 
be detected. The general spin glass framework by Reichardt and Bornholdt (\cite{reichardt06} and Section~\ref{sub_sec6_01}) is a typical example,
where $\gamma$ is the resolution parameter. The extension of the method to weighted graphs has been recently discussed~\cite{heimo08}.

Pons has proposed a method~\cite{pons06} consisting of the optimization of multiscale quality functions, including
the {\it multiscale modularity}
\begin{equation}
Q^{\cal M}_{\alpha}=\sum_{c=1}^{n_c}\Big[\alpha\frac{l_c}{m}-(1-\alpha)\left(\frac{d_c}{2m}\right)^2\Big],
\label{eq:mult0}
\end{equation}
and two other additive quality functions, derived from the {\it performance} (Eq.~\ref{eq10_perf}) and a measure based on the 
similarity of vertex pairs. In Eq.~\ref{eq:mult0}
$0\leq\alpha\leq 1$ is the resolution parameter and the notation is otherwise the same as in Eq.~\ref{eq:mod}. We see that, for $\alpha=1/2$,
one recovers standard modularity. However, since multiplicative factors in $Q^{\cal M}_{\alpha}$ do not change the results of the optimization,
we can divide $Q^{\cal M}_{\alpha}$ by $\alpha$, recovering the same quality function as in Eq.~\ref{eqr13}, with $\gamma=(1-\alpha)/\alpha$, 
up to an irrelevant multiplicative constant. To evaluate the relevance of the partitions, for any given multiscale quality function,
Pons suggested that the 
length of the $\alpha$-range $[\alpha_{min}({\cal C}),\alpha_{max}({\cal C})]$, for which a community ${\cal C}$ ``lives'' in the maximum modularity partition,
is a good indicator of the stability of the community. He then defined the {\it relevance function} of a community ${\cal C}$ at scale $\alpha$ as 
\begin{eqnarray}
\nonumber
R_{\alpha}({\cal C})&=&\frac{\alpha_{max}({\cal C})-\alpha_{min}({\cal C})}{2}\\
&+&
\frac{2(\alpha_{max}({\cal C})-\alpha)(\alpha-\alpha_{min}({\cal C}))}{\alpha_{max}({\cal C})-\alpha_{min}({\cal C})}.
\label{eq:mult01}
\end{eqnarray}
The relevance $R(\alpha)$ of a partition ${\cal P}$ at scale $\alpha$ is the average of the relevances of the clusters of the partition, weighted by
the cluster sizes. Peaks in $\alpha$ of $R(\alpha)$ reveal the most meaningful partitions.

\begin{figure}
\begin{center}
\includegraphics[width=\columnwidth]{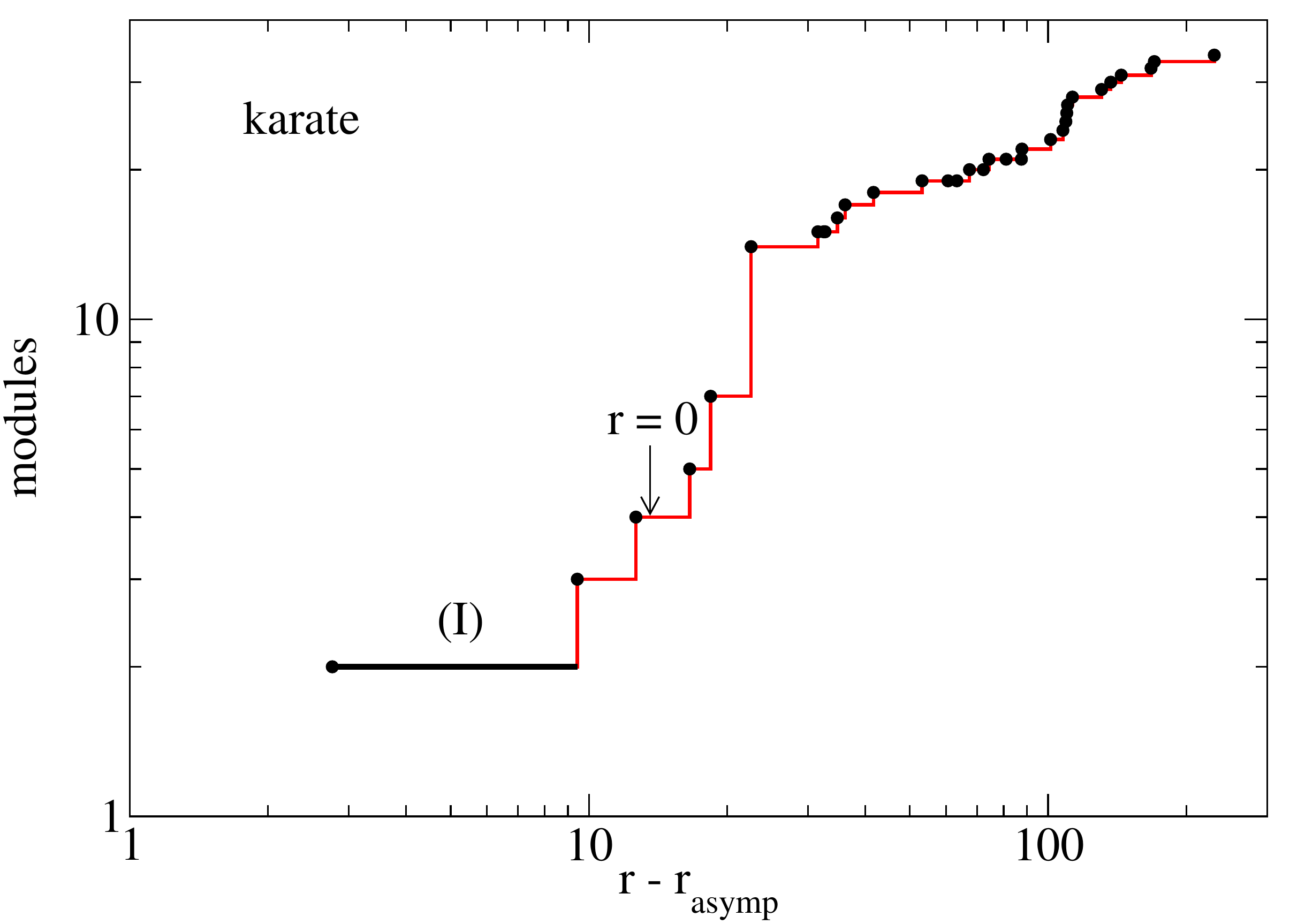}
\caption {\label{figarenasplat} Analysis of Zachary's karate club with the multiresolution method by Arenas et al.~\cite{arenas08b}. 
The plot shows the number of 
clusters obtained in correspondence of the resolution parameter $r$. The longest plateau $(I)$ indicates the most stable partition, which exactly matches 
the social fission observed by Zachary. The partition obtained with straight modularity optimization ($r=0$) consists of four clusters and 
is much less stable with respect to $(I)$, as suggested by the much shorter length of its plateau. Reprinted figure with permission from 
Ref.~\cite{arenas08b}. \copyright 2008 by IOP Publishing.}
\end{center}
\end{figure}

Another interesting technique has been devised by Arenas et al.~\cite{arenas08b}, and consists of 
a modification of the original expression of modularity. The idea is to make vertices contribute as well to the computation of the 
edge density of the clusters, by adding a self-loop of strength $r$ to each vertex. 
Arenas et al. remarked that the parameter $r$ does not affect the structural
properties of the graph in most cases, which are usually determined by an adjacency matrix without diagonal elements. 
With the introduction of the vertex strength $r$, modularity reads
\begin{equation}
Q_r=\sum_{c=1}^{n_c}\Big[\frac{2W_c+N_cr}{2W+nr}-\left(\frac{S_c+N_cr}{2W+nr}\right)^2\Big],
\label{eq:mult1}
\end{equation}
for the general case of a weighted graph. The notation is the same as in Eq.~\ref{eq:mod20}, $N_c$ is the number of vertices in cluster $c$.
We see that now the relative importance of the two terms in each summand depends on $r$, which can take any value in
$]-2W/n, \infty[$. Arenas et al. made a sweep in the range of $r$, and determined for each $r$ the maximum modularity 
with extremal optimization (Section~\ref{sub_sec6_0_3}) and {\it tabu search}\footnote{Tabu search consists in 
moving single vertices from one community to another, chosen at random, 
or to new communities, starting from some initial partition. After a sweep over all vertices, the best move, i. e. the one 
producing the largest increase of modularity, is accepted and applied, yielding a new partition. The procedure is repeated 
until modularity does not increase further. To escape local optima, a list of recent accepted moves is kept and updated, so that 
those moves are not accepted in the next update of the configuration (tabu list). 
The cost of the procedure is about the same of other stochastic optimization techniques like, e. g.,
simulated annealing.}~\cite{glover86}. 
Meaningful cluster structures 
correspond to plateaus in the plot of the number of clusters versus $r$ (Fig.~\ref{figarenasplat}). The length of a plateau gives a measure 
of the {\it stability} of the partition against the variation of $r$. 
The procedure is able to disclose the community structure of a number of 
real benchmark graphs. As expected, the most relevant partitions can be found in intervals of $r$ not including the value $r=0$,
which corresponds to the case of standard modularity (Fig.~\ref{figarenasplat}). 
A drawback of the method is that it is very slow, as one has to compute the modularity maximum for many values of 
$r$ in order to discriminate between relevant and irrelevant partitions. If the modularity maximum is computed with precise methods like simulated
annealing and/or extremal optimization, as in Ref.~\cite{arenas08b}, only graphs with a few hundred vertices can be analyzed on a single processor.
On the other hand the algorithm can be trivially parallelized by running the optimization for different values of $r$ on different processors. 
This is a common feature
of all multiresolution methods discussed in this Section.
In spite of the different formal expressions of modularity, the methods by Arenas et al.
and Reichardt and Bornholdt are somewhat related to each other and yield similar results~\cite{kumpula07b} on Zachary's karate club~\cite{zachary77} 
(Section~\ref{sec6_1}), synthetic graphs \'a la Ravasz-Barab\'asi~\cite{ravasz03} and on a model graph with the properties of real 
weighted social networks\footnote{Related does not mean equivalent, though. Arenas et al. have shown that their method is better than that by 
Reichardt and Bornholdt when the graph at hand includes communities of different sizes~\cite{arenas08b}.}. In fact,
their modularities can be both recovered from the continuous-time version of the stability of clustering
under random walk, introduced by Delvenne et al.~\cite{delvenne08} (Section~\ref{sec44_2}).
 
Lancichinetti et al. have designed a multiresolution method which is capable of detecting both the hierarchical structure of graphs 
and overlapping communities~\cite{lancichinetti09}.
It is based on the optimization of a fitness function, which estimates the strength of a cluster and entails a
resolution parameter $\alpha$. The function could in principle be arbitrary,
in their applications the authors chose a simple ansatz based on the tradeoff between the internal and the total degree of the cluster.
The optimization procedure starts from a cluster with a single vertex, arbitrarily selected. Given a cluster core, 
one keeps adding and removing neighboring vertices of the cluster as long as its fitness increases. The fitness is recalculated after each 
addition/removal of a vertex.
At some point one reaches a local maximum and the cluster is ``closed''. Then, another vertex is chosen at random, 
among those not yet assigned to a cluster, a new cluster is built, and so on, until all vertices have been assigned to clusters. 
During the buildup of a cluster, vertices already assigned to other clusters may be included, i.~e. communities may overlap. 
The computational complexity of the algorithm, estimated on sparse Erd\"os-R\'enyi
random graphs, is $O(n^\beta)$, with $\beta\sim 2$ for small values of the resolution parameter $\alpha$, and 
$\beta\sim 1$ if $\alpha$ is large. For a complete analysis, the worst-case computational complexity is 
$O(n^2\log n)$, where the factor $\log n$ comes from the minimum number of different $\alpha$-values which are needed to resolve the actual community
structure of the graph. Relevant partitions are revealed by pronounced spikes in 
the histogram of the fitness values of covers obtained for different $\alpha$-values, where the fitness of a cover is defined as the 
average fitness of its clusters.

A technique based on the Potts model, similar to that of Reichardt and Bornholdt~\cite{reichardt06}, has been suggested by
Ronhovde and Nussinov~\cite{ronhovde08}. The energy of their spin model is 
\begin{equation}
{\cal H}(\{\sigma\})=-\frac{1}{2}\sum_{i\neq j}[A_{ij}-\gamma(1-A_{ij})]\delta(\sigma_i,\sigma_j).
\label{eq:mult2}
\end{equation}
The big difference with Eq.~\ref{eqr13} is the absence of a null model term. The model considers pairs of vertices 
in the same community: edges between vertices are energetically rewarded, whereas 
missing edges are penalized. The parameter $\gamma$ fixes the tradeoff between the two 
contributions. The energy is minimized by sequentially shifting single vertices/spins to the communities which yield the largest decrease of the system's energy, 
until convergence. If, for each vertex, one just examines the communities of its neighbors, the energy is minimized in a time $O(m^\beta)$, where $\beta$
turns out to be slightly above $1$ in most applications, allowing for the analysis of large graphs. 
This essentially eliminates the problem of limited resolution, as the criterion 
to decide about the merger or the split of clusters only depends on local parameters. Still, for the detection of possible hierarchical levels
tuning $\gamma$ is mandatory. In a successive paper~\cite{ronhovde09}, the authors have introduced a new stability criterion for the partitions,
consisting of the computation of the similarity of partitions obtained for the same $\gamma$ and different initial conditions. The idea is that,
if a partition is robust in a given range of $\gamma$-values, most replicas delivered by the algorithm will be very similar. On the other hand, if 
one explores a region of resolutions in between two strong partitions, the algorithm will deliver the one or the other partition and the 
individual replicas will be, on average, not so similar to each other. So, by plotting the similarity as a function of the resolution
parameter $\gamma$, stable communities are revealed by peaks. Ronhovde and Nussinov adopted similarity measures borrowed from information theory
(Section~\ref{sec6_2}). Their criterion of stability can be adopted 
to determine the relevance of partitions obtained with any multiresolution algorithm.

A general problem of multiresolution methods is how to assess the stability of partitions for large graphs. The rapidly increasing number of
partitions, obtained by minimal shifts of vertices between clusters, introduces a large amount of noise, that blurs 
signatures of stable partitions like plateaus, spikes, etc. that one can observe in small systems. In this respect, it seems far more reliable
focusing on correlations
between partitions (like the average similarity used by Ronhovde and Nussinov~\cite{ronhovde08,ronhovde09}) 
than on properties of the individual partitions (like the measures of occurrence used by Arenas et al.~\cite{arenas08b} and by
Lancichinetti et al.~\cite{lancichinetti09}).

\subsection{Hierarchical methods}
\label{sub_sec6_202}

The natural procedure to detect the hierarchical structure of a graph is hierarchical clustering, that we have discussed in Section~\ref{sec4_2}.
There we have emphasized the main weakness of the procedure, which consists of the necessity to introduce 
a criterion to identify relevant partitions (hierarchical levels) out of the full dendrogram produced by the given algorithm. Furthermore,
there is no guarantee that the results indeed reflect the actual hierarchical structure of the graph, and that they are not mere artefacts of 
the algorithm itself. Scholars have just started to deal with these problems. 

Sales-Pardo et al. have proposed a top-down approach~\cite{sales07}. Their method consists of two steps: 1) measuring
the similarity between vertices; 2) deriving the hierarchical structure of the graph from the similarity matrix.  
The similarity measure, named {\it node affinity}, is based on Newman-Girvan modularity.
Basically the affinity between two vertices is the
frequency with which they coexist in the same community in partitions corresponding to 
local optima of modularity. The latter are configurations for which modularity is stable, i.~e. it cannot increase
if one shifts one vertex from one cluster to another or by merging or splitting clusters.
The set of these partitions is called ${\cal P}_{max}$.
Before proceeding with the next step, one verifies whether the graph has a significant community structure or not. This is 
done by calculating the $z$-score (Eq.~\ref{eq:zscore}) for the average modularity of the partitions in ${\cal P}_{max}$ with respect to the 
average modularity of partitions with local modularity optima of the equivalent ensemble of null model graphs, obtained as usual 
by randomly rewiring the edges of the original graph under the condition that the expected degree sequence is the same as the degree sequence of the graph.
Large $z$-scores indicate meaningful cluster structure: Sales-Pardo et al. used a threshold corresponding to the $1\%$ significance level\footnote{We remind
that the significance of the $z$-score
has to be computed with respect to the actual distribution of the maximum modularity for the null model graphs, as the latter 
is not Gaussian (Section~\ref{sub_sec6_1}).}.
If the graph has a relevant cluster structure, one proceeds with the second step, which consists in putting 
the affinity matrix in a form as close as possible to block-diagonal, by minimizing a 
cost function expressing the average distance of connected vertices from the diagonal. The blocks correspond to the communities
and the recovered partition represents the uppermost organization level. To determine lower levels,
one iterates the procedure for each subgraph identified at the previous level, which
is treated as an independent graph. The procedure stops when all blocks found do not have a relevant cluster structure, i.~e. their
$z$-scores are lower than the threshold. 
The partitions delivered by the method are hierarchical by construction, as 
communities at each level are nested within communities at higher levels. However, 
the method may find no relevant partition (no community structure), a single partition 
(community structure but no hierarchy) or more (hierarchy) and in this respect it is better
than most existing methods.
The algorithm is not fast,
as both the search of local optima for modularity and the rearrangement of the 
similarity matrix are performed with simulated annealing\footnote{The reordering of the matrix is by far the most time-consuming
part of the method. The situation improves if one adopts faster optimization strategies than simulated annealing, at the cost of less accurate results.}, 
but delivers good results for computer generated networks, and meaningful partitions for some real networks, like 
the world airport network~\cite{barrat04}, an email exchange network of a Catalan university~\cite{guimera03}, 
a network of electronic circuits~\cite{itzkovitz05} and metabolic networks of {\it E. coli}~\cite{guimera07c}.

\begin{figure}
\begin{center}
\includegraphics[width=\columnwidth]{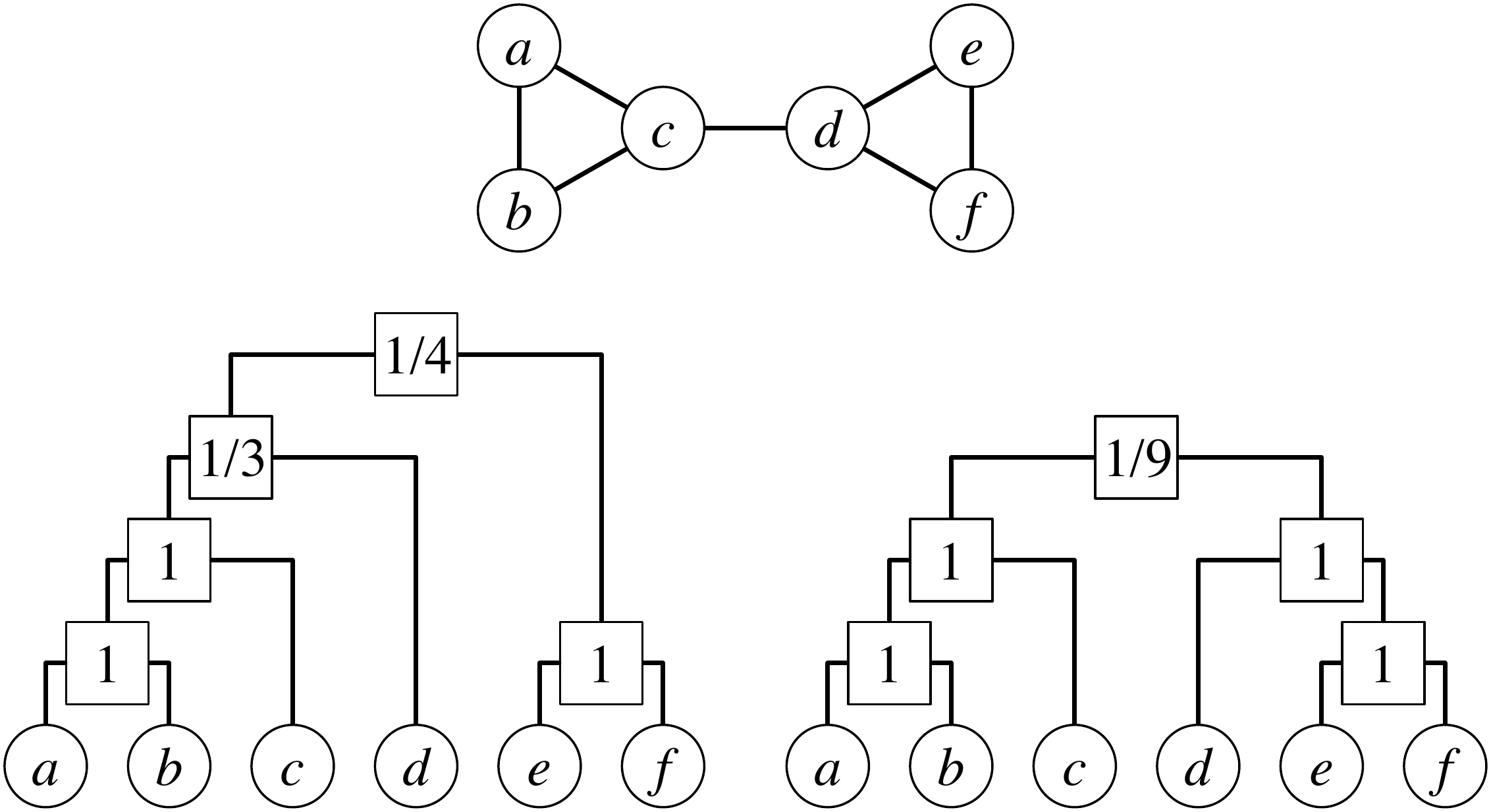}
\caption {\label{clausethier} Hierarchical random graphs by Clauset et al.~\cite{clauset08}. The picture shows two possible dendrograms
for the simple graph on the top. The linking probabilities on the internal nodes of the dendrograms yield 
the best fit of the model graphs to the graph at study. Reprinted figure with 
permission from Ref.~\cite{clauset08}. \copyright 2008 by the Nature
Publishing Group.}
\end{center}
\end{figure}

Clauset et al.~\cite{clauset07,clauset08} described the hierarchical organization of a graph by
introducing a class of {\it hierarchical random graphs}. A hierarchical random graph is defined by a dendrogram $\cal D$, which is the natural
representation of the hierarchy, and by a set of probabilities $\{p_r\}$ associated to the $n-1$ internal nodes of the dendrogram.
An {\it ancestor} of a vertex $i$ is any internal node of the dendrogram that is encountered by starting from the ``leaf'' vertex $i$ and 
going all the way up to the top of the dendrogram. The probability that vertices $i$ and $j$ are linked to each other is given by the 
probability $p_r$ of the lowest common ancestor of $i$ and $j$. Clauset et al. searched for the model $({\cal D}, \{p_r\})$ that best fits the 
observed graph topology, by using Bayesian inference (Section~\ref{subsec_statinf1}). The probability that the model fits the graph is 
proportional to the likelihood 
\begin{equation}
{\cal L}({\cal D}, \{p_r\})=\prod_{r\in {\cal D}}p_r^{E_r}(1-p_r)^{L_rR_r-E_r}.
\label{hier1}
\end{equation}
Here, $E_r$ is the number of edges connecting vertices whose lowest common ancestor is $r$, $L_r$ and $R_r$ are the numbers of 
graph vertices in the left and right subtrees descending from the dendrogram node $r$, and the product runs over all internal 
dendrogram nodes. For a given dendrogram $\cal D$, the maximum likelihood ${\cal L}({\cal D})$ corresponds to the set of probabilities $\{\bar{p}_r\}$,
where ${\bar{p}_r}$ equals the actual density of edges $E_r/(L_rR_r)$ between the two subtrees of $r$ (Fig.~\ref{clausethier}). 
One can define
the statistical ensemble of hierarchical random graphs describing a given graph ${\cal G}$, by assigning to each model graph $({\cal D}, \{\bar{p}_r\})$
a probability proportional to the maximum likelihood ${\cal L}({\cal D})$. The ensemble can be sampled by a Markov chain Monte Carlo method~\cite{newman99}.
The procedure suggested by Clauset et al. seems to converge to equilibrium roughly in a time $O(n^2)$, although the actual complexity may be much higher.
Still, the authors were able to investigate graphs with a few thousand vertices. From sufficiently large sets of model configurations sampled at equilibrium,
one can compute average properties of the model, e. g. degree distributions, clustering coefficients. etc., and compare them with 
the corresponding properties of the original graph. Tests on real graphs reveal that the model is indeed capable to describe closely the graph properties.
Furthermore, the model enables one to predict missing connections between vertices of the original graph. This is 
a very important problem~\cite{liben03}: edges of real graphs are the result of observations/experiments, that may fail to discover
some relationships between the units of the system. From the ensemble of the hierarchical random graphs one can derive the average
linking probability between all pairs of graph vertices. By ranking the probabilities corresponding to vertex pairs which are disconnected in the original
graph, one may expect that the pairs with highest probabilities are likely to be connected in the system, even if such connections are not observed.
Clauset et al. pointed out that their method does not deliver a sharp hierarchical organization for a given graph, but a class of possible
organizations, with well-defined probabilities. It is certainly reasonable to assume that many structures are compatible with a given graph topology.
In the case of community structure, it is not clear which information one can extract from averaging over the ensemble of 
hierarchical random graphs. Moreover, since the hierarchical structure is represented by a dendrogram, it is impossible to rank partitions
according to their relevance. In fact, the work by Clauset et al. questions the concept of ``relevant partition'', and 
opens a debate in the scientific community about the meaning itself of graph clustering. The software of the method can be found at
{\tt http://www.santafe.edu/$\sim$aaronc/hierarchy/}.

\section{Detection of dynamic communities}
\label{sec7_1_2}

\begin{figure}
\begin{center}
\includegraphics[width=\columnwidth]{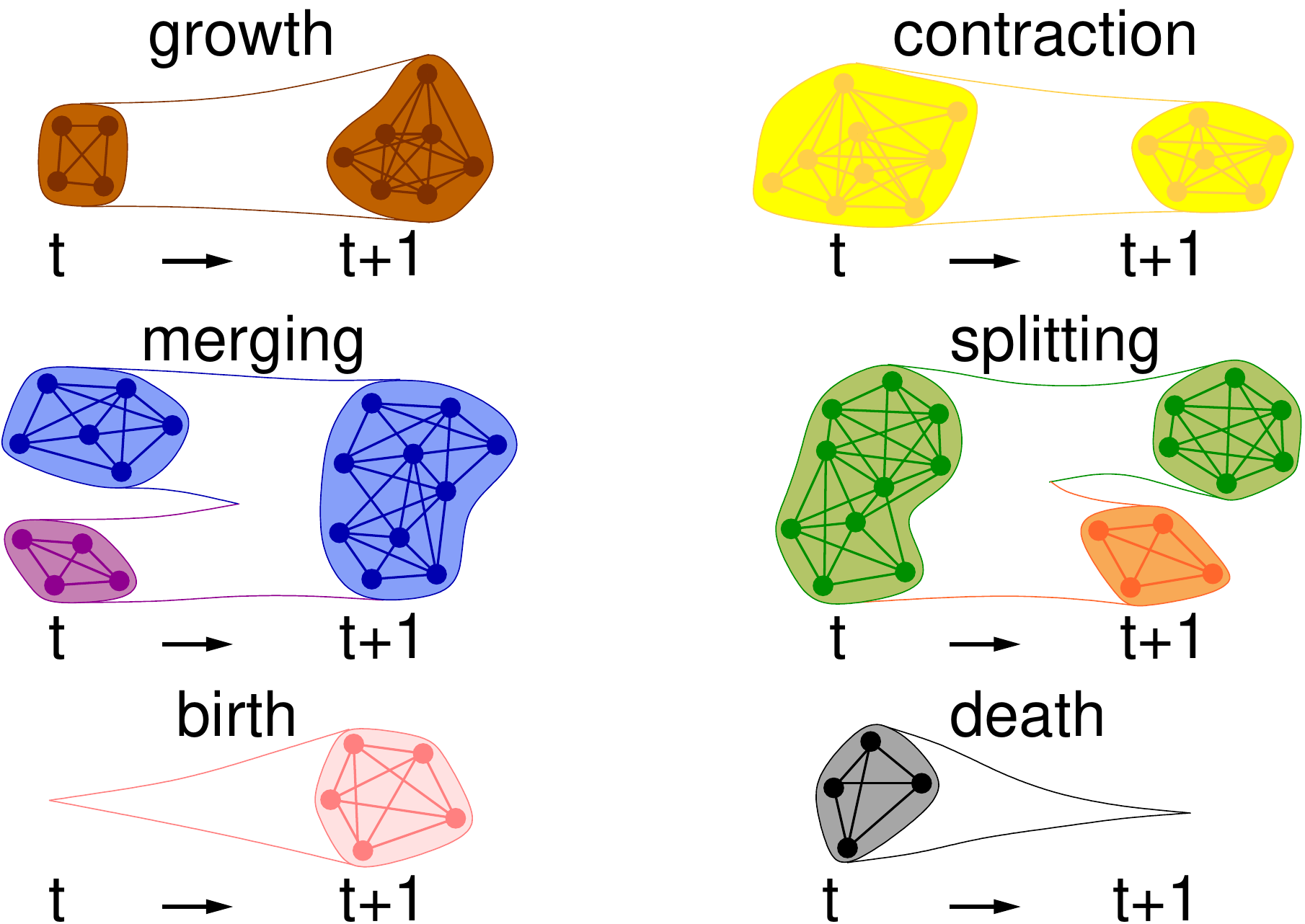}
\caption {\label{commevol} Possible scenarios in the evolution of communities. Reprinted figure with permission from 
Ref.~\cite{palla07}. \copyright 2007 by the Nature Publishing Group.}
\end{center}
\end{figure}

The analysis of dynamic communities is still in its infancy. Studies in this direction have been mostly
hindered by the fact that the problem of graph clustering
is already controversial on single graph realizations, so it is understandable that most efforts still concentrate on the ``static'' version of the problem.
Another difficulty is represented by the dearth of timestamped data on real graphs. 
Recently, several data sets have become available, enabling to monitor the evolution in time of real systems~\cite{kumar03,kumar06,leskovec05,leskovec08}.
So it has become possible to investigate how communities form, evolve and die. 
The main phenomena occurring in the lifetime of a community are (Fig.~\ref{commevol}): birth, growth, contraction, merger with other
communities, split, death.

The first study was carried out by Hopcroft et al.~\cite{hopcroft04}, who analyzed several snapshots of the citation graph 
induced by the NEC CiteSeer Database~\cite{giles98}. The snapshots cover the period from $1990$ to $2001$. Communities are detected 
by means of (agglomerative) hierarchical clustering (Section~\ref{sec4_2}), where the similarity between vertices is the {\it cosine similarity}
of the vectors describing the corresponding papers, a well known measure used in information retrieval~\cite{baeza99}. In each snapshot Hopcroft et al.
identified the {\it natural communities}, defined as those communities of the hierarchical tree that are only slightly affected by minor
perturbations of the graph, where the perturbation consists in removing a small fraction of the vertices (and their edges).
Such natural communities are conceptually similar to the {\it stable} communities we will see in Section~\ref{sec6_4}. 
Hopcroft et al. found the best matching natural communities across different snapshots, and in this way they could follow the history of communities.
In particular they could see the emergence of new communities, corresponding to new research topics. The main drawback of the method 
comes from the use of hierarchical clustering, which is unable to sort out meaningful communities out of the hierarchical tree, which
includes many different partitions of the graph.

More recently, Palla et al. performed a systematic analysis of 
dynamic communities~\cite{palla07}. They studied two social systems: 1) a graph of phone calls between customers of a mobile
phone company in a year's time; 2) a collaboration network between scientists, describing the coauthorship 
of papers in condensed matter physics from the electronic e-print archive (cond-mat) maintained by Cornell University Library,
spanning a period of 142 months. 
The first problem is identifying the image of a community ${\cal C}(t+1)$ at time $t+1$ among 
the communities of the graph at time $t$. A simple criterion, used in other works, is to measure the relative overlap (Eq.~\ref{eqt20})
of ${\cal C}(t+1)$ with all communities at time $t$, and pick the community which has the largest overlap with ${\cal C}(t+1)$. This
is intuitive, but in many cases it may miss the actual evolution of the community. For instance, if ${\cal C}(t)$
at time $t+1$ grows considerably and overlaps with another community ${\cal B}(t+1)$ (which at the previous
time step was disjoint from ${\cal C}(t)$), the relative overlap between ${\cal C}(t+1)$ and ${\cal B}(t)$
may be larger than the relative overlap between ${\cal C}(t+1)$ and ${\cal C}(t)$. It is not clear whether there 
is a general prescription to avoid this problem. Palla et al. solved it by exploiting the features of the 
Clique Percolation Method (CPM) (Section~\ref{sec45_1}), that they used to detect communities. The idea is to analyze the graph ${\cal G}(t, t+1)$,
obtained by merging the two snapshots ${\cal G}(t)$ and ${\cal G}(t+1)$ of the evolving graph, 
at times $t$ and $t+1$ (i. e., by putting together all their vertices and edges). 
Any CPM community of ${\cal G}(t)$ and ${\cal G}(t+1)$ does not get lost, as it is included within one of the CPM communities of ${\cal G}(t, t+1)$. 
For each CPM community ${\cal V}_k$ of ${\cal G}(t, t+1)$, one finds the CPM communities $\{{\cal C}_k^t\}$ and 
$\{{\cal C}_k^{t+1}\}$ (of ${\cal G}(t)$ and ${\cal G}(t+1)$, respectively) which are contained
in ${\cal V}_k$. The image of any community in $\{{\cal C}_k^{t+1}\}$ at time $t$ is the community of $\{{\cal C}_k^t\}$ that has the largest
relative overlap with it. 

\begin{figure}
\begin{center}
\includegraphics[width=\columnwidth]{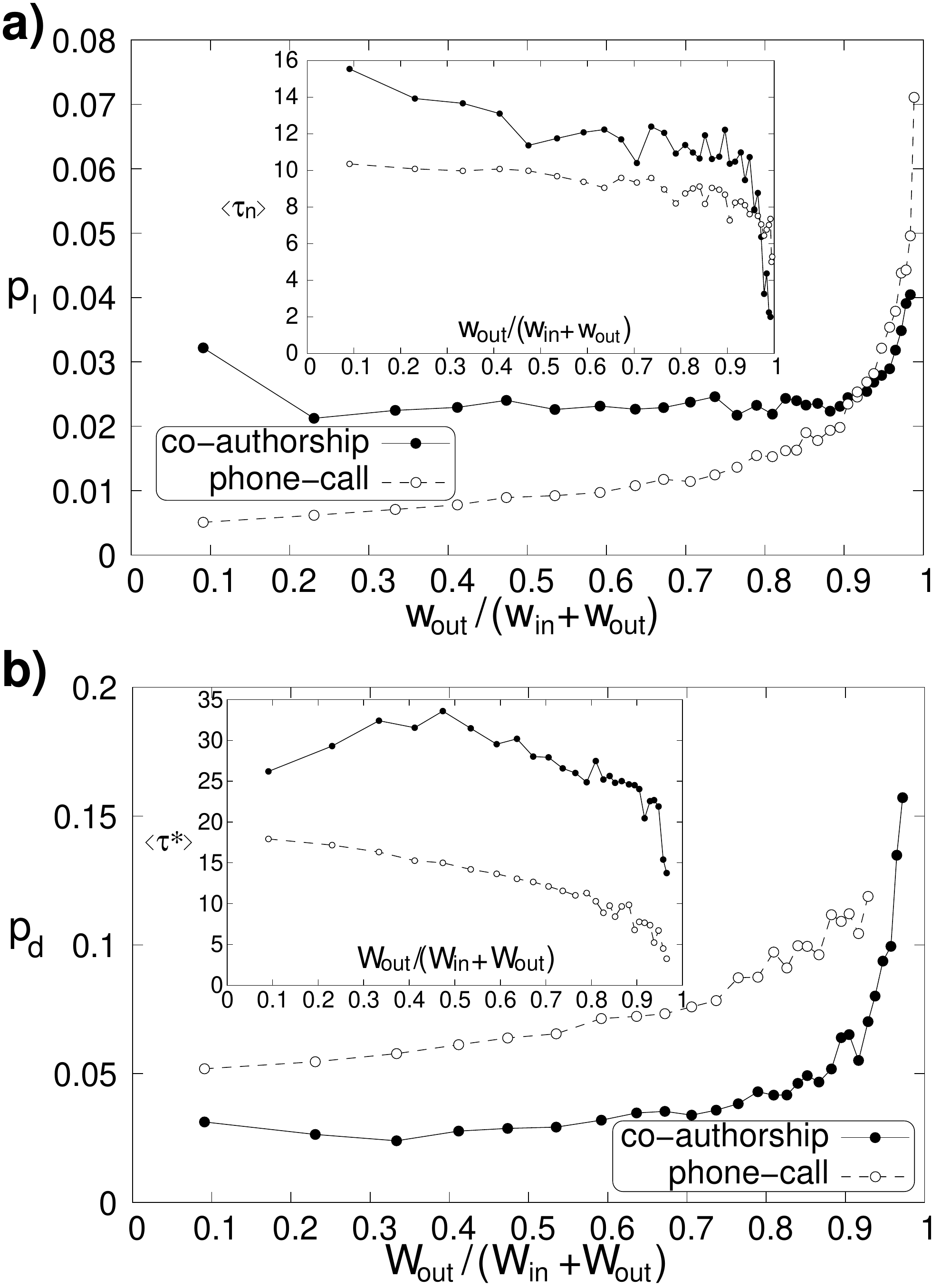}
\caption {\label{evolpred} Relation between structural features and evolution of a community. a) Relation between the probability 
that a vertex will abandon the community in the next time step and its relative external strength. b) Relation between the 
probability of disintegration of a community in the next time step and its relative external strength. Reprinted figure with permission from 
Ref.~\cite{palla07}. \copyright 2007 by the Nature Publishing Group.}
\end{center}
\end{figure}

The age $\tau$ of a community is the time since its birth. It turns out that the age 
of a community is positively correlated with its size $s(\tau)$, i. e.
that older communities are also larger (on average). The time evolution of a community ${\cal C}$ can be described by means of the
relative overlap $C(t)$ between states of the community separated by a time $t$:
\begin{equation}
C(t)=\frac{|{\cal C}(t_0)\bigcap {\cal C}(t_0+t)|}{|{\cal C}(t_0)\bigcup {\cal C}(t_0+t)|}.
\label{eqtrol3}
\end{equation}
One finds that, in both data sets, $C(t)$ decays faster for larger communities, so the composition of large communities is 
rather variable in time, whether small communities are essentially static. Another important question is whether it is possible to 
predict the evolution of communities from information on their structure or on their vertices. In Fig.~\ref{evolpred}a  
the probability $p_l$ that a vertex will leave the community in the next step of the evolution is plotted as a function of the 
relative external strength of the vertex, indicating how much of the vertex strength lies on edges connecting it to vertices outside 
its community. The plot indicates that there is a clear positive correlation: vertices which are only loosely connected to vertices of their community
have a higher chance (on average) to leave the community than vertices which are more ``committed'' towards the other community members.
The same principle holds at the community level too. 
Fig.~\ref{evolpred}b shows that the probability $p_d$ that a community
will disintegrate in the next time step is positively correlated with the relative external strength of the community. Finally,
Palla et al. have found that the probability for two communities to merge increases with the community sizes much more than what one expects from
the size distribution, which is consistent with the faster dynamics observed for large communities. Palla et al. analyzed two different real systems,
a network of mobile phone communications and a coauthorship network,
to be able to infer general properties of community evolution. However, communities were only found with the CPM, so their results
need to be cross-checked by employing other clustering techniques.

Asur et al.~\cite{asur07} explored the dynamic relationship between vertices and communities.
Communities were found with
the MCL method by Van Dongen~\cite{vandongen00} (Section~\ref{sec44_2}), by analyzing the graph at different timestamps. Asur et al. distinguished 
events involving communities and events involving the vertices. 
Events involving communities are {\it Continue} 
(the community keeps most of its vertices in consecutive time steps), {\it $\kappa$-Merge} (two clusters merge into another),
{\it $\kappa$-Split} (two clusters split in two parts), {\it Form} (no pair of vertices of the cluster at time $t+1$ were in 
the same cluster at time $t$) and {\it Dissolve} (opposite of {\it Form}). Events involving vertices are {\it Appear} 
(if a vertex joins the graph for the first time), {\it Disappear} (if a vertex of the graph at time $t$ is no longer there 
at time $t+1$), {\it Join} (if a vertex of a cluster at time $t+1$ was not in that cluster at time $t$) and {\it Leave} 
(if a vertex which was in a cluster at time $t$ is not in that cluster at time $t+1$). Based on such events,
four measures are defined in order to catch the behavioral tendencies of vertices contributing to the evolution of the graph: 
the {\it stability index} (measuring the tendency of a vertex to interact with the same vertices over time), the {\it sociability index}
(measuring the number of different interactions of a vertex, basically the number of {\it Join} and {\it Leave} events), the
{\it popularity index} (measuring the number of vertices attracted by a cluster in a given time interval) and the
{\it influence index} (measuring the influence a vertex has on the others, which is computed from the number of 
vertices that leave or join a cluster together with the vertex). Applications on a coauthorship network of computer scientists and on a 
network of subjects for clinical trials show that the behavioral measures above enable one to make reliable predictions about the
time evolution of such graphs (including, e. g., the inference of missing links~\cite{liben03}). 

Dynamic communities can be as well detected with methods of information compression, such as some of those we have seen in Section~\ref{subsec_statinf2}.
Sun et al.~\cite{sun07} applied the Minimum Description Length (MDL) principle~\cite{rissanen78,grunwald05} to find the minimum encoding cost 
for the description of a time sequence of graphs and their partitions in communities. The method is quite similar to that successively developed by
Rosvall and Bergstrom~\cite{rosvall07}, which is however defined only for static graphs 
(Section~\ref{subsec_statinf2}). Here one considers bipartite graphs evolving in time. 
The time sequence of graphs can be separated in segments, each containing some number of consecutive snapshots of the system.
The graphs of each segment
are supposed to have the same modular structure (i. e. they represent the same phase in the history of the system), 
so they are characterized by the same partition of the two vertex classes.
For each graph segment it is possible to define an encoding cost, which combines the encoding cost of 
the partition of the graphs of the segment with the entropy of compression of the segment in the subgraph segments induced by the partition.
The total encoding
cost $C$ of the graph series is given by the sum of the encoding costs of its segments. Minimizing $C$ enables one to find not only the 
most modular partition for each graph segment (high modularity\footnote{We stress that here by modularity we mean 
the feature of a graph having community structure,
not the modularity of Newman and Girvan.} 
corresponds to low encoding costs for a partition), but also the most compact subdivision of the snapshots into segments, such that
graphs in the same segment are strongly correlated with each other. The latter feature allows to identify 
{\it change points} in the time history of the system, i. e. short periods in which the dynamics produces big changes in the graph structure
(corresponding to, e.g., extreme events).
The minimization of $C$ is ${\bf NP}$-hard, so the authors propose an approximation method called {\it GraphScope}, which consists of two steps:
first, one looks for the best partition of each graph segment; second, one looks for the best division in segments. In both cases the ``best'' result
corresponds to the minimal encoding cost. The best partition within a graph segment is found by local search. GraphScope has the big
advantage not to require any input, like the number and sizes of the clusters. It is also suitable to operate in a streaming environment, in which new 
graph configurations are added in time, following the evolution of the system: the computational complexity required to process a snapshot 
(on average) is stable over time.
Tests on real evolving data sets show that GraphScope is able to find meaningful
communities and change points.

Since keeping track of communities in different time steps
is not a trivial problem, as we have seen above, it is perhaps easier to adopt a vertex-centric perspective, in which one monitors
the community of a given vertex at different times. For any method, given a vertex $i$ and a time $t$, the community 
to which $i$ belongs at time $t$ is well defined. Fenn et al.~\cite{fenn09} used the multiresolution method 
by Reichardt et al.~\cite{reichardt06} (Section~\ref{sub_sec6_01})
and investigated a fully connected graph with time-dependent weights, representing the correlations of time series of hourly exchange rate returns.
The resolution parameter $\gamma$ is fixed to the value that occurs in most stability plateaus of the system at different time steps. 
Motivated by the work of Guimer\`a and Amaral~\cite{guimera05} (Section~\ref{sec7_1}), Fenn et al. identify
the role of individual vertices in their community through the pair $(z^{in},z^b)$, where $z^{in}$ is the $z$-score
of the internal strength (weighted degree, Section~\ref{sec1_1}), defined in Eq.~\ref{eqtrol1}, and $z^b$ the $z$-score of the site betweenness,
defined by replacing the internal degree with the site betweenness of Freeman~\cite{freeman77} in Eq.~\ref{eqtrol1}. 
We remind that the site betweenness is a measure of the number of shortest paths running through a vertex. The variable 
$z^b$ expresses the importance of a vertex in processes of information diffusion with respect to the other members of its community.
Another important issue regards the {\it persistence} of communities in time, i. e. how stable they are during the evolution.
As a measure of persistence, Fenn et al. introduced a vertex-centric version of the relative overlap of Eq.~\ref{eqtrol3}
\begin{equation}
a_i^t(\tau)=\frac{|{\cal C}_i(t)\bigcap {\cal C}_i(t+\tau)|}{|{\cal C}_i(t)\bigcup {\cal C}_i(t+\tau)|},
\label{eqtrol4}
\end{equation}
where $i$ is the vertex and ${\cal C}_i(t)$, ${\cal C}_i(t+\tau)$ the communities of $i$ at times $t$, $t+\tau$, respectively.
The decay of $a_i^t(\tau)$ depends on the type of vertex. In particular, if the vertex is 
strongly connected to its community ($z^{in}$ large), $a_i^t(\tau)$ decays quite slowly, meaning that it tends to stay attached to a stable core of vertices.

The methods described above are basically two-stage approaches, in which clusters are detected at each timestamp of the graph evolution, 
independently of the results at different times, and relationships between partitions at different times are inferred successively.
However, this often produces significant variations between partitions close in time, especially when, as it usually happens, the 
datasets are noisy. In this case one would be forced to introduce {\it ad hoc} hypotheses on the graph evolution 
to justify the variability of the modular structure, whereas such variability is mostly an artefact of the approach. 
It would be desirable instead to have a unified framework, in which clusters are deduced both from the current structure 
of the graph and from the knowledge of the cluster structure at previous times. This is the principle of {\it evolutionary
clustering}, a framework introduced by Chakrabarti et al.~\cite{chakrabarti06} and successively adopted and refined by other authors.
Let ${\cal C}_t$ be the partition of the graph at time $t$. The {\it snapshot quality} of ${\cal C}_t$ measures the goodness of the
partition with respect to the graph structure at time $t$. The {\it history cost} is a measure of the distance/dissimilarity
of ${\cal C}_t$ with respect to the partition ${\cal C}_{t-1}$ at the previous time step. The overall quality of ${\cal C}_t$ 
is given by a combination of the snapshot quality and the history cost at each time step. Ideally, a good partition should have high
snapshot quality (i. e. it should cluster well the data at time $t$) and low history cost (i. e. it should be similar to the partition 
at the previous time step). In order to find ${\cal C}_t$ from ${\cal C}_{t-1}$ and the relational data at time $t$
Chakrabarti et al. suggested to minimize 
the difference between the snapshot quality and the history cost, with a relative
weight $cp$ that is a tunable parameter. The input of the procedure consists in the sequence of adjacency/similarity matrices at different time steps.
In practice, one could use modified versions of such matrices, obtained by performing (weighted) averages of the data over some time window,
in order to make the relational data more robust against noise and the results of the clustering procedure more reliable.
One can adopt arbitrary measures to compute the snapshot quality and the historical cost.
Besides, several known clustering techniques used for static graphs can be reformulated within this evolutionary framework. 
Chakrabarti et al. derived evolutionary versions of hierarchical clustering (Section~\ref{sec4_2}) and $k$-means clustering
(Section~\ref{sec4_3}), whereas Chi et al.~\cite{chi07} designed two implementations for
spectral clustering (Section~\ref{sec4_4}). Based on evolutionary clustering, Lin et al.~\cite{lin08} introduced a framework, 
called {\it FacetNet}, that allows vertices to belong to more communities at the same time. Here the 
snapshot cost\footnote{Lin et al. used the cost and not the quality to evaluate the fit of the 
partition to the data. The two estimates are obviously related: the lower the cost, the higher the quality.} 
is the Kullback-Leibler (KL) divergence~\cite{kullback51} between the 
adjacency/similarity matrix at time $t$ and the matrix describing the community structure of the graph at time $t$; the 
history cost is the KL divergence between the matrices describing the community structure of the graph at times $t-1$ and $t$.
FacetNet can be extended to handle
adding and removing of vertices as well as variations of the number of clusters in consecutive time steps. However, it is not able to
account for the creation and the disintegration of communities and not scalable to large systems due to the high number of 
iterations necessary for the matrix computations to reach convergence. These issues have been addressed in a recent approach
by Kim and Han~\cite{kim09b}.

Naturally, what one hopes to achieve at the end of the day is to see how real groups form and evolve in time.
Backstrom et al.~\cite{backstrom06} have carried out an analysis of group dynamics in the free online community of LiveJournal 
({\tt http://www.livejournal.com/})
and in a coauthorship network of computer scientists. Here the groups are identified through the declared memberships of users (for LiveJournal)
and conferences attended by computer scientists, respectively. Backstrom and coworkers have found that the probability that an individual 
joins a community grows with the number of friends/coauthors who are already in the community and (for LiveJournal) with their degree of interconnectedness.
Moreover, the probability of growth of LiveJournal communities is positively correlated to a combination of factors including the community size, the number of 
friends of community members which are not in the community and the ratio of these two numbers. A high density of triads within a community appears
instead to hinder its growth.

\section{Significance of clustering}
\label{sec6_4}

Given a network, many partitions could represent meaningful clusterings
in some sense, and it could be difficult for some methods to discriminate between them. Quality functions
evaluate the goodness of a partition (Section~\ref{sec3_2_2}), so one could say that high quality corresponds to meaningful partitions.
But this is not necessarily true. In Section~\ref{sub_sec6_1} we have seen that high values of the modularity of Newman and Girvan
do not necessarily indicate that a graph has a definite cluster structure. In particular we have seen that partitions of random graphs  
may also achieve considerably large values of $Q$, although we do not expect them to have community structure, due to the lack of 
correlations between the linking probabilities of the vertices. The optimization of quality functions, like modularity, delivers
the best partition according to the criterion underlying the quality function. But is the optimal clustering also {\it significant}, i.~e.
a relevant feature of the graph, or is it  
just a byproduct of randomness and basic structural properties like, e. g., 
the degree sequence? Little effort has been devoted to this crucial issue, that we discuss here. 

In some works the concept of significance has been related to that of {\it robustness} or {\it stability} of a partition against
random perturbations of the graph structure. The basic idea is that, if a partition is significant, it will be recovered
even if the structure of the graph is modified, as long as the modification is not too extensive. Instead, if a partition is not
significant, one expects that minimal modifications of the graph will suffice to disrupt the partition, so other clusterings are recovered.
A nice feature of this approach is the fact that it can be applied for any clustering technique.
Gfeller et al.~\cite{gfeller05} considered the general case of weighted graphs. A graph is modified, in that its edge weights are increased or decreased by a
relative amount $0<\sigma< 1$. This choice also allows to account for the possible effects of uncertainties in the values of the edge weights,
resulting from measurements/experiments carried out on a given system. After fixing $\sigma$ (usually to $0.5$), multiple realizations of 
the original graph are generated. The best partition for each realization is identified and, for each pair of adjacent vertices $i$ and $j$, 
the {\it in-cluster} probability $p_{ij}$ is computed, i.~e. the fraction of realizations in which $i$ and $j$ were classified in the same cluster. Edges
with in-cluster probability smaller than a threshold $\theta$ (usually $0.8$) are called {\it external edges}.  
The stability of a partition is estimated through the {\it clustering entropy}
\begin{equation}
S=-\frac{1}{m}\sum_{(i,j):A_{ij}=1}[p_{ij}\log_2 p_{ij}-(1-p_{ij})\log_2(1-p_{ij})],
\label{sign1}
\end{equation}
where $m$ is, as usual, the number of graph edges, and the sum runs over all edges. The most stable partition has $p_{ij}=0$ along
inter-cluster edges and $p_{ij}=1$ along intra-cluster edges, which yields $S=0$; the most unstable partition has $p_{ij}=1/2$ on all edges,
yielding $S=1$. The absolute value of $S$ is not meaningful, though, and needs to be compared with the corresponding value for a null model graph,
similar to the original graph, but with supposedly no cluster structure. Gfeller et al. adopted the same null model of Newman-Girvan modularity, i.~e.
the class of graphs with expected degree sequence coinciding with that of the original graph. Since the null model is defined on unweighted graphs, the significance of $S$ 
can be assessed only in this case, although it would not be hard to think of a generalization to weighted graphs.
The approach enables one as well to identify {\it unstable vertices}, i.~e. vertices
lying at the boundary between clusters. In order to do that, 
the external edges are removed and  
the connected components of the resulting disconnected graph are associated with 
the clusters detected in the original graph, based on their relative overlap (computed through Eq.~\ref{eqt20}). 
Unstable vertices end up in components that are not associated to any of the initial clusters. A weakness of the method by Gfeller et al. 
is represented by the two parameters $\sigma$ and $\theta$, whose values are in principle arbitrary.

More recently, Karrer et al.~\cite{karrer08} adopted a similar strategy to unweighted graphs. Here one performs a sweep over all edges: the perturbation consists
in removing each edge with a probability $\alpha$ and replacing it with another edge between a pair of vertices $(i,j)$, chosen at random 
with probability $p_{ij}=k_ik_j/2m$, where $k_i$ and $k_j$ are the degrees of $i$ and $j$. We recognize the probability of the null model
of Newman-Girvan modularity. Indeed, by varying the probability $\alpha$ from $0$ to $1$ one smoothly interpolates between 
the original graph (no perturbation) and the null model (maximal perturbation). The degree sequence of the graph remains invariant (on average) 
along the whole process, by construction. The idea is that the perturbation affects solely the organization of the vertices, keeping the 
basic structural properties. For a given value of $\alpha$, many realizations of the perturbed graph are generated, their cluster structures 
are identified with some method (Karrer et al. used modularity optimization) and compared with the partition obtained from the original unperturbed
graph. The partitions are compared by computing the variation of information $V$ (Section~\ref{sec6_2}). From the plot of the average
$\langle V\rangle$ versus $\alpha$ one can assess the stability of the cluster structure of the graph. If $\langle V(\alpha)\rangle$ 
changes rapidly for small values of $\alpha$
the partition is likely to be unstable. As in the approach by Gfeller et al. the behaviour of the function $\langle V(\alpha)\rangle$
does not have an absolute meaning, but needs to be compared with the corresponding curve obtained for a null model. 
For consistency, the natural choice is again the null model of modularity, which is already used in the process of graph perturbation.
The approaches by Gfeller et al. and Karrer et al., with suitable modifications, can also be used to check for the stability of the cluster structure 
in parts of a graph, up to the level of individual
communities. This is potentially important as it may happen that some parts of the graph display a strong community structure and 
other parts weak or no community structure at all. 

Rosvall and Bergstrom~\cite{rosvall08b} defined the significance of clusters with the bootstrap method~\cite{efron93}, which is a standard
procedure to check for the accuracy of a measurement/estimate based on resampling from the empirical data. The graph at study 
is supposed to be generated by a parametric model, which is used to create many samples. This is done by assigning
to each edge a weight taken by a Poisson distribution with mean equal to the original edge weight. For the initial graph and each sample
one identifies the community structure with some method, that can be arbitrary. For each cluster of the partition of the 
original graph one determines the largest subset of vertices that are classified in the same cluster in at least $95\%$ of all bootstrap samples.
Identifying such cluster cores enables one to track the evolution in time of the community structure, as we explained in Section~\ref{sec7_1_2}.

A different approach has been proposed by Massen and Doye~\cite{massen06}. They analyzed an equilibrium canonical ensemble 
of partitions, with $-Q$ playing the role of the energy, $Q$ being Newman-Girvan modularity. This means that the 
probability of occurrence of a partition at temperature $T$ is proportional to $\exp(Q/T)$. The idea is that, if a graph has 
a significant cluster structure, at low temperatures one would recover essentially the same partition, corresponding to the modularity maximum, which 
is separated by an appreciable gap from the modularity values of the other partitions. 
On the contrary, graphs with no community structure, e. g. random graphs, have many competing (local) maxima,  
and the corresponding configurations will emerge already at low temperatures, since their modularity 
values are close to the absolute maximum\footnote{As we have seen in Section~\ref{sub_sec6_1}, Good et al.~\cite{good09} have 
actually shown that the modularity landscape has a huge degeneracy of states with
high modularity values, close to the global maximum, especially on graphs with community structure. So the results of the method by Massen and Doye may
be misleading.}. These distinct behaviors can manifest themselves in various ways. For instance, if one
considers the variation of the specific heat $C=-dQ/dT$ with $T$, the gap in the modularity landscape is associated to a sharp peak
of $C$ around some temperature value, like it happens in a phase transition. If the gap is small and there are many partitions
with similar modularity values, the peak of $C$ becomes broad. Another strategy to assess the significance of the maximum modularity partition
consists of the investigation of the similarity between partitions recovered at a given temperature $T$. This similarity can be expressed by the
{\it frequency matrix}, whose element $f_{ij}$ indicates the relative number of times vertices $i$ and $j$ have been classified in the same cluster.
If the graph has a clear community structure, at low temperatures the frequency matrix can be put in block-diagonal form, with the blocks corresponding to the 
communities of the best partition; if there is no significant community structure, the frequency matrix is rather homogeneous. 
The Fiedler eigenvalue~\cite{fiedler73}
$\lambda_2$, the second smallest eigenvalue of the Laplacian matrix associated to the frequency matrix, allows to estimate how ``block-diagonal'' the
matrix is (see Section~\ref{sec4_1}). At low temperatures $\lambda_2\sim 0$ if there is one (a few) partitions with maximum or near to maximum modularity; if there are many
(almost) degenerate partitions, $\lambda_2$ is appreciably different from zero even when $T\rightarrow 0$. A sharp transition from low to high values 
of $\lambda_2$ by varying temperature indicates significant community structure. Another clear signature of significant community structure
is the observation of a rapid drop
of the average community size with $T$, as ``strong'' communities break up in many small pieces for a modest temperature increase, while 
the disintegration of ``weak'' communities takes place more slowly. In scale-free graphs (Section~\ref{sec1_3}) 
clusters are often not well separated, due to the presence of the hubs; in these cases the above-mentioned transitions of 
ensemble variables are not so sharp and take place over a broader temperature range.  
The canonical ensemble of partitions is generated through 
single spin heatbath simulated annealing~\cite{reichardt06}, combined with parallel tempering~\cite{earl05}. The approach 
by Massen and Doye could be useful to recognize graphs without cluster structure, if the modularity landscape is characterized by many maxima
with close values (but see Footnote).
However, it can happen that gaps between the absolute modularity maximum and the rest of the modularity values are created 
by fluctuations, and the method is unable to identify these situations. Furthermore, the approach heavily relies on modularity and on a costly
technique like simulated annealing: extensions to other quality functions and/or optimization procedures do not appear straightforward.

In a recent work by Bianconi et al.~\cite{bianconi09} the notion of entropy of graph ensembles~\cite{bianconi08,bianconi08a} is employed
to find out how likely it is for a cluster structure
to occur on a graph with a given degree sequence. The entropy is computed from the number of graph configurations which are compatible
with a given classification of the vertices in $q$ groups. The clustering is quantitatively described by fixing the number of edges $A(q_1,q_2)$ 
running between clusters $q_1$ and $q_2$, for all choices of $q_1\neq q_2$. Bianconi et al. proposed the following indicator of clustering significance
\begin{equation}
\Theta_{\vec{k},\vec{q}}=\frac{\Sigma_{\vec{k},\vec{q}}-\langle\Sigma_{\vec{k},\pi(\vec{q})}\rangle_{\pi}}
{\sqrt{\langle\delta\Sigma_{\vec{k},\pi(\vec{q})}^2\rangle_{\pi}}},
\label{eq:mod13}
\end{equation}
where $\Sigma_{\vec{k},\vec{q}}$ is the entropy of the graph configurations with given degree sequence $\vec{k}$ and 
clustering $\vec{q}$ (with fixed numbers of
inter-cluster edges $A(q_1,q_2)$), and $\langle\Sigma_{\vec{k},\pi(\vec{q})}\rangle_{\pi}$ is the average entropy of the configurations with the same
degree sequence and a random permutation $\pi(\vec{q})$ of the cluster labels. The absolute value of the entropy $\Sigma_{\vec{k},\vec{q}}$ 
is not meaningful, so the comparison of $\Sigma_{\vec{k},\vec{q}}$ and  
$\langle\Sigma_{\vec{k},\pi(\vec{q})}\rangle_{\pi}$ is crucial, as it tells how relevant the actual cluster structure is with respect to 
a random classification of the vertices. However, different permutations of the assignments $\vec{q}$ yield different values of the
entropy, which can fluctuate considerably. Therefore one has to compute the standard deviation $\langle\delta\Sigma_{\vec{k},\pi(\vec{q})}^2\rangle_{\pi}$
of the entropy corresponding to all random permutations $\pi(\vec{q})$, to estimate how significant the 
difference between $\Sigma_{\vec{k},\vec{q}}$ and $\langle\Sigma_{\vec{k},\pi(\vec{q})}\rangle_{\pi}$ is. In this way, if $\Theta_{\vec{k},\vec{q}}\leq 1$, the
entropy of the given cluster structure is of the same order as the entropy of some random permutation of the cluster labels, so it is 
not relevant. Instead, if $\Theta_{\vec{k},\vec{q}}\gg 1$, the cluster structure is far more likely 
than a random classification of the vertices, so the clustering is relevant. The indicator $\Theta_{\vec{k},\vec{q}}$ can be simply generalized
to the case of directed and weighted graphs. 

\begin{figure*}
\begin{center}
\includegraphics[width=\textwidth]{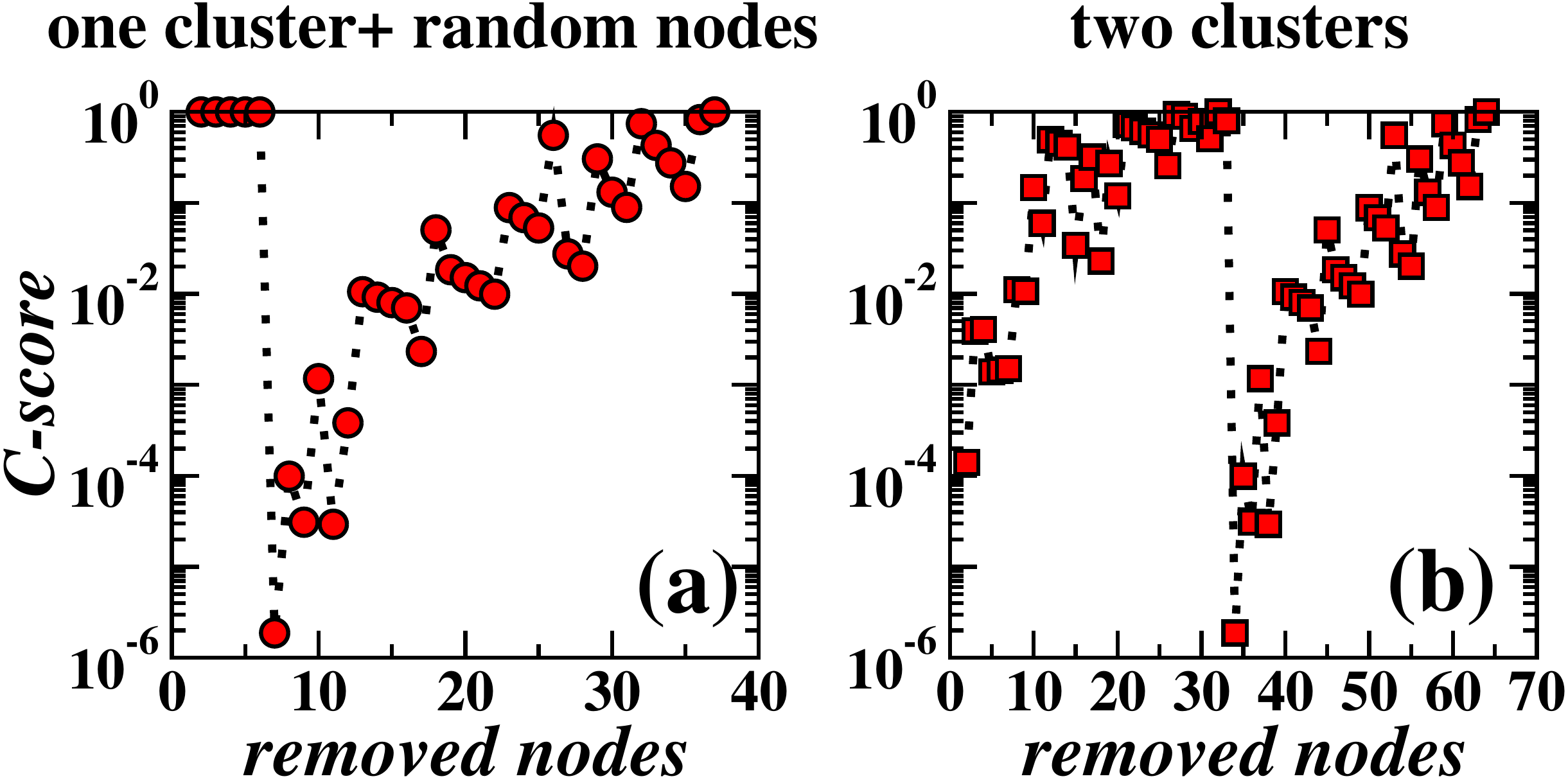}
\caption {\label{cscore} Application of the ${\cal C}$-score by Lancichinetti et al.~\cite{lancichinetti09c}
to identify modules within subgraphs. In (a) the subgraph consists of a compact cluster 
(generated with the LFR benchmark~\cite{lancichinetti08,lancichinetti09b}) plus some randomly added vertices. In (b) 
the subgraph consists of two compact clusters interconnected by a few edges. Vertices are removed from each subgraph in increasing 
order of their internal degree. The ${\cal C}$-score displays sharp drops after all the spurious vertices (a) and
all the vertices of one of the two clusters (b) are removed. We notice that the first subgraph (a) is not significant 
(high ${\cal C}$-score) until the noise represented by the randomly added vertices disappears, whereas the second subgraph (b) is 
a community at the very beginning, as it should be, it loses significance when one of the clusters is heavily damaged
(because the remainder of the cluster appears as noise, just like the spurious vertices in (a)), and becomes significant again when the damaged 
cluster is totally removed.
Reprinted figure with permission from Ref.~\cite{lancichinetti09c}.}
\end{center}
\end{figure*}

Lancichinetti et al.~\cite{lancichinetti09c} as well addressed the issue by comparing the cluster structure of the graph
with that of a random graph with similar properties. An important
novelty of this approach is the fact that it estimates the significance of single communities, not
of partitions. In fact, not all communities are equally significant, in general, so it makes
a lot of sense to check them individually. 
In particular, it may happen that real networks are not fully modular, due to their particular history or generating mechanisms,
and that only portions of them display community structure. 
The main idea is to verify how likely it is that a community ${\cal C}$ 
is a subgraph of a random graph with the same degree sequence of the original graph. This likelihood is 
called ${\cal C}$-score, and is computed by examining the
vertex $w$ of ${\cal C}$, with the lowest internal degree $k_{w}^{in}$ in ${\cal C}$ (the ``worst'' vertex).
The ${\cal C}$-score is defined as the probability that the internal degree of the worst vertex in the corresponding community 
of the null model graphs is larger than or equal to $k_{w}^{in}$. This probability is computed by using tools from
Extreme and Order Statistics~\cite{david03,beirlant04}. A low value of the ${\cal C}$-score ($\leq 5\%$) is a strong
indication that the group of vertices at study is a community and not the product of random fluctuations.
In addition, the measure can be used to check whether a subgraph has an internal modular structure.
For that, one removes the vertices of the subgraph one at a time, starting from the worst
and proceeding in increasing order of the internal degree, and observes how the ${\cal C}$-score varies at each vertex removal:
sharp drops indicate the presence of dense subgraphs (Fig.~\ref{cscore}). Therefore, one could think of using
the ${\cal C}$-score as ingredient of new clustering techniques.
As we have seen, the ${\cal C}$-score is based on the behavior of the vertex with lowest internal degree of the subgraph.
Real networks are characterized by noise, which could strongly
affect the structural relationships between vertices and clusters. For this reason, relying on the properties of a single vertex 
to evaluate the significance of a subgraph could be a problem for applications to real networks. Lancichinetti et al. 
have shown that the 
${\cal C}$-score can be easily extended to consider the $t$ vertices with lowest internal degree, with $t>1$ (${\cal B}$-score).
The main limit of the ${\cal C}$-score is the fact that its null model is the same as that of Newman-Girvan modularity. 
According to this null model, each vertex can in principle be connected to any other, no matter how large the system is.
This is however not realistic, especially for large graphs, where it is much more reasonable to assume that each vertex has its own
``horizon'', i.e. a subset of other vertices with which it can interact, which is usually much smaller than the 
whole system (see Section~\ref{sub_sec6_1}).
How to define such ``horizons'' and, more in general, realistic null models is still an open problem. However, the 
${\cal C}$-score could be easily reformulated with any null model, so one could readily derive more reliable definitions.

We conclude with a general issue which is related to the significance of community structure. The question is:
given a cluster structure in a graph, can it be recovered {\it a priori} by an algorithm? 
In a recent paper~\cite{reichardt08}, Reichardt and Leone studied under which conditions a special built-in cluster structure 
can be recovered. The clusters have equal size and a pair of vertices 
is connected with probability $p$ if they belong to the same cluster, with probability $r<p$ otherwise. In computer science
this is known as the {\it planted partitioning problem}~\cite{condon01}. The goal is to propose algorithms that recover the planted
partition for any choice of $p$ and $r$. For dense graphs, i.~e. graphs whose average degree grows with the number $n$ of vertices, 
algorithms can be designed that find the solution
with a probability which equals $1$ minus a term that vanishes in the limit of infinite graph size, regardless of the 
difference $p-r$, which can then be chosen arbitrarily small. Since many real networks are not dense graphs, as their average degree $\langle k \rangle$
is usually much smaller than $n$ and does not depend on it, Reichardt and Leone investigated the problem in the case of fixed
$\langle k \rangle$ and infinite graph size. We indicate with $q$ the number of clusters and with $p_{in}$ 
the probability that a randomly selected edge of the graph lies within any of the $q$ clusters. In this way, if $p_{in}=1/q$, the
inter-cluster edge density matches the intra-cluster edge density (i.~e. $p=r$), and the planted partition would not
correspond to a recoverable clustering, whereas for $p_{in}=1$, there are no inter-cluster edges and the partition can be trivially 
recovered. The value of $p_{in}$ is in principle unknown, so one has to detect the cluster structure ignoring this information. Reichardt and Leone
proposed to look for a minimum cut partition, i.~e. for the partition that minimizes the number of inter-cluster edges, 
as it is usually done in the graph partitioning problem (discussed in Section~\ref{sec4_1}). Clearly, for $p_{in}=1$ the minimum cut partition 
trivially coincides with the planted partition, whereas for $1/q<p_{in}<1$ there should be some overlap, which is expected
to vanish in the limit case $p_{in}=1/q$. The minimum cut partition corresponds to the minimum 
of the following ferromagnetic Potts model Hamiltonian
\begin{equation}
{\cal H}_{part}=-\sum_{i<j}J_{ij}\delta_{\sigma_i,\sigma_j},
\label{eq:mod12}
\end{equation}
over the set of all spin configurations with zero magnetization. Here the spin $\sigma_i$ indicates the cluster
vertex $i$ belongs to, and the coupling matrix $J_{ij}$ is just the adjacency matrix of the graph. The constraint of zero magnetization
ensures that the clusters have all the same size, as required by the planted partitioning problem. The energy of a spin configuration, expressed
by Eq.~\ref{eq:mod12}, is the negative of the number of edges that lie within clusters: the minimum energy corresponds to the maximum 
number of intra-cluster edges, which is coupled to the minimum number of inter-cluster edges. The minimum energy can be computed with 
the cavity method, or belief propagation, at zero temperature~\cite{mezard03}. The accuracy of the solution with respect to the planted partition
is expressed by the fraction of vertices which are put in the same class in both partitions. The analysis yields a striking result: 
the planted clustering is accurately recovered for $p_{in}$ larger than a critical threshold $p_{in}^c>1/q$. So, there is a range of 
values of $p_{in}$,  $1/q<p_{in}<p_{in}^c$, in which the clustering is not recoverable, as the minimum cut partition is uncorrelated with it. The threshold 
$p_{in}^c$ depends on the degree distribution $p(k)$ of the graph.

\section{Testing Algorithms}
\label{sec6}

When a clustering algorithm is designed, it is necessary to
test its performance, and compare it with that of other methods. In the previous sections we have said very little about
the performance of the algorithms, other than their computational complexity. Indeed, the issue of testing 
algorithms has received very little attention in the literature on graph clustering. This is a serious limit of the field.
Because of that, it is still impossible to state which method (or subset of methods) is the most reliable in applications, and people rely
blindly on some algorithms instead of others for reasons that have nothing to do with the actual performance of the algorithms, like. e.g. 
popularity (of the method or of its inventor). This lack of control is also the main reason for the  
proliferation of graph clustering techniques in the last few years. Virtually in any paper, where a new method is introduced, the part about testing
consists in applying the method to a small set of simple benchmark graphs, whose cluster structure is fairly easy
to recover. Because of that, the freedom in the design of a clustering algorithm is basically infinite, whereas it is not clear what a new 
procedure is adding to the field, if anything.

In this section we discuss at length the issue of testing. First, we describe the fundamental ingredients of any testing procedure, i.~e.
benchmark graphs with built-in community structure, that methods have to identify (Section~\ref{sec6_1}). We proceed by reviewing 
measures to compare graph partitions with each other (Section~\ref{sec6_2}). In Section~\ref{sec6_3} we present the comparative evaluations
of different methods that have been performed in the literature. 
 
\subsection{Benchmarks}
\label{sec6_1}

Testing an algorithm essentially means applying it to a specific problem whose solution is known and comparing such solution with that
delivered by the algorithm. In the case of graph clustering,
a problem with a well-defined solution is a graph with a clear community structure. This concept is not trivial, however. 
Many clustering algorithms are based on similar intuitive notions of what a community is, but different implementations. So it is crucial 
that the scientific community agrees on a set of reliable benchmark graphs. This mostly applies to computer-generated graphs,
where the built-in cluster structure can be arbitrarily designed. In the literature real networks are used as well,
in those cases in which communities are well defined because of information about the system. 

\begin{figure*}
\begin{center}
\includegraphics[width=\textwidth]{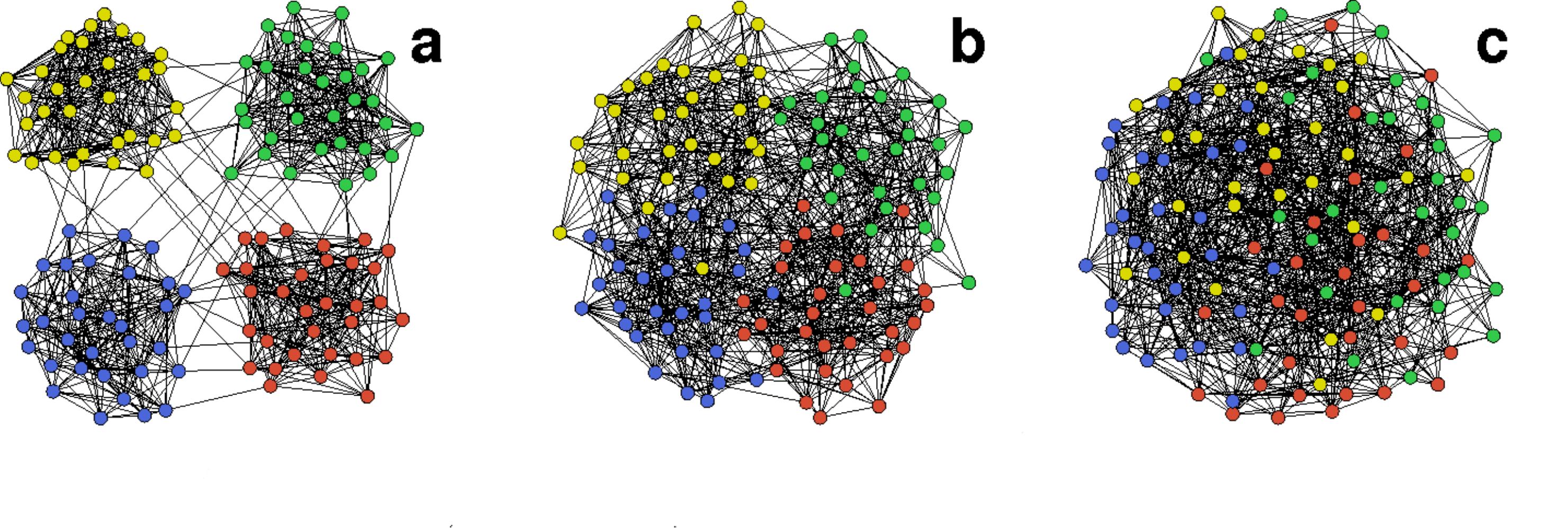}
\caption {\label{Figure11} Benchmark of Girvan and Newman. The three pictures correspond 
to $z_{in}=15$ (a), $z_{in}=11$ (b) and $z_{in}=8$ (c). In (c) the four groups are hardly 
visible. Reprinted figure with permission from Ref.~\cite{guimera05}. \copyright 2005 by the
Nature Publishing Group.}
\end{center}
\end{figure*}

We start our survey from computer-generated benchmarks.
A special class of graphs has become
quite popular in the last years. They are generated with the so-called {\it planted $\ell$-partition model}~\cite{condon01}.
The model partitions a graph with $n=g\cdot {\ell}$ vertices in $\ell$ groups with $g$ vertices each. Vertices
of the same group are linked with a probability $p_{in}$, whereas vertices of different groups are linked with a probability $p_{out}$.
Each subgraph corresponding to a group is then a random graph \'a la Erd\"os-R\'enyi with connection probability
$p=p_{in}$ (Section~\ref{sec1_3}). The average degree of a vertex is $\langle k\rangle=p_{in}(g-1)+p_{out}g(\ell-1)$.
If $p_{in}>p_{out}$ the intra-cluster edge density exceeds the inter-cluster edge density and the graph has a community 
structure. This idea is quite intuitive and we have encountered it in several occasions in the previous sections. 
Girvan and Newman considered a special case of the planted $\ell$-partition model~\cite{girvan02}. They set $\ell=4$, $g=32$ (so the 
number of graph vertices is $n=128$)
and fixed the average total degree $\langle k\rangle$ to $16$. This implies that $p_{in}+3p_{out}\approx 1/2$, so
the probabilities $p_{in}$ and $p_{out}$ are not independent parameters. In calculation it is common to use as parameters
$z_{in}=p_{in}(g-1)=31p_{in}$ and $z_{out}=p_{out}g(\ell-1)=96p_{out}$, indicating the expected internal and external degree of a vertex, respectively.
These particular graphs have by now gained the status of standard benchmarks~\cite{girvan02} (Fig.~\ref{Figure11}).
In the first applications of the graphs one assumed that communities are well defined when $z_{out}<8$, corresponding to the situation
in which the internal degree exceeds the external degree. However, the threshold $z_{out}=z_{in}=8$ implies $p_{in}\approx 1/4$ and $p_{out}=1/12$,
so it is not the actual threshold of the model, where $p_{in}=p_{out}=1/8$, corresponding to $z_{out}\approx 12$. 
So, one expects\footnote{However, we stress that, even if communities are there, 
methods may be unable to detect them. The reason is that, due to fluctuations
in the distribution of links in the graphs, already before the limit imposed by the planted partition model it may be impossible
to detect the communities and the model graphs may look similar to random graphs.}
to be able to detect the planted partition up until $z_{out}\approx 12$. 

Testing a method against the Girvan-Newman benchmark 
consists in calculating the similarity between the partitions determined by the
method and the natural partition of the graph in the four equal-sized groups. Several measures of partitions' similarity
may be adopted; we describe them in Section~\ref{sec6_2}. 
One usually builds many graph realizations for a particular value of 
$z_{out}$ and computes the average similarity between the solutions of the method and the built-in solution.
The procedure is then iterated for different values of $z_{out}$.
The results are usually represented in a plot, where the average similarity is drawn as a function of $z_{out}$. 
Most algorithms usually do a good job 
for small $z_{out}$ and start to fail when $z_{out}$ approaches $8$.
Fan et al.~\cite{fan07} have designed a weighted version of the benchmark 
of Girvan and Newman, in that one gives different weights to edges inside and between communities. One could pick just two values, one for
intra- and the other for inter-community edges, or uniformly distributed values in two different ranges. For this benchmark there are then two 
parameters that can be varied: $z_{out}$ and the relative importance of the internal and the external weights. Typically one fixes the topological
structure and varies the weights. This is particularly insightful when $z_{out}=4$, which delivers graphs without topological cluster structure: 
in this case, the question whether there are clusters or not depends entirely on the weights.

\begin{figure*}
\begin{center}
\includegraphics[width=\textwidth]{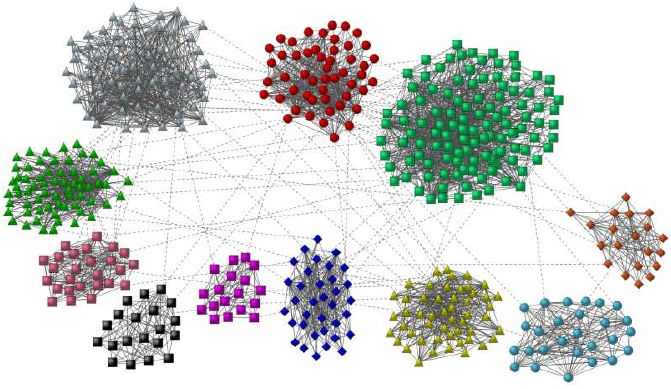}
\caption{\label{fignewbench} A realization of the LFR benchmark graphs~\cite{lancichinetti08}, with $500$ vertices.
The distributions of the vertex degree and of the community size are both power laws. Such benchmark is a more faithful approximation
of real-world networks with community structure than simpler benchmarks like, e. g., that by Girvan and Newman~\cite{girvan02}.
Reprinted figure with permission from 
Ref.~\cite{lancichinetti08}. \copyright 2008 by the American Physical Society.}
\end{center}
\end{figure*}

As we have remarked above, the planted $\ell$-partition model generates mutually interconnected random graphs \'a la Erd\"os-R\'enyi.
Therefore, all vertices have approximately the same degree. Moreover, all communities have exactly the same size by construction. 
These two features are at odds with what is observed in graph representations of real systems. Degree distributions are usually skewed,
with many vertices with low degree coexisting with a few vertices with high degree. A similar heterogeneity is also observed in 
the distribution of cluster sizes, as we shall see in Section~\ref{sec7_1}. So, the planted $\ell$-partition model is not a good description 
of a real graph with community structure. However, the model can be modified to account for the heterogeneity of degrees and community sizes. A modified version
of the model, called {\it Gaussian random partition generator}, was designed by
Brandes et al.~\cite{brandes03}. Here the cluster sizes have a Gaussian distribution, so they are not the same, although they do not differ much 
from each other. The heterogeneity of the cluster sizes introduces a heterogeneity in the degree distribution as well, as the 
expected degree of a vertex depends on the number of vertices of its cluster. Still, the variability of degree and cluster size is not appreciable. Besides,
vertices of the same cluster keep having approximately the same degree.
A better job in this direction has been recently done by Lancichinetti et al. (LFR benchmark)~\cite{lancichinetti08}. They assume that the distributions
of degree and community size are power laws, with exponents $\tau_1$ and $\tau_2$, respectively.
Each vertex shares a fraction $1-\mu$ of its edges with the other vertices of its community and a fraction $\mu$ with the vertices
of the other communities; $0\leq \mu\leq 1$ is the {\it mixing parameter}. 
The graphs are built as follows: 
\begin{enumerate}
\item{A sequence of community sizes obeying the prescribed power law distribution
is extracted. This is done by picking random numbers from a power law distribution with exponent
$\tau_2$.} 
\item{Each vertex $i$ of a community receives an internal degree $(1-\mu)k_i$, where $k_i$ is the degree of vertex $i$, which is
taken by a power law distribution with exponent $\tau_1$. In this way, each vertex $i$ has a number of stubs $(1-\mu)k_i$.}
\item{All stubs of vertices of the same community are randomly attached to each other, until no more stubs are ``free''. 
In this way the sequence of internal degrees of each vertex in its community is maintained.}
\item{Each vertex $i$ receives now an additional number of stubs, equal to $\mu k_i$ (so that the final degree of the vertex is $k_i$),
that are randomly attached to vertices of different communities, until no more stub is ``free''.}
\end{enumerate}
Numerical tests show that this procedure has a complexity $O(m)$, where $m$ is as usual the number of edges of the graph, 
so it can be used to create graphs of sizes spanning several orders of magnitude. Fig.~\ref{fignewbench} shows an example of a LFR benchmark graph. 
Recently the LFR benchmark has been extended to directed and weighted graphs with overlapping communities~\cite{lancichinetti09b}. The software to create the 
LFR benchmark graphs can be freely downloaded at {\tt http://santo.fortunato.googlepages.com/inthepre}\\{\tt ss2}.

A class of benchmark graphs with power law degree distributions had been previously introduced by Bagrow~\cite{bagrow08}.
The construction process starts from a graph with a power-law degree
distribution. Bagrow used Barab\'asi-Albert scale free graphs~\cite{barabasi99}. Then, vertices are randomly assigned to one of four
equally-sized communities. Finally, pairs of edges between two communities are rewired so that either edge ends up within the same community,
without altering the degree sequence (on average). This is straightforward: suppose that the edges join the vertex pairs $a_1$, $b_1$
and $a_2$, $b_2$, where $a_1$, $a_2$ belong to community $A$ and $b_1$, $b_2$ to community $B$. It the edges are replaced by 
$a_1$-$a_2$ and $b_1$-$b_2$ (provided they do not exist already), all vertices keep their degrees. With this rewiring procedure one
can arbitrarily vary the edge density within and, consequently, between clusters. In this class of benchmarks, however,
communities are all of the same size by construction, although one can in principle relax this condition.

A (seemingly) different benchmark is represented by the class of {\it relaxed caveman graphs}, which were originally introduced to explain
the clustering properties of social networks~\cite{watts03}. 
The starting point is a set of disconnected cliques. With some probability edges are rewired to link different cliques. Such model
graphs are interesting as they are smooth variations of the ideal graph with ``perfect'' communities, i.~e. disconnected cliques. On the other hand
the model is equivalent to the planted $\ell$-partition model, where $p_{in}=1-p$ and $p_{out}$ is proportional to $p$, with coefficient
depending on the size of the clusters.

Benchmark graphs have also been introduced to deal with special types of graphs and/or cluster structures. For instance,
Arenas et al.~\cite{arenas06} have introduced a class of benchmark graphs with embedded hierarchical structure, which extends the 
class of graphs by Girvan and Newman. Here there are $256$ vertices and two hierarchical levels, corresponding to a partition in $16$ groups
(microcommunities) with $16$ vertices and a partition in $4$ larger groups of $64$ vertices (macrocommunities), 
comprising each $4$ of the smaller groups. The edge densities
within and between the clusters are indicated by three parameters $z_{in_1}$, $z_{in_2}$ and $z_{out}$: $z_{in_1}$ is the expected internal degree
of a vertex within its microcommunity; $z_{in_2}$ is the expected number of edges that the vertex shares with the
vertices of the other microcommunities within its macrocommunity; $z_{out}$ is the expected 
number of edges connecting the vertex with vertices of the other three macrocommunities. The average degree 
$\langle k\rangle=z_{in_1}+z_{in_2}+z_{out}$ of a vertex is fixed to $18$. Fig.~\ref{fig8} shows an example of hierarchical graph 
constructed based on the same principle, with $512$ vertices and an average degree of $32$. 

Guimer\`a et al.~\cite{guimera07b} have proposed a model of bipartite graphs with built-in communities. 
They considered a bipartite graph of actors and teams, here we describe how to build the benchmarks for general bipartite graphs.
One starts from 
a bipartite graph whose vertex classes $A$ and $B$ are partitioned into $n_c$ groups, ${\cal C}^A_i$ and ${\cal C}^B_i$ ($i=1, 2, ..., n_c$). 
Each cluster ${\cal C}_i$ comprises 
all vertices of the subgroups ${\cal C}^A_i$ and ${\cal C}^B_i$, respectively. With probability 
$p$ edges are placed between vertices of subgroups ${\cal C}^A_i$ and ${\cal C}^B_i$ ($i=1, 2, ..., n_c$), i.~e. within clusters.
With probability $1-p$, edges are placed between vertices of subgroups ${\cal C}^A_i$ and ${\cal C}^B_j$, where $i$ and $j$ are chosen at random, so they
can be equal or different. By construction, a non-zero value of the probability $p$ indicates a preference by vertices to share links with vertices of the
same cluster, i.~e. for $p>0$ the graph has a built-in community structure. For $p=1$ there would be edges only within clusters,
i. e. the graph has a perfect cluster structure. 

Finally, Sawardecker et al. introduced a general model, that accounts for the possibility that clusters overlap~\cite{sawardecker09}. 
The model is based on the reasonable assumption that the probability $p_{ij}$ 
that two vertices are connected by an edge grows with the number $n_0$ of communities both vertices belong to. For vertices in different clusters,
$p_{ij}=p_0$, if they are in the same cluster (and only in that one) $p_{ij}=p_1$, if they belong to the same two clusters $p_{ij}=p_2$, etc..
By hypothesis, $p_0<p_1\leq p_2\leq p_3...$. The planted $\ell$-partition model is recovered when $p_1=p_2=p_3...$.

As we have seen, nearly all existing benchmark graphs are inspired by the planted $\ell$-partition model, to some extent.
However, the model needs to be refined to provide a good description of real graphs with community structure. 
The hypothesis that the linking probabilities of each vertex with the vertices of its community or
of the other communities are constant is not realistic. It is more plausible that each pair of vertices $i$ and $j$ 
has its own linking probability $p_{ij}$, and that such probabilities are correlated for vertices in the same cluster.

Tests on real networks usually focus on a very limited number of examples,
for which one has precise information about the vertices and their properties. 

In Section~\ref{sec2} we have introduced two popular real networks with known community structure, i. e. 
the social network of Zachary's karate club and the social network of bottlenose dolphins 
living in Doubtful Sound (New Zealand), studied by Lusseau.
Here, the question is whether the actual separation
in two social groups could be predicted from the graph topology. Zachary's karate club is by far the most investigated system.
Several algorithms are actually able to identify the two classes, modulo a few intermediate vertices, which may be misclassified.
Other methods are less successful: for instance, the maximum of Newman-Girvan modularity corresponds
to a split of the network in four groups~\cite{duch05,donetti04}.
Another well known example is the network of American 
college football teams, derived by Girvan and Newman~\cite{girvan02}. 
There are $115$ vertices, representing the 
teams, and two vertices are connected if their teams play against each other. 
The teams are divided into $12$ conferences. Games between teams in the same 
conference are more frequent than games between teams of different conferences, 
so one has a natural partition where the communities correspond to the 
conferences.

When dealing with real networks, it is useful to resolve their community structure with different clustering techniques,
to cross-check the results and make sure that they are consistent with each other, as in some cases the answer may strongly depend 
on the specific algorithm adopted.
However, one has to keep in mind that there is no guarantee that
``reasonable'' communities, defined on the basis of non-structural information,
must coincide with those detected by methods based only on the graph structure.

\subsection{Comparing partitions: measures}
\label{sec6_2}

Checking the performance of an algorithm involves defining a criterion to establish how
``similar'' the partition delivered by the algorithm is to the partition one wishes to recover.
Several measures for the similarity of partitions exist. In this section we present and discuss the most popular measures.   
A thorough introduction of similarity measures for graph partitions has been given by 
Meil\u{a}~\cite{meila07} and we will follow it closely.

Let us consider two generic partitions ${\cal X}=(X_1, X_2, ... , X_{n_X})$ and ${\cal Y}=(Y_1, Y_2, ... , Y_{n_Y})$ of a graph $\cal G$, 
with $n_X$ and $n_Y$ clusters, respectively. We indicate with $n$ the number of graph vertices, 
with $n_i^X$ and $n_j^Y$ the number of vertices in clusters $X_i$ and $Y_j$ and with 
$n_{ij}$ the number of vertices shared by clusters $X_i$ and $Y_j$: $n_{ij}=|X_i\bigcap Y_j|$. 

In the first tests using the benchmark graphs by Girvan and Newman (Section~\ref{sec6_1}) scholars used a measure proposed by
Girvan and Newman themselves, the {\it fraction of correctly classified vertices}. A vertex is correctly classified if it is in the same 
cluster with at least half of its ``natural'' partners.
If the partition found by the algorithm has clusters given by the merging of two or more natural groups, all vertices of the cluster
are considered incorrectly classified. The number of correctly classified vertices is then divided by the total size of the 
graph, to yield a number between $0$ and $1$. The recipe to label vertices as 
correctly or incorrectly classified is somewhat arbitrary, though.

Apart from the fraction of correctly classified vertices, which is somewhat {\it ad hoc} and distinguishes the roles
of the natural partition and of the algorithm's partition, most similarity measures can be divided in three categories: measures
based on {\it pair counting}, {\it cluster matching} and {\it information theory}.

Measures based on pair counting depend on the number of pairs of vertices which are classified in the same (different) clusters 
in the two partitions. In particular $a_{11}$ indicates the number of pairs of vertices which are in the same community in both partitions, $a_{01}$
($a_{10}$) the number of pairs of elements which are put in the same community in ${\cal X}$ (${\cal Y}$) and in
different communities in ${\cal Y}$ (${\cal X}$) and $a_{00}$ the number of pairs of vertices that are in different communities in both partitions. 
Wallace~\cite{wallace83} proposed the two indices
\begin{equation}
W_I=\frac{a_{11}}{\sum_kn_k^X(n_k^X-1)/2}; \hskip1cm W_{II}=\frac{a_{11}}{\sum_kn_k^Y(n_k^Y-1)/2}.
\label{eqt01}
\end{equation}
$W_I$ and $W_{II}$ represent the probability that vertex pairs in the same cluster of ${\cal X}$ are also 
in the same cluster for ${\cal Y}$, and viceversa. These indices are asymmetrical, as the role of the two partitions is not the same.
Fowlkes and Mallows~\cite{fowlkes83} suggested to use the geometric mean of $W_I$ and $W_{II}$, which is symmetric.

The {\it Rand index}~\cite{rand71} is the ratio of the number of vertex pairs correctly classified in both partitions (i.~e. either in the same
or in different clusters), by the total number of pairs
\begin{equation}
R({\cal X},{\cal Y})=\frac{a_{11}+a_{00}}{a_{11}+a_{01}+a_{10}+a_{00}}.
\label{eqt02}
\end{equation}
A measure equivalent to the Rand index is the {\it Mirkin metric}~\cite{mirkin96}
\begin{equation}
M({\cal X},{\cal Y})=2(a_{01}+a_{10})=n(n-1)[1-R({\cal X},{\cal Y})].
\label{eqt03}
\end{equation}
The {\it Jaccard index} is the ratio of the number of vertex pairs classified in the same cluster in both partitions, by the
number of vertex pairs which are classified in the same cluster in at least one partition, i.~e.
\begin{equation}
J({\cal X},{\cal Y})=\frac{a_{11}}{a_{11}+a_{01}+a_{10}}.
\label{eqt04}
\end{equation}
Adjusted versions of both the Rand and the Jaccard index exist, in that a null model is introduced, corresponding to the hypothesis of 
independence of the two partitions~\cite{meila07}. The null model expectation
value of the measure is subtracted from the unadjusted version, and the result is normalized by the range of this difference, yielding
$1$ for identical partitions and $0$ as expected value for independent partitions (negative values are possible as well).
Unadjusted indices have the drawback that they are not local, i.~e. the result depends on how the whole graph is partitioned, even when
the partitions differ only in a small region of the graph.

Similarity measures based on cluster matching aim at finding the largest overlaps between pairs of clusters of different partitions.
For instance, the {\it classification error} $H({\cal X},{\cal Y})$ is defined as~\cite{meila01}
\begin{equation}
H({\cal X},{\cal Y})=1-\frac{1}{n}\max_{\pi}\sum_{k=1}^{n_X}n_{k\pi(k)},
\label{eqt05}
\end{equation}
where $\pi$ is an injective mapping from the cluster indices of partition ${\cal Y}$ to the cluster indices of partition ${\cal X}$.
The maximum is taken over all possible injections $\{\pi\}$. In this way one recovers the maximum overlap 
between the clusters of the two partitions. An alternative measure is the {\it normalized Van Dongen metric}, defined as~\cite{vandongen00b}
\begin{equation}
D({\cal X},{\cal Y})=1-\frac{1}{2n}\left[\sum_{k=1}^{n_X}\max_{k^\prime}n_{kk^{\prime}}+\sum_{k^{\prime}=1}^{n_Y}\max_{k}n_{kk^{\prime}}\right].
\label{eqt06}
\end{equation}
A common problem of this type of measures is that some clusters may not be taken into account, if their overlap with clusters of the
other partition is not large enough. Therefore if we compute the similarity between two partitions $\cal X$ and $\cal X^\prime$ 
and partition $\cal Y$, with $\cal X$ and $\cal X^\prime$
differing from each other by a different subdivision
of parts of the graph that are not used to compute the measure, one would obtain the same score. 

The third class of similarity measures is based on reformulating the problem of comparing 
partitions as a problem of message decoding within the framework of information theory~\cite{mackay03}.
The idea is that, if two partitions are similar, one needs very little information to infer one partition given the other. 
This extra information can be used as a measure of dissimilarity. To evaluate the Shannon information content~\cite{mackay03} of a partition,
one starts by considering the community assignments $\{x_i\}$ and $\{y_i\}$, where $x_i$ and $y_i$ indicate the cluster labels of 
vertex $i$ in partition ${\cal X}$ and ${\cal Y}$, respectively. One assumes that the labels $x$ and $y$ are values of two random
variables $X$ and $Y$, with joint distribution $P(x,y)=P(X=x,Y=y)=n_{xy}/n$, which implies that $P(x)=P(X=x)=n_x^X/n$ 
and $P(y)=P(Y=y)=n_y^Y/n$. 
The {\it mutual information} $I(X,Y)$ of two random variables has been previously defined [Eq.~(\ref{eqstinf8})], and can be 
applied as well to partitions ${\cal X}$ and ${\cal Y}$, since they are described by random variables.
Actually $I(X,Y)=H(X)-H(X|Y)$, where $H(X)=-\sum_xP(x)\log P(x)$ is the Shannon entropy of $X$ and $H(X|Y)=-\sum_{x,y}P(x,y)\log P(x|y)$
is the conditional entropy of $X$ given $Y$. The mutual information is not ideal as a similarity measure: in fact, given a partition
${\cal X}$, all partitions derived from ${\cal X}$ by further partitioning (some of) its clusters would all have 
the same mutual information with ${\cal X}$, even though they could be very different from each other. In this case the mutual information
would simply equal the entropy $H(X)$, because the conditional entropy would be systematically zero. To avoid that, Danon et al.
adopted the {\it normalized mutual information}~\cite{danon05}
\begin{equation}
I_{norm}({\cal X}, {\cal Y})=\frac{2I(X,Y)}{H(X)+H(Y)},
\label{eqt08}
\end{equation}
which is currently very often used in tests of graph clustering algorithms. 
The normalized mutual information equals $1$ if the partitions are identical, whereas it has an expected value of $0$ if the partitions are independent.
The measure, defined for standard partitions, in which each vertex belongs to only one cluster, has been recently extended to the 
case of overlapping clusters by Lancichinetti et al.~\cite{lancichinetti09}. 
The extension is not straightforward as the community assignments of a partition
are now specified by a vectorial random variable, since each vertex may belong to more clusters simultaneously. 
In fact, the definition by Lancichinetti et al. is not a proper extension
of the normalized mutual information, in the sense that it does not recover exactly the same value 
of the original measure for the comparison of proper partitions
without overlap, even though the values are close.

Meil\u{a}~\cite{meila07} introduced the {\it variation of information} 
\begin{equation}
V({\cal X}, {\cal Y})=H(X|Y)+H(Y|X),
\label{eqt09}
\end{equation}
which has some desirable properties with respect to the normalized mutual information and other measures. In particular, it defines
a metric in the space of partitions as it has the properties of distance. It is also a local measure, i.~e. the similarity of 
partitions differing only in a small portion of a graph depends on the differences of the clusters in that region, and not 
on the partition of the rest of the graph. The maximum value of the variation of information is $\log n$, so similarity values
for partitions of graphs with different size cannot be compared with each other. For meaningful comparisons one could divide $V({\cal X}, {\cal Y})$
by $\log n$, as suggested by Karrer et al.~\cite{karrer08}.

A concept related to similarity is that of {\it distance}, which indicates basically how many operations need to be performed in order to
transform a partition to another. Gustafsson et al. defined two distance measures for partitions~\cite{gustafsson06}. They are both based on
the concept of {\it meet} of two partitions, which is defined as
\begin{equation}
{\cal M}=\bigcup_{i=1}^{n_A}\bigcup_{j=1}^{n_B}\left[X_i\bigcap Y_j\right].
\label{eqt1}
\end{equation}
The distance measures are $m_{moved}$ and $m_{div}$. In both cases they are determined by 
summing the distances of ${\cal X}$ and ${\cal Y}$ from the meet ${\cal M}$. For $m_{moved}$ 
the distance of ${\cal X}$ (${\cal Y}$) from the meet is
the minimum number of elements that must be moved between ${\cal X}$ and ${\cal Y}$
so that ${\cal X}$ (${\cal Y}$) and ${\cal M}$ coincide~\cite{gusfield02}. For $m_{div}$ the distance of ${\cal X}$ (${\cal Y}$) from the meet is
the minimum number of divisions that must be done in ${\cal X}$ (${\cal Y}$) so that ${\cal X}$ (${\cal Y}$) and ${\cal M}$ coincide~\cite{stanley97}.
Such distance measures can easily be transformed in similarity measures, like 
\begin{equation}
I_{moved}=1-m_{moved}/n,\hskip1cm I_{div}=1-m_{div}/n.
\label{eqt2}
\end{equation}
Identical partitions have zero mutual distance and similarity $1$ based on Eqs.~\ref{eqt2}.

Finally an important problem is how to define the similarity between clusters. If two partitions ${\cal X}$ and ${\cal Y}$ of a graph are similar, 
each cluster of ${\cal X}$ will be very similar to one cluster of ${\cal Y}$, and viceversa, and it is important to identify the pairs
of corresponding clusters. For instance, if one has information about the time evolution of a graph, one could monitor 
the dynamics of single clusters as well, by keeping track of each cluster at different time steps~\cite{palla07}. Given clusters $X_i$ and $Y_j$,
their similarity can be defined through the {\it relative overlap} $s_{ij}$
\begin{equation}
s_{ij}=\frac{|X_i\bigcap Y_j|}{|X_i\bigcup Y_j|}.
\label{eqt20}
\end{equation}
In this way, looking for the cluster of ${\cal Y}$ corresponding to $X_i$ means finding the cluster $Y_j$ that maximizes $s_{ij}$. The index
$s_{ij}$ can be used to define similarity measures for partitions as well~\cite{zhang06b,fan07}. An interesting discussion on 
the problem of comparing partitions, along with further definitions of similarity measures not discussed here, can be found in Ref.~\cite{traud08}.

\subsection{Comparing algorithms}
\label{sec6_3}

The first systematic comparative analysis of graph clustering techniques has been carried out by Danon et al.~\cite{danon05}.  
They compared the performances of various algorithms on the benchmark graphs by Girvan and Newman (Section~\ref{sec6_1}). 
The algorithms examined are listed in Table~\ref{Table_Orders}, along with their complexity. 
\begin{table*}
  \centering

\begin{tabular}{|c|c|c|c|}


  \hline

  Author &Ref. & Label & Order \\

  \hline

  \hline

  Eckmann \& Moses&\cite{eckmann02}&  EM & $O(m\langle k^2\rangle)$ \\

  Zhou \& Lipowsky &\cite{zhou04} & ZL & $O(n^3)$ \\

  Latapy \& Pons & \cite{latapy05} & LP & $O(n^3)$ \\

  Clauset et al. &\cite{clauset04} & NF & $O(n\log^2n)$ \\

  Newman \& Girvan &\cite{newman04b} &  NG & $O(nm^2)$ \\

  Girvan \& Newman &\cite{girvan02} &  GN & $O(n^2m)$ \\

  Guimer\`{a} et al. & \cite{guimera04,guimera05} & SA & parameter dependent\\

  Duch \& Arenas &\cite{duch05} & DA & $O(n^2\log n)$ \\

  Fortunato et al. &\cite{fortunato04} & FLM & $O(m^3n)$ \\

  Radicchi et al. &\cite{radicchi04} & RCCLP & $O(m^4/n^2)$\\

  Donetti \& Mu\~noz&\cite{donetti04,donetti05} & DM/DMN & $O(n^3)$ \\

  Bagrow \& Bollt &\cite{bagrow05}&  BB & $O(n^3)$ \\

  Capocci et al. &\cite{capocci05}& CSCC & $O(n^2)$ \\

  Wu \& Huberman &\cite{wu04}& WH & $O(n+m)$ \\

  Palla et al. & \cite{palla05} & PK & $O(\exp(n))$\\

  Reichardt \& Bornholdt &\cite{reichardt04} & RB & parameter dependent\\

  \hline

\end{tabular}

\caption{\label{Table_Orders}List of the algorithms used in the comparative analysis of Danon et al.~\cite{danon05}. The first column indicates
the names of the algorithm designers, the second the original reference of the work, the third the symbol used to indicate the algorithm
and the last the computational complexity of the technique. Adapted from Ref.~\cite{danon05}.}
\end{table*}
\begin{figure}
\begin{center}
\includegraphics[width=\columnwidth]{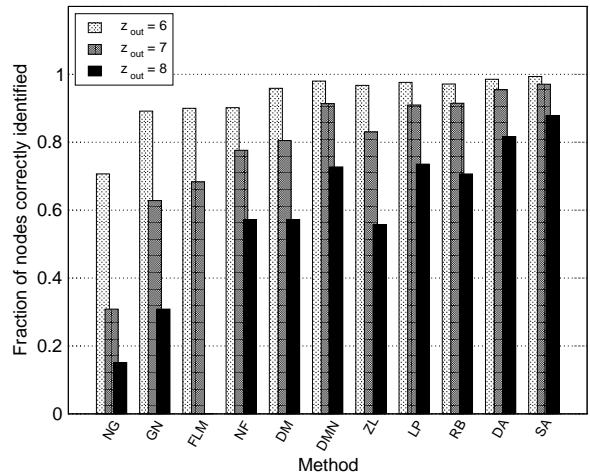}
\caption{\label{figdanon} Relative performances of the algorithms listed in Table~\ref{Table_Orders} on the Girvan-Newman benchmark, for
three values of the expected average external degree $z_{out}$. Reprinted figure with permission from Ref.~\cite{danon05}.
\copyright 2005 by IOP Publishing and SISSA.}
\end{center}
\end{figure}
\begin{table*}
  \centering

\begin{tabular}{|c|c|c|c|}


  \hline

  Author &Ref. & Label & Order \\

  \hline

  \hline

  Girvan \& Newman &\cite{girvan02,newman04b}&  GN & $O(nm^2)$ \\

  Clauset et al. &\cite{clauset04} & Clauset et al. & $O(n\log^2n)$ \\

  Blondel et al. &\cite{blondel08} & Blondel et al. & $O(m)$ \\

  Guimer\`{a} et al. & \cite{guimera04,guimera05} & Sim. Ann. & parameter dependent\\

  Radicchi et al. &\cite{radicchi04} & Radicchi et al. & $O(m^4/n^2)$ \\

  Palla et al. & \cite{palla05} & Cfinder & $O(\exp(n))$\\

  Van Dongen &\cite{vandongen00} & MCL & $O(nk^2)$, $k<n$ parameter \\

  Rosvall \& Bergstrom &\cite{rosvall07} & Infomod & parameter dependent \\

  Rosvall \& Bergstrom &\cite{rosvall08} & Infomap & $O(m)$ \\

  Donetti \& Mu\~noz&\cite{donetti04,donetti05} & DM & $O(n^3)$ \\

  Newman \& Leicht & \cite{newman07} & EM & parameter dependent \\

  Ronhovde \& Nussinov & \cite{ronhovde09} & RN & $O(m^\beta\log n)$, $\beta\sim 1.3$ \\

  \hline

\end{tabular}

\caption{\label{Table2} List of the algorithms used in the comparative analysis of Lancichinetti and 
Fortunato~\cite{lancichinetti09d}. The first column indicates
the names of the algorithm designers, the second the original reference of the work, the third the symbol used to indicate the algorithm
and the last the computational complexity of the technique.}
\end{table*}
Fig.~\ref{figdanon} shows the performance of all algorithms. Instead of showing the 
whole curves of the similarity versus $z_{out}$ (Section~\ref{sec6_1}), which would display 
a fuzzy picture with many strongly overlapping curves, 
difficult to appreciate, Danon et al.
considered three values for $z_{out}$ ($6$, $7$ and $8$), and represented the result for each algorithm as a group of three columns,
indicating the average value of the similarity between the planted partition and the partition found by the method for each of the three
$z_{out}$-values. The similarity was measured in terms of the fraction of correctly classified vertices (Section~\ref{sec6_1}). The comparison
shows that modularity optimization via simulated annealing (Section~\ref{sub_sec6_0_2}) yields the best results, although it is a rather slow procedure, that 
cannot be applied to graphs of size of the order of $10^5$ vertices or larger.
On the other hand, we have already pointed out that the benchmark by Girvan and Newman is not a good representation
of real graphs with community structure, which are characterized by heterogeneous distributions of degree and community sizes.
In this respect, the class of graphs designed by Lancichinetti et al. (LFR benchmark)~\cite{lancichinetti08} (Section~\ref{sec6_1}) 
poses a far more severe test to clustering techniques.
For instance, many methods have problems to detect clusters of very different sizes (like most methods listed
in Table~\ref{Table_Orders}). 
For this reason, Lancichinetti and Fortunato have carried out a careful comparative analysis
of community detection methods on the much more restrictive LFR benchmark~\cite{lancichinetti09d}. The algorithms
chosen are listed in Table~\ref{Table2}.
In Fig.~\ref{companalysis} the performances of the algorithms on the LFR benchmark are compared.
Whenever possible, tests on the versions of the LFR benchmark with directed edges, weighted edges 
and/or overlapping communities~\cite{lancichinetti09b} were carried out.
Lancichinetti and Fortunato also tested the methods on random graphs, to check whether they are able
to notice the absence of community structure. 
From the results of all tests, the Infomap method by Rosvall and Bergstrom~\cite{rosvall08} appears to be the best, but also the algorithms by
Blondel et al.~\cite{blondel08} and by Ronhovde and Nussinov~\cite{ronhovde09} have a good performance. These three methods
are also very fast, with a complexity which is essentially linear in the system size, so they can be applied to large systems.
On the other hand, modularity-based methods (with the exception of the method by Blondel et al.)
have a rather poor performance, which worsens for larger systems and smaller communities, due to the well known resolution limit
of modularity~\cite{fortunato07}. The performance of the remaining methods worsens considerably if one increases the system size 
(DM and Infomod) or the community size (Cfinder, MCL and method by Radicchi et al.).
\begin{figure}
\begin{center}
\includegraphics[width=8cm]{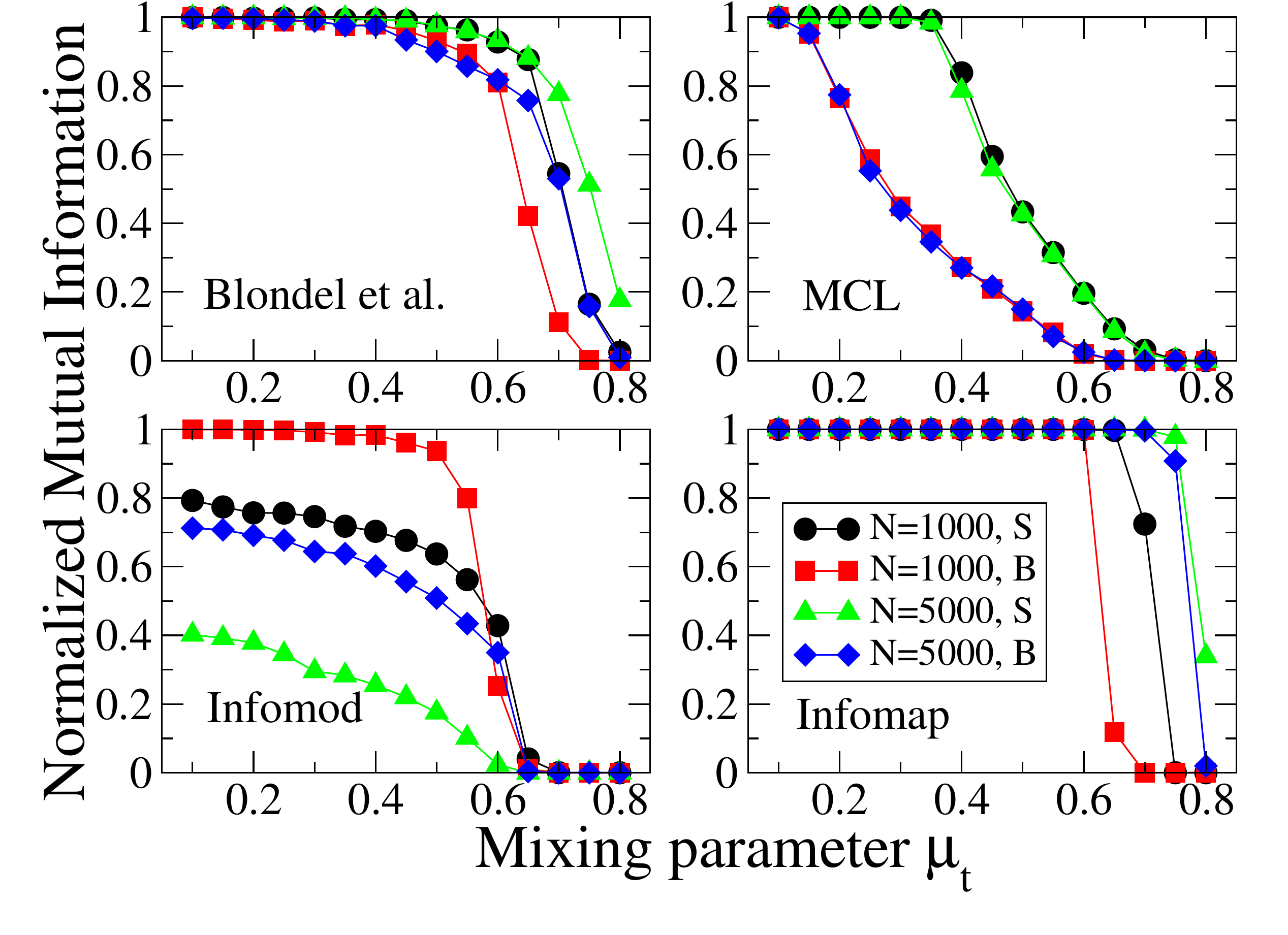}
\includegraphics[width=8cm]{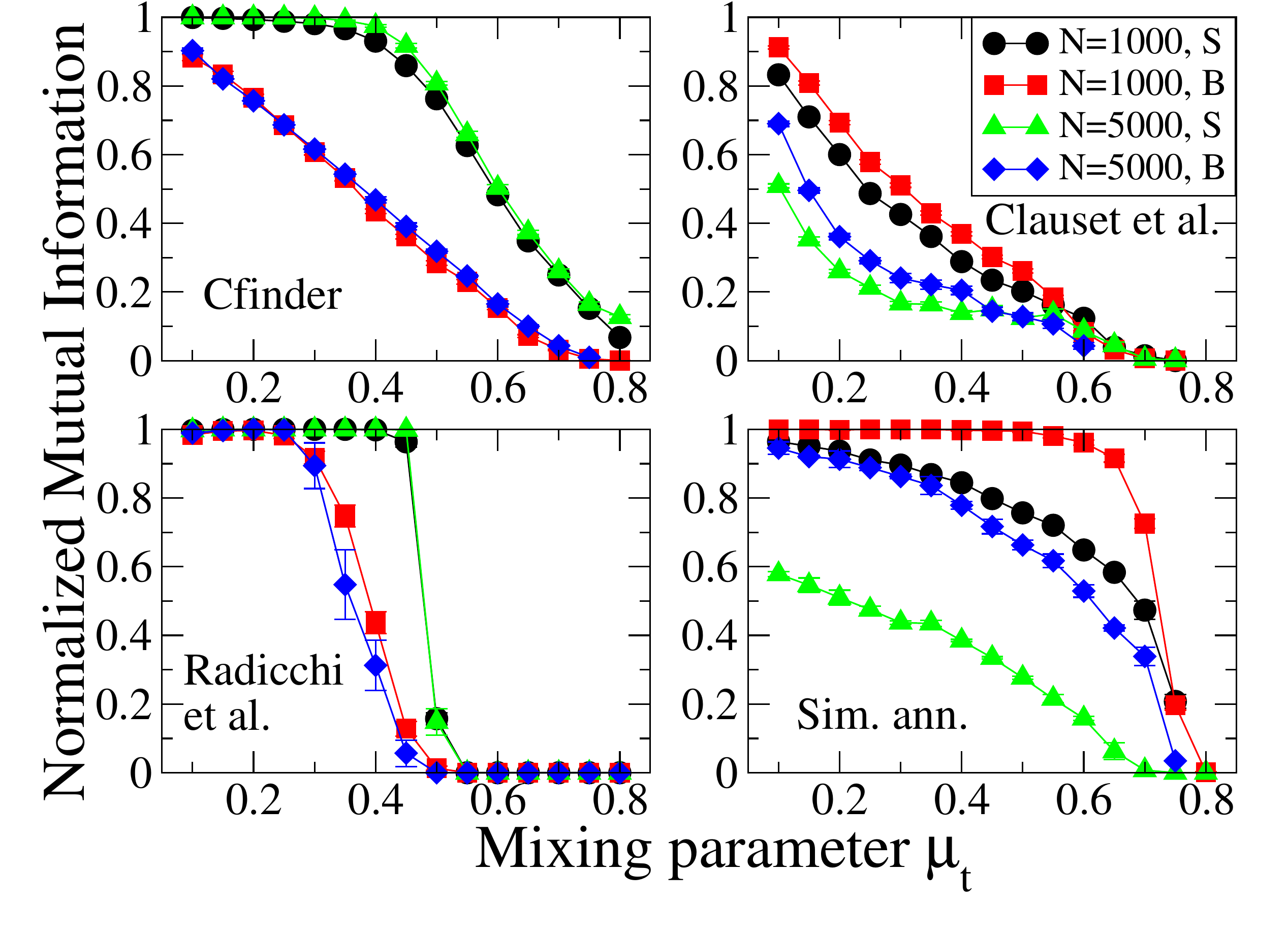}
\includegraphics[width=8cm]{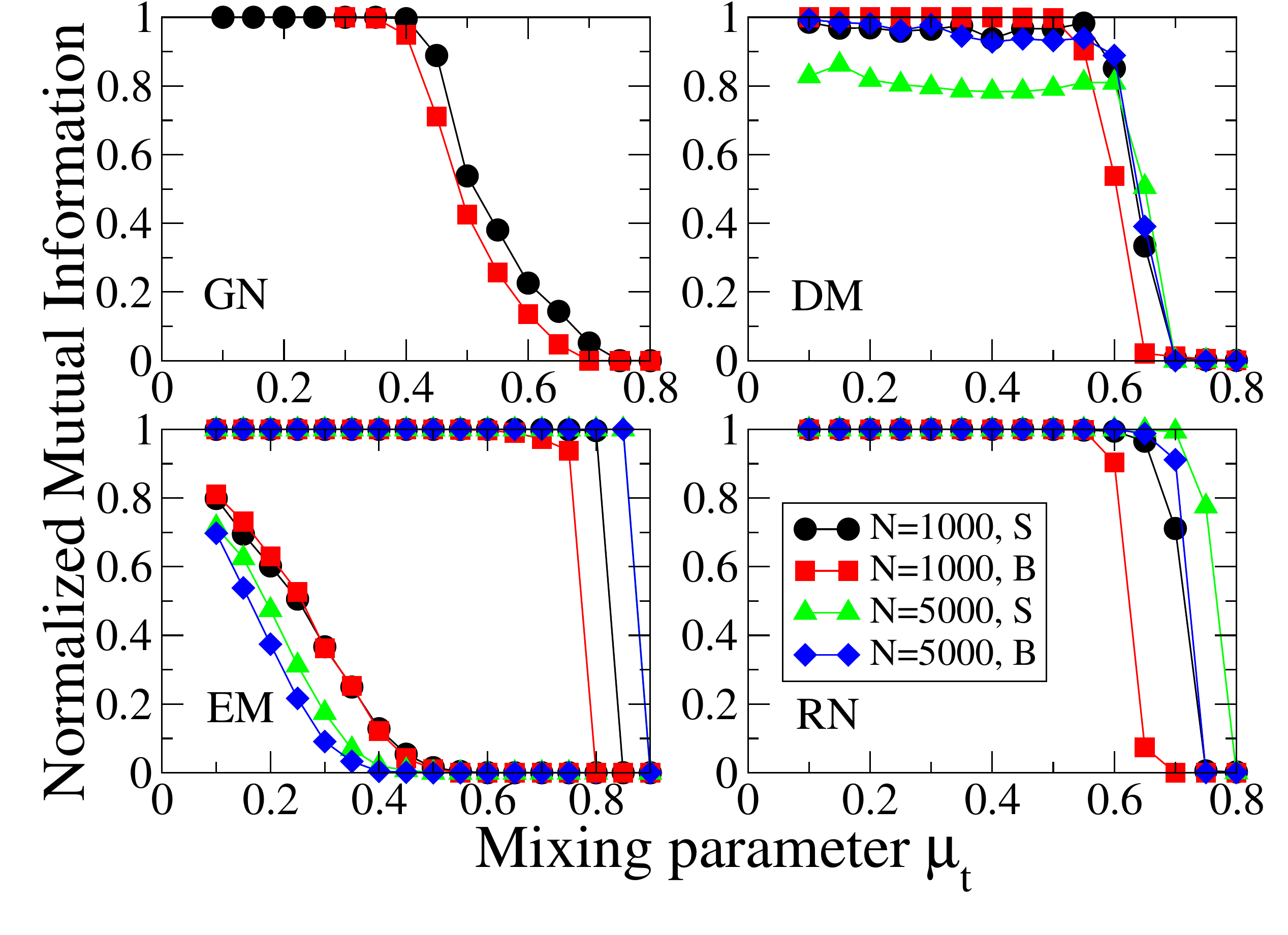}
\caption{\label{companalysis} Performances of several algorithms on the LFR benchmark~\cite{lancichinetti09d}. The plots show
the normalized mutual information (in the version proposed in Ref.~\cite{lancichinetti09}) as a function of the mixing parameter
of the benchmark graphs. The different curves for each method refer to different system sizes ($1000$ and $5000$ vertices)
and community size ranges [(S)=from $10$ to $50$ vertices, (B)=from $20$ to $100$ vertices]. For the GN algorithm only the smaller
graph size was adopted, due to the high complexity of the method, whereas for the EM method there are eight curves instead of four because
for each set of benchmark graphs the algorithm was run starting from two different initial conditions. 
Reprinted figure with permission from 
Ref.~\cite{lancichinetti09d}. \copyright 2009 by the American Physical Society.}
\end{center}
\end{figure}

\begin{figure}
\begin{center}
\includegraphics[width=\columnwidth]{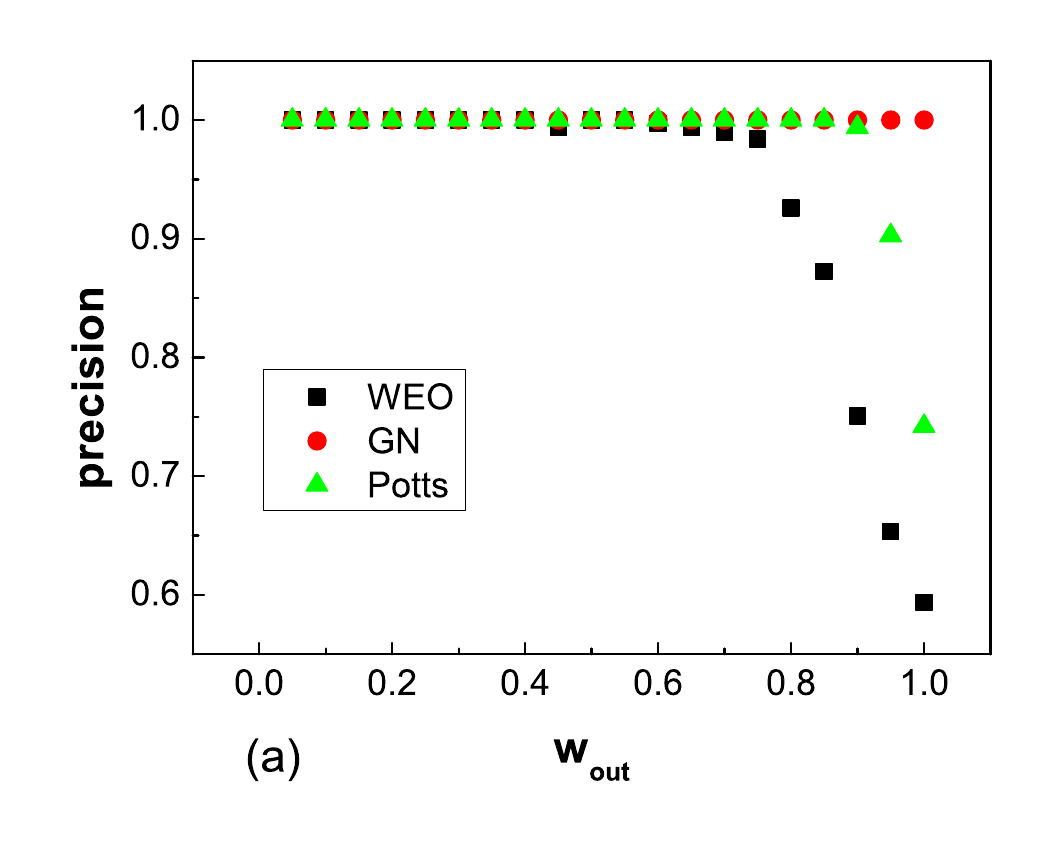}
\includegraphics[width=\columnwidth]{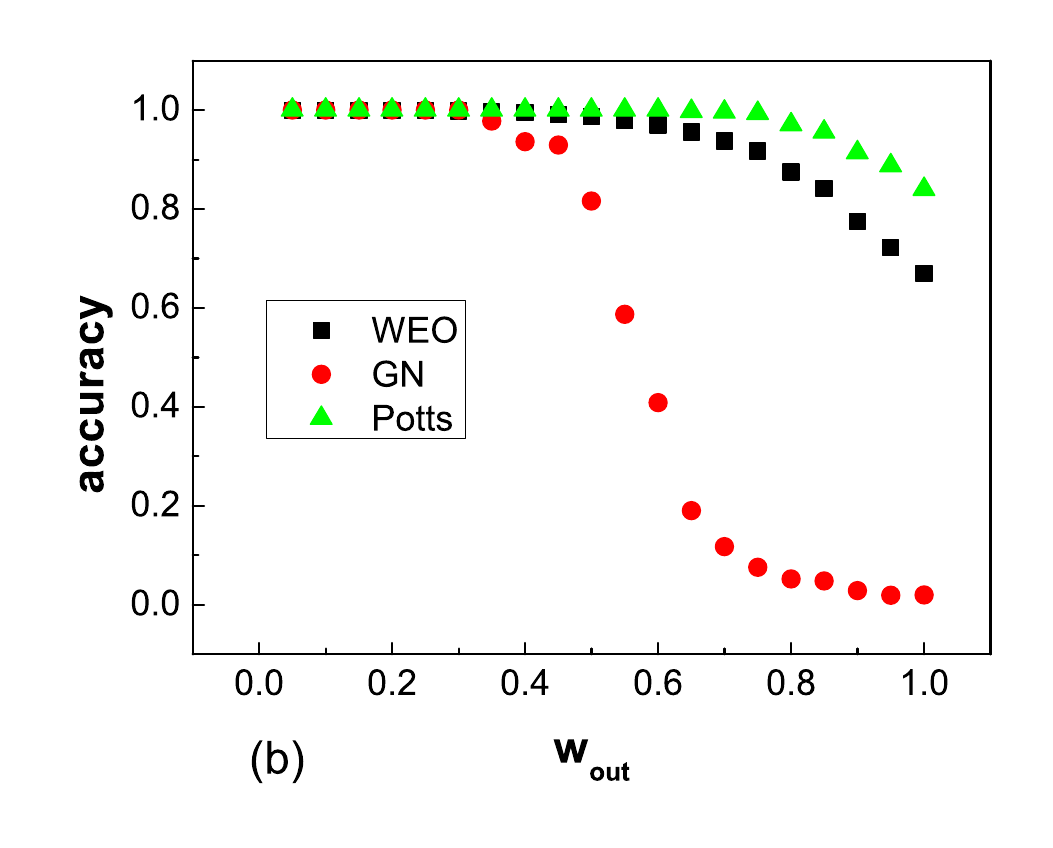}
\caption{\label{fantest} Comparative evaluation of the performances of algorithms to find communities in weighted graphs. Tests are carried out on
a weighted version of the benchmark of Girvan and Newman. The two plots show how good the algorithms are in terms of the precision and accuracy with which
they recover the planted partition of the benchmark. Precision indicates how close the values of similarity between the planted and the model partition are 
after repeated experiments with the same set of parameters; accuracy indicates how close the similarity values are to the ideal result (1) after
repeated experiments with the same set of parameters. The similarity measure adopted here is based on the relative overlap of clusters
of Eq.~\ref{eqt20}. We see that the maximization of modularity with extremal optimization (WEO) and the Potts model
algorithm (Potts) are both precise and accurate as long as the weight of the inter-cluster edges $w_{inter}$ remains lower than the weight of the
intra-cluster edges ($w_{inter}<1$). Reprinted figures with permission from 
Ref.~\cite{fan07}. \copyright 2007 by Elsevier.}
\end{center}
\end{figure}

Fan et al. have evaluated the performance of some algorithms to detect communities on weighted graphs~\cite{fan07}. The algorithms are:
modularity maximization, carried out with extremal optimization (WEO) (Section~\ref{sub_sec6_0_3}); the Girvan-Newman algorithm (WGN)
(Section~\ref{subsec5_1}); the Potts model algorithm by Reichardt and Bornholdt (Potts) (Section~\ref{sec44_1}). 
All these techniques have been originally introduced
for unweighted graphs, but we have shown that they can easily be extended to weighted graphs. The algorithms
were tested on the weighted version of the benchmark of Girvan and Newman, that we discussed in Section~\ref{sec6_1}.
Edge weights have only two values: $w_{inter}$ for inter-cluster edges and $w_{intra}$ for intra-cluster edges. Such values are
linked by the relation $w_{intra}+w_{inter}=2$, so they are not independent. For testing one uses realizations of the benchmark with fixed topology
(i.~e. fixed $z_{out}$) and variable weights.
In Fig.~\ref{fantest} the comparative performance of the three algorithms is illustrated. 
The topology of the benchmark graphs 
corresponds to $z_{out}=8$, i.~e. to graphs in which each vertex has approximately the same number of neighbors inside and outside its
community. By varying $w_{inter}$ from $0$ to $2$ one goes smoothly from a situation in which most of the weight is concentrated inside
the clusters, to a situation in which instead the weight is concentrated between the clusters. From Fig.~\ref{fantest} we see that WEO and Potts are 
more reliable methods.  

Sawardecker et al. have tested methods to detect overlapping communities~\cite{sawardecker09}. 
They considered three algorithms: modularity optimization,
the Clique Percolation Method (CPM) (Section~\ref{sec45_1}) and the 
modularity landscape surveying method by Sales-Pardo et al.~\cite{sales07} (Section~\ref{sub_sec6_202}).
For testing, Sawardecker et al. defined a class of benchmark graphs in which the linking probability between vertices
is an increasing function of the number of clusters the vertices belong to. We have described this benchmark in Section~\ref{sec6_1}.
It turns out that the modularity landscape surveying method is able to identify overlaps between communities, as long as the 
fraction of overlapping vertices is small. Curiously, the CPM, designed to find overlapping communities, has a poor performance,
as the overlapping vertices found by the algorithm are in general different from the overlapping vertices of the planted partition of the benchmark.
The authors also remark that, if the overlap between two clusters is not too small, it may be hard (for any method) to recognize whether
the clusters are overlapping or hierarchically organized, i.~e. loosely connected clusters within a large cluster.

We close the section with some general remarks concerning testing. We have seen that a testing procedure
requires two crucial ingredients: benchmark graphs with built-in community structure and clustering algorithms that try to recover it.
Such two elements are not independent, however, as they are both based on the concept of community. If the underlying notions
of community for the benchmark and the algorithm are very different, one can hardly expect that the algorithm will do a good job on the
benchmark. Furthermore, there is a third element, i.~e. the quality of a partition. All benchmarks start from
a situation in which communities are clearly identified, i.~e. connected components of the graph, and introduce some amount of noise, that eventually leads
to a scenario where clusters are hardly or no longer detectable. It is then important to keep track of how the quality of the natural partition of the 
benchmark worsens as the amount of noise increases, in order to distinguish configurations in which the graphs have a cluster structure,
that an algorithm should then be able to resolve, from configurations in which the noise prevails and the natural clusters are not meaningful.
Moreover, quality functions are important to evaluate the performance of an algorithm on graphs whose community structure is unknown.
Quality functions are strongly related to the concept of community as well, as they 
are supposed to evaluate the goodness of the clusters, so they require a clear quantitative concept of what a cluster is. It is very important
for any testing framework to check for the mutual dependencies between the benchmark, the quality function used to evaluate 
partitions, and the clustering algorithm to be tested. This issue has so far received very little attention~\cite{delling07}.
Finally, empirical tests are also very important, as one ultimately wishes to apply clustering techniques to real graphs. Therefore, 
it is crucial to collect more data sets of graphs whose community structure is known or deducible from information on the vertices and their edges. 

\section{General properties of real clusters}
\label{sec7_1}

What are the general properties of partitions and clusters of real graphs? In many papers on graph clustering 
applications to real systems are presented. In spite of the variety of clustering methods that one could employ, in many cases
partitions derived from different techniques are rather similar to each other, so the general properties of clusters
do not depend much on the particular algorithm used. The analysis of clusters and their properties
delivers a {\it mesoscopic description} of the graph, where the communities, and not the vertices, are the elementary 
units of the topology. The term mesoscopic is used because the relevant scale here lies between the scale of the vertices and
that of the full graph. 

One of the first issues addressed was whether the communities 
of a graph are usually about of the same size or whether the community sizes have some special distribution.
Most clustering techniques consistently find skewed distributions of community sizes, with a tail 
described with good approximation by a power law (at least, a sizeable portion of the curve) with exponents in the range between 
$1$ and $3$~\cite{palla05,newman04, danon07, clauset04, radicchi04}. So, there seems to be no characteristic
size for a community: small communities usually coexist with large ones.
\begin{figure}
\begin{center}
\includegraphics[width=8cm]{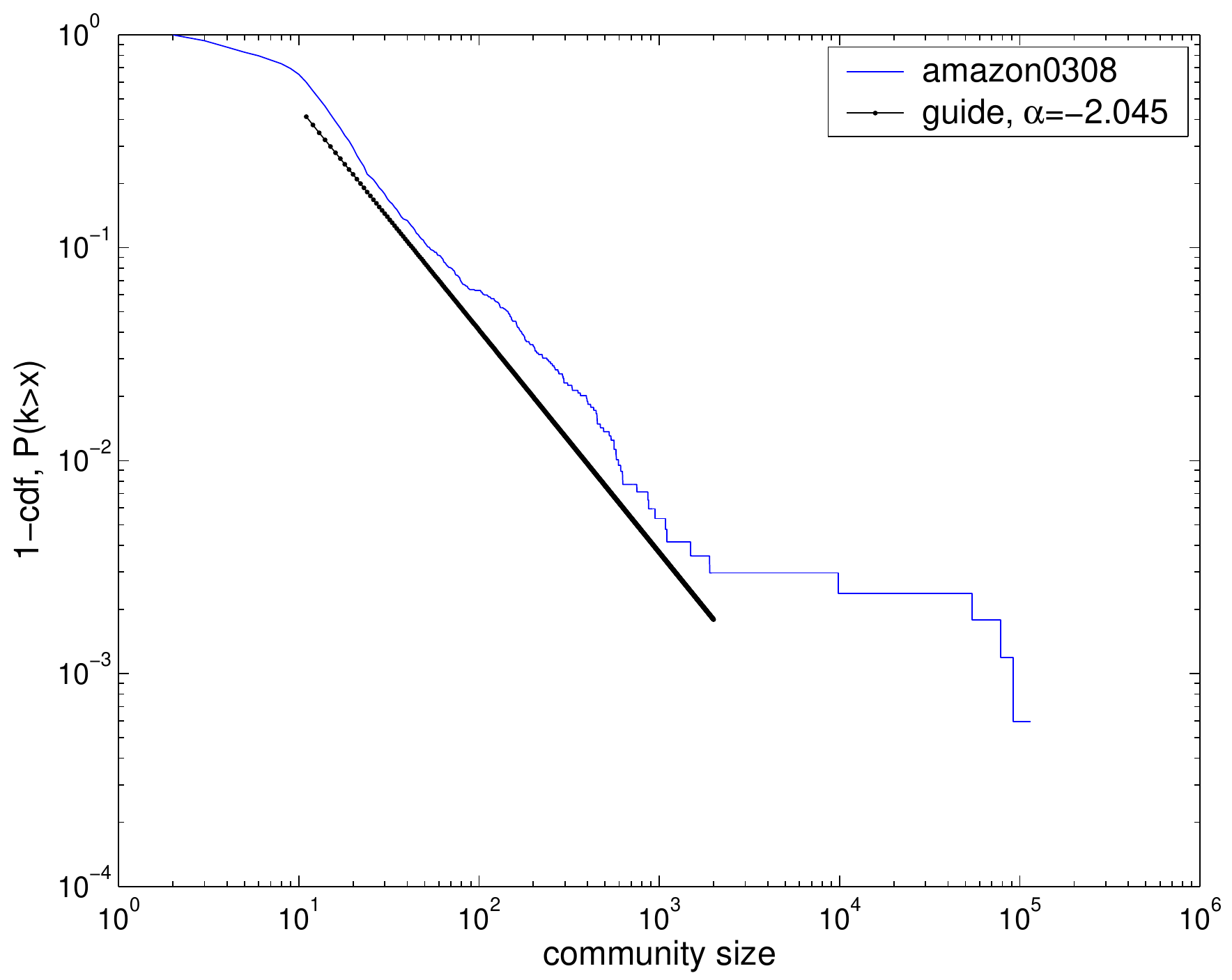}
\caption {\label{Figure13} Cumulative distribution of community sizes for the 
Amazon purchasing network. The partition is derived by greedy modularity optimization.  
Reprinted figure with permission 
from Ref.~\cite{clauset04}. \copyright 2004 by the American Physical Society.}
\end{center}
\end{figure}
As an example, Fig.~\ref{Figure13} shows the cumulative distribution of community sizes for 
a recommendation network of the online vendor Amazon.com. Vertices are products and 
there is a connection between item $A$ and $B$ if $B$ was frequently purchased by buyers of $A$.
Recall that the cumulative distribution is the integral of the probability distribution: if the 
cumulative distribution is a power law $s^{-\alpha}$, the probability distribution is also 
a power law with exponent $-(\alpha+1)$. 

\begin{figure}
\begin{center}
\includegraphics[width=\columnwidth]{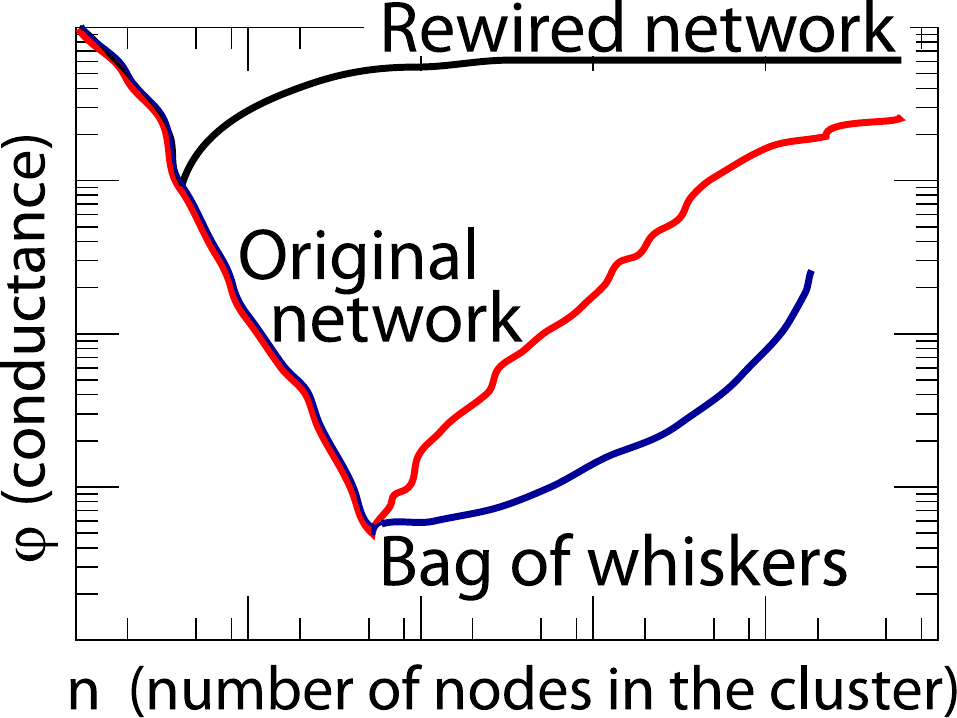}
\vskip0.5cm
\includegraphics[width=\columnwidth]{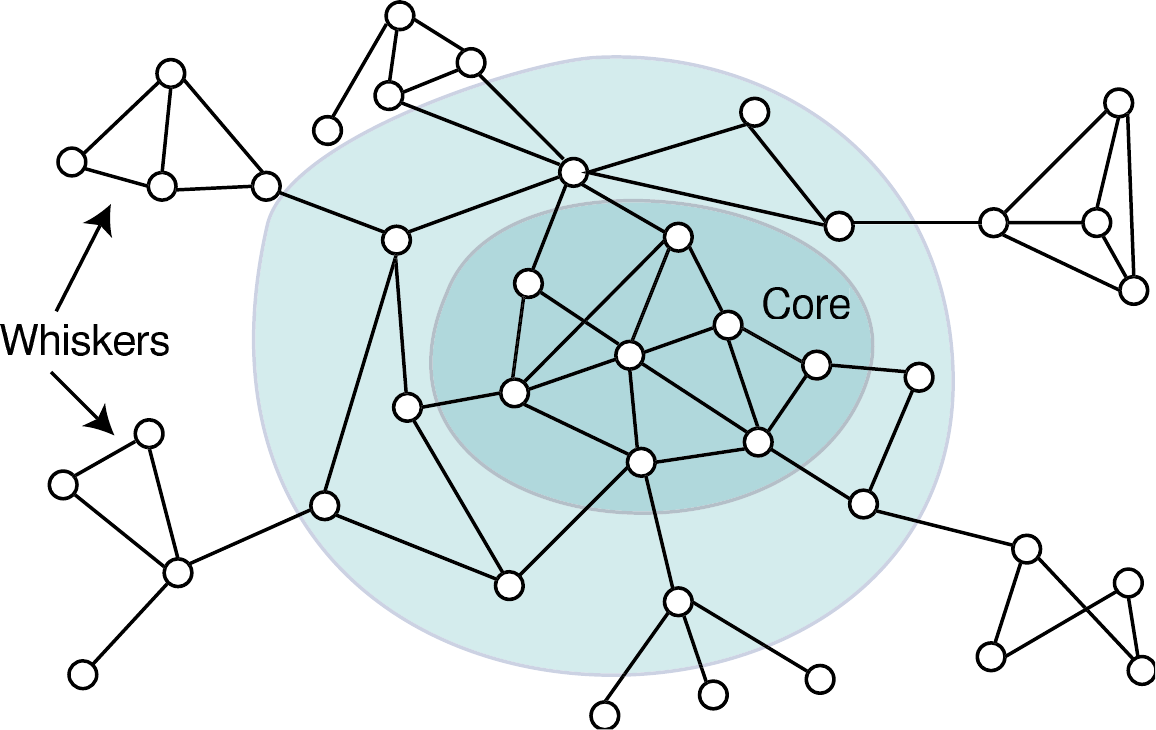}
\caption {\label{figlesk} Analysis of communities in large real networks by Leskovec et al.~\cite{leskovec08b}.
(Left) Typical shape of the network community profile plot (NCPP), showing how the minimum conductance of subgraphs of size $n$ 
varies with $n$. The plot indicates that the ``best'' communities have a size of about $100$ vertices (minimum of the curve), whereas communities of larger
sizes are not well-defined. In the plot two other NCPPs are shown: the one labeled {\it Rewired network}
corresponds to a randomized version of 
the network, where edges are randomly rewired by keeping the degree distribution; the one labeled {\it Bag of whiskers} 
gives the minimum conductance scores of clusters composed of disconnected pieces.
(Right) Scheme of the core-periphery
structure of large social and information networks derived by Leskovec et al. based on the results of their empirical analysis.
Most of the vertices are in a central core, which does not have a clear community structure, whereas the
best communities, which are rather small, are weakly connected to the core. 
Reprinted figure with permission from 
Ref.~\cite{leskovec08b}.}
\end{center}
\end{figure}

Leskovec et al.~\cite{leskovec08b} have gone one step further. They
carried out a systematic analysis of communities in large real networks, including traditional and on-line
social networks, technological, information networks and web graphs. The main goal was to 
assess the quality of communities at various sizes. As a quality function the {\it conductance} of the cluster was chosen.
We remind that the conductance of a cluster is the ratio between the cut size of the cluster and the minimum between the total degree
of the cluster and that of the rest of the graph (Section~\ref{sec4_1}). So, if the cluster is much smaller than the whole graph, the 
conductance equals the ratio between the cut size and the total degree of the cluster. Since a ``good'' cluster is characterized by 
a low cut size and a large internal density of edges, low values of the conductance indicate good clusters. For each 
real network Leskovec et al. derived the {\it network community profile plot} (NCPP), showing the minimum conductance score
among subgraphs of a given size as a function of the size. Interestingly, they found that the NCPPs of all networks they studied
have a characteristic shape: they go downwards up until subgraphs with about $100$ vertices, and then they rise monotonically for larger subgraphs
(Fig.~\ref{figlesk}). This seems to suggest that communities are well defined only when they are fairly small in size.
Such small clusters are weakly connected to the rest of the graph,
often by a single edge (in this case they are called {\it whiskers}), and form the {\it periphery} of the network. 
The other vertices form a big {\it core}, in which 
larger clusters are well connected to each other, and are therefore barely distinguishable (Fig.~\ref{figlesk}).
Leskovec et al. performed low-conductance cuts with several methods, to ensure that the result is not a simple artefact 
of a particular chosen technique. Moreover, they have also verified that, for large real networks with known 
community structure (such as, e.g., the social network of the on-line blogging site 
{\it LiveJournal}, with its user groups), the NCPP has the same qualitative shape
if one takes the real communities instead of low-conductance subgraphs. The analysis by Leskovec et al.
may shed new light on our understanding of community structure and its detection in large networks.
The fact that the ``best'' communities appear to have a characteristic size of about $100$ vertices is consistent
with Dunbar conjecture that $150$ is the upper size limit for a working human community~\cite{dunbar98}. 
On the other hand, if large communities are very mixed with each other, as Leskovec et al. claim,
they could hardly be considered communities, and the alleged ``community structure'' of large networks would be limited
to their peripheral region. The results by Leskovec et al. may be affected by the properties of  
conductance, and need to be validated with alternative approaches.  
In any case, whatever the value of the quality score of a cluster may be (low or high), it is necessary to 
estimate the significance of the cluster (Section~\ref{sec6_4}), before deciding whether it is a meaningful structure or not.

\begin{figure}
\begin{center}
\includegraphics[width=\columnwidth]{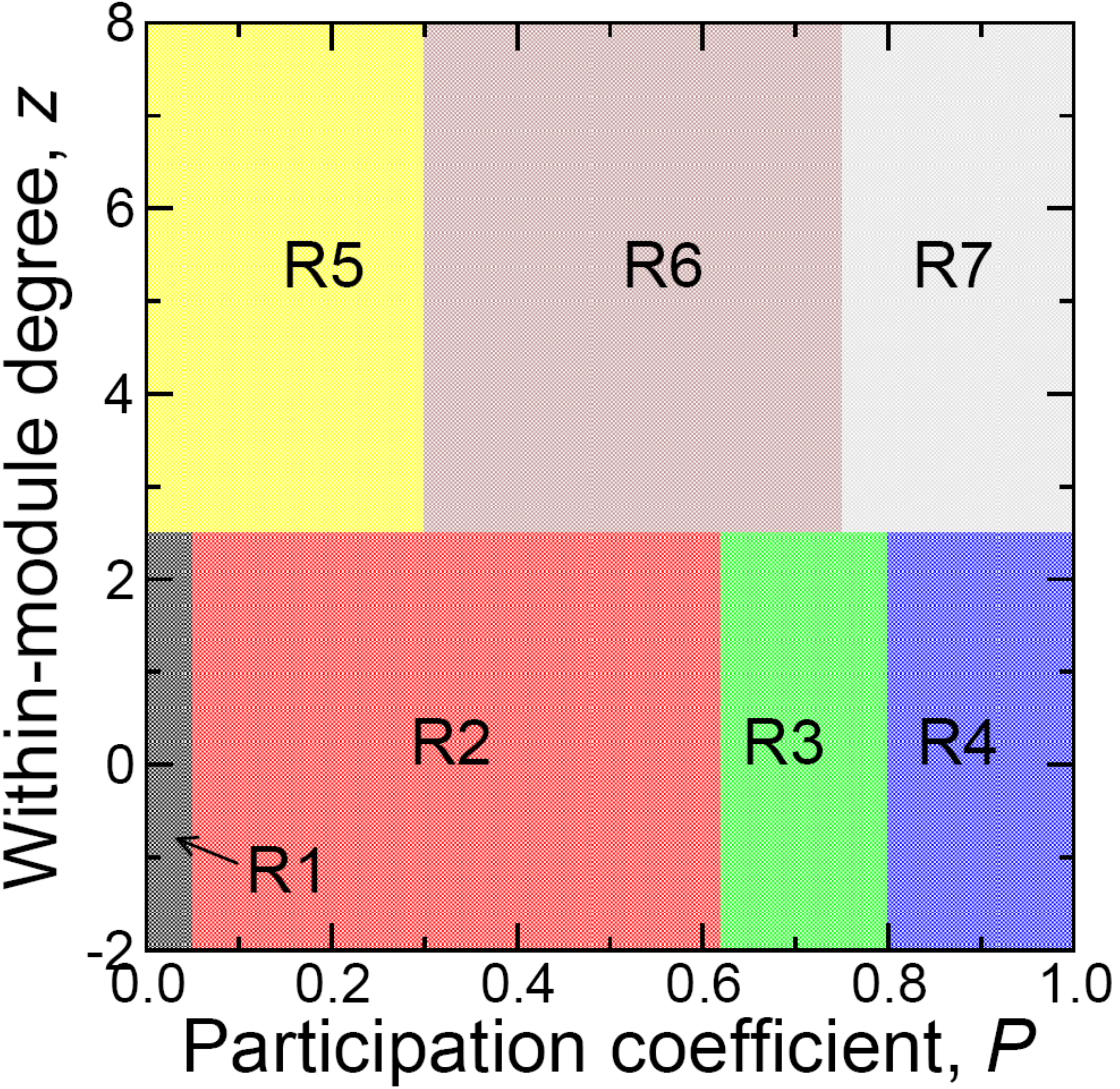}
\caption {\label{figroles} Regions of the $z-P$ plane defining the roles of vertices in the modular structure of a graph, 
according to the scheme of Guimer\`a and Amaral~\cite{guimera05, guimera05b}. Reprinted figure with permission from 
Ref.~\cite{guimera05}. \copyright 2005 by the Nature Publishing Group.}
\end{center}
\end{figure}

If the community structure of a graph is known, it is possible to classify vertices according
to their roles within their community, which may allow to infer individual properties of the
vertices. A promising classification has been proposed by Guimer\`a and 
Amaral~\cite{guimera05,guimera05b}. The role of a vertex depends on the values of 
two indices, the {\it within-module degree} and the {\it participation ratio} (though other variables may be chosen, in principle).
The within-module degree $z_i$ of vertex $i$ is defined as 
\begin{equation}
z_i=\frac{\kappa_i-{\bar{\kappa}}_{s_i}}{\sigma_{\kappa_{s_i}}},
\label{eqtrol1}
\end{equation}
where $\kappa_i$ is the internal degree of $i$ in its cluster $s_i$, ${\bar{\kappa}}_{s_i}$ and $\sigma_{\kappa_{s_i}}$
the average and standard deviation of the internal degrees for all vertices of cluster $s_i$. The within-module degree
is then defined as the $z$-score of the internal degree $\kappa_i$. Large values of $z$ indicate
that the vertex has many more neighbors within its community than most other vertices of the community.  
Vertices with $z\geq 2.5$ are classified as {\it hubs},
if $z<2.5$ they are {\it non-hubs}. The participation ratio $P_i$ of vertex $i$ is defined as
\begin{equation}
P_i=1-\sum_{s=1}^{n_c}\left(\frac{\kappa_{is}}{k_i}\right)^2.
\label{eqtrol2}
\end{equation}
Here $\kappa_{is}$ is the internal degree of $i$ in cluster $s$, $k_i$ the degree of $i$. Values of $P$ close to $1$
indicate that the neighbors of the vertex are uniformly distributed among all clusters; if all neighbors are within 
the cluster of the vertex, instead, $P=0$. Based on the values of the pair $(z, P)$, Guimer\`a and Amaral distinguished seven roles for
the vertices. Non-hub vertices can be {\it ultra-peripheral} ($P\approx 0$), {\it peripheral} ($P<0.625$), {\it connectors}
($0.625<P<0.8$) and {\it kinless vertices} ($P>0.8$). Hub vertices are classified in {\it provincial hubs} 
($P<\sim 0.3$), {\it connector hubs} ($0.3<P<0.75$) and 
{\it kinless hubs} ($P>0.75$). The regions of the $z-P$ plane corresponding to the seven roles are highlighted in Fig.~\ref{figroles}. We stress that
the actual boundaries of the regions can be chosen rather arbitrarily.
On graphs without community structure, like Erd\"os-R\'enyi~\cite{erdos59} random graphs and Barab\'asi-Albert~\cite{barabasi99} graphs (Section~\ref{sec1_3}), 
non-hubs are mostly kinless vertices. In addition, if there are hubs, like in Barab\'asi-Albert graphs, they are kinless hubs.
Kinless hubs (non-hubs) vertices have less than half (one third) of their neighbors inside any cluster, so they are not clearly associated to a cluster.
On real graphs, the topological roles can be correlated to functions of vertices: in metabolic networks, for instance,
connector hubs, which share most edges with vertices of other clusters than their own, are often
metabolites which are more conserved across species than other metabolites,  
i.~e. they have an evolutionary advantage~\cite{guimera05}. 

If communities are overlapping, one can explore other statistical properties, like the distributions
of the overlaps and of the vertex memberships. The overlap is defined as 
the number of vertices shared by each pair of overlapping clusters; the membership of a vertex
is the number of communities including the vertex. Both distributions turn out to be skewed, so there 
seem to be no characteristic values for the overlap and the membership. Moreover, 
one could derive a network, where the
communities are the vertices and pairs of vertices are connected if their corresponding communities
overlap~\cite{palla05}. Such networks seem to have some special properties. For instance, 
the degree distribution is a particular function, with an initial exponential
decay followed by a slower power law decay\footnote{This holds for the networks considered by Palla et al.~\cite{palla05} like, e. g.,
the word association network (Section~\ref{sec2}) and a coauthorship network of physicists. There is no {\it a priori} reason to believe that this 
result is general.}. 
We stress that the above results have been obtained with the Clique Percolation
Method by Palla et al. (Section~\ref{sec45_1}) and it is not clear whether other techniques would confirm them or not.
In a recent analysis it has been shown that the degree distribution of the network of communities
can be reproduced by assuming that the graph grows 
according to a simple preferential attachment mechanism, where communities
with large degree have an enhanced chance to interact/overlap with new communities~\cite{pollner06}. 

\section{Applications on real-world networks}
\label{sec7_2}

The ultimate goal of clustering algorithms is trying to infer properties of and relationships between vertices, that are not 
available from direct observation/measurement. If the scientific community agrees on a set of reliable techniques, one could then proceed with 
careful investigations of systems in various domains. So far, most works in the literature on graph clustering focused on the 
development of new algorithms, and applications were limited to those few benchmark graphs that one typically uses for testing (Section~\ref{sec6_1}).
Still, there are also applications aiming at understanding real systems. Some
results have been actually mentioned in the previous sections. This section is supposed to give a flavor of what 
can be done by using clustering algorithms. Therefore, the list of works presented here is by no means exhaustive.
Most studies focus on biological and social networks. We mention a few applications to other
types of networks as well.

\subsection{Biological networks}
\label{sec7_2_1}

The recent abundance of genomic data has allowed us to explore the cell at an unprecedented depth. A wealth of information 
is available on interactions involving proteins and genes, metabolic processes, etc. In order to study cellular systems, 
the graph representation is regularly used. Protein-protein interaction networks (PIN), gene regulatory networks (GRN) and metabolic networks (MN)
are meanwhile standard objects of investigation in biology and bioinformatics~\cite{junker08}. 

Biological networks are characterized by a remarkable modular organization, reflecting functional associations between their components.  
For instance, proteins tend to be associated in two types of cellular modules: {\it protein complexes} and {\it functional modules}. 
A protein complex is a group of proteins that mutually interact at the same time and space, forming a sort of physical object. Examples are 
transcription factor complexes, protein transport and export complexes, etc. Functional modules instead are groups of proteins taking place in the same
cellular process, even if the interactions may happen at different times and places. Examples are the CDK/cyclin module, responsible for cell-cycle
progression, the yeast pheromone response pathway, etc.. Identifying cellular modules is fundamental to uncover the organization and dynamics of 
cell functions. However, the information on cell units (e. g. proteins, genes) and their interactions is often incomplete, or even incorrect, 
due to noise in the data
produced by the experiments. Therefore, inferring modules from the topology of cellular networks enables one to restrict the set of possible
scenarios and can be a safe guide for future experiments. 

Rives and Galitski~\cite{rives03} studied the modular organization of a subset of the PIN of the yeast 
({\it Saccharomyces cerevisiae}), consisting
of the (signaling) proteins involved in the processes leading the microorganism to a filamentous form. 
The clusters were detected with a hierarchical clustering technique. Proteins mostly interacting with members of their 
own cluster are often essential proteins; edges between modules are important points of communication. Spirin and Mirny~\cite{spirin03}
identified protein complexes and functional modules in yeast with different techniques: clique detection, 
superparamagnetic clustering~\cite{blatt96} and optimization of cluster edge density. They estimated the statistical significance
of the clusters by computing the $p$-values of seeing those clusters in random graphs with the same expected degree sequence as the original network.
From the known functional annotations of yeast genes one can see that the modules usually group proteins with the same or consistent biological
functions. Indeed, in many cases, the modules exactly coincide with known protein complexes. The results appear robust if noise is introduced
in the system, to simulate the noise present in the experimental data. Functional modules in yeast were also found by Chen and Yuan~\cite{chen06}, who
applied the algorithm by Girvan and Newman with a modified definition of edge betweenness (Section~\ref{subsec5_1}). The standard 
Girvan-Newman algorithm has proved to be reliable to detect functional modules in PINs~\cite{dunn05}. 
The novelty of the work by Chen and Yuan is its focus on weighted PINs, 
where the weights come from information derived through microarray expression profiles. Weights add information about the system
and should lead to a more reliable modular structure.
By knocking out genes in the same structural cluster similar phenotypes appeared, suggesting that the genes have similar biological roles. Moreover, 
the clusters often contained known protein complexes, either entirely or to a large extent. Finally, Chen and Yuan were able to make predictions
of the unknown function of some genes, based on the structural module they belong to: gene function prediction is the most promising
outcome deriving from the application of clustering techniques to PINs. Farutin et al.~\cite{farutin06} have adopted a local concept of community, and 
derived a hierarchical decomposition of PINs, in that the modules identified at some level become the vertices of a network at the higher level. 
Communities are overlapping, to account for the fact that proteins (and whole modules) may have diverse biological functions.
High level structures detected in a human PIN correspond to general biological concepts like signal transduction, regulation of gene expression,
intercellular communication. Sen et al.~\cite{sen06} identified protein clusters for yeast from the eigenvectors of the 
Laplacian matrix (Section~\ref{sec1_2}), computed via Singular Value Decomposition. In a recent analysis, Lewis et al.~\cite{lewis09}
carefully explored the relationship between structural communities of PINs and their biological function. 
Communities were detected with the multiresolution approach 
by Reichardt and Bornholt~\cite{reichardt06} (Section~\ref{sub_sec6_01}). A community is considered biologically homogeneous if 
the functional similarity between protein pairs of the community (extracted through the Gene Ontology database~\cite{ashburner00}) is larger than
the functional similarity between all protein pairs of the network. Lewis et al. also specified the comparison to interacting and non-interacting
protein pairs. As a result, many communities turn out to be biologically homogeneous, especially if they are not too small. Moreover,
some topological attributes of communities, like  
the within-community clustering coefficient (i.e. the average value of the clustering coefficients 
of the vertices of a community, computed by considering just 
the neighbors belonging to the community) and link density
(density of internal edges), are good indicators of biological homogeneity: the former 
is strongly correlated with 
biological homogeneity, independently of the community size, whereas for the latter the correlation is strong for large communities.  

Metabolic networks have also been extensively investigated. We have already discussed the ``functional cartography'' designed by Guimer\`a
and Amaral~\cite{guimera05,guimera05b}, which applies to general types of networks, not necessarily metabolic. A hierarchical decomposition 
of metabolic networks has been derived by Holme et al.~\cite{holme03}, by using a hierarchical clustering technique inspired by the
algorithm by Girvan and Newman (Section~\ref{subsec5_1}). Here, vertices are removed based on their betweenness values, which are obtained by dividing
the standard site betweenness scores~\cite{freeman77} by the indegree of the respective vertices. A picture of metabolic network emerges, in which 
there are core clusters centered at hub-substances, surrounded by outer shells of less connected substances, and a few other clusters at intermediate scales.
In general, clusters at different scales seem to be meaningful, so the whole hierarchy should be taken into account.

Wilkinson and Huberman~\cite{wilkinson04} analyzed a network of gene co-occurrence to find groups of related genes. 
The network is built by connecting pairs of genes that are mentioned together in the abstract of articles of the Medline database
({\tt http://medline.cos.com/}). 
Clusters were found with a modified version of the algorithm by Girvan and Newman, in which 
edge betweenness is computed by considering the shortest paths of a small subset of all vertex pairs, to gain computer time (Section~\ref{subsec5_1}).
As a result, genes belonging to the same cluster turn out to be functionally related to each other.
Co-occurrence of terms is also used to extract associations between genes and diseases, to find out which
genes are relevant for a specific disease. Communities of genes related to colon
cancer can be helpful to identify the function of the genes.

\subsection{Social networks}
\label{sec7_2_2}

Networks depicting social interactions between people have been studied for decades~\cite{wasserman94,scott00}.
Recently the modern Information and Communication Technology (ICT) has opened new 
interaction modes between individuals, like mobile phone communications
and online interactions enabled by the Internet. Such
new social exchanges can be accurately monitored for very large systems, 
including millions of individuals, whose study
represents a huge opportunity for social science. Communities of social networks
can be friendship circles, groups of people sharing common interests and/or activities, etc.. 

\begin{figure}
\begin{center}
\includegraphics[width=8cm]{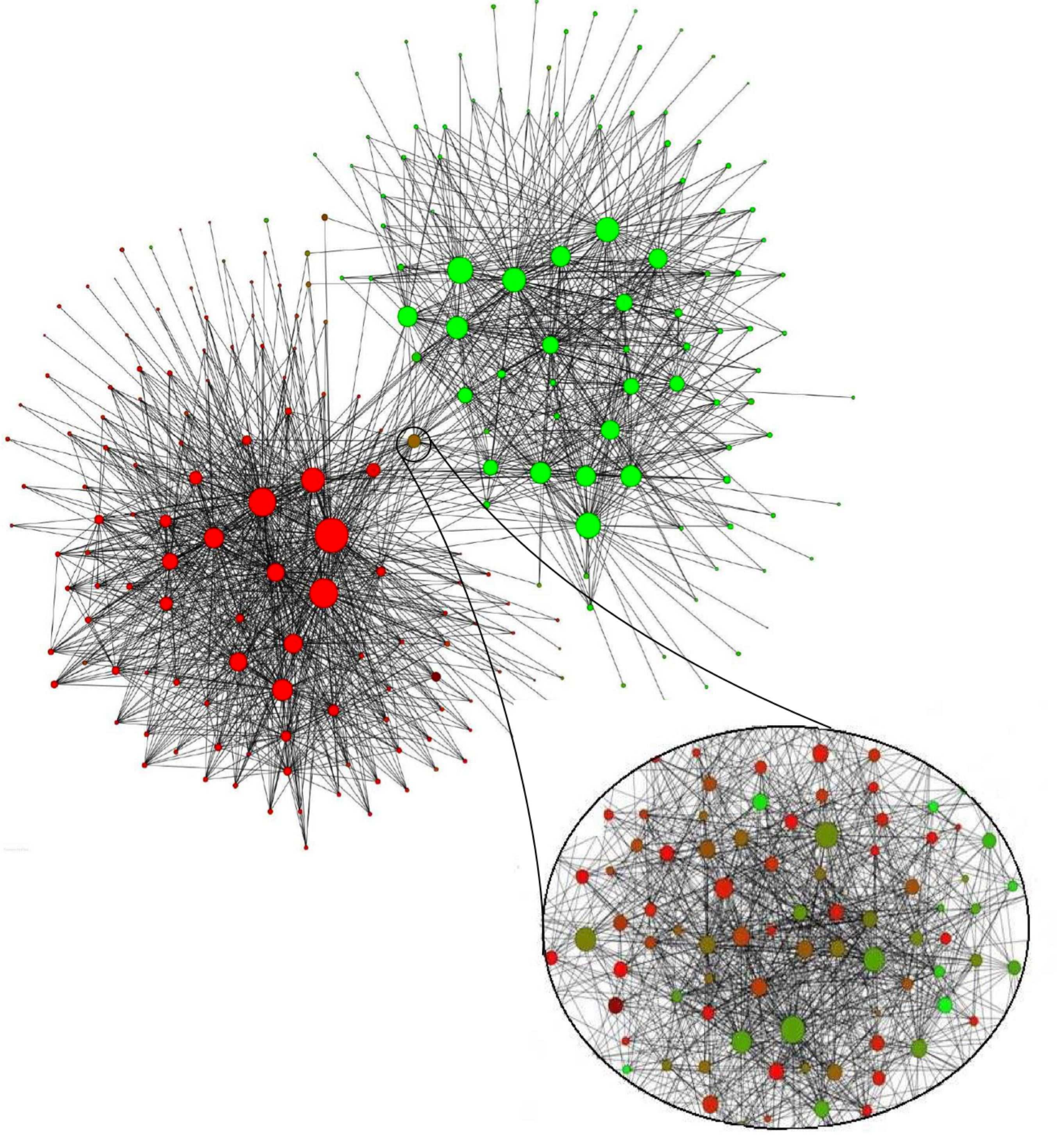}
\caption {\label{blondphone} Community structure of a social network of mobile phone communication in Belgium. Dots indicate subcommunities at
the lower hierarchical level (with more than $100$ people) and are colored in a red-green scale to represent the level of representation of the two
main languages spoken in Belgium (red for French and green for Dutch).
Communities of the two larger groups are linguistically homogeneous, with more than $85\%$ of people
speaking the same language. Only one community (zoomed), which lies at the border between the two main aggregations, has a more balanced
distribution of languages. Reprinted figure with permission from Ref.~\cite{blondel08}.
\copyright 2008 by IOP Publishing and SISSA.}
\end{center}
\end{figure}

Blondel et al. have analyzed a network of mobile phone communications between users of a Belgian phone operator~\cite{blondel08}.
The vertices of the graph are 2.6 millions and the edges are weighted by the cumulative duration of phone calls between users in the observation time frame.
The clustering analysis, performed with a fast hierarchical modularity optimization technique developed
by the authors (discussed in Section~\ref{sub_sec6_0_1}), delivers six hierarchical levels. 
The highest level consists of $261$ groups with more than $100$ vertices, which are clearly 
arranged in two main groups, linguistically homogeneous,
reflecting the linguistic split of Belgian population (Fig.~\ref{blondphone}).
Tyler et al.~\cite{tyler03} studied a network of e-mail exchanges between people working at HP Labs. They applied the same modified
version of Girvan-Newman algorithm that two of the authors have used to find communities of related genes~\cite{wilkinson04} (Section~\ref{sec7_2_1}).
The method enables one to measure the degree of membership of each vertex in a community and allows for overlaps between communities. 
The detected clusters matched quite closely the organization of the Labs in departments and project groups, as confirmed by interviews
conducted with researchers. 
\begin{figure}
\begin{center}
\includegraphics[width=7cm]{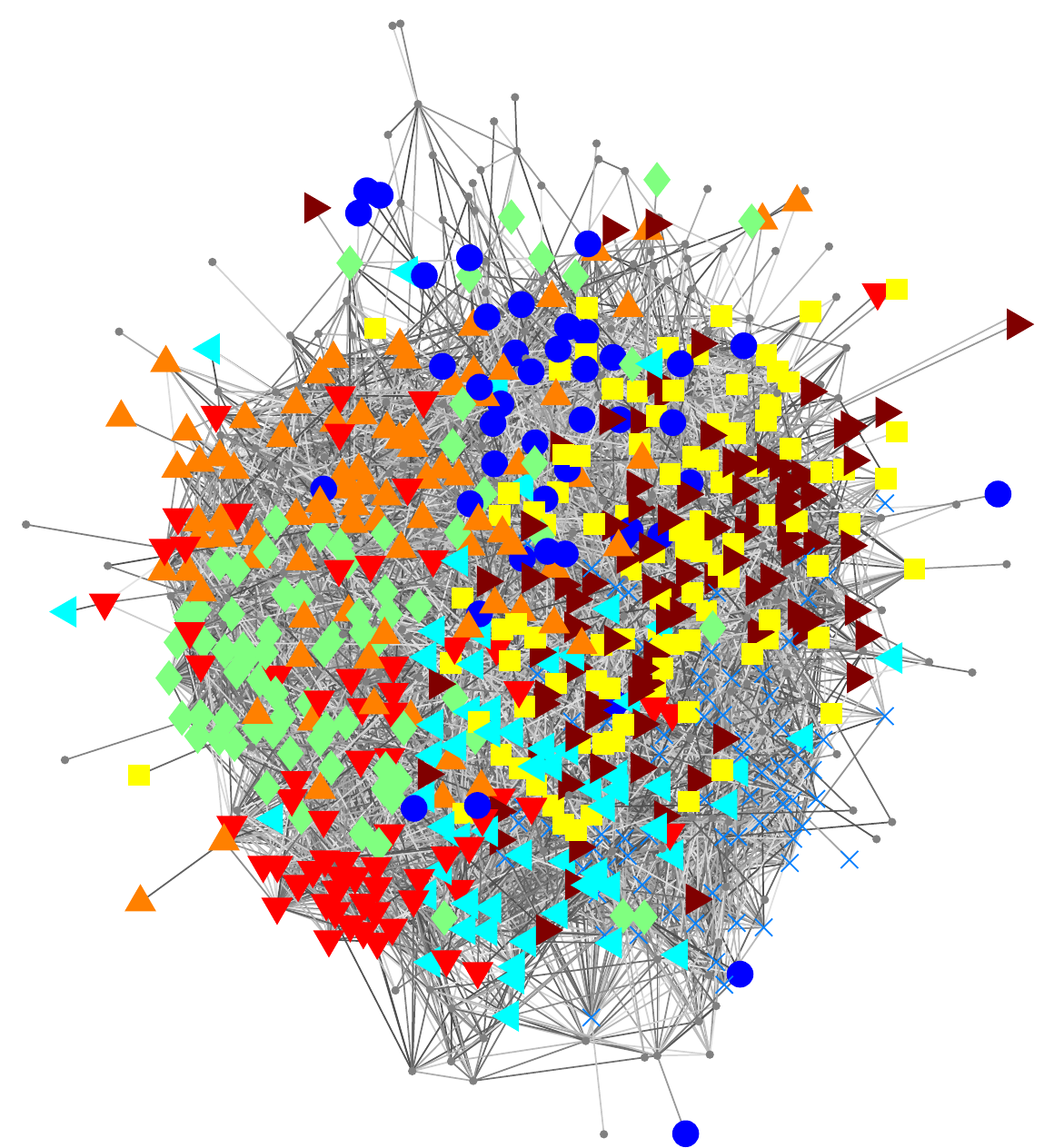}
\includegraphics[width=7cm]{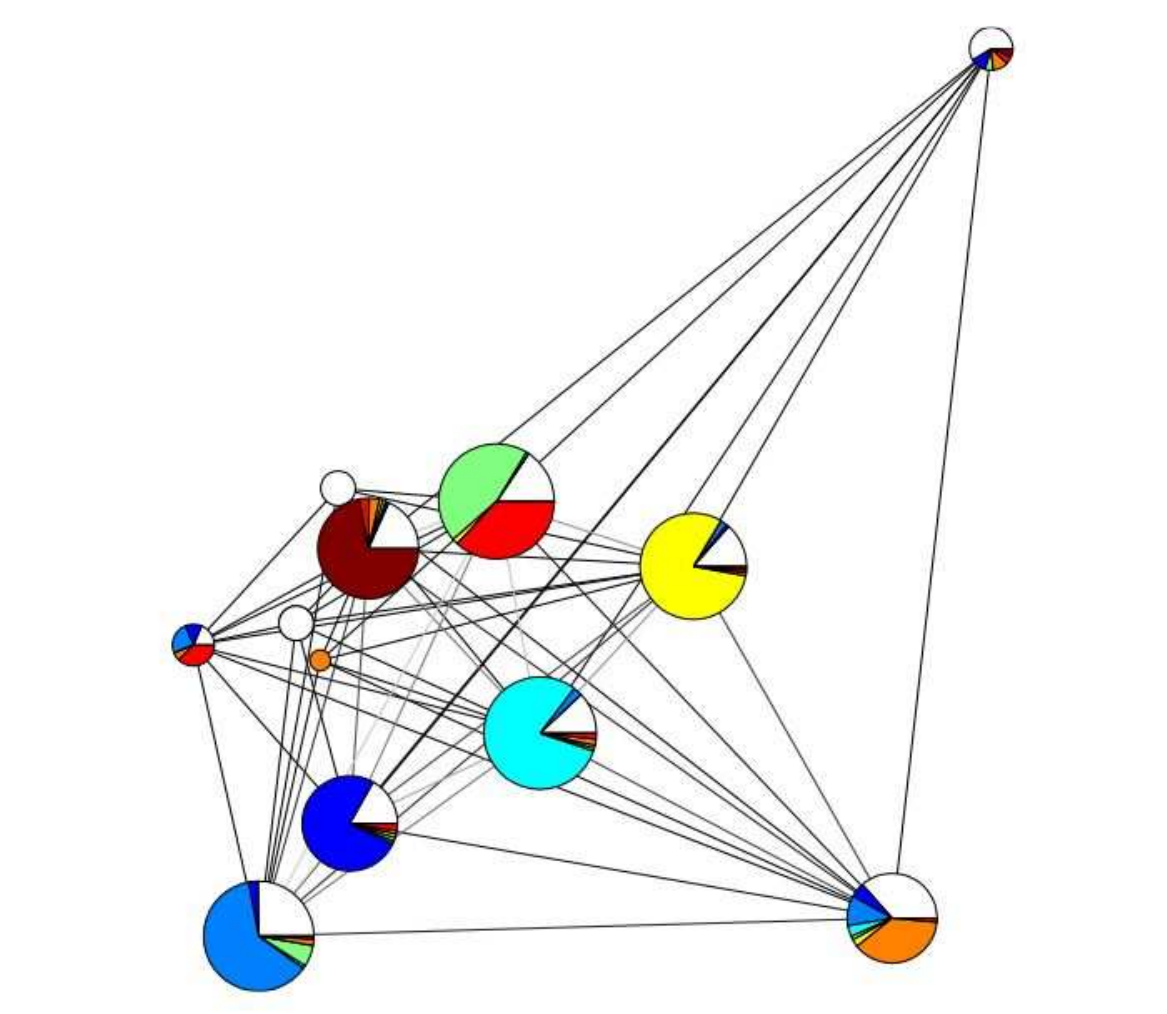}
\caption {\label{figcaltech} Communities in social networking sites. (Top) Visualization of a network of friendships between students at Caltech, 
constructed from Facebook data (September 2005). The colors/shapes indicate the dormitories (Houses) of the students. (Bottom) Topological communities of the network,
which are quite homogeneous with respect to House affiliation. Reprinted figures with permission from Refs.~\cite{porter09} and \cite{traud08}.}
\end{center}
\end{figure}

Social networking sites, like {\it Myspace} ({\tt www.myspace.com}), {\it Friendster}
({\tt www.friendster.com}), {\it Facebook} ({\tt www.facebook.com}), etc. have become extremely popular in the last years.
They are online platforms that allow people to communicate with friends, send e-mails, solicit opinions on specific issues, spread ideas and/or fads, etc.
Traud et al.~\cite{traud08} used anonymous Facebook data to create networks of friendships between students of different American universities,
where vertices/students are connected if they are friends on Facebook. Communities were detected by applying a variant of 
Newman's spectral optimization of modularity (Section~\ref{sub_sec6_0_4}): the results
were further refined through additional steps \'a la Kernighan-Lin (Section~\ref{sec4_1}).
One of the goals of the study was to infer relationships between the online and offline lives of the students. By using demographic information on
the students' populations, one finds that communities are organized by class year or by House (dormitory) affiliation, depending on the 
university (Fig.~\ref{figcaltech}).
Yuta et al.~\cite{yuta07} observed a gap in the community size distribution 
of a friendship network extracted from {\it mixi} ({\tt mixi.jp}), the largest social networking
site in Japan (as of December 2006). Communities were identified with the fast greedy 
modularity optimization by Clauset et al.~\cite{clauset04}. The gap occurs in the intermediate range of sizes between $20$ and $400$, where but
a few communities are observed. Yuta et al. introduced a model where people form new friendships both by 
``closing'' ties with people who are friends of friends, and by setting new links with individuals having similar interests. In this way
most groups turn out to be either small or large, and medium size groups are rare.

Collaboration networks, in which individuals are linked if they are (were) involved in a common activity,
have been often studied because they embed an implicit objective concept of acquaintance, that is not easy to capture 
in direct social experiments/interviews. For instance, somebody may consider another individual a friend, while the latter may disagree.
A collaboration instead is a proof of a social relationship between individuals. The analysis of the structure of scientific collaboration
networks~\cite{newman01} has exerted a big influence on the development of the modern network science. Scientific collaboration is associated to
coauthorship: two scientists are linked if they have coauthored at least one paper together. Information about coauthorships can be extracted from different
databases of research papers. Communities indicate groups of people with common research interests, i. e. topical or disciplinary groups.
In the seminal paper by Girvan and Newman~\cite{girvan02}, the authors applied their method on a collaboration network of scientists working at the Santa Fe Institute, 
and were able to discriminate between research divisions (Fig.~\ref{fig3}b). The community structure of scientific collaboration networks 
has been investigated by many
authors~\cite{zhou03b,newman04c,radicchi04,donetti04,palla05,duch05,white05,danon06,pujol06,reichardt06,son06,vragovic06,newman06b,palla07,farkas07,
gregory07,lehmann07,richardson09,nepusz08,shen09,noack09b}.
Other types of collaboration networks have been studied too. Gleiser and Danon~\cite{gleiser03} considered a collaboration network
of jazz musicians. Vertices are either musicians, connected if they played in the same band, or bands, connected if they have a musician in common.
By applying the algorithm of Girvan and Newman they found that communities reflect both racial segregation (with two main groups comprising only
black or white players) and geographical separation, due to the different recording locations.

\subsection{Other networks}
\label{sec7_2_3}

\begin{figure*}
\begin{center}
\includegraphics[width=\textwidth]{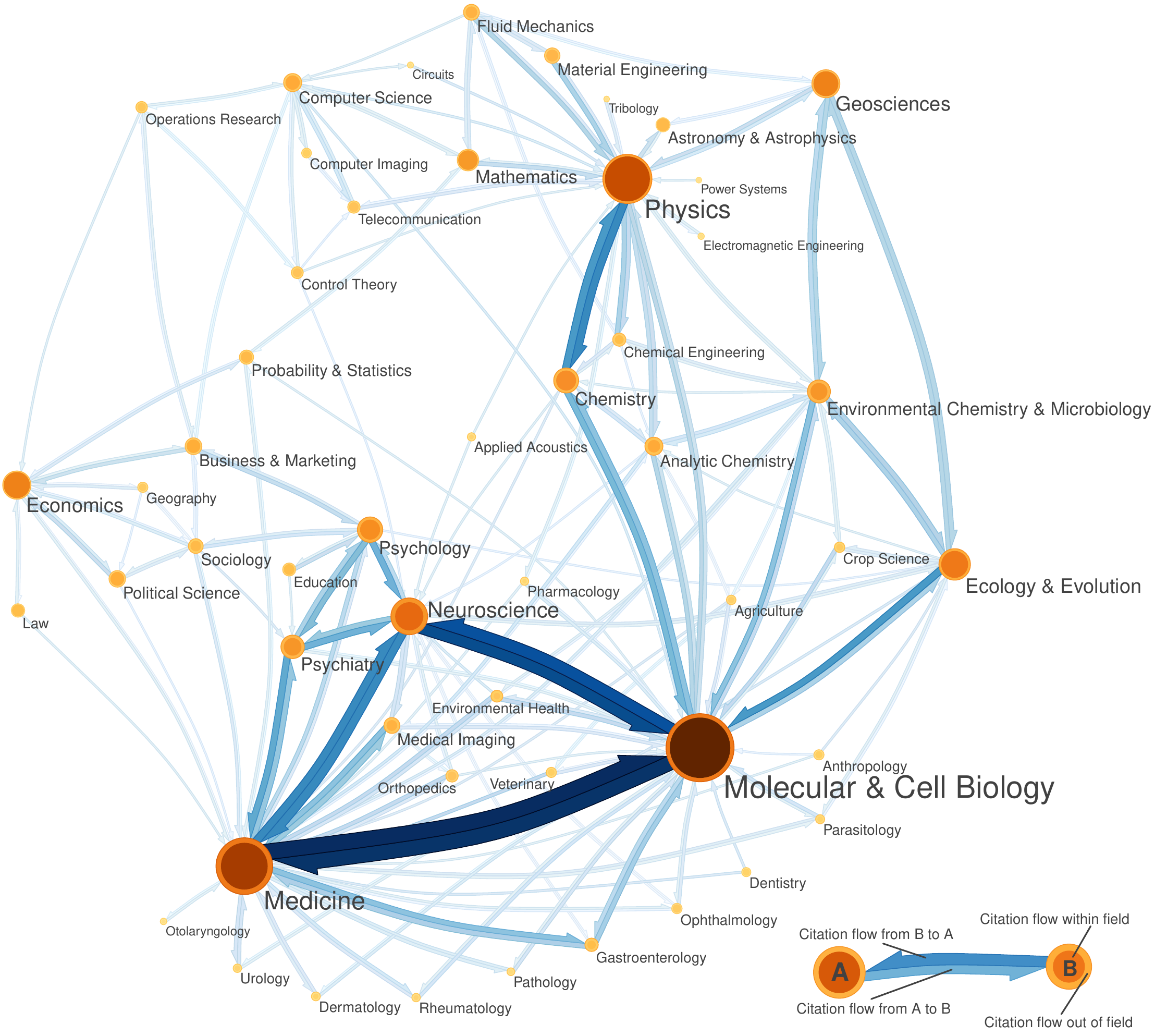}
\caption {\label{maprosvall} Map of science derived from a clustering analysis of a citation network comprising more than $6000$ journals.
Reprinted figure with permission from 
Ref.~\cite{rosvall08}. \copyright 2008 by the National Academy of Science of the USA.}
\end{center}
\end{figure*}

Citation networks~\cite{price65} have been regularly used to understand the citation patterns of authors and to 
disclose relationships between disciplines. Rosvall and Bergstrom~\cite{rosvall08} used a citation network of over $6000$ scientific journals
to derive a map of science. They used a clustering technique based on compressing the information on random walks taking place on the graph 
(Section~\ref{subsec_statinf2}). A random walk follows the flow of citations from one field to another, and the fields emerge naturally
from the clustering analysis (Fig.~\ref{maprosvall}). The structure of science resembles the letter ${\bf U}$, with the social sciences
and engineering at the terminals, joined through a chain including medicine, molecular biology, chemistry and physics.

Reichardt and Bornholdt~\cite{reichardt07c} performed a clustering analysis on a network built from bidding data taken from the German version of
Ebay ({\tt www.ebay.de}), the most
popular online auction site. The vertices are bidders and two vertices are connected if the corresponding bidders have expressed interest for
the same item. Clusters were detected with the multiresolution modularity optimization developed by the authors 
themselves~\cite{reichardt06} (Section~\ref{sub_sec6_01}). In spite of the variety of items that it is possible to 
purchase through Ebay, about $85\%$ of bidders were classified into a few major clusters, reflecting 
bidders' broad categories of interests. Ebay data were also examined by
Jin et al.~\cite{jin07}, who considered bidding networks where the vertices are the individual auctions and edges are placed between auctions
having at least one common bidder. Communities, detected with greedy modularity optimization~\cite{newman04c} (Section~\ref{sub_sec6_0_1}),
allow to identify substitute goods, i. e. products that have value for the same bidder, so that
they can be purchased together or alternatively.

Legislative networks enable one to deduce associations between politicians through their parliamentary activity, 
which may be related or not to party affiliation. Porter and coworkers
have carried out numerous studies on the subject~\cite{porter05,porter07,zhang08}, by using data on the Congress of the United States.
In Refs.~\cite{porter05,porter07}, they examined the community structure of networks of 
committees in the US House of Representatives. Committees sharing common members are connected by edges, which are weighted by
dividing the number of common members by the number one would expect to have if committee memberships were randomly assigned.
Hierarchical clustering (Section~\ref{sec4_2}) reveals close connections between some of the committees. In another work~\cite{zhang08}, 
Zhang et al. analyzed networks of legislation cosponsorship, in which vertices are legislators and two legislators are linked if
they support at least one common bill. Communities, identified with a modification of 
Newman's spectral optimization of modularity (Section~\ref{sub_sec6_0_4}), are correlated with party affiliation, but also with
geography and committee memberships of the legislators.

Networks of correlations between time series of stock returns have received a growing attention in the past few years~\cite{mantegna99}.
In early studies, scholars found clusters of correlated stocks by computing the 
{\it maximum spanning tree} of the network~\cite{bonanno00,bonanno03,onnela02,onnela03} (Section~\ref{sec1_1}), and 
realized that such clusters match quite well the economic sectors of the stocks. More recently, the community structure of the networks has been
investigated by means of proper clustering algorithms. Farkas et al.~\cite{farkas07} have applied the weighted version of the Clique
Percolation Method (Section~\ref{sec45_1}) and found that the presence of two strong (i. e. carrying high correlation)
edges in triangles is usually accompanied by the presence of a strong third edge.
Heimo et al.~\cite{heimo08} used the weighted version of the multiresolution method by Reichardt and Bornholdt~\cite{reichardt06} (Section~\ref{sub_sec6_01}).
Clusters correspond to relevant business sectors, as indicated by Forbes classification; moreover, smaller clusters at 
lower hierarchical levels seem to correspond to (economically) meaningful substructures of the main clusters.

\section{Outlook}
\label{sec8}

Despite the remote origins and the great popularity of the last years, research on graph clustering has not yet 
given a satisfactory solution of the problem and leaves us with a number of important open issues.
From our exposition it appears that the field has grown in a rather chaotic way, without a precise direction or 
guidelines. In some cases, interesting new ideas and tools have been presented,  
in others existing methods have been improved, becoming more accurate and/or faster.

What the field lacks the most is a theoretical framework that defines precisely what clustering algorithms are supposed to do.
Everybody has his/her own idea of what a community is, and most ideas are consistent with each other, but, as long as there is still disagreement,
it remains impossible to decide which algorithm does the best job and there will be no control on the creation of new methods. 
Therefore, we believe that the first and foremost task that the
scientific community working on graph clustering has to solve in the future 
is defining a set of reliable benchmark graphs, against which algorithms should be tested (Section~\ref{sec6_1}). 
Defining a benchmark goes far beyond the issue of testing. It means designing practical examples of 
graphs with communities, and, in order to do that, one has to agree on the fundamental concepts
of community and partition. Clustering algorithms have to be devised consistently with such definitions, in order to 
give the best performance on the set of designated benchmarks, which represent a sort of ground truth. 
The explosion in the number of algorithms we have witnessed in recent times is due precisely to the present lack 
of reliable mechanisms of control of their quality and comparison of their performances. If the community agrees on a 
benchmark, the future development of the field will be more coherent and the progress brought by new methods can be
evaluated in an unbiased manner.
The planted $\ell$-partition model~\cite{condon01}
is the easiest recipe one can think of when it comes to defining clusters, and is the criterion underlying well-known benchmarks, 
like that by Girvan and Newman. We believe that the new benchmarks have to be defined along the same lines. The benchmark graphs recently introduced  
by Lancichinetti et al.~\cite{lancichinetti08,lancichinetti09b} and by Sawardecker et al.~\cite{sawardecker09} are an important step in this direction.

Defining a benchmark implies specifying the ``natural'' partition of a graph, the one that any algorithm should find.
This issue in turn involves the concept of quality of a partition, that has characterized large part of the development of the field, 
in particular after the introduction of Newman-Girvan modularity (Section~\ref{sec3_2_2}). Estimating the quality of a partition 
allows to discriminate among the large number of partitions of a graph. In some cases this is not difficult. For instance, in the
benchmark by Girvan and Newman there is a single meaningful partition, and it is hard to argue with that. But most 
graphs of the real world have a hierarchical structure, with communities including smaller communities and so on. Hence there are several
meaningful partitions, corresponding to different hierarchical levels, and discriminating among them is hard, as they may be all relevant, in a sense.
If we consider the human body, we cannot say that the organization in tissues of the cells is more important than the organization in organs.
We have seen that there are recent methods dealing with the problem of finding meaningful hierarchical levels (Section~\ref{sub_sec6_20}). 
Such methods rank the hierarchical partitions based on some criterion and one can assess their relevance through the ranking.
One may wonder whether it makes sense sorting out levels, which means introducing a kind of threshold on the quality index chosen to 
rank partitions (to distinguish ``good'' from ``bad'' partitions), 
or whether it is more appropriate to keep the information given by the whole set of hierarchical partitions. The work 
by Clauset et al. on hierarchical random graphs~\cite{clauset07,clauset08}, discussed in Section~\ref{sub_sec6_202}, indirectly raises this issue.
There it was shown that the ensemble of model graphs, represented by dendrograms, encodes most of the information on the structure of the graph
at study, like its degree distribution, transitivity and distribution of shortest path lengths. At the same time, by construction, the model reveals 
the whole hierarchy of communities, without any distinction between good and bad partitions. The information given by a dendrogram 
may become redundant and confusing when the graph is large, as then there is a big number of partitions. This is actually the reason  
why quality functions were originally introduced. However, in that case, one was dealing with artificial hierarchies, produced
by techniques that systematically yield a dendrogram as a result of the analysis
(like, e. g., hierarchical clustering), regardless of whether the graph actually has a hierarchical structure or not.
Here instead we speak of real hierarchy, which is a fundamental element of real graphs and, as such, it must be 
considered in any serious approach to graph clustering. Any good clustering method must be able to tell 
whether a graph has community structure or not, and, in the first 
case, whether the community structure is hierarchical (i. e. with two or more levels) or flat (one level).
We expect that the concept of hierarchy will become 
a key ingredient of future clustering techniques. 
In particular, assessing the consistence of the concepts of partitions' quality and hierarchical structure
is a major challenge.

A precise definition of null models, i. e. of graphs without community structure, is also missing. This aspect is extremely important, though,
as defining communities also implies deciding whether or not they exist in a specific graph.
At the moment, it is generally accepted
that random graphs have no communities. The null model of modularity (Section~\ref{sec3_2_2}), by far the most popular, comprises all graphs 
with the same expected degree sequence of the original graph and random rewiring of edges. This class of graphs is characterized, by construction, by the
fact that any vertex can be linked to any other, as long as the constraint on the degree sequence is satisfied. But this is by no means the only possibility.
A community can be generically defined as a subgraph whose vertices have a higher probability to be connected to the other vertices of the subgraph than
to external vertices. The planted $\ell$-partition model~\cite{condon01} is based on this principle, as we have seen. However, this does not mean that 
the linking probabilities of a vertex with respect to the other vertices in its community or in different communities 
be constant (or simply proportional to their degrees, as in the
configuration model~\cite{luczak92,molloy95}). In fact, in large networks it is reasonable
to assume that the probability that a vertex is linked to most vertices is zero, as the vertex ``ignores'' their existence. This does not 
exclude that the probability that the vertex gets connected to the ``known'' vertices is the same (or proportional to their degrees), in which
case the graph would still be random and have no communities. We believe that we are still far from a precise definition 
and a complete classification of
null models. This represents an important research line for the future of the field,
for three main reasons: 1) to better disentangle ``true'' communities from byproducts of random fluctuations; 2) to pose a stringent test
to existing and future clustering algorithms, whose reliability would be questionable if they found ``false positives'' in null model graphs; 
3) to handle ``hybrid'' scenarios, where
a graph displays community structure on some portions of it, while the rest is essentially random and has no communities.

In the previous chapters we have seen a great number of clustering techniques. Which one(s) shall we use? 
At the moment the scientific community is unable to tell. Modularity optimization is probably the most popular   
method, but the results of the analysis of large graphs are likely to be unreliable 
(Section~\ref{sub_sec6_1}). Nevertheless, people have become
accustomed to use it, and there have been several attempts to improve the measure.
A newcomer, who wishes to find clusters in a given network and is not familiar with clustering techniques, would not know,
off-hand, which method to use, and he/she would hardly 
find indications about good methods in any single paper on graph clustering, except perhaps 
on the method presented in the paper. So, people keep using algorithms because they have heard of them, or because 
they know that other people are using them, or because of the reputation of the scientists who designed them.
Waiting for future reliable benchmarks, that may give an objective assessment of the quality of the algorithms, 
there are at the moment hardly solid reasons to prefer an algorithm to another: the comparative analyses by Danon et al.~\cite{danon05} and by
Lancichinetti and Fortunato~\cite{lancichinetti09d} (Section~\ref{sec6_3}) represent a first serious assessment of this issue.
However, we want to stress that there is no such thing as
the perfect method, so it is pointless to look for it. Among the other things, if one tries to look for a very general method, that should give
good results on any type of graphs, one is inevitably forced to make very general assumptions on the structure of the graph and on
the properties of communities. In this way one neglects a lot of specific features of the system, that may lead to a more accurate
detection of the clusters. Informing a method with features characterizing some types of graphs makes it far more
reliable to detect the community structure of those graphs than a general method, even if its applicability may be limited. Therefore
in the future we envision the development of domain-specific clustering techniques. The challenge here is 
to identify the peculiar features of classes of graphs, which are bound to become crucial ingredients  
in the design of suitable algorithms.
Some of the methods available today are actually based on assumptions that 
hold only for some specific categories of graphs. The Clique Percolation Method by Palla et al.~\cite{palla05}, for instance, 
may work well for graphs characterized by a large number of cliques, like certain social networks, whereas it may give poor results 
otherwise.

Moving one step further, one should learn how to use specific information about a graph, 
whenever available, e. g. properties of vertices and/or partial information
about their classification. For instance, it may be that one has some information on a subset of vertices,
like demographic data on people of a social network, and such data may highlight relationships between 
people that are not obvious from the network of social interactions. In this case, using only the social network 
may be reductive and ideally one should exploit both the structural and the non-structural information 
in the search of clusters, as the latter should be consistent with both inputs. How to do this is an open problem.
The scientific community has just begun to study this aspect~\cite{allahverdyan09}.   

Most algorithms in the literature deal with the ``classical'' case of a graph with undirected and unweighted edges.
This is certainly the simplest case one could think of, and graph clustering is already a complex task on such
types of graphs. We know that real networks may be directed, have weighted connections, be bipartite.  
Methods to deal with such systems have been developed, as we have seen, especially in the most recent literature,
but they are mostly preliminary attempts and there is room for improvement. Another situation that 
may occur in real systems is the presence of edges with positive and negative weights, 
indicating attractive and repulsive interactions, respectively.
This is the case, for instance, of correlation data~\cite{mantegna99}. In this case, ideal partitions
would have positively weighted intracluster edges and negatively weighted intercluster edges. We have  
discussed some studies in this direction~\cite{traag09,gomez09,kaplan08}, but we are just at the beginning of this endeavour.
Instead, there are no algorithms yet which are capable to deal with graphs in which there are edges of several types, indicating 
different kinds of interactions between the vertices ({\it multigraphs}). 
Agents of social networks, for instance, may be joined by working relationships, friendship, family ties, etc. 
At the moment there are essentially two ways of proceeding in these instances:  1) keeping edges of one type and
forgetting the others, repeating the analysis for each type of edges and eventually comparing the results obtained; 2) analyzing
a single (weighted) graph, obtained by ``combining'' the contributions of the different types of edges in some way.
Finally, since most real networks are built through the results of experiments, which carry errors in their estimates, it would be useful
to consider as well the case in which edges have not only associated weights, but also errors on their values.

Since the paper by Palla et al.~\cite{palla05}, overlapping communities have received a lot of attention (Section~\ref{sec45}). 
However, there is still no consensus about a quantitative definition of the concept of overlapping community,
and most definitions depend on the method adopted. Intuitively, one would expect that clusters share vertices 
lying at their borders, and this idea has inspired most algorithms. However, clusters detected with the Clique Percolation Method
(Section~\ref{sec45_1}) often share central vertices of the clusters, which makes sense in specific instances, especially
in social networks. So, it is still unclear how to characterize overlapping vertices. Moreover, the concept of overlapping clusters
seems at odds with that of hierarchical structure. No dendrogram can be drawn if there are
overlapping vertices, at least in the standard way. Due to the relevance of both features in real networks, 
it is necessary to adapt them to each other in a consistent way. Overlapping vertices pose problems as well when it comes to comparing
the results of different methods on the same graph. Most similarity measures are defined only in the case of partitions, where
each vertex is assigned to a single cluster (Section~\ref{sec6_2}). It is then necessary to extend such definitions
to the case of overlapping communities, whenever possible.

Another issue that is getting increasingly more popular is the study of graphs evolving in time. This 
is now possible due to the availability of timestamped network data sets. Tracking the evolution of 
community structure in time is very important, to uncover how communities are generated and how they interact  
with each other. Scholars have just begun to study this 
problem~\cite{hopcroft04,palla07,fenn09,asur07,sun07,chakrabarti06,chi07,lin08,kim09b} (Section~\ref{sec7_1_2}).
Typically one analyzes separately snapshots at different times and checks what happened at time $t+1$ to the communities 
at time $t$. It would be probably better to use simultaneously the whole dynamic data set, and future work shall aim at 
defining proper ways to do that. In this respect, the evolutionary clustering framework by Chakrabarti et al.~\cite{chakrabarti06} 
is a promising starting point.

The computational complexity of graph clustering algorithms has improved by at least one power in the 
graph size (on average) in just a couple of years.
Due to the large size of many systems one wishes to investigate, the ultimate goal would be to 
design techniques with linear or even sublinear complexity.
Nowadays partitions in graphs with up to millions of vertices can be found. However, the results 
are not yet very reliable, as they are usually obtained by greedy optimizations, which yield rough approximations of 
the desired solution. In this respect the situation could improve by focusing on the development of efficient local methods, for two reasons: 
1) they enable analyses of portions of the graph, independently of the rest; 2) they
are often suitable for parallel implementations, which may speed up considerably the computation.

Finally, if there has been a tremendous effort in the design of clustering algorithms, basically nothing 
has been done to make sense of their results. What shall we do with communities? What can they tell us about a system?
The hope is that they will enable one to disclose ``hidden'' relationships between vertices, due to features that are not known, because 
they are hard to measure, for instance. It is quite possible that the scientific community will converge sooner or later to a
definition {\it a posteriori} of community. Already now, most algorithms yield similar results in practical applications.
But what is the relationship between the vertex classification given by the algorithms and real classifications? 
This is the main question beneath the whole endeavor.  

\vskip0.5cm

{\centerline{\bf Acknowledgments}}

I am indebted to these people for giving useful suggestions and advice to improve this manuscript at various stages: 
A. Arenas, J. W. Berry, A. Clauset, P. Csermely,
S. G\'omez, S. Gregory, V. Gudkov, R. Guimer\`a, Y. Ispolatov, R. Lambiotte, A. Lancichinetti, 
J.-P. Onnela, G. Palla, M. A. Porter, F. Radicchi, J. J. Ramasco, C. Wiggins.
I gratefully acknowledge ICTeCollective, grant number 238597 of the European Commission.

\begin{appendix}
\section{Elements of Graph Theory}
\label{sec1}

\subsection{Basic Definitions}
\label{sec1_1}

\begin{figure}
\begin{center}
\includegraphics[width=\columnwidth]{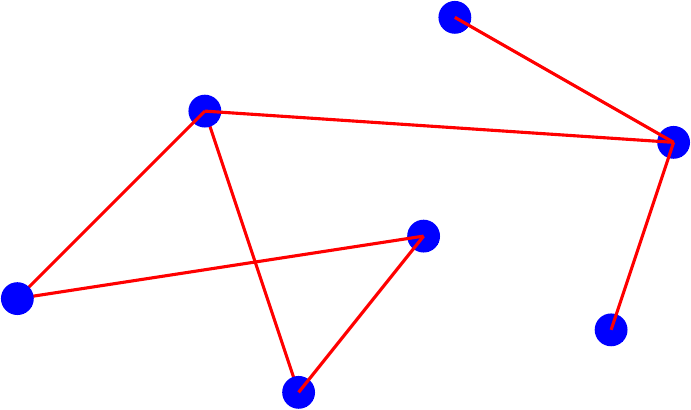}
\caption{\label{fig1} A sample graph with seven vertices and seven edges.}
\end{center}
\end{figure}

A {\it graph} $\cal G$ is a pair of sets $(V,E)$, where $V$ is a set of {\it vertices} or {\it nodes} and $E$ is a subset
of $V^2$, the set of unordered pairs of elements of $V$. The elements of $E$ are called {\it edges} or {\it links}, the two vertices
that identify an edge are called {\it endpoints}. An edge is {\it adjacent} to each of its endpoints. 
If each edge is an ordered pair of vertices one has a {\it directed} graph (or {\it digraph}). In this
case an ordered pair $(v,w)$ is an edge directed from $v$ to $w$, or an edge beginning at $v$ and ending at $w$.
A graph is visualized as a set of points connected by lines, as shown in Fig.~\ref{fig1}. In many real examples, graphs are {\it weighted}, i.~e. 
a real number is associated to each of the edges. Graphs do not include {\it loops}, i.~e. edges
connecting a vertex to itself, nor multiple edges, i.~e. several edges joining the same pair of vertices. Graphs with loops and multiple edges are
called {\it multigraphs}. Generalizations of graphs admitting edges between any number of vertices (not necessarily two) are called {\it hypergraphs}.    

A graph ${\cal G}^\prime=(V^\prime,E^\prime)$ is a {\it subgraph} of ${\cal G}=(V,E)$ if $V^\prime\subset V$ and 
$E^\prime\subset E$. If $\cal G^\prime$ contains
all edges of $\cal G$ that join vertices of $V^\prime$ one says that the subgraph $\cal G^\prime$ is induced or spanned by $V^\prime$. A partition
of the vertex set $V$ in two subsets $S$ and $V-S$ is called a {\it cut}; the {\it cut size} is the
number of edges of $\cal G$ joining vertices of $S$ with vertices of $V-S$.  

We indicate the number of vertices and edges of a graph with $n$ and $m$, respectively.
The number of vertices is the {\it order} of the graph, the number of edges its {\it size}. The maximum size of a graph equals the total number of 
unordered pairs of vertices, $n(n-1)/2$. If $|V|=n$ and $|E|=m=n(n-1)/2$, the graph is a {\it clique} (or {\it complete graph}), and is indicated as
$K_n$. 
Two vertices are {\it neighbors} (or {\it adjacent}) if they are connected by an edge. The set of neighbors of a vertex $v$ is called {\it neighborhood},
and we shall denote it with $\Gamma(v)$. The {\it degree} $k_{v}$ of a vertex $v$ is the number of its neighbors. The {\it degree sequence} is the 
list of the degrees of the graph vertices, $k_{v_1},k_{v_2}, ..., k_{v_n}$.
On directed graphs, one distinguishes two types of degree for a vertex $v$: the {\it indegree}, i.~e.
the number of edges beginning at $v$ and the {\it outdegree}, i.~e. the number of edges ending at $v$. The analogue of degree 
on a weighted graph is the {\it strength}, i.~e. the sum of the weights of the edges adjacent to the vertex.
Another useful local property of graphs is {\it transitivity} or {\it clustering}~\cite{watts98}, which indicates the level of cohesion
between the neighbors of a vertex~\footnote{The term clustering is commonly adopted to indicate community detection in some disciplines, like
computer science, and we often used it in this context throughout the manuscript. We paid attention to disambiguate the occurrences in which
clustering indicates instead the local property of a vertex neighborhood described here.}. 
The clustering coefficient $c_v$ of vertex $v$ is the ratio between the number of edges joining pairs of neighbors
of $v$ and the total number of possible edges, given by $k_v(k_v-1)/2$, $k_v$ being the degree of $v$. According to this definition,
$c_v$ measures the probability that a pair of neighbors of $v$ are connected.
Since all neighbors of $v$ are connected to $v$
by definition, edges connecting pairs of neighbors of $v$ form triangles with $v$. This is why the definition is often given in terms of number of triangles.

A {\it path} is a graph ${\cal P}=(V({\cal P}), E({\cal P}))$, with $V({\cal P})=\{x_0,x_1,...,x_l\}$  
and $E({\cal P})=\{x_0x_1,x_1x_2,...,x_{l-1}x_l\}$. The vertices $x_0$ and $x_l$ are the {\it endvertices} of $\cal P$, whereas $l$ is its {\it length}. 
Given the notions of vertices, edges and paths, one can define the concept of {\it independence}. A set of vertices (or edges) of a graph are independent
if no two elements of them are adjacent. Similarly, two paths are independent if they only share the endvertices.
A {\it cycle} is a closed path whose vertices and edges are all distinct.
Cycles of length $l$ are indicated with $C_l$. The smallest non-trivial cycle is the {\it triangle},
$C_3$.

Paths allow to define the concept of connectivity and distance in graphs. A graph is {\it connected} if, given any pair of vertices, there is at least
one path going from one vertex to the other. In general, there may be multiple paths connecting two vertices, with different lengths. A {\it shortest path},
or {\it geodesic}, between two vertices of a graph, is a path of minimal length. Such minimal length is the {\it distance} between the two vertices. 
The {\it diameter} of a connected graph is the maximal
distance between two vertices.
If there is no path between two vertices, the graph is divided in at least two connected subgraphs. Each maximal connected subgraph of a graph is called
{\it connected component}. 

A graph without cycles is a {\it forest}. A connected forest is a {\it tree}. Trees are very important in graph theory and deserve some attention.
In a tree, there can be only one path from a vertex to any other. In fact, if there were at least two paths between the same pair of vertices
they would form a cycle, while the tree is an acyclic graph by definition. Further, the number of edges of a tree with $n$ vertices is $n-1$.
If any edge of a tree is removed, it would get disconnected in two parts; if a new edge is added, there would be at least one cycle. This is why a tree
is a minimally connected, maximally acyclic graph of a given order. Every connected graph contains a {\it spanning tree}, i.~e. a tree 
sharing all vertices of the graph. On weighted graphs, one can define a {\it minimum} ({\it maximum}) {\it spanning tree}, i.~e. a spanning tree such that the sum
of the weights on the edges is minimal (maximal). Minimum and maximum spanning trees are often used in graph optimization problems, including clustering.

A graph $\cal G$ is {\it bipartite} if the vertex set $V$ is separated in two disjoint subsets $V_1$ and $V_2$, or {\it classes},  
and every edge joins a vertex of $V_1$ with a vertex of $V_2$. The definition can be extended to that of
{\it $r$-partition}, where the vertex classes are $r$ and no edge joins vertices within the same class. In this case one speaks of 
{\it multipartite} graphs.

\subsection{Graph Matrices}
\label{sec1_2}

The whole information about the topology of a graph of order $n$ is entailed in the {\it adjacency matrix} {\bf A}, 
which is an $n\times n$ matrix whose element
$A_{ij}$ equals $1$ if there is an edge joining vertices $i$ and $j$, otherwise it is zero. Due to the absence of loops
the diagonal elements of the adjacency matrix are all zero. For an undirected graph 
{\bf A} is a symmetric matrix. The sum of the elements of the $i$-th row or column yields the degree of node $i$. If the edges are weighted,
one defines the {\it weight matrix} $\bf W$, whose element $W_{ij}$ expresses the weight of the edge between vertices $i$ and $j$.

The {\it spectrum} of a graph $\cal G$ is the set of eigenvalues of its adjacency matrix {\bf A}.
Spectral properties of graph matrices play an important role in the study of graphs.
For instance, the {\it stochastic matrices} rule the process of diffusion (random walk) on a graph. The 
{\it right stochastic matrix} $\bf R$ is
obtained from {\bf A} by dividing the elements of each row $i$ by the degree of vertex $i$. The {\it left stochastic matrix} $\bf T$,
or {\it transfer matrix}, is the transpose of $\bf R$.
The spectra of stochastic matrices allow to evaluate, for instance, the mixing time of the random walk, i.~e. the time it takes to reach 
the stationary distribution of the process. The latter is obtained by computing the eigenvector of the transfer matrix
corresponding to the largest eigenvalue.

Another important matrix is the {\it Laplacian} ${\bf L}={\bf D}-{\bf A}$, where {\bf D} is the diagonal matrix  
whose element $D_{ii}$ equals the degree of vertex $i$. The matrix ${\bf L}$ is usually referred to as {\it unnormalized Laplacian}.
In the literature one often uses {\it normalized Laplacians}~\cite{chung97}, of which there are two main forms: 
${\bf L_{sym}}={\bf D}^{-1/2}{\bf L}{\bf D}^{-1/2}$ and ${\bf L_{rw}}={\bf D}^{-1}{\bf L}={\bf I}-{\bf D}^{-1}{\bf A}={\bf I}-{\bf T}$.
The matrix ${\bf L_{sym}}$ is symmetric; ${\bf L_{rw}}$ is not symmetric and is closely related to a random walk taking place on the graph.
All Laplacian matrices have a straightforward extension to the case of weighted graphs. 
The Laplacian is one of the most studied matrices and finds application
in many different contexts, like graph connectivity~\cite{bollobas98}, synchronization~\cite{barahona02,nishikawa03}, 
diffusion~\cite{chung97} and 
graph partitioning~\cite{pothen97a}. 
By construction, the sum of the elements of each row of the Laplacian (normalized or unnormalized) is zero. This implies that 
{\bf L} always has at least one zero eigenvalue, corresponding to the eigenvector with all equal components, such as $(1, 1,..., 1)$.
Eigenvectors corresponding to different eigenvalues 
are all orthogonal to each other.
Interestingly, {\bf L} has as many zero eigenvalues as there are connected components in the graph. So, the Laplacian of a connected graph
has but one zero eigenvalue, all others being positive. 
Eigenvectors of Laplacian matrices are regularly used in {\it spectral clustering} (Section~\ref{sec4_4}).
In particular, the eigenvector corresponding to the second smallest eigenvalue, called
{\it Fiedler vector}~\cite{fiedler73,fiedler75}, is used for graph bipartitioning, as described in Section~\ref{sec4_1}.

\subsection{Model graphs}
\label{sec1_3}

In this section we present the most popular models of graphs introduced to describe real systems, at least to some extent.
Such graphs are useful null models in community detection, as they do not have community structure, so they can be used for negative tests
of clustering algorithms. 
\begin{figure}
\begin{center}
\includegraphics[width=9cm]{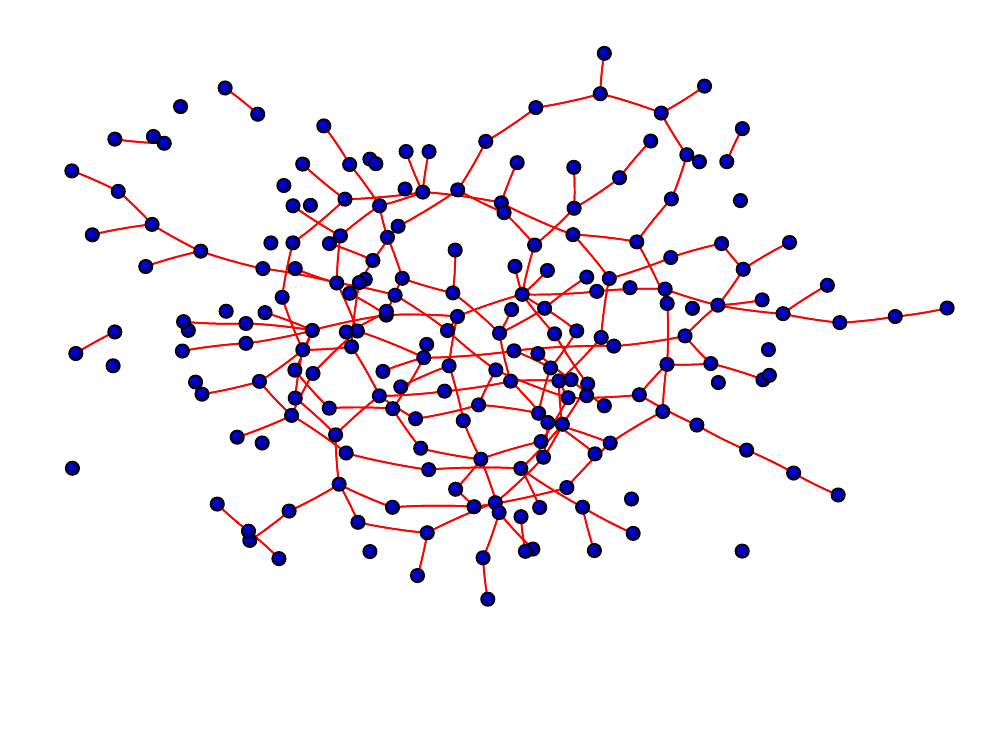}
\includegraphics[width=9cm]{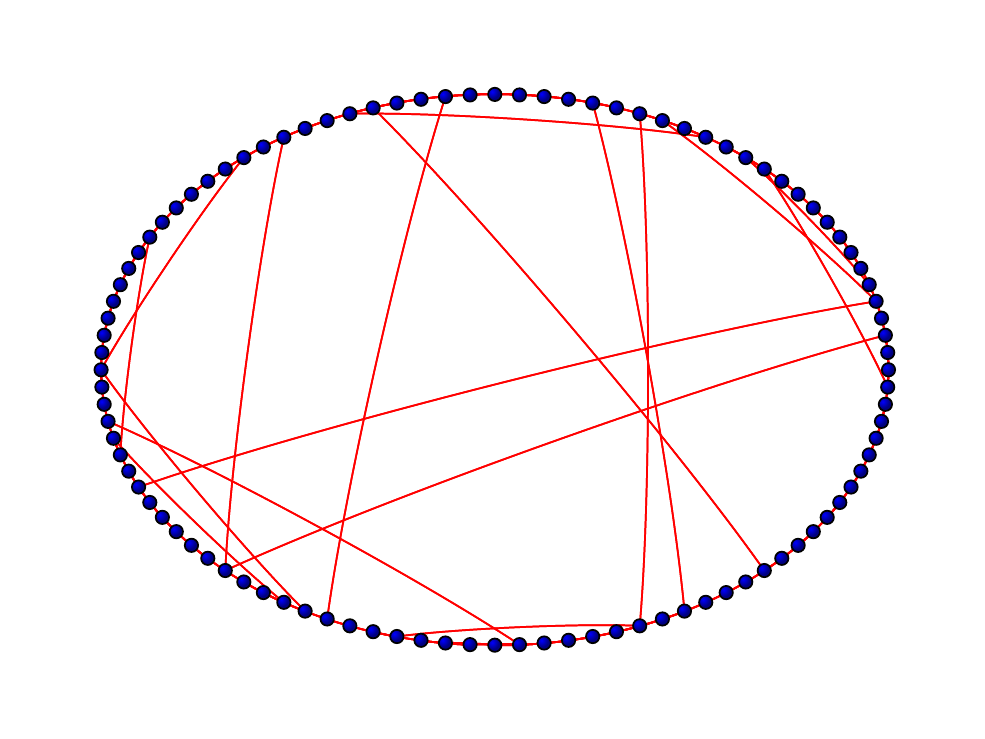}
\includegraphics[width=9cm]{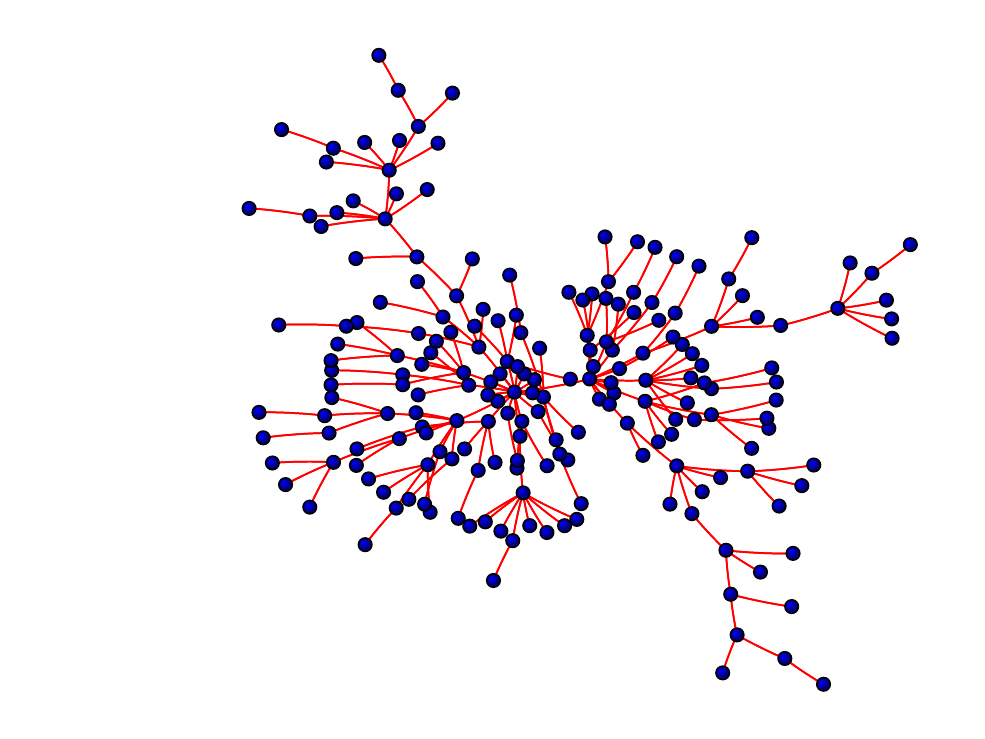}
\caption{\label{fig2} Basic models of complex networks. (Top) Erd\"os-R\'enyi random graph with $100$ vertices and a 
link probability $p=0.02$. (Center) Small world graph \'a la Watts-Strogatz, with $100$ vertices and a rewiring probability $p=0.1$.
(Bottom) Barab\'asi-Albert scale-free network, with $100$ vertices and an average degree of $2$. Courtesy by J.~J. Ramasco.}
\end{center}
\end{figure}

The oldest model is that of {\it random graph}, proposed by Solomonoff and Rapoport~\cite{solomonoff51} 
and independently by Erd\"os and R\'enyi~\cite{erdos59}. There are two parameters: the number of vertices $n$ and the connection probability $p$.
Each pair of vertices is connected with equal probability $p$ independently of the other pairs. The expected number of 
edges of the graph is $pn(n-1)/2$, and the expected mean degree $\langle k\rangle=p(n-1)$. The degree distribution of the vertices of a random graph 
is binomial, and in the limit $n\rightarrow\infty$, $p\rightarrow 0$ for fixed $\langle k\rangle$ it converges to a Poissonian.
Therefore, the vertices have all about the same degree, close to $\langle k\rangle$ (Fig.~\ref{fig2}, top). 
The most striking property of this class of graphs is 
the phase transition observed by varying $\langle k\rangle$ in the limit $n\rightarrow\infty$. For $\langle k\rangle<1$, the
graph is separated in connected components, each of them being microscopic, i.~e. occupying but a vanishing portion of the system size.
For $\langle k\rangle>1$, instead, one of the components becomes macroscopic ({\it giant component}), 
i.~e. it occupies a finite fraction of the graph vertices. 

The diameter of a random graph with $n$ vertices is very small, growing 
only logarithmically with $n$. This property ({\it small-world effect}) 
is very common in many real graphs. The first evidence that social networks are characterized by paths of small length was provided by a series of 
famous experiments conducted by the phychologist Stanley Milgram~\cite{milgram67, travers69}.
The expected clustering coefficient of a vertex of a random graph is $p$,
as the probability for two vertices to be connected is the same whether they are neighbors of the same vertex or not. 
Real graphs, however, are characterized by far higher values of the clustering coefficient as compared to random graphs of the same size.
Watts and Strogatz~\cite{watts98} showed that the small world property and high clustering coefficient can coexist in the same system.
They designed a class of graphs which result from an interpolation between a regular lattice, which has high clustering coefficient, and a random
graph, which has the small-world property. One starts from a ring lattice in which each vertex has degree $k$, and with a probability $p$
each edge is rewired to a different target vertex (Fig.~\ref{fig2}, center).
It turns out that low values of $p$ suffice
to reduce considerably the length of shortest paths between vertices, because rewired edges act as shortcuts between initially remote
regions of the graph. On the other hand, 
the clustering coefficient remains high, since few rewired edges do not perturb appreciably the local structure of the graph, which
remains similar to the original ring lattice.
For $p=1$ all edges are rewired and the resulting structure is a random graph \'a la Erd\"os and R\'enyi. 

The seminal paper of Watts and Strogatz triggered a huge interest towards the graph representation of real systems.
One of the most important discoveries was that the distribution of the vertex degree of real graphs is very 
heterogeneous~\cite{albert99},
with many vertices having few neighbors coexisting with some vertices with many neighbors. In several cases the tail of this distribution
can be described as a power law with good approximation\footnote{The power law is however not necessary to explain the
properties of complex networks. It is enough that the tails of the degree distributions 
are ``fat'', i. e. spanning orders of magnitude in degree. They may or may not
be accurately fitted by a power law.}, 
hence the expression {\it scale-free networks}.  
Such degree heterogeneity is responsible for a number of 
remarkable features of real networks, such as resilience to random failures/attacks~\cite{albert00}, and the absence of 
a threshold for percolation~\cite{cohen00} and epidemic spreading~\cite{pastor01}. The most popular model 
of a graph with a power law degree distribution is the model by Barab\'asi and Albert~\cite{barabasi99}.  
A version of the model for directed graphs had been proposed much earlier by de Solla Price~\cite{price76}, building up on previous
ideas developed by Simon~\cite{simon55}.
The graph is created with a dynamic procedure, where vertices are added one by one to an initial core.
The probability for a new vertex to set an edge with a preexisting vertex is proportional to the degree
of the latter. In this way, vertices with high degree have large probability of being selected as neighbors by 
new vertices; if this happens, their degree further increases so they will be even more likely to be chosen
in the future.
In the asymptotic limit of infinite number of vertices, this {\it rich-gets-richer} strategy generates a graph with 
a degree distribution characterized by a power-law tail with exponent $3$. In Fig.~\ref{fig2} (bottom) we show an example of Barab\'asi-Albert (BA)
graph. The clustering coefficient of a BA graph decays with the size of the graph, and it is much lower than in
real networks. Moreover, the power law decays of the degree distributions observed in real networks are characterized by a range of exponents' values
(usually between $2$ and $3$), whereas the BA model yields a fixed value. However, many refinements of the BA model
as well as plenty of different models have been later introduced to account more closely for the features observed in real systems 
(for details see~\cite{albert02,mendes03,newman03,pastor04, boccaletti06,barrat08}).

\end{appendix}

\end{document}